\documentclass[aj,numberedappendix,usenatbib,usegraphicx]{emulateapj}

\usepackage{morefloats,epsf,apjfonts,amsmath,subfigure}
\usepackage{natbib, placeins}
\bibpunct{(}{)}{;}{a}{}{,}

\newcommand{\kms}{\ensuremath{\,\mbox{km}\,\mbox{s}^{-1}}}

\newcommand{\HI}{H {\sc i}}
\newcommand{\degree}{\ensuremath{^{\circ}}}

\newcommand{\dx}{\ensuremath{\delta}{\sc X}}
\newcommand{\dy}{\ensuremath{\delta}{\sc Y}}

\newcommand{\VROT}{{\sc vrot}}
\newcommand{\A}{\ensuremath{\tilde{A}_r}}
\newcommand{\Am}{\ensuremath{\tilde{A}_m}}
\newcommand{\Ar}{\ensuremath{A_r(r)}}
\newcommand{\rotcur}{{\sc rotcur}}
\newcommand{\reswri}{{\sc reswri}}
\newcommand{\vmax}{{\ensuremath{V_{\mathrm{tot}}}}}

\slugcomment{Accepted for publication in the AJ special THINGS issue}
\begin{document}

\title{Dynamical Centers and Non-Circular Motions in THINGS Galaxies:
Implications for Dark Matter Halos}
\shorttitle{Non-Circular Motions in THINGS}
\shortauthors{Trachternach et al.}

\author{C. Trachternach\altaffilmark{1}, W.J.G. de
  Blok\altaffilmark{2}, F. Walter\altaffilmark{3},
  E. Brinks\altaffilmark{4}, R.C. Kennicutt, Jr.\altaffilmark{5}}

\email{trachter@astro.rub.de}
\email{edeblok@ast.uct.ac.za}
\email{walter@mpia-hd.mpg.de}
\email{E.Brinks@herts.ac.uk}
\email{robk@ast.cam.ac.uk}
\altaffiltext{1}{Astronomisches Institut, Ruhr-Universit\"at
  Bochum, Universit\"atsstra{\ss}e 150, 44780 Bochum, Germany}
\altaffiltext{2}{Department of Astronomy, University of Cape Town, Private Bag
X3, Rondebosch 7701, South Africa}
\altaffiltext{3}{Max Planck Institut f\"ur Astronomie, K\"onigstuhl 17, 69117
  Heidelberg, Germany}
\altaffiltext{4}{Centre for Astrophysics Research, University of
  Hertfordshire, College Lane, Hatfield, AL10 9AB, United Kingdom}
\altaffiltext{5}{Institute of Astronomy, University of Cambridge, Madingley
  Road, Cambridge CB3 0HA, United Kingdom}

\begin{abstract}
We present harmonic decompositions of the velocity fields of 19
galaxies from THINGS (The \HI\ Nearby Galaxy Survey) which quantify
the magnitude of the non-circular motions in these galaxies and yield
observational estimates of the elongations of the dark matter halo
potentials. Additionally, we present accurate dynamical center
positions for these galaxies.  We show that the positions of the
kinematic and photometric centers of the large majority of the
galaxies in our sample are in good agreement.  The median absolute
amplitude of the non-circular motions, averaged over our sample, is
$6.7\ \kms$, with $\sim 90$ percent of the galaxies having median
non-circular motions of less than $\sim 9\kms$.  As a fraction of the
total rotation velocity this translates into $4.5$ percent on average.
The mean elongation of the gravitational potential, after a
statistical correction for an unknown viewing angle, is $0.017\pm
0.020$, i.e., consistent with a round potential.  Our derived
non-circular motions and elongations are smaller than what is needed
to bring Cold Dark Matter (CDM) simulations in agreement with the
observations. In particular, the amplitudes of the non-circular
motions are not high enough to hide the steep central mass-density
profiles predicted by CDM simulations.  We show that the amplitudes of
the non-circular motions decrease towards lower luminosities and later
Hubble types.
\end{abstract}


\section{Introduction}
Dark Matter is one of the fundamental ingredients of our current
cosmological paradigm \citep[e.g.,][]{spergel-2007}. It is the
dominant mass component in most galaxies, and as such determines the
properties and evolution of these objects. Measuring or modeling the
properties of these dark matter halos can thus help improve our
knowledge of galaxy evolution and its relation with cosmology. Dark
matter cannot be observed directly, so its characteristics must be
derived observationally by studying the gravitational effect it has on
tracer particles (be it atoms or photons). Alternatively, its
properties can be studied using sophisticated computer
simulations. The latter have now converged on a picture where dark
matter consists of massive, non-relativistic particles, also called
Cold Dark Matter (CDM).

Simulations adopting this framework find galaxy halos to be tri-axial
\citep[e.g.,][]{frenk-1988, dubinski-1994, hayashi-2004,hayashi-2007,
  capuzzo-2007}, and to have a characteristic mass density
profile. Its most distinghuishing feature is the steep increase in
dark matter density towards the center, usually described as a
power-law $\rho \sim r^\alpha$ with $\alpha \leq -1$
\citep[e.g.,][]{navarro-1996, navarro-1997}. Much effort has been
dedicated to obtaining observational confirmation of these
predictions, concentrating mainly on dark matter dominated galaxies.
Most observational determinations of the distribution of dark matter
in these galaxies (as deduced from their kinematics) seem, however, to
indicate that it is not characterized by a $r^{-1}$ ``cusp'', but by a
central kpc-sized ``core'' with a constant dark matter density
\citep[][]{mcgaugh-1998b, deblok-2001a, deblok-2001b, marchesini-2002,
  deblok-2002, gentile-2004, simon-2003, simon-2005, deblok-2005,
  zackrisson-2006, gentile-2007, spano-2007, naray-2006, naray-2008}.

Whether this core is real or a result of unrecognized systematics in
the observations has been the subject of some debate. For example, it
is conceivable that the photometric and kinematic centers could be
physically offset from each other \citep[see, e.g,][]{matthews-2002},
which would lead to an artifical flattening of the inferred mass
density distribution if the ``wrong'' center is chosen \citep[see,
e.g.,][]{swaters-2003a}.  A second effect which has been discussed by
many authors is the effect of non-circular motions. These are a
natural consequence of tri-axial halos
\citep[][]{hayashi-2004,hayashi-2006,hayashi-2007}, but can also be
caused by bars, spiral arms, or other perturbations to the potential
\citep[cf.][]{rhee-2004,valenzuela-2007}. As an example of how they
could affect the observations, \cite{hayashi-2006} show that the
ellipticity of the dark matter halo can induce large non-circular
motions in the inner parts of galaxies (up to $\sim 15\,$ percent of
the total rotation velocity), thus making an intrinsic \emph{cuspy}
density profile appear \emph{cored}.  Using simulated rotation curves,
\citet{deblok-2003} show that non-circular motions of the order of
$\sim 20\ \mathrm{km\,s^{-1}}$ over a large fraction of the disk are
needed to make them consistent with CDM halos \citep[see also][ who
reach similar conclusions derived from actual
observations]{gentile-2005}.

As mentioned above, most of these studies have concentrated on dark
matter dominated galaxies, and more specifically on extremely
late-type spirals as well as late-type dwarf galaxies (a group of
galaxies commonly known as Low Surface Brightness [LSB] galaxies). The
baryonic components of these galaxies, being a minor contribution to the total mass, can be used in these systems as a tracer for the
dark matter component. However, a question
which naturally arises is how representative these galaxies are for
the galaxy population at large.  The shallow potentials and low mass
surface densities potentially allow the possibility that their
photometric and kinematic centers have particularly large offsets,
and/or that the importance of non-circular motions in their disks is
anomalously high.  Systematic observational studies of center
positions and non-circular motions in late-type spiral and dwarf
galaxies are, however, still comparatively rare \citep[but see,
e.g.,][]{matthews-2002,simon-2003,simon-2005,gentile-2005,naray-2006,naray-2008}.

One can also take a broader approach and study non-circular
motions and center offsets across the Hubble sequence, with the aim of,
firstly, tying the existing studies of dark matter-dominated galaxies into
a larger, established scheme of galaxy properties changing with Hubble
type and luminosity, and, secondly, providing a baseline for future
studies.

For this reason, we have embarked on a major study of a large sample
spanning a major part of the Hubble sequence to measure possible
offsets between positions of the dynamical and photometric centers, as
well as the importance of any non-circular motions. We will be using
the THINGS \citep[The HI Nearby Galaxy Survey;][see Section
2]{walter-07} sample which covers a large range in physical
properties, and enables us to address aspects such as the importance
of dark matter as a function of Hubble type, the applicability of dark
matter models, the shapes of rotation curves, etc. Many of these
topics are covered in some detail in \cite{deblok-07} and
\cite{se-heon}.  The THINGS sample furthermore contains a number of dark
matter dominated dwarf galaxies that should provide direct
clues on the behaviour of the centers and the importance of
non-circular motions in extremely late-type galaxies and  enable
a comparison with earlier Hubble types.


Deviations from purely circular motions in galaxies in general can
have different causes, including chaotic non-circular motions induced
for example by star formation \citep[as investigated in detail for two
THINGS dwarf galaxies in][]{se-heon}, or systematic non-circular
motions that relate to the potential (e.g., spiral arms, tri-axiality
of the halo). Here, we will focus on the systematic non-circular
motions.  One way to quantify these is to make a harmonic
decomposition of the velocity field. An extensive description of the
harmonic decomposition technique is given in
\citet{schoenmakers-thesis}. Several groups have measured non-circular
motions in disk galaxies using this technique
\citep[e.g.,][]{schoenmakers-1997, wong-2004, gentile-2005}. However,
none of these works report non-circular motions as large as those
proposed by \cite{hayashi-2006}.  \cite{simon-2003} used a different
technique to measure the non-circular motions in NGC 2976. They
conclude that although the inner 300 pc of NGC 2976 contain relatively
large non-circular motions, the slope of its (cored) density profile
is not significantly affected by that. \cite{swaters-2003b} on the
contrary argue that the non-circular motions in DDO 39 might prevent
the derivation of a well-constrained density profile.

As for the center offsets, \citet{deblok-2001a} and
\citet{deblok-2004} show that significant differences 
between the mass density profiles of cored and cuspy halos become
prominent interior to a radius of $\sim 1$ kpc. This radius follows
directly from the halo models and is fairly independent of (realistic)
assumptions on halo parameters of either model. Center offsets
significantly smaller than a kpc are thus unlikely to have a major
impact on the shapes of the rotation curves.

A previous study by \citet{matthews-2002} directly addresses the issue
of center offsets using high-resolution rotation curves of a sample of
21 extreme late-type, low-luminosity, low-surface brightness spiral
galaxies. They compare the systemic velocities as derived from HI and
H$\alpha$ observations and in general find excellent agreement (for
two-thirds of their sample the velocities agree to better than 10 km
s$^{-1}$), leading them to conclude that ``most extreme late-type
spirals seem to have well-defined centers in spite of their often
diffuse stellar disks and shallow central potentials.''

In this paper we will quantify the central positions of the THINGS
galaxies, as well as quantify the strength of the non-circular motions
in these galaxies, and investigate their dependence on global galaxy
properties, such as luminosity. We will also check whether the mass
models of the THINGS galaxies as presented in \cite{deblok-07} and
\cite{se-heon} are affected by these non-circular motions. 

The paper is organized as follows.  After a brief description of the
sample in \S 2, we extensively discuss the various ways in which one
can define the center of a galaxy in \S 3, and present our best
determinations of the positions of the true kinematic centers of our
sample galaxies. These center positions are used as inputs for our
harmonic decompositions of the observed velocities, as described in \S
4. We illustrate the procedure and show the results for one case-study
galaxy in detail in \S 5. Results for the rest of the sample are
presented in the Appendix.  Our results and several quality checks are
presented and discussed in \S 6, and the conclusions are summarized in
\S 7.

\section{Sample and data}
For our analysis, we used THINGS (The \HI\ Nearby Galaxy
Survey), an \HI\ spectral line survey of 34 nearby disk galaxies obtained at
the NRAO\footnote{The National Radio Astronomy Observatory is a
  facility of the National Science Foundation operated under
  cooperative agreement by Associated Universities, Inc.} VLA in B, C,
and D arrays. THINGS contains a wide range of galaxy types,
i.e., high and low surface brightness galaxies, grand
design spirals and dwarf irregulars, barred and non--barred
  galaxies, all observed at high spatial ($\sim 10\arcsec$ for the natural weighted data cubes) and velocity ($\le
5.2\ \mathrm{km\,s^{-1}}$) resolution \citep[for a detailed description of
THINGS, see][]{walter-07}. The great
advantage of THINGS is that all galaxies have been observed, reduced, and
analyzed in a homogeneous manner. The large range in physical
properties enables one to study trends with, e.g., Hubble type. 
Furthermore, as THINGS was designed to overlap with the SINGS survey
\citep[Spitzer Infrared Nearby Galaxies Survey,][]{SINGS}, we can use its IRAC
3.6\,$\mu$m images (which give a 
virtually dust-free view of the stellar disk) to directly compare
photometric and kinematic centers.

In this paper, we will study the kinematics of galaxies. We therefore
use the sub-sample of THINGS for which \citet{deblok-07} derived
accurate rotation curves. This sample includes THINGS galaxies with
inclinations larger than 40\degree\ that are not undergoing obvious
strong interactions, as well as NGC 6946, which, despite its lower
inclination proved suitable for the derivation of a rotation
curve. Basic properties of the galaxies of our sample are given in
Table~\ref{table:sample-properties}.  For our analysis, we used the
velocity fields which were created by fitting hermite polynomials to
the velocity profiles from the natural weighted data cubes \citep[as
  described in detail in][]{deblok-07}. Additionally, we made use of
the radio continuum maps from THINGS \citep{walter-07}, as well as the
3.6\,$\mu$m images from SINGS \citep{SINGS}.

\begin{deluxetable*}{lrrrrrrrrr}
\tablewidth{0pt}
\tabletypesize{\footnotesize}  
\tablecaption{Basic properties for the galaxies in our sample.\label{table:sample-properties}}
\tablehead{
\colhead{Name} & \colhead{D} & \colhead{$r_{25}$} & \colhead{$M_{\mathrm{B}}$} & \colhead{$i$} & \colhead{PA} & \colhead{$V_{\mathrm{tot}}$} & \colhead{$\Delta r$} & \colhead{$\mathrm{M_{HI}}$} & \colhead{Type}\\
& \colhead{Mpc} & \colhead{kpc} & \colhead{mag} & \colhead{$\degree$} & \colhead{$\degree$} & \colhead{$\kms$} & \colhead{$\arcsec$} & \colhead{$10^8\ \mathrm{M}_{\odot}$} & \\
\colhead{(1)} & \colhead{(2)} & \colhead{(3)} & \colhead{(4)} & \colhead{(5)} & \colhead{(6)} & \colhead{(7)}& \colhead{(8)} & \colhead{(9)} & \colhead{(10)} \\
}
\startdata
NGC 925  & 9.2  & 14.2 & $-$20.04 & 66 & 287 & 115 & 3.0 & 45.8   & 7 \\ 
NGC 2366 & 3.4  & 2.2  & $-$17.17 & 64 & 40  & 55  & 6.0 & 6.5    & 10\\ 
NGC 2403 & 3.2  & 7.4  & $-$19.43 & 63 & 124 & 135 & 4.0 & 25.8   & 6 \\ 
NGC 2841 & 14.1 & 14.2 & $-$21.21 & 74 & 153 & 260 & 5.0 & 85.8   & 3 \\ 
NGC 2903 & 8.9  & 15.2 & $-$20.93 & 65 & 204 & 190 & 7.0 & 43.5   & 4 \\ 
NGC 2976 & 3.6  & 3.8  & $-$17.78 & 65 & 335 & 80  & 3.5 & 1.4    & 5 \\ 
NGC 3031 & 3.6  & 11.6 & $-$20.73 & 59 & 330 & 200 & 6.0 & 36.4   & 2 \\ 
NGC 3198 & 13.8 & 13.0 & $-$20.75 & 72 & 215 & 150 & 6.0 & 101.7  & 5 \\ 
IC 2574  & 4.0  & 7.5  & $-$18.11 & 53 & 56  & 70  & 6.0 & 14.8   & 9 \\ 
NGC 3521 & 10.7 & 12.9 & $-$20.94 & 73 & 340 & 210 & 6.0 & 80.2   & 4 \\ 
NGC 3621 & 6.6  & 9.4  & $-$20.05 & 65 & 345 & 140 & 6.5 & 70.7   & 7 \\ 
NGC 3627 & 9.3  & 13.9 & $-$20.74 & 62 & 173 & 133 & 5.0 & 8.2    & 3 \\ 
NGC 4736 & 4.7  & 5.3  & $-$19.80 & 41 & 296 & 120 & 5.0 & 4.0    & 2 \\ 
DDO 154  & 4.3  & 1.2  & $-$14.23 & 66 & 230 & 48  & 6.5 & 3.6   & 10\\ 
NGC 4826 & 7.5  & 11.4 & $-$20.63 & 65 & 121 & 150 & 5.0 & 5.5    & 2 \\ 
NGC 5055 & 10.1 & 17.4 & $-$21.12 & 59 & 102 & 190 & 5.0 & 91.0   & 4 \\ 
NGC 6946 & 5.9  & 9.8  & $-$20.61 & 33 & 243 & 220 & 3.0 & 41.5   & 6 \\ 
NGC 7331 & 14.7 & 19.6 & $-$21.67 & 76 & 168 & 233 & 3.0 & 91.3   & 3 \\ 
NGC 7793 & 3.9  & 6.0  & $-$18.79 & 50 & 290 & 130 & 6.0 & 8.9    & 7 \\ 
\enddata
\tablecomments{(1): the name of the galaxy; (2): distance as given in \cite{walter-07}; (3): radius of the major axis of the galaxy at the $\mu_\mathrm{B} = 25\, \mathrm{mag\,arcsec^{-2}}$ isophote level, taken from LEDA; (4): absolute \emph{B}-band magnitude as given in \cite{walter-07}; (5): average inclination as given in \cite{deblok-07}; (6): average position angle as given in \cite{deblok-07}; (7): total rotation velocity (used for the normalization of the non-circular motions); (8): adopted spacing of the tilted-rings; (9): \HI\ mass as listed in \cite{walter-07}; (10): morphological type from LEDA} 
\end{deluxetable*}

\section{Estimating galaxy centers}
For a correct appraisal of the non-circular motions in a galaxy (as
derived from a harmonic decomposition) it is important to accurately
determine the position of its center
\citep[cf.][]{schoenmakers-thesis}.  There are several ways in which
one can determine the position of the center of a galaxy. As central
activity in galaxies is likely to coincide with the bottom of the
galaxy potential well, a central compact radio source is a good
indicator of the center of a galaxy. Similarly, nuclear star clusters
can be used to locate the center \citep{matthews-2002}. One can also
fit ellipses to the surface brightness distribution as well as
construct a tilted-ring model using the kinematic data to obtain
additional, independent estimates.

Ideally, these different determinations should all coincide. Large
discrepancies can indicate strong disturbances, caused, e.g., by
spiral arms or strong bars, or a genuine offset between the kinematic
and photometric center. For the majority of our galaxies, multiple
center estimates can be derived, namely kinematic centers derived from
the velocity fields, as well as photometric centers as deduced from
the 20-cm radio continuum maps taken from THINGS and from the {\it
  Spitzer} 3.6\,$\mu$m images.

As will be shown in the subsequent analysis, for most galaxies the
kinematic center position agrees with the position of the radio
continuum source and/or the center as derived from the IRAC image
within the uncertainties. In those cases where the positions of the
photometric and kinematic centers could be determined accurately, we
have generally adopted one of the photometric centers as our best
center position, given that the photometric centers usually have
smaller uncertainties.  Note that this particular choice affects our
determination of the non-circular motions. The position of the
dynamical center is to first order that position which minimizes the
non-circular motions in a tilted-ring fit and adopting this (or a
variable) center position would thus results in non-circular motions
slightly smaller than the ones derived here for the photometric
centers.  For most of our galaxies the difference will be minimal, but
the general effect is important to keep in mind.

Also note that it is still important to evaluate any possible
differences between the two types of centers, even when both are
well-defined. This will enable a better understanding of possible
systematic effects inherent in the method, which is important for the
interpretation of the results from galaxies where the centers may not
be as well-defined.
 
Below follows a description of the methods used to determine the various
center estimates. An application of these methods is given for
one ``case-study'' galaxy in Section~\ref{sec:test-candidate}. Detailed
and more technical descriptions are then given for all galaxies in
Appendix~\ref{sec:indiv-gal}.

\subsection{Radio continuum}
A nuclear point source in the radio continuum is usually
associated with a central compact object, which naturally should be at
(or very close to) the bottom of the potential well of the
galaxy. For all galaxies which show such a source in the THINGS radio
continuum maps, 
its position was determined by fitting a Gaussian to the central
source. The uncertainties for the center positions estimated in this way are
all similar ($\le 1\arcsec$) and we therefore do not show
individual uncertainties for center estimates deduced from the radio continuum.

\subsection{{\it Spitzer}/IRAC 3.6\,$\mu$m image}
We make use of the high-resolution 3.6\,$\mu$m
images from SINGS \citep{SINGS}\footnote{A small number of galaxies in our sample were not part of SINGS. For these galaxies, the data were retrieved from the \emph{Spitzer} archive.}. These allow an almost dust-free view of the
predominantly old stellar populations, though we note that the 3.6 
$\mu$m band can also contain some trace emission from hot dust, PAHs and AGB stars. 

For all galaxies in our sample, we determined the central 3.6\,$\mu$m position by fitting ellipses using the GIPSY\footnote{GIPSY, the Groningen Image Processing SYstem
  \citep{vanderhulst-1992}} task {\sc ellfit}, taking care that the
ellipse fits were not affected by small scale structures.
For those galaxies which also show a well-defined nuclear source in the 3.6\,$\mu$m 
image, we additionally derived the central position by fitting a
Gaussian to the central source. The two different center estimates generally agree very well. 
However, as the center position derived by fitting a Gaussian to the central source is usually better constrained as the center from {\sc ellfit}, we only list the former (where available) in Table~\ref{table:center-pos}.
Because of the homogeneous and consistently small positional uncertainty of less than 1\arcsec, we
do not list these here.

\subsection{Kinematic center}
In addition to the photometric centers mentioned above, we also derive kinematic
centers using the GIPSY task \rotcur. This task fits a set of tilted
rings of a given width 
to the velocity field of a galaxy and determines their central positions, rotation
and systemic velocities, inclinations and position angles. We use
the best available center position
(i.e., a central continuum source, if present, otherwise a nuclear
source in the 3.6\,$\mu$m image
and as a last resort the center as derived using {\sc ellfit}) as an initial center estimate for
\rotcur\ and make a fit with all parameters left free (including the
center). By averaging the central positions over a radial range unaffected by
spiral arms or other large-scale disturbances, we 
derive the position of the kinematic center for each galaxy.

The determination of the positions of all centers are described in Appendix~\ref{sec:indiv-gal}, and all center estimates are summarized in Table~\ref{table:center-pos}, where our adopted best center positions are shown in bold face.

\section{Harmonic decomposition}\label{sec:harm-decomp-introduction}
We perform a harmonic decomposition of the velocity fields by decomposing the velocities found along the tilted-rings into multiple terms of sine and cosine.

Following \citet{schoenmakers-thesis}, we describe the line-of-sight velocity,
$v_{\mathrm{los}}$, as:
\begin{equation}
v_{\mathrm{los}}(r)=v_{\mathrm{sys}}(r)+\sum_{m=1}^N c_m(r)\,\cos\, m\,\psi\,
+\,s_m(r)\, \sin\, m\,\psi,
\label{eq:vlos}
\end{equation}
where $N$ is the maximum fit order used, $r$ is the radial distance
from the dynamical center, $\psi$ is the azimuthal angle in the plane
of the disk, and $v_{\mathrm{sys}}$ is the $\mathrm{0^{th}}$ order 
harmonic component, $c_0$.
Initial tests showed that a decomposition of the velocity fields up to third
order (i.e., $N=3$) is sufficient to capture most of the non-circular signal,
as is described in Section~\ref{sec:reswri-rotcur-residuals}.
 
The usual description of the apparent velocity, under the assumption of
purely circular motion can be 
retrieved by only including $m=0$ and $m=1$ terms in Eq.~\ref{eq:vlos}, i.e.,
\begin{equation}
v_{\mathrm{los}}(r)=v_{\mathrm{sys}}(r)+c_1(r)\,\cos\, \psi\,
+\,s_1(r)\, \sin\, \psi,
\label{eq:vlos-circ}
\end{equation}
and by ignoring streaming (radial) motions (i.e., $s_1=0$).
The circular rotation velocity corresponds therefore to $c_1$. Note that the dependence 
on inclination is included in the $c_m$ and $s_m$ terms.
For the following discussion it is worthwhile to repeat a few rules of
thumb which apply to harmonic decompositions as given in \citet{schoenmakers-1997} and \citet{schoenmakers-thesis}:

{\bf (1)} A perturbation of the gravitational potential of order $m$ will cause
  $m+1$ and 
  $m-1$ harmonics in the velocity field (so an $m=2$ two-armed spiral
  component will cause $m=1$ and $m=3$ harmonics in the velocity field).

{\bf (2)} Perturbations in the gravitational potential are independent
  and can therefore be 
  added. The same holds for velocity perturbations.

{\bf (3)} The elongation of the potential $\epsilon_{\mathrm{pot}}$ in the plane of the disk of the galaxy
can be calculated at each radius as follows:
\begin{equation}
\epsilon_{\mathrm{pot}}\,\sin\,2\varphi_2=(s_3-s_1)\frac{1+2q^2+5q^4}{c_1(1-q^4)},
\label{eq:epot}
\end{equation}
where $q=\cos i$. The only remaining unknown quantity is $\varphi_2$, the unknown angle in the plane of the ring between the minor axis of the elongated ring and the observer.

{\bf (4)} Velocities induced by a global elongation of the potential will result in a constant offset in
  $\epsilon_{\mathrm{pot}}\,\sin\,2\varphi_2$. Velocities induced by spiral arms occur on much smaller scales, and will therefore only lead to perturbations (``wiggles'') around this offset.

{\bf (5)} If the fitted inclination is close to the intrinsic inclination of the disk, then
  $c_3=0$. Small offsets of a few \kms\ result in only small (1-2 degree) inclination offsets.

We use the GIPSY
task \reswri. This task performs a tilted-ring fit
assuming circular rotation, creates a model velocity field, subtracts
this from the original velocity field, and does a harmonic expansion of the
residuals. \reswri\ does not down-weight velocities along the minor axis as is
usually done in standard rotation curve analysis, and thus emphasizes the non-circular motions. 
For the width of the
annuli (cf. Col (8) of Table~\ref{table:sample-properties}), we chose half the beam width; neighboring rings are thus not
independent. 

We calculate the quadratically added amplitude (``power'') for each order of the harmonic
decomposition using 
\begin{equation}
A_1(r)=\sqrt{s_1^2(r)},\label{eq:A(m)1}
\end{equation}
for $m=1$ (note that $c_1$ corresponds to the circular velocity and is not
included in the calculation of the amplitude of $A_1(r)$), and
\begin{equation}
A_m(r)=\sqrt{c_m^2(r)+s_m^2(r)},\label{eq:A(m)}
\end{equation}
for $m>1$.\\
Additionally, we calculate the quadratically added amplitude of all (i.e., up to $N=3$) non-circular harmonic components (``total power''):
\begin{equation}
A_r(r) = \sqrt{s_1^2(r)+c_2^2(r)+s_2^2(r)+c_3^2(r)+s_3^2(r)}.\label{eq:A(r)}
\end{equation}

The radial variation of $A_m(r)$ and $A_r(r)$ can be checked for
coincidence with visible features in the galaxies. To quantify the
power of the non-circular motions in a compact way, we also use \Am,
the median of $A_m(r)$ defined for each value of $m$, and \A, the
median of $A_r(r)$.  We derive two values for \Am\ and \A, one for the
entire radial range and one for the inner 1 kpc (for those galaxies
where sufficiently high signal-to-noise \HI\ is present in the inner
parts).  The 1 kpc value is not arbitrarily chosen but follows from
both the core and cusp models for reasonable choices of halo
parameters. The distinction between a cusp and a core can be made most
clearly interior to this radius \citep{deblok-2001a, deblok-2004}, and
it is therefore important to separately quantify the non-circular
motions in this inner region.  Values determined for the entire radial
range are thus determined over the entire extent of the \HI\ disk,
except for radii where \reswri\ failed to converge properly or produce
stable results. These radii are usually characterized by very small
filling factors and are mostly found in the outer parts of the
galaxies.

In the following we will discuss the absolute non-circular motions, as
well as the same motions expressed as a fraction of the local and
total rotation velocity. The latter is defined as the rotation
velocity of the flat part of the rotation curve, or as the maximum
rotation velocity if the rotation curve is still rising at the
outermost point.

\reswri\ creates a residual velocity field which can be used as an
indicator for those non-circular motions that are not captured with
the harmonic decomposition (as will be shown in
Section~\ref{sec:reswri-rotcur-residuals}). As the residual velocity
fields contain only values which scatter around a mean zero level, we
use \emph{absolute} residual velocity fields in our analysis.

Finally, as mentioned in \cite{schoenmakers-thesis}, the center
position should be kept fixed during the harmonic decomposition. This
is because a galaxy which has real, physical $c_2$ and $s_2$ terms in
its velocity field would, in the case of an unconstrained center
position, appear to have a center which drifts in such a way as to
minimize these terms. Moreover, rapidly varying center positions at
small radii (with offsets larger than the relevant ring radii) have no
physical basis.

\section{NGC 3198 --- A case--study}\label{sec:test-candidate}
Here we present the derivation of the central position and the
harmonic decomposition in some detail for one galaxy. 
The purpose of this section is to explain the conventions and notations used
and demonstrate our methods for one galaxy in our sample.
A complete description and discussion of all galaxies from our sample is
given in the Appendix.\\

\subsection{Center estimates}
The 3.6\,$\mu$m IRAC image of NGC 3198 shows two well-defined spiral arms, 
emanating from a prominent bulge. The central component has a nuclear point
source embedded, which has a counterpart in the radio continuum. The
IRAC and continuum centers agree to within 1\arcsec.
For ease of reference, and in order to have a compact notation for the center positions, we will in our discussion frequently refer to positions with respect to the pointing center \citep[as listed in][]{walter-07}. This pointing center has no physical meaning, and merely provides a convenient zero-point, unrelated to any particular choice of the central position. 
We will refer to the offsets from the pointing center as \dx\ (positive in the direction of decreasing right ascension) and \dy\ (positive in the direction of increasing declination). Both are expressed in arcseconds. Note that we list the full coordinates
of all center positions we derive in Table~\ref{table:center-pos} (with the position used for further analysis shown in bold face).

In our determination of the kinematic center of NGC 3198, we start with an unconstrained \rotcur\
fit with the position of the radio continuum 
center (\dx= $-1$\farcs 4, \dy= $-0$\farcs 1) as an initial estimate.
\begin{figure*}[]
\begin{center}
\includegraphics[angle=0,width=0.80\textwidth,bb=18 144 592 520,clip=]{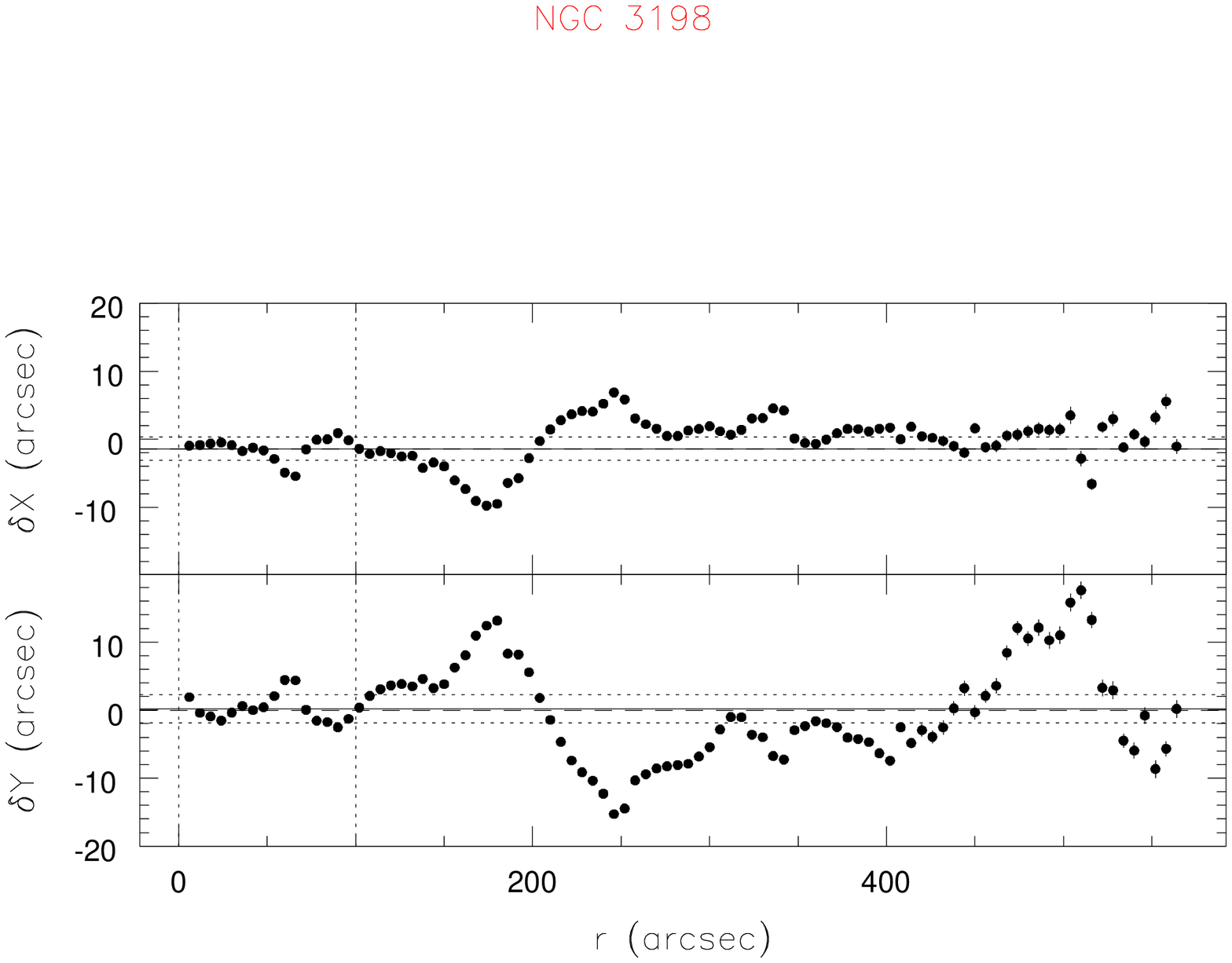}
\caption[Estimating the dynamical center of NGC 3198]{Radial variation of the center positions of the individual
  tilted-rings derived in an unconstrained fit with \rotcur. The center
  positions are given as an offset from the pointing center (in
  arcseconds). The two vertical lines at $r=0\arcsec$ and $r=100\arcsec$
  denote the radial range over which the center positions were averaged in
  order to derive a kinematic center. The resulting kinematic center
  (together 
  with its standard deviation) is indicated by the black solid (and dotted)
  lines. Our best center position (in this case from the radio continuum) is
  indicated by the dashed horizontal line, which is however barely
  distinguishable from the kinematic estimate because of the good agreement
  between these two.}\label{fig:n3198-1}
\end{center}
\end{figure*}
Fig.~\ref{fig:n3198-1} shows the variation of the center position from the
\rotcur\ fit over the radial range of the galaxy, together with our best
photometric center. As can
be seen by the variation of \dx\ and \dy, the outer parts ($r \ge
150\arcsec$) of 
NGC 3198 are strongly affected by the spiral arms. For the derivation of the
dynamical center, we therefore restrict the averaging of \dx\ and \dy\ to
radii with $r\le 100\arcsec$ (indicated in Fig.~\ref{fig:n3198-1} by the
vertical lines at these radii).  
The dynamical center derived in such a way is offset from the pointing
center by $\dx = -1\farcs 4 \pm 1\farcs 7$, $\dy=-0\farcs 2 \pm
2\farcs 1$ (i.e., to the south-east).
Averaging \dx\ and \dy\ over all tilted-rings results in a similar center position, though with a larger scatter.

\begin{figure*}[]
\begin{center}
\includegraphics[angle=0,width=0.85\textwidth,bb=18 79 520 520,clip=]{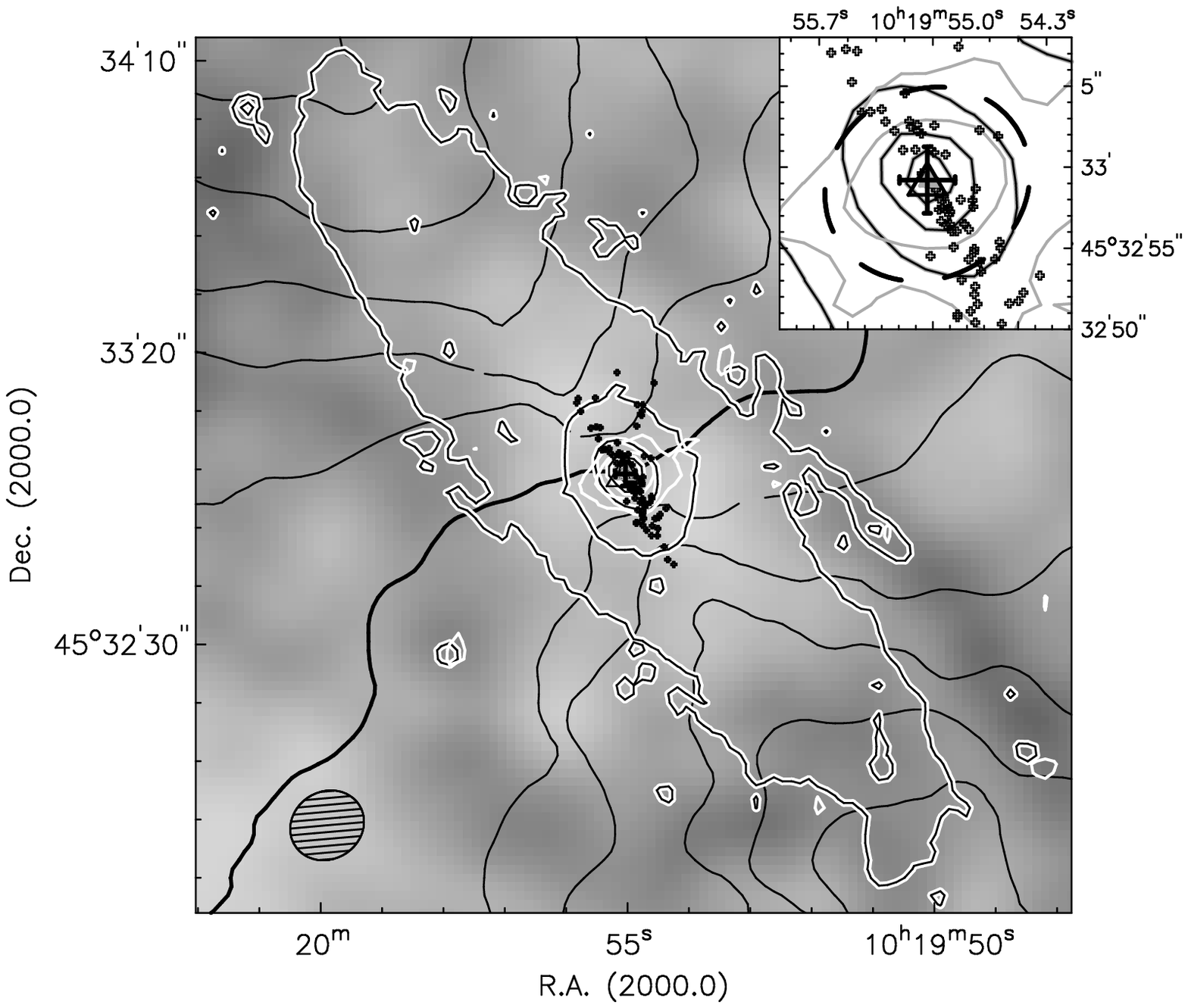}
\caption[Comparison of different center estimates of NGC 3198]{This figure shows the inner $150\arcsec\,\times\,150\arcsec$ of
  NGC~3198. The integrated natural-weighted \HI\ map is shown in grayscales. The
  beam is indicated in the bottom-left corner. The black contours show
  iso-velocity contours from the hermite velocity field. The thick black
  contour is shown at 663\,\kms\ and represents the systemic velocity given in
  \citet{deblok-07}. The other velocity contours are spaced by 25\,\kms. The
  thin black contours overlaid on the thick white contours represent the
  3.6\,$\mu$m IRAC image. They are drawn at 2, 5, 10, 20, and 50 percent of
  the maximum intensity level. 
The white contours represent the THINGS radio continuum
map, and are drawn at 10, 20, and 50 percent of the peak
  intensity. The black dots indicate the individual center
positions from 
\rotcur\ and the black cross represents the derived dynamical center together
with its uncertainty. The derived center from the 3.6\,$\mu$m image is shown as a
gray, filled triangle, whereas the one from the radio continuum is shown as a
black, open triangle.
{\bf Inset:} To better highlight the different
center estimates, we show an inset of the innermost 18\arcsec\ in the
upper-right corner. For clarity reasons we 
do not show the \HI\ grayscale and velocity field contours here. The
contours from the 3.6\,$\mu$m image are shown in black and are given at the same
intensity levels as in the main plot. The same holds for the radio continuum
contours, which are shown here in gray. The individual center estimates from
\rotcur\ are shown as small crosses.
In the inset, the beam is indicated by the thick black dashed ellipse and is centered on
our best center position.} \label{fig:n3198-2}
\end{center}
\end{figure*}

To put our estimates for the center positions in context, we show them
together with the IRAC and radio continuum map (where available) overlaid on
the central $150\arcsec\,\times\,150\arcsec$ of the \HI\ total intensity map
(cf. Fig.~\ref{fig:n3198-2}). Also shown are the central
positions of the individual tilted-rings. 
As can be seen, all center positions agree well within the uncertainties and to within one natural-weighted beam (hereafter referred to
as ``the beam'') and we therefore adopt the center as
derived from the radio continuum map as our best center position.\\

\subsection{Harmonic expansion}
\begin{figure*}[]
\begin{center}
\includegraphics[angle=0,width=1\textwidth,bb=19 404 592 697,clip=]{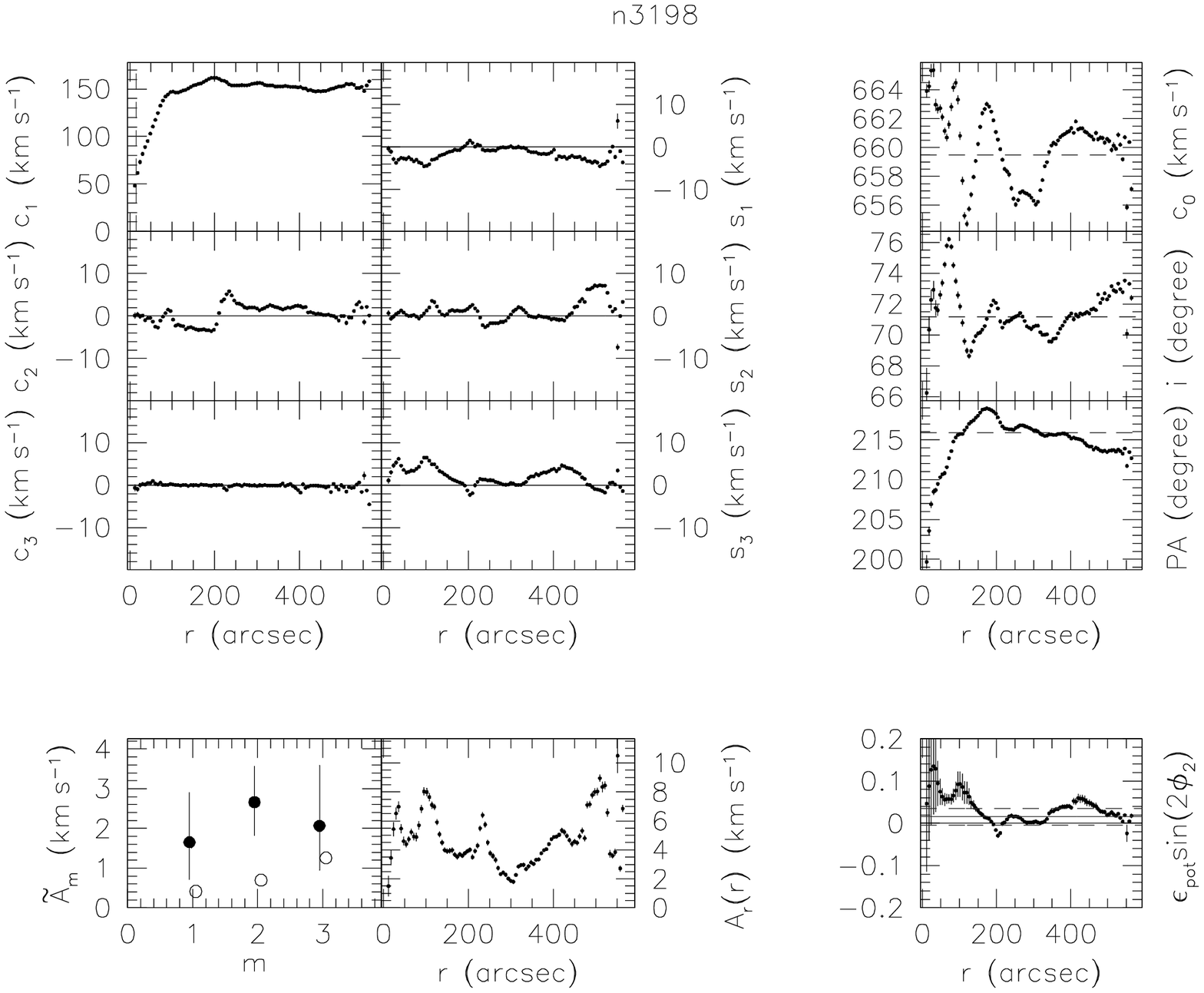}
\caption[Harmonic decomposition of NGC 3198: individual components]{{\bf Left:} circular ($c_1$), and non-circular ($c_2$, $c_3$, $s_1$,
  $s_2$, $s_3$) harmonic components (derived according to Eq.~\ref{eq:vlos}), all corrected for inclination and plotted
  {\it vs.} radius. The dashed vertical line in the panel showing the
  distribution of $c_1$ indicates the radius corresponding to 1 kpc.
 {\bf Right:} systemic velocity $c_0$, inclination and position angle, plotted
  {\it vs.} radius. The dashed horizontal lines represent the error weighted
  means. The inclination and position angle are from the tilted-ring fit assuming circular rotation. 
The error bars shown in all panels of this figure are the formal
  uncertainties from \reswri.}
\label{fig:n3198-3}
\end{center}
\end{figure*}

The radial distribution of all fitted parameters from the harmonic decomposition of NGC 3198 are shown in Fig.~\ref{fig:n3198-3}.
The PA of NGC 3198 rises swiftly within the inner $200\arcsec$, and then
slowly declines. The inclination varies in the inner parts
over a range of about five degrees, but shows a steady increase beyond
$r\sim 450\arcsec$, 
indicating that the outer disk is warped. The $c_3$ term is 
small for all radii, meaning that the fitted inclination is close to
the intrinsic inclination of the disk. 
Although there is no global offset from zero for $c_2$ and
$s_2$, they show small deviations at radii coinciding with the locations of 
spiral arms in the 3.6\,$\mu$m image
of NGC 3198, or in the total \HI\ map presented in \cite{walter-07}.
The $s_1$ and $s_3$ terms are best described as wiggles caused by spiral arms
on top of 
a slight offset. The amplitudes of all non-circular components ($c_2$, 
$c_3$, $s_1$, $s_2$, $s_3$) are generally only a few \kms.

\begin{figure*}[]
\begin{center}
\includegraphics[angle=0,width=1\textwidth,bb=19 235 592 375,clip=]{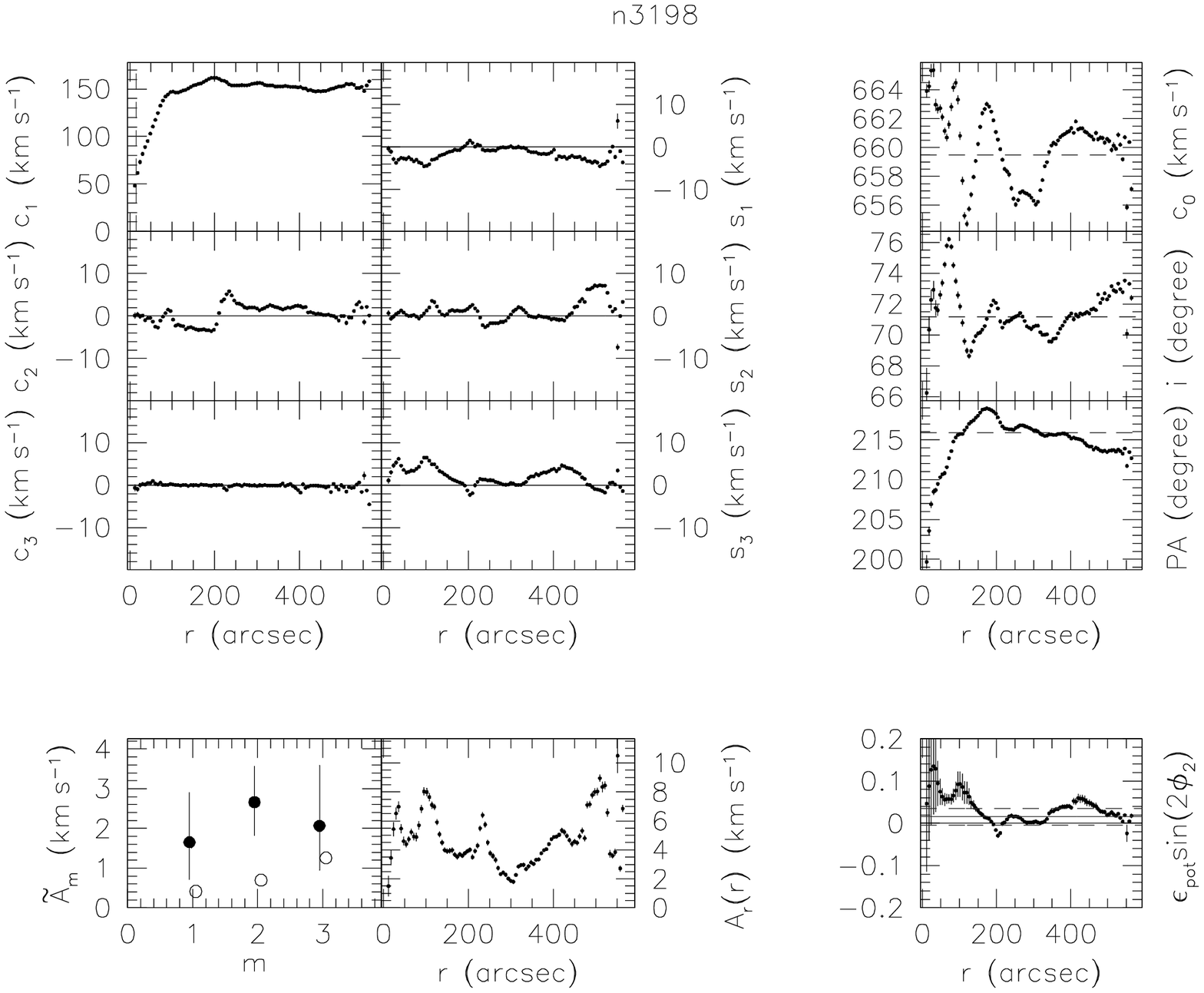}
\caption[Harmonic decomposition of NGC 3198: quadratically added amplitudes
and elongation of the potential]{{\bf Left:} \Am, the median amplitudes of the individual harmonic
components, derived following Eqs.~\ref{eq:A(m)1} and \ref{eq:A(m)}, plotted
{\it vs.} harmonic number $m$. The error bars denote the upper and lower
quartile of the distribution of $A_m(r)$. We have calculated the
median amplitudes twice: once for the entire 
radial range (filled circles), and once for the inner 1 kpc (open
circles). Note that in the case of NGC 3198, the median for the inner
1 kpc has no 
error bars, as the inner 1 kpc contains only one data point.  
 {\bf Middle:} \Ar, the
quadratically added amplitude of all non-circular components (derived
following Eq.~\ref{eq:A(r)}) {\it vs.} radius. The errors were estimated
assuming Gaussian error propagation. For most radii, the amplitude
is $\Ar \le 8 \kms$.
 {\bf Right:} the radial distribution of the elongation of
the potential, $\epsilon_{\mathrm{pot}}\,\sin(2\varphi_2)$, derived according
to Eq.~\ref{eq:epot}. The uncertainties are also estimated assuming Gaussian
error propagation. The elongation is fairly unconstrained in the inner $\sim
30\arcsec$, mainly because of the large uncertainty in the inclination, which
enters into the uncertainty of $\epsilon_{\mathrm{pot}}$ as a fourth
power. The weighted mean elongation (and its standard deviation) are
represented by the horizontal dotted (and dashed) lines. The weighted mean
elongation of NGC 3198 is $ \langle \epsilon_{\mathrm{pot}}\,\sin(2\varphi_2)
\rangle =0.017 \pm 0.020$, and thus consistent with a round potential.}
\label{fig:n3198-4}
\end{center}
\end{figure*}

In Fig.~\ref{fig:n3198-4}, we show derived parameters which were
calculated according to Eqs.~\ref{eq:epot}-\ref{eq:A(r)}.
The median amplitudes of the individual harmonic
components, derived following Eqs.~\ref{eq:A(m)1} and \ref{eq:A(m)}, are 
similar in amplitude, ranging from $\Am \sim 2-3$ km\,s$^{-1}$ 
(when averaged over the entire radial range), or $\le$~2 percent of \vmax. For the 
inner 1 kpc, the amplitudes are even smaller ($\Am < 1.5 \kms$). 
The distribution of \Ar\ (cf. Eq. \ref{eq:A(r)}) shows that the amplitude of the
non-circular motions is $\Ar \le 8\ \mathrm{km\,s^{-1}}$ for most 
radii. The median amplitude is $\A \sim 4.5\,\kms$ when averaged over the
entire radial range, and $\A \sim 1.5\, \kms$ when averaged over the
inner 1 kpc only.

The elongation of the potential, $\epsilon_{\mathrm{pot}}\,\sin(2\varphi_2)$, 
derived according to 
Eq.~\ref{eq:epot} is small over most of the radial range, but shows
traces of the same spiral arms which cause the variation in, e.g., the 
$s_3$ component. Inwards of $r\sim 30\arcsec$, the elongation is
rather unconstrained --- mainly because of the larger uncertainty in
the derived inclination, which enters into the uncertainty in
$\epsilon_{\mathrm{pot}}$ as a fourth power.
The weighted mean elongation of the potential is
fairly small with $ \langle \epsilon_{\mathrm{pot}}\,\sin(2\varphi_2)
\rangle =0.017 \pm 0.020$, and within
the uncertainties consistent with zero.

As mentioned in Section~\ref{sec:harm-decomp-introduction}, \reswri\ creates residual
velocity fields, which can be used to quantify the signal which was
not captured in the harmonic expansion. For NGC 3198, we find a median absolute
value of $\sim 2.6\,\kms$, showing 
that a harmonic decomposition up to third order is capable of capturing most
non-circular motions. This is addressed more fully in
Section~\ref{sec:reswri-rotcur-residuals}.\\

The results from the harmonic decompositions are summarized (for all
galaxies) in Table~\ref{table:harm-decomp} and detailed descriptions for
all galaxies are given in Appendix~\ref{sec:indiv-gal}. 

\begin{deluxetable*}{lllllll}
\tablewidth{0pt}
\tabletypesize{\footnotesize}  
\tablecaption{Center positions for the galaxies in our sample.\label{table:center-pos}}
\tablehead{
\colhead{Name} & 
  \multicolumn{2}{c}{kinematic (\rotcur)
  center} & \multicolumn{2}{c}{3.6\,$\mu$m center} & \multicolumn{2}{c}{radio continuum center} \\
& \colhead{$\alpha_{2000}$} & \colhead{$\delta_{2000}$}&
\colhead{$\alpha_{2000}$} & \colhead{$\delta_{2000}$}&
\colhead{$\alpha_{2000}$} & \colhead{$\delta_{2000}$} \\
& \colhead{(h m s)} & \colhead{($\circ\,\,\, \arcmin\,\,\, \arcsec$)}&
\colhead{(h m s)} & \colhead{($\circ\,\,\, \arcmin\,\,\, \arcsec$)}&
\colhead{(h m s)} & \colhead{($\circ\,\,\, \arcmin\,\,\, \arcsec$)} \\
}
\startdata
NGC 925  & {\bf 02 27 16.5$\pm$0.7} & {\bf $+$33 34 43.5$\pm$4.1} & 02 27 17.0$^a$
& $+$33 34 42.4\footnote{Derived by fitting ellipses with {\sc ellfit}} & \nodata & \nodata \\
NGC 2366 & {\bf 07 28 53.9$\pm$0.7} & {\bf $+$69 12 37.4$\pm$7.8} & 07 28 53.4$^a$ & $+$69 12 40.3$^a$ & \nodata & \nodata \\
NGC 2403 & {\bf 07 36 51.1$\pm$0.9} & {\bf $+$65 36 02.9$\pm$4.2} & 07 36 51.0$^a$ & $+$65 36 02.1$^a$ & \nodata & \nodata \\
NGC 2841 & 09 22 02.6$\pm$0.1 & $+$50 58 35.3$\pm$1.0 & 09 22 02.7 & $+$50 58 35.4 & {\bf 09 22 02.7} & {\bf $+$50 58 35.4} \\
NGC 2903 & 09 32 10.0$\pm$0.2 & $+$21 30 02.5$\pm$2.2 & 09 32 10.1 & $+$21 30 04.9 & {\bf 09 32 10.1} & {\bf $+$21 30 04.3} \\
NGC 2976 & 09 47 14.9$\pm$0.5 & $+$67 55 00.8$\pm$1.4 & {\bf 09 47 15.3} & {\bf $+$67 55 00.0} & \nodata & \nodata \\
NGC 3031 & 09 55 33.5$\pm$0.6 & $+$69 03 52.0$\pm$3.9 & 09 55 33.3 & $+$69 03 54.6 & {\bf 09 55 33.1} & {\bf $+$69 03 54.7} \\ 
NGC 3198 & 10 19 55.0$\pm$0.2 & $+$45 32 59.2$\pm$2.0 & 10 19 55.0 & $+$45 32 59.1 & {\bf 10 19 55.0} & {\bf $+$45 32 58.9} \\
IC 2574\footnote{The center coordinates of IC2574 were derived using a bulk velocity field cleared of non-circular motions as presented in \citet{se-heon}} & {\bf 10 28 27.5$\pm$2.7} & {\bf $+$68 24 58.7$\pm$10.4} & \nodata & \nodata & \nodata & \nodata\\
NGC 3521 & 11 05 48.6$\pm$0.1 & $-$00 02 08.4$\pm$1.4 & {\bf 11 05 48.6} & {\bf $-$00 02 09.2} & \nodata & \nodata \\
NGC 3621 & 11 18 16.6$\pm$0.2 & $-$32 48 48.5$\pm$6.5 & {\bf 11 18 16.5} & {\bf $-$32 48 50.9} & \nodata & \nodata \\
NGC 3627 & 11 20 15.3$\pm$0.2 & $+$12 59 22.7$\pm$4.8 & 11 20 15.0 & $+$12 59 29.2 & {\bf 11 20 15.0} & {\bf $+$12 59 29.6} \\
NGC 4736 & 12 50 53.0$\pm$0.2 & $+$41 07 14.2$\pm$2.0 & 12 50 53.1 & $+$41 07 11.9 & {\bf 12 50 53.0} & {\bf $+$41 07 13.2} \\
DDO 154 & {\bf 12 54 05.9$\pm$0.2} & {\bf $+$27 09 09.9$\pm$3.4} & \nodata & \nodata & \nodata & \nodata \\
NGC 4826 & 12 56 43.6$\pm$0.1 & $+$21 40 59.3$\pm$0.8 & 12 56 43.6 & $+$21 40 59.2 & {\bf 12 56 43.6} & {\bf $+$21 41 00.3} \\
NGC 5055 & 13 15 49.3$\pm$0.2 & $+$42 01 45.1$\pm$1.3 & 13.15 49.3 & $+$42 01 45.5 & {\bf 13 15 49.2} & {\bf $+$42 01 45.3} \\
NGC 6946 & 20 34 52.4$\pm$0.6 & $+$60 09 11.8$\pm$5.9 & 20 34 52.3 & $+$60 09 14.3 & {\bf 20 34 52.2} & {\bf $+$60 09 14.4} \\
NGC 7331 & 22 37 04.1$\pm$0.1 & $+$34 24 54.4$\pm$2.2 & {\bf 22 37 04.1} & {\bf $+$34 24 56.5} & \nodata & \nodata \\
NGC 7793 & 23 57 49.8$\pm$0.2 & $-$32 35 25.2$\pm$2.1 & {\bf 23 57 49.7} & {\bf $-$32 35 27.9} & \nodata & \nodata \\
\enddata
\tablecomments{The uncertainties in the \rotcur\ centers are given in
  units of seconds (for right ascension) and arcseconds (for
  declination). The center position chosen for subsequent analysis is
  shown in bold face.} 
\end{deluxetable*}

\begin{deluxetable*}{lrrrrrr}
\tablewidth{0pt}
\tabletypesize{\footnotesize}  
\tablecaption{Derived quantities from the harmonic decompositions.\label{table:harm-decomp}}
\tablehead{
\colhead{Name} & 
  \colhead{$\tilde{A}_r$} & \colhead{$\tilde{A}_{r, 1\rm{kpc}}$} & \colhead{$\tilde{A}_r/V_{\rm{max}}$} & \colhead{$\langle \epsilon_{\mathrm{pot}}\,\sin(2\varphi_2) \rangle$} & \colhead{$\tilde{M}_{\rm{resid}}$} & \colhead{$r_{\mathrm{max}}$} \\
& \colhead{km\,s$^{-1}$} & \colhead{km\,s$^{-1}$}& percent& &
  \colhead{km\,s$^{-1}$} & \colhead{\arcsec} \\
\colhead{(1)} & \colhead{(2)} & \colhead{(3)} & \colhead{(4)} & \colhead{(5)} & \colhead{(6)} & \colhead{(7)} \\
}
\startdata
NGC 925  &  $6.30_{-1.74}^{+1.63}$ & $9.45_{-2.98}^{+0.64}$  & 5.5 &  0.000 $\pm$ 0.046 & 3.0 & 282 \\[1.2ex]
NGC 2366 &  $2.94_{-1.24}^{+1.46}$ & $1.17_{-0.40}^{+0.12}$  & 5.3 &  0.004 $\pm$ 0.066 & 2.4 & 252 \\[1.2ex]
NGC 2403 &  $4.03_{-1.33}^{+1.39}$ & $2.60_{-0.48}^{+0.59}$  & 3.0 & $-$0.022 $\pm$ 0.025 & 2.9 & 950 \\[1.2ex]
NGC 2841 &  $6.71_{-3.54}^{+4.67}$ & \nodata                  & 2.6 & $-$0.001 $\pm$ 0.014 & 3.5 & 635 \\[1.2ex]
NGC 2903 &  $6.10_{-2.68}^{+3.64}$ & $13.55_{-6.18}^{+2.70}$ & 3.2 &  0.006 $\pm$ 0.028 & 2.8 & 602  \\[1.2ex]
NGC 2976 &  $2.81_{-1.18}^{+0.75}$ & $2.18_{-0.55}^{+0.77}$  & 3.5 & $-$0.010 $\pm$ 0.018 & 2.1 & 147 \\[1.2ex]
NGC 3031 &  $9.14_{-3.05}^{+2.08}$ & \nodata                  & 4.6 &  0.007 $\pm$ 0.045 & 3.0 & 840 \\[1.2ex]
NGC 3198 &  $4.49_{-0.91}^{+1.00}$ & $1.50$\footnote{This value has no upper or lower quartile as the inner 1 kpc of NGC 3198 contain only one data point.}  & 3.0 &  0.016 $\pm$ 0.020 & 2.6 & 565 \\[1.2ex]
IC 2574  &  $3.75_{-1.04}^{+1.78}$ & $1.36_{-0.06}^{+0.49}$  & 5.4 &  0.012 $\pm$ 0.047 & 2.7 & 505 \\[1.2ex]
NGC 3521 &  $8.80_{-4.62}^{+3.45}$ & $3.12_{-1.77}^{+12.67}$ & 4.2 &  0.017 $\pm$ 0.019 & 4.5 & 415 \\[1.2ex]
NGC 3621 &  $3.36_{-1.09}^{+1.00}$ & $5.52_{-3.21}^{+0.94}$  & 2.4 &  0.002 $\pm$ 0.022 & 2.3 & 600 \\[1.2ex]
NGC 3627 & $28.49_{-5.87}^{+10.91}$ & \nodata                  & 14.7& $-$0.024 $\pm$ 0.071 & 3.6 & 165 \\[1.2ex]
NGC 4736 & $10.01_{-2.14}^{+3.63}$ & $8.79_{-1.56}^{+1.87}$  & 8.3 & $-$0.055 $\pm$ 0.149 & 2.5 & 400 \\[1.2ex]
DDO 154  &  $1.61_{-0.65}^{+0.42}$ & $1.43_{-0.53}^{+0.14}$  & 3.4 &  0.024 $\pm$ 0.033 & 1.2 & 325 \\[1.2ex]
NGC 5055 &  $4.11_{-0.73}^{+1.61}$ & $8.38_{-2.60}^{+11.14}$ & 2.2 & $-$0.003 $\pm$ 0.025 & 3.1 & 450\\[1.2ex]
NGC 6946 &  $7.28_{-2.88}^{+3.12}$ & \nodata                  & 3.6 &  0.004 $\pm$ 0.069 & 3.4 & 420 \\[1.2ex]
NGC 7331 &  $5.94_{-1.21}^{+1.65}$ & \nodata                  & 2.6 & $-$0.003 $\pm$ 0.017 & 4.2 & 297 \\[1.2ex]
NGC 7793 &  $5.08_{-1.67}^{+0.90}$ & $3.41_{-0.48}^{+0.64}$  & 3.9 &
$-$0.067 $\pm$ 0.085 & 2.2 & 372 \\[1.2ex]
\hline
{\bf Sample mean} & {\bf 6.72} & {\bf 4.80} & {\bf 4.5} &{\bf 0.011} & {\bf 2.9}\\
{\bf Sample rms} & {\bf 5.91} & {\bf 3.99} & {\bf 2.9} & {\bf 0.013} & {\bf 0.8}
\enddata
\tablecomments{(1): the name of the galaxy; (2): \A, the median of the
  quadratically added amplitude of the non-circular motions, averaged over the
  entire radial range. Uncertainties indicate the lower and upper quartile; (3): same 
  as (2) but averaged over the inner 1 kpc only; (4): the percentage
  the non-circular motions contribute to the total rotation velocity; (5): the weighted
  mean elongation of the potential; (6): the median of the absolute residual
  velocity field after the harmonic decomposition; (7): maximum radius for the averaging of \A\ and $\epsilon_{\mathrm{pot}}$. The bottom two rows
  contain the mean values and their rms over the entire sample. For
  $\epsilon_{\mathrm{pot}}$, this value represents the weighted mean.}
\end{deluxetable*}

\section{Results}
In the previous section and in the detailed notes on the individual galaxies
in the Appendix, we show how the center positions are derived and
present the results of the harmonic decompositions.
The different center estimates are summarized in
Table~\ref{table:center-pos}, where our adopted positions are shown in bold
face. The results from the harmonic decomposition are summarized in
Table~\ref{table:harm-decomp}.

In this section, we will put 
these results in context and address some of the astrophysical questions
discussed in the Introduction. Section \ref{sec:center-results} will deal with the
results of the center estimates, whereas Section \ref{sec:harm-decomp} deals with
the results of the harmonic decompositions. In Section \ref{sec:theory}, we compare
the latter with predictions from CDM simulations. 
In Section \ref{sec:sanity-checks}, we
present and discuss several consistency checks which we applied to test our methods.

\subsection{Quality of the center estimates}\label{sec:center-results}

In this section, we discuss the quality and reliability of our center
positions.  An intrinsically cuspy density profile can be mistaken for
a flat, constant-density one, if possible offsets between the
kinematic and the photometric center of a galaxy are ignored. Using
the photometric center to derive a rotation curve or a mass model in
the presence of such an offset will result in a less steep rotation
curve and density profile. A potential cusp could then appear as a
core-like density profile.

As we have determined the kinematic and photometric centers of the
galaxies in our sample, we can directly test if such offsets exist.
This is shown in Fig.~\ref{fig:kpc-offset}, where we show the
offsets between the dynamical and photometric centers for the galaxies
in our sample.  The offsets are generally less than $\sim 0.1$ kpc,
with only two galaxies reaching offsets of $\sim 0.3$ kpc.

These results are similar to those derived by \citet{matthews-2002}
for a sample of 21 extreme late-type spiral galaxies. They
compare the systemic velocites as derived from the rotation curves of
these galaxies with the central velocities of the corresponding global
HI profiles. For over two-thirds of their sample they find offsets less
than 10 km s$^{-1}$ and from this they conclude that their sample
galaxies, despite their late-type, diffuse stellar disks and shallow
potentials, do have well-defined centers. A similar comparison of the
systemic velocities and central HI profile velocities (measured at 20
percent of the peak level) of our sample galaxies (using the values as listed in
\citealt{walter-07} and \citealt{deblok-07}) shows a similarly good
agreement with an absolute velocity difference $\Delta V = 1.7 \pm
1.7$ km s$^{-1}$, the largest offset being 6.2 km s$^{-1}$.

\begin{figure}[]
\begin{center}
\includegraphics[angle=0,width=0.45\textwidth,bb=23 346 330 523,clip=]{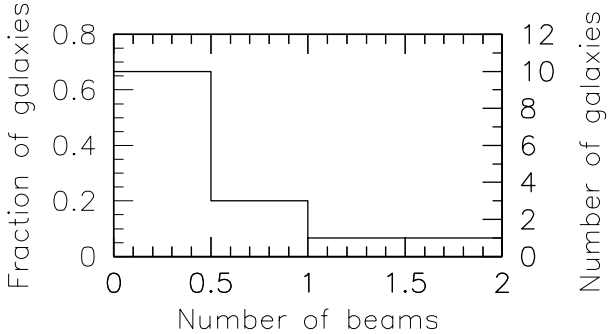}
\caption{The histogram shows the offset between the dynamical and the best
  (photometric) center for those 15 galaxies in our sample which have a well-constrained photometric center position. The offset is shown
 in terms of beam size. For 13 of the 15 galaxies, the different center estimates agree 
to within one beam (typical beam size: $\sim 10\arcsec$), showing that
  there is no indication for a genuine and general offset 
 between kinematic and photometric centers in our sample. } \label{fig:histo-offset}
\end{center}
\end{figure}

\begin{figure}[]
\begin{center}
\includegraphics[angle=0,width=0.45\textwidth,bb=291 436 572 700,clip=]{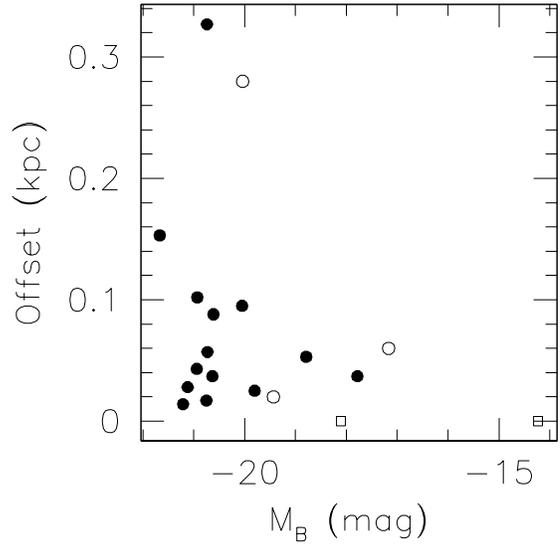}
\caption{Offsets between kinematic and photometric centers of the galaxies in our sample. Filled circles indicate galaxies for which we adopted the position of the photometric center as our best center estimate. Open circles represent galaxies for which we adopted the position of the kinematic center. Open squares indicate the two galaxies for which no reliable photometric center position could be derived. } \label{fig:kpc-offset}
\end{center}
\end{figure}

We also show the distribution of offsets between the kinematic and the
best (photometric) center estimates for the 15 galaxies in our sample
with well-constrained photometric centers.  For 13 out of the 15
galaxies shown in Fig.~\ref{fig:histo-offset}, the dynamical center
differs by less than the size of one beam from the best (photometric)
center --- and for 10 galaxies the agreement is even better than half
the size of the beam. Only two galaxies (NGC 3627 and NGC 6946) show
moderate offsets of between one and two beam sizes. Given that the
former galaxy is extremely asymmetric (due to its interactions in the
Leo group) and the latter has a low inclination which makes fitting
tilted-rings more difficult, these outliers can be understood.

\begin{figure}[]
\begin{center}
\includegraphics[angle=0,width=0.45\textwidth,bb=23 346 346 536,clip=]{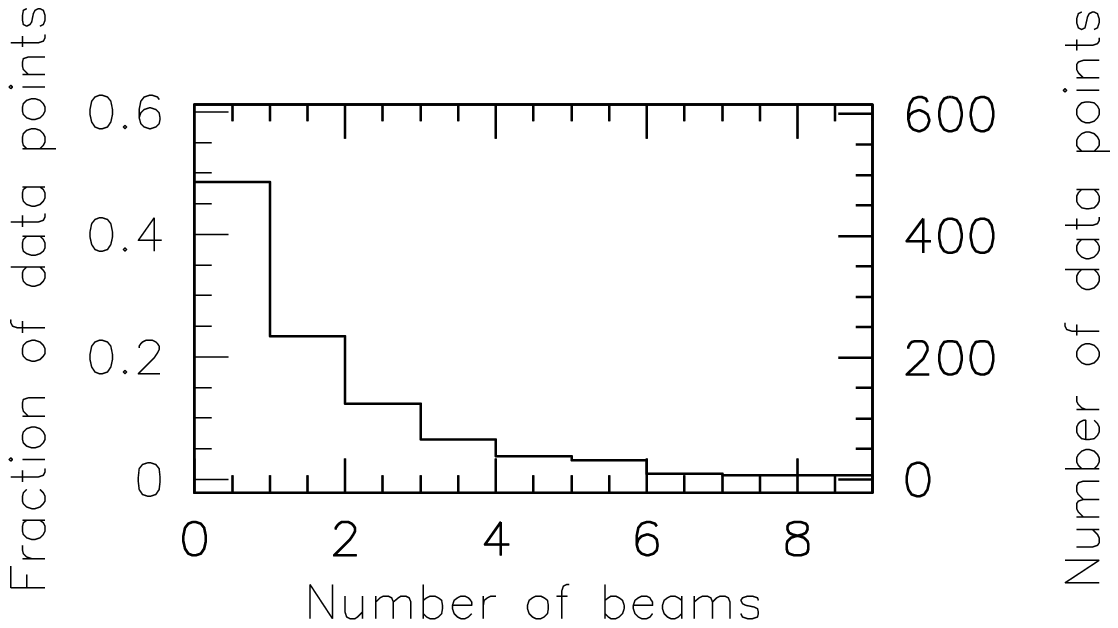}
\caption{The histogram shows (for all galaxies in our sample) the offset between the 
individual tilted-ring center estimates and the corresponding
  best center position in terms of beam size. The typical beam size
  is $\sim 10\arcsec$ (this corresponds to a physical size
  of 170 pc $-$ 750 pc). Only 
  tilted-rings inwards of $r=0.5\, r_{\rm{max}}$ were included. The histogram
  demonstrates that our dynamical center estimates are well-constrained and in good
agreemend with the photometric center positions.} \label{fig:histo-beam}
\end{center}
\end{figure}

The strength of the agreement we find does depend on how
well-constrained the dynamical center estimates are.  A tightly
constrained kinematic center position which agrees with a
well-determined photometric center allows one to draw strong
conclusions.  A weakly constrained kinematic center less so, as the
increased uncertainties allow agreement with a whole range of
photometric centers, as long as they are located somewhere near the
kinematic center. To test the strength of our conclusions we therefore
determine for each galaxy the offsets between the centers of
individual tilted-rings and the best center estimate.  Because of the
sparsely filled rings in the very outer parts, as well as the presence
of warps and asymmetries, we restrict this analysis to the data points
in the inner half of each galaxy (i.e., those with $r<0.5\,
r_{\rm{max}}$), with $r_{\rm{max}}$ as listed in
Table~\ref{table:harm-decomp}.  In Fig.~\ref{fig:histo-beam}, we show
a histogram of the distribution of these offsets in terms of beam size
for all galaxies in our sample.  Approximately 50 percent of the
center positions of the $\sim$1000 individual tilted-rings differ less
than one beam from our best center position. Another 25 percent show
an offset between one and two beam sizes and only a small fraction
shows large offsets.  Note that the center positions of some of the
tilted-rings can be affected by the presence of spiral arms and other
features. Large offsets for individual rings do therefore not
necessarily imply intrinsic offsets between center positions, but need
to be regarded within the context of the results for the whole
galaxy. The radial variation of the center positions are shown in
Fig.~\ref{fig:n3198-1} for NGC 3198 and in the top panels of
Figs.~\ref{fig:ngc-925}-\ref{fig:ngc-7793} for the other galaxies in
our sample.

Our results show that for the large majority of the galaxies studied
here, the photometric and kinematic center positions agree within
their uncertainties and also to within a beam. See also
Sec.~\ref{sec:center-offsets} where we evaluate the impact of a shift
in center position on our results.

\subsection{Results of the harmonic decomposition}\label{sec:harm-decomp}

Here, we discuss the results of the harmonic decompositions in a more
general way. For a description of the results for individual galaxies
we refer to Appendix~\ref{sec:indiv-gal}. Because of the low
filling-factor of its velocity field \citep[cf.][]{deblok-07}, we did
not perform a harmonic decomposition for NGC 4826.

\begin{figure}[]
\begin{center}
\includegraphics[angle=0,width=0.45\textwidth,bb=19 160 586 697,clip=]{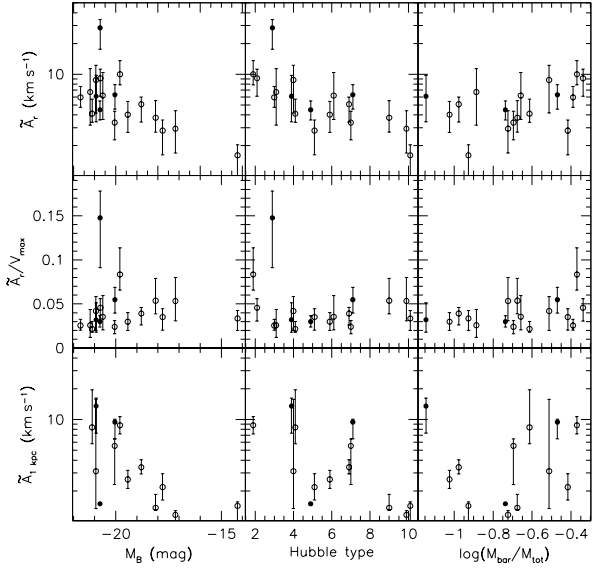}
\caption[Amplitude of the non-circular motions]{
The non-circular motions measured in different ways for our entire sample.
Barred galaxies are shown as filled circles, non--barred
  as open circles. {\bf Top panel:} Median amplitude of the
  non-circular motions (\A) on a logarithmic scale {\it vs}. absolute magnitude (left), Hubble type
  (center) and baryonic to total mass ratio (right). Error bars indicate the upper and lower quartile of the radial distribution of \A. {\bf
    Middle panel:} The percentage the non-circular motions 
  contribute to the total rotation velocity. {\bf Bottom
  panel:} Like top panel, but the averaging was restricted to the inner 1 kpc of
the galaxies. Note that for five galaxies, the data do not show significant amounts of \HI\ in the inner kpc, and they are therefore not shown in the two bottom panels.} \label{fig:amplitude} 
\end{center}
\end{figure}

In Fig.~\ref{fig:amplitude} we show plot the harmonic decomposition
results as a function of absolute magnitude, Hubble type and dark
matter dominance.  In all panels, we distinguish between barred and
non--barred galaxies. The Hubble types are listed in
Table~\ref{table:sample-properties} and the bar classifications are
based on NED. The dark matter dominance is expressed as the ratio of
total baryonic mass to total dynamical mass $M_{\rm bar}/M_{\rm
  tot}$. The baryonic mass is calculated using the mass models
presented in \citet{deblok-07}, where we have used the Kroupa IMF
results (their Table 4). The total dynamical mass is calculated using
the radius of the outermost point of the rotation curve and the
velocity is taken from Table~\ref{table:sample-properties}.  We note
that for NGC 3627 we cannot show $M_{\rm bar}/M_{\rm tot}$ as the
non-circular motions in this galaxies were too large for
\citet{deblok-07} to derive a mass model.

In the upper panels of Fig.~\ref{fig:amplitude}, we show the results
for $\tilde{A}_r$.  It is clear that the magnitude of the non-circular
motions decreases towards lower luminosities and later Hubble types
(top panel of Fig.~\ref{fig:amplitude}). A more tentative trend is
visible for $M_{\rm bar}/M_{\rm tot}$ where the dark-matter dominated
galaxies have on average the lowest non-circular motions.  When these
motions are expressed as a fraction of \vmax, we see that their
contribution is roughly independent of luminosity, Hubble type or dark
matter dominance (Fig.~\ref{fig:amplitude}, middle panel).  For 16 out
of the 18 galaxies, the non-circular contribution is smaller than 6
percent of \vmax\ and it has a mean value (averaged over the entire
sample) of 4.5 percent.  Both in the top and in the middle panel of
Fig.~\ref{fig:amplitude}, barred galaxies do not stand out, except for
NGC~3627. This is different if we look only at the non-circular
motions in the inner 1~kpc (Fig.~\ref{fig:amplitude}, bottom
panel). As expected, most barred galaxies have rather high
non-circular motions in the innermost region. Note that some galaxies
are not shown in these diagrams, as we have no data above a detection
limit of 3 $\sigma$ in their inner 1~kpc.

\begin{figure}[t]
\begin{center}
\includegraphics[angle=0,width=0.45\textwidth,clip=]{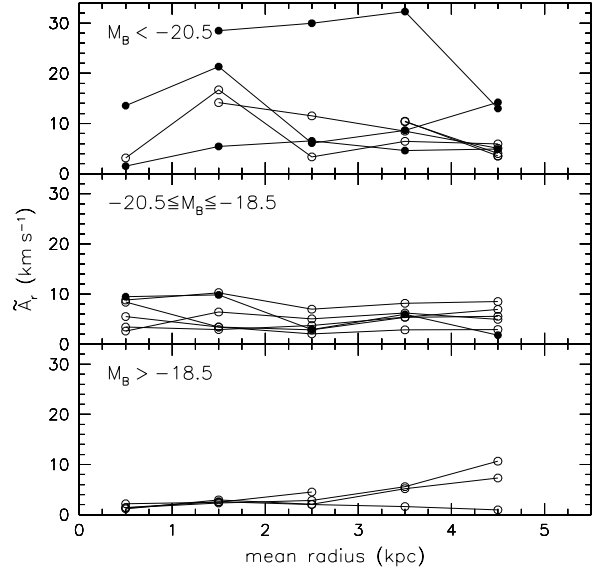}
\caption[Absolute amplitudes of the non-circular motions within $0<r<1$, $1<r<2$, ..., $4<r<5$ kpc]{The amplitude of the non-circular motions within rings of 1 kpc width (i.e., $0<r<1$ kpc, $1<r<2$ kpc, ..., $4<r<5$ kpc) for each galaxy (if
  data available). All measurements of a 
  specific galaxy are connected by a line. Filled symbols correspond to barred
  galaxies, open symbols to non--barred galaxies. We divide the plot into three
  subpanels. The top panel contains all galaxies 
  with $M_B<-20.5$, the middle panel those with $-20.5\le M_B \le -18.5$ and
  the bottom panel all galaxies with $M_B>-18.5$.\\
} \label{fig:12345kpc_V2}
\end{center}
\end{figure}

\begin{figure}[t]
\begin{center}
\includegraphics[angle=0,width=0.45\textwidth,clip=]{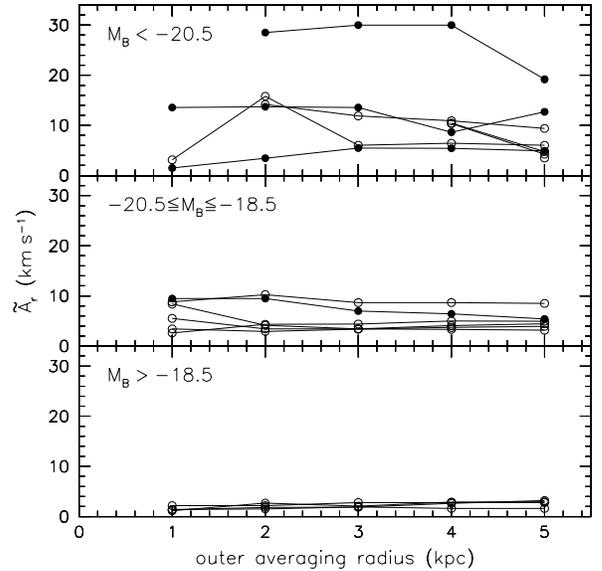}
\caption[Absolute amplitudes of the non-circular motions within $0<r<1$, $0<r<2$, ..., $0<r<5$ kpc]{Like Fig.~\ref{fig:12345kpc_V2}, but the amplitude of the non-circular motions 
were averaged within rings of increasing radius (i.e., $0<r<1$ kpc, $0<r<2$ kpc, ..., $0<r<5$ kpc).
As the region over which the non-circular motions are averaged increases outwards, the amplitudes shown are less affected by local features such as the effects of star formation. } \label{fig:12345kpc}
\end{center}
\end{figure}

As a comparison, we also calculated the amplitude of the non-circular
motions within the inner 2, 3, 4, and 5 kpc, using two different
methods.  The approach shown in Fig.~\ref{fig:12345kpc_V2} measures
the non-circular motions by taking the median of the \Ar\ values
within rings of 1 kpc width, as measured over radii $0<r<1$ kpc,
$1<r<2$ kpc, ..., $4<r<5$ kpc. Our second approach, shown in
Fig.~\ref{fig:12345kpc}, simply increases the radial range over which
the amplitudes are averaged ($0<r<1$ kpc, $0<r<2$ kpc, ..., $0<r<5$
kpc). In both cases, we have divided our sample into three absolute
magnitude bins.  It is apparent from the two figures that galaxies
with low luminosity have the lowest amplitudes of non-circular motions
in the inner parts, regardless of the chosen method.  Since the
approach shown in Fig~\ref{fig:12345kpc_V2} measures the non-circular
motions more locally than the method shown in Fig~\ref{fig:12345kpc},
it is also more affected by local features like effects of star
formation. For instance, the two relatively high amplitudes in the
$4<r<5$ kpc bin of the bottom panel of Fig.~\ref{fig:12345kpc_V2}
belong to IC 2574 and NGC 2366. The former has a supergiant shell at
those radii \citep[see][]{walter-1998}, whereas the latter has a large
star forming region in its outer parts (see \citealt{se-heon} for an
in-depth analysis of the small-scale non-circular motions in these two
galaxies).  The only other low-luminosity galaxy in our sample which
extends out to 5 kpc radius (DDO 154) is completely quiescent in
contrast. In the analysis shown in Fig.~\ref{fig:12345kpc}, these
effects have ``averaged out'', due to the larger area used for the
averaging.

\begin{figure}[t]
\begin{center}
\includegraphics[angle=0,width=0.45\textwidth,clip=]{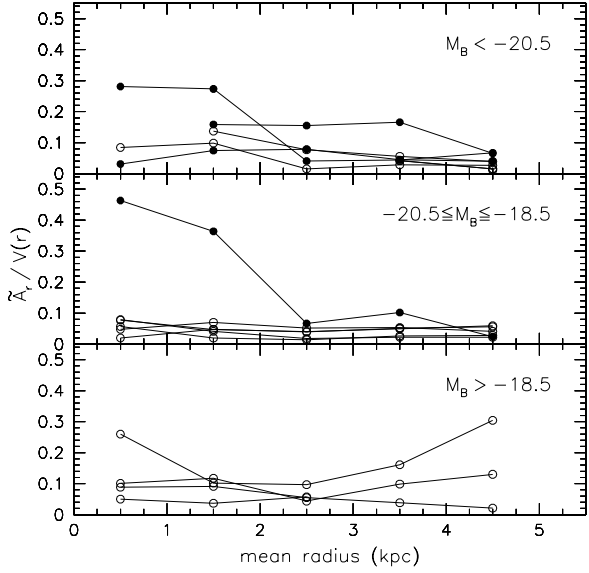}
\caption[Amplitudes of the non-circular motions relative to the local rotation velocity within $0<r<1$, $1<r<2$, ..., $4<r<5$ kpc]{Like Fig.~\ref{fig:12345kpc_V2}, but the amplitudes of the non-circular motions are normalized by the local rotation velocity. Except for a few galaxies, the non-circular motions in the inner few kpc have an amplitude of  $\sim 10$ percent of the local rotation velocity, irrespective of the galaxies' luminosity. Only three galaxies have non-circular motions in their centers which are larger than 20 percent of the local rotation velocity. 
Note also that large non-circular motions (in relative terms) are predominantly found in barred galaxies.} \label{fig:12345kpc_V3}
\end{center}
\end{figure}

\begin{figure}[t]
\begin{center}
\includegraphics[angle=0,width=0.45\textwidth,clip=]{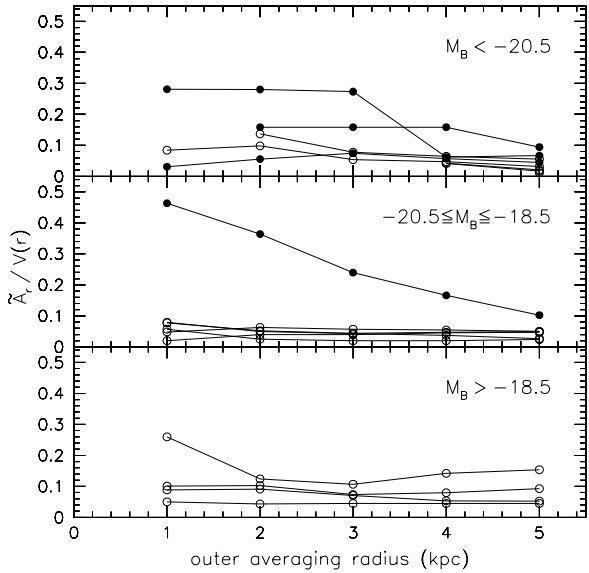}
\caption[Amplitudes of the non-circular motions relative to the local rotation velocity within $0<r<1$, $0<r<2$, ..., $0<r<5$ kpc]{Like Fig.~\ref{fig:12345kpc}, but the amplitudes of the non-circular motions are normalized by the local rotation velocity. Analogous to Fig.~\ref{fig:12345kpc}, the increasing region for the averaging of the non-circular motions causes the amplitudes to be somewhat less affected by local features such as the effects of star formation.
} \label{fig:12345kpc_V4}
\end{center}
\end{figure}

However, no matter how they are binned, in absolute terms the measured
amplitudes of the non-circular motions are small.  In order to account
for the different rotation velocities, we have normalized the
amplitudes shown in Figs.~\ref{fig:12345kpc_V2} and \ref{fig:12345kpc}
by the local rotation velocity. The results are shown in
Figs.~\ref{fig:12345kpc_V3} and \ref{fig:12345kpc_V4}, respectively.
For the large majority of the galaxies in our sample, the non-circular
motions in the inner few kpc contribute approximately ten percent to
the local rotation velocity. Three galaxies (NGC 925, NGC 2903, and IC
2574) contain non-circular motions larger than 20 percent of the local
rotation velocity.  For NGC 925 and NGC 2903, it is likely that the
large non-circular motions are associated with their stellar bars,
especially as Figs.~\ref{fig:12345kpc_V3} and \ref{fig:12345kpc_V4}
indicate that large non-circular motions in the central parts are
predominantly found in barred galaxies.  For IC 2574, the large ratio
between non-circular and circular motions is mainly caused by the slow
rise of its rotation curve in the inner kpc. Over the same radial
range the \emph{absolute} non-circular motions are generally less than
2.3\kms.

\begin{figure}[t]
\begin{center}
\includegraphics[angle=0,width=0.45\textwidth,bb=19 160 586 697,clip=]{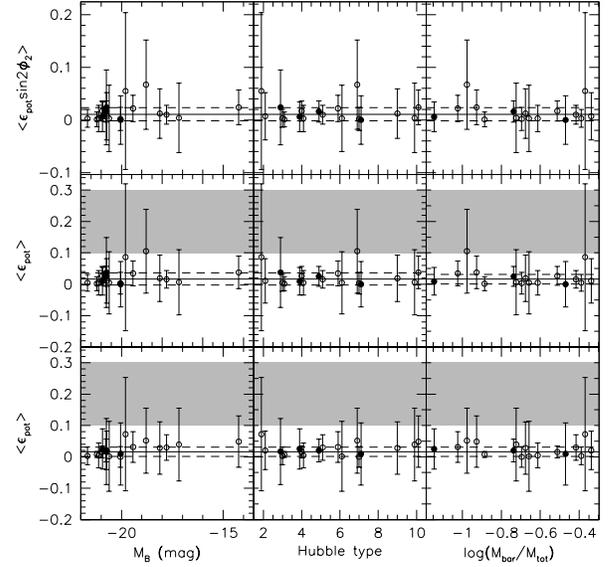}
\caption[Elongation of the potential]{{\bf Top panel:} Elongation of
  the potential {\it vs}. absolute magnitude (left), Hubble type
  (center) and baryonic to total mass ratio (right). The horizontal lines represent the weighted mean and its
  standard deviation ($ \langle
  \epsilon_{\mathrm{pot}}\,\sin(2\varphi_2) \rangle =0.011\pm
  0.013$). Barred galaxies are shown as filled circles, non--barred
  ones as open circles. The distribution of $ \langle
  \epsilon_{\mathrm{pot}}\,\sin(2\varphi_2) \rangle$ shows neither a
  trend with luminosity, nor with Hubble type.  {\bf Middle panel:}
  Like top panel, but a statistical correction for the unknown viewing
  angle was applied to the individual elongation measurements and to
  their mean and standard deviation. The hatched area indicates the
  CDM predictions by \cite{hayashi-2006}.  The elongations of the
  large majority of the galaxies shown here are systematically lower
  (rounder) than what is predicted by CDM simulations, although the
  measurements for some galaxies have large enough error bars to make
  them marginally consistent with the lower end of the CDM
  predictions.  {\bf Bottom panel:} Like middle panel, but we averaged
  the elongation of the potential not over the entire radial range,
  but only out to $r<0.5\, r_{\rm{max}}$ ($r_{\rm{max}}$ is the
  maximum radius to which our analysis extends).  This has generally
  only little effect on the individual elongation measurements.  The
  mean value has not changed ($ \langle \epsilon_{\mathrm{pot}}
  \rangle =0.017\pm 0.020$ {\it vs.}  $ \langle
  \epsilon_{\mathrm{pot}} \rangle =0.016\pm 0.015$) and we therefore
  do not see an indication that the elongation of the potentials
  increase inwards.  } \label{fig:elongation}
\end{center}
\end{figure}

\begin{figure}[t]
\begin{center}
\includegraphics[angle=0,width=0.45\textwidth,bb=19 475 570 700 clip=]{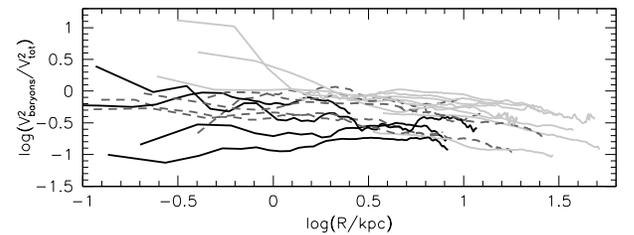}
\caption[Local baryon to total mass ratios]{ The radial distribution
  of $M_{\rm bar}/M_{\rm tot}$ in the THINGS galaxies, derived from
  the Kroupa IMF mass models presented in
  \citet{deblok-07}. Light-gray full curves represent galaxies
  brighter than $M_B=20.5$. Dashed, dark-gray curves show galaxies
  with $-20.5 \le M_B < -18.5$. Black curves show galaxies fainter
  than $M_B = -18.5$.  } \label{fig:localdm}
\end{center}
\end{figure}

We will now discuss the elongation of the potential (see
Fig.~\ref{fig:elongation}).  The weighted average elongation and its
standard deviation are $ \langle
\epsilon_{\mathrm{pot}}\,\sin(2\varphi_2) \rangle =0.011\pm 0.013$.
This elongation and the individual elongation measurements still
contain an unknown viewing angle $\varphi_2$, for which one can apply
a statistical correction by dividing the elongation by the expectation
value of $\varphi_2$, $2/\pi$ (middle panel of
Fig.~\ref{fig:elongation}). The average, corrected elongation is $
\langle \epsilon_{\mathrm{pot}} \rangle =0.017\pm 0.020$, i.e.,
consistent with a round potential.

The elongations show neither a trend with absolute magnitude, nor with
Hubble type, nor with dark matter dominance.  If the influence of the
disk decreases with baryonic mass, and if the dark matter halos are
indeed tri-axial \citep[or more precisely, have an elliptical
potential distortion in the plane of their disk, cf.][]{hayashi-2006},
one would expect the measured elongations to increase with decreasing
(baryonic) mass. However, no such trend is seen in our data.

One explanation could be that the baryons dominate the total potential
sufficiently so that the elongation values we measure are really the
elongations of the potentials of the baryonic disks. As the
self-interaction (dissipation) of the (gas component) baryons tends to
circularize the orbits, any (kinematical) traces of the tri-axial dark
matter potential would thus be wiped out. With a variation in total
$M_{\rm bar}/M_{\rm tot}$ of close to an order of magnitude, this
means the circularization process ought to be extremely effective over
a large range of mass ratios.  The relevant quantity here, however, is
not the \emph{total} dark matter dominance but the \emph{local}
one. We therefore show in Fig.~\ref{fig:localdm} a measure for the
local ratio of $M_{\rm bar}/M_{\rm tot}$ as a function of radius,
based on the Kroupa IMF mass models presented in \citet{deblok-07}. We
see that whilst for the high luminosity galaxies the local importance
of dark matter and baryons in the inner parts is roughly equal, it
drops to values of $M_{\rm bar}/M_{\rm tot} \sim 1/3$ in the outer
parts. For the low luminosity galaxies the importance of the baryons
is, however, close to an order of magnitude lower. Again, this implies
that the circularization process should be equally effective over a
large range in local mass ratios, and already manage to wipe out the
tri-axial signal when the baryons contribute only $\sim 10$ percent to
the local mass. Note that in the above we have assumed that both stars
and gas contribute to the circularization process. It is however
likely that only the gas component is relevant, as the stars will
not necessary dissipate as strongly. If this is the case, the relevant
quantity is the $M_{\rm gas}/M_{\rm tot}$ ratio, which can be another
order of magnitude lower than the $M_{\rm bar}/M_{\rm tot}$ ratios
discussed here.  In summary, attributing the circular potentials to
dissipation requires that the circularization process is already
effective when the baryons contribute only $\sim 10$ percent to the
total mass.  Alternatively, the dissipation may not be relevant at
such low mass ratios, in which case the potentials are truly
round. The well-known dark matter dominance of LSB galaxies may 
imply that similar conclusions also hold for them; direct and equally
detailed measurements should provide the additional information needed
to draw definitive conclusions.

\subsection{Comparison with predictions from simulations}\label{sec:theory}
One of the clearest outcomes of simulations is that CDM halos are
triaxial \citep[e.g.,][]{frenk-1988, dubinski-1994, hayashi-2004,
  moore-2004, kasun-2005,hayashi-2006, hayashi-2007}.  We compare our
results with predictions based on cosmological $\Lambda$CDM
simulations.  \cite{hayashi-2004, hayashi-2006} present models where
they use an elliptical distortion of the gravitational potential
(mimicking a tri-axial halo) to explain the observed solid-body
rotation curves of dwarfs and LSB galaxies within a CDM context. This
distortion affects the orbits and thus also the rotation curve. In
this picture, the major axis rotation curve depends on the viewing
angle of the elliptical distortion. This viewing angle can be chosen
in such a way that the major axis rotation curve in the elliptical
potential looks identical to what it would have been, had the gas been
on circular orbits (i.e., without elliptical distortion and without
non-circular motions). It would therefore hide the presence of the
disturbance due to tri-axiality and look like a solid-body rotation
curve.

However, even if the angle is adjusted in such a way, the minor axis
rotation curve of the tri-axial case would still show large streaming
motions. These can be up to 15 percent of the total rotation velocity.
As we show in Fig.~\ref{fig:amplitude}, the observed non-circular
motions in our sample are generally much smaller and never reach 15
percent of the total rotation \citep[see also the minor-axis
position-velocity diagrams presented in][]{deblok-07}.  In
Figs.~\ref{fig:12345kpc_V3} and \ref{fig:12345kpc_V4}, we normalize
the amplitudes of the non-circular motions by the \emph{local}
(circular) rotation velocity.  At a radius of 1 kpc,
\cite{hayashi-2004} report non-circular motions as high as 50 percent
of the local rotation velocity. At smaller radii, the contribution of
the non-circular motions is even larger. The results shown in
Figs. \ref{fig:12345kpc_V3} and \ref{fig:12345kpc_V4}, however, show
that for the large majority of the galaxies in our sample, the
non-circular motions are only as high as ten percent of the local
rotation velocity. Only in one barred galaxy (NGC 925) do the
non-circular motions reach an amplitude close to 50 percent of the
local rotation velocity in the inner 1 kpc.  The dwarf galaxies
(bottom panels of Figs.~\ref{fig:12345kpc_V3} and
\ref{fig:12345kpc_V4}) are of most interest here, as these galaxies
are the ones most dominated by dark matter. The non-circular motions
in three out of the four dwarfs in our sample (NGC 2366, NGC 2976, and
DDO 154) contribute only ten percent to the local rotation
velocity. Therefore, the non-circular motions in these galaxies do not
significantly affect the mass models presented in \cite{deblok-07} and
\cite{se-heon} and preclude the possibility that their cored density
profiles (cf. aforementioned papers) are intrinsic cuspy profiles
which have been artificially flattened by large non-circular motions.
The non-circular motions in the inner kpc of IC 2574 amount to $\sim
26$ percent of the local rotation velocity, a factor two smaller than
predicted in \cite{hayashi-2004}. We refer to \cite{se-heon} for a
full treatment of the IC 2574 non-circular motions.

The gravitational potential in the simulations of \citet{hayashi-2006}
is elongated in the inner parts as $ \langle \epsilon_{\mathrm{pot}}
\rangle =0.2 \pm 0.1$ (their Fig.~3).  All but two of the individual
elongation measurements presented here are systematically lower than
those predictions and consistent with a round potential, although some
have large enough error bars to make them marginally consistent with
the lower end of the CDM predictions.

If DM halos do indeed have an elliptical distortion in their inner
parts, we would expect to find higher elongations there.  To test this
idea, we have also determined the elongations of the potential by
using only data out to $r=0.5\, r_{\rm{max}}$ (bottom panel of
Fig.~\ref{fig:elongation}). The individual elongation measurements do
not change significantly if averaged only over the inner half of the
disk.  Their weighted mean ($ \langle \epsilon_{\mathrm{pot}} \rangle
=0.016\pm 0.015$) is not significantly different from the value
obtained by averaging over the entire disk ($ \langle
\epsilon_{\mathrm{pot}} \rangle =0.017\pm 0.020$) and we therefore see
no indication that the gravitational potential is more elongated
towards the center of
the galaxies.

One caveat relating to the above discussion is that the model
presented in \citet{hayashi-2006} consists of only a dark matter
component, while real galaxies contain baryons as well. Baryonic
processes, such as dissipation in the gas component might affect the
dynamics within the visible disk of observed galaxies, thus
invalidating any comparisons. As we showed in
Sect.~\ref{sec:harm-decomp}, however, in some of the galaxies in our
sample the local baryon to total mass ratio is only $\sim 10$
percent. If dissipation or circularization of the orbits affects the
results found for our galaxies, this implies that the 10 percent of
baryons are able to wipe out the dynamical signature of the 90 percent
of dark matter. If dissipation is really this efficient, it would also
imply that analyses of, e.g., the scatter in the Tully-Fisher relation
in order to put limits on the tri-axiality of haloes (e.g.,
\citealt{franx-92}) are only relevant for the elongations of the
baryonic disks, but not for those of dark matter haloes.  In summary,
unless the effects of baryonic dissipation are extremely efficient, it
is difficult to reconcile the tri-axialities found in simulations with
those inferred for the low-luminosity galaxies in our sample.  More
extensive observations and more detailed simulations ought to show
whether these results hold for the general disk galaxy population.

\subsection{Consistency Checks}\label{sec:sanity-checks}
In this section, we compare the results of our harmonic
decomposition with the traditional tilted-ring analysis
presented in \citet{deblok-07}. Additionally, we present quality controls
which show the limitations of our methods, test and validate our current
results as well as indicate room for improvement.

\subsubsection{\rotcur\ {\it vs}. \reswri: residual velocity fields}\label{sec:reswri-rotcur-residuals}
\begin{figure}[t]
\begin{center}
\includegraphics[angle=0,width=0.45\textwidth,clip=]{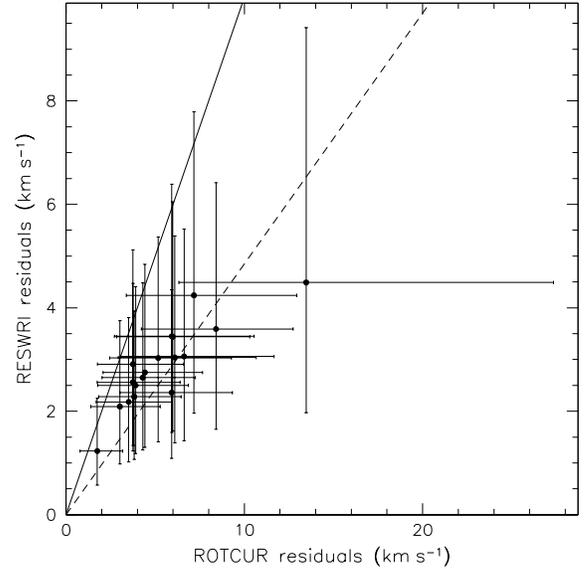}
\caption[\rotcur\ residuals {\it vs.} \reswri\ residuals]{Median of the absolute residual velocity field from the rotation
  curve analysis of \cite{deblok-07} {\it vs}. the one from the harmonic
  decomposition presented in this paper. The error bars indicate the lower and
  upper quartile, the solid line represents a one--to--one relation,
  and the dashed 
  line the unweighted least square fit through the data points. The residuals
  in the traditional \rotcur\ analysis are 
  all larger than those from the \reswri\ fits, as expected. The 
  difference between the solid line and dashed line is due to the
  non-circular motions quantified in this paper.} \label{fig:residuals-median}
\end{center}
\end{figure}

Here, we compare the residual velocity fields from the
\rotcur\ 
analysis by \cite{deblok-07} with those from our analysis with
\reswri. As mentioned in Section~\ref{sec:harm-decomp-introduction}, the residual
velocity fields are derived by subtracting a model using the
final parameter estimates from the original data. 
For both types of residual fields, we calculate the median and the lower and
upper quartile of the {\it absolute} residual velocity fields. 
These are compared in Fig.~\ref{fig:residuals-median}. It is clear that all
galaxies have larger residuals in the \rotcur\ 
analysis than they have in the \reswri\ analysis. This is of course to be
expected, given that \rotcur\
considers only circular motion, while \reswri\ also takes non-circular motions
into account. 
Nevertheless, it is a quality test which shows that our results are behaving in the
expected manner. The largest median
amplitude of the residual \reswri\ velocity fields is 4.5\,\kms\ (for
NGC~3521). The average
value of the sample is 2.9\,\kms, which clearly shows that a harmonic
expansion up to third order is capable of capturing the majority of the
non-circular motions in most galaxies.

\begin{figure}[t]
\begin{center}
\includegraphics[angle=0,width=0.45\textwidth,clip=]{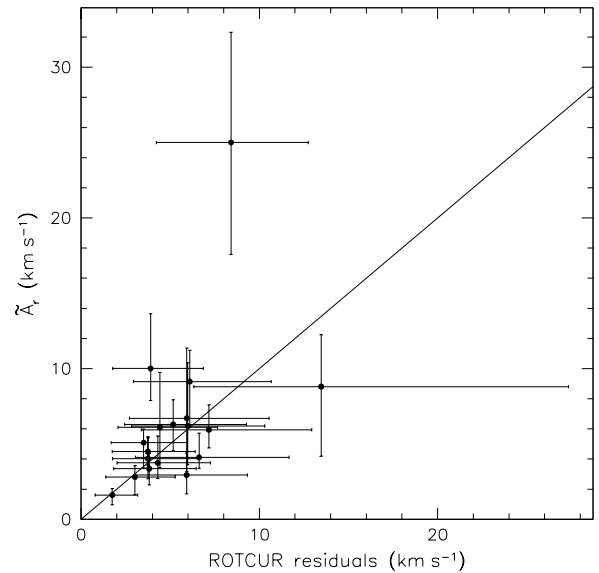}
\caption[\rotcur\ residuals {\it vs.} amplitude of the non-circular motions]{Median of the absolute residual velocity field from the rotation
  curve analysis in \cite{deblok-07} {\it vs}. \A, the median of the (radially
  averaged) quadratically added amplitudes of the
  non-circular motions quantified in this paper. The error bars indicate the
  lower and upper quartile. Galaxies with large residuals in the \rotcur\
  fits show generally also larger non-circular motions. The outlier in the
  upper part of the panel is NGC\ 3627, which shows large
  non-circular motions, but 
  only moderate residuals in the \rotcur\ analysis by \citet{deblok-07}.} \label{fig:residuals-A}
\end{center}
\end{figure}

Looking at Fig.~\ref{fig:residuals-median}, it becomes also clear that galaxies
with small \rotcur\ residuals also have small \reswri\ residuals. Small
\rotcur\ residuals, however, indicate that a model assuming circular rotation already provides a good approximation. Therefore, it is to be expected that these galaxies show only small non-circular motions.
This expectation is tested in Fig.~\ref{fig:residuals-A}, where we plot the 
median values of the
absolute residual velocity fields from \cite{deblok-07} {\it vs.} the median
amplitudes \A\ of the non-circular motions from our harmonic
decompositions. Most galaxies are located near the line of unity, showing that
our expectation is indeed correct.
The
only real outlier in Fig.~\ref{fig:residuals-A} is NGC\ 3627, which is clearly offset from all other
galaxies as it has --- despite its large non-circular
motions ($ \A \sim 28.5\,\kms$) --- only moderate \rotcur\
residuals. The results for the other galaxies, however, show that even a simple model considering only circular rotation can fit most galaxies quite well. These small residuals therefore demonstrate (independently from our \reswri\ analysis) that non-circular motions are generally small.

\subsubsection{Making prior assumptions during the \reswri\ runs}
The rotation curves of all galaxies in our sample were derived and discussed
in detail by \cite{deblok-07}. The general procedure for the derivation of a
rotation curve includes keeping some parameters fixed in the
tilted-ring fits, thus 
reducing the number of free parameters for each individual fit with
\rotcur. 
In the current analysis we derive all tilted-ring
parameters in a single fit with all parameters left free 
and do not attempt to correct for the motions induced by, e.g., star formation
or spiral arms. Thus, it is expected that the tilted-ring parameters derived
by us differ to some extent from the 
ones derived by \cite{deblok-07}. In order to check this and to estimate the
impact on our results, we compare the derived values for our unconstrained
harmonic decompositions with those from constrained decompositions with PA and
$i$ fixed to the final values from \citet{deblok-07}. We apply this test to
two representative galaxies, NGC~3198 and DDO~154, a spiral galaxy and a dwarf irregular.

\begin{figure*}[t]
\begin{center}
\includegraphics[angle=0,width=0.65\textwidth,bb=19 235 592 697,clip=]{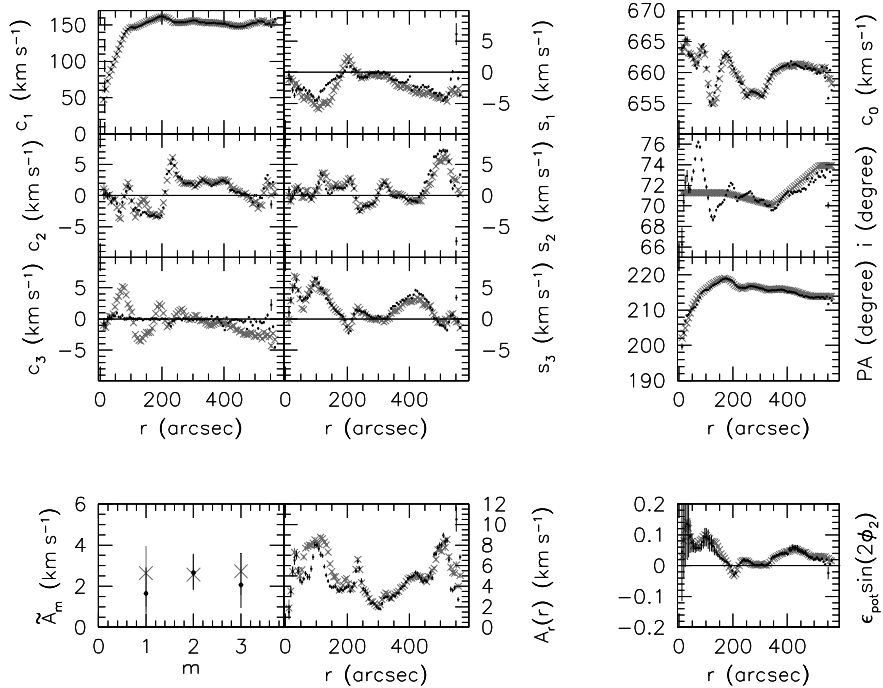}
\caption[Constrained {\it vs.} unconstrained harmonic decomposition of NGC 3198]{The results of the unconstrained (black dots) and constrained
  (gray crosses) 
  harmonic decomposition of NGC\ 3198. For the constrained fit, the
  inclination and position angle were fixed to the values from
  \citet[][]{deblok-07}. The layout of the figure is identical to that of
  Figs.~\ref{fig:n3198-3} and \ref{fig:n3198-4}, except that we do not show any weighted means in the panels on the right-hand side. The parameters that are plotted in this figure are defined in Sec.~\ref{sec:test-candidate}.
The constant inclination in the inner parts of 
  NGC~3198 
  cause a non-zero $c_3$ term which results in slightly higher non-circular
  motions in the inner parts.} \label{fig:fixed-free-ngc3198} 
\end{center}
\end{figure*}

In the case of NGC\ 3198 (Fig.~\ref{fig:fixed-free-ngc3198}), most values agree
remarkably well between the constrained and the unconstrained fit.
Because \cite{deblok-07} have assumed a constant inclination in the inner
parts (in order to compensate for effects caused by spiral arms), our unconstrained
values differ slightly in this region. Note though that the effect on the rotation curve is negligible. 
Fixing the inclination has the predictable effect that the $c_3$ term
reaches values of a few \kms\ as this term tries to compensate for the
effect the spiral arms have on the velocity field.
The impact of this, however, is
minor, as one can see in the distribution of \Ar\ and \Am\, as well as in the
negligible difference in the derived rotation velocity. 
The elongation of the potential is also largely unaffected by constraining $i$
and PA. The only noticeable differences are the
different-sized error bars. As the inclination contributes to the
fourth power to 
the error of the elongation of the potential, the error bars are naturally
smaller in the case of a fixed inclination. 

\begin{figure*}[t]
\begin{center}
\includegraphics[angle=0,width=0.65\textwidth,bb=19 235 592 697,clip=]{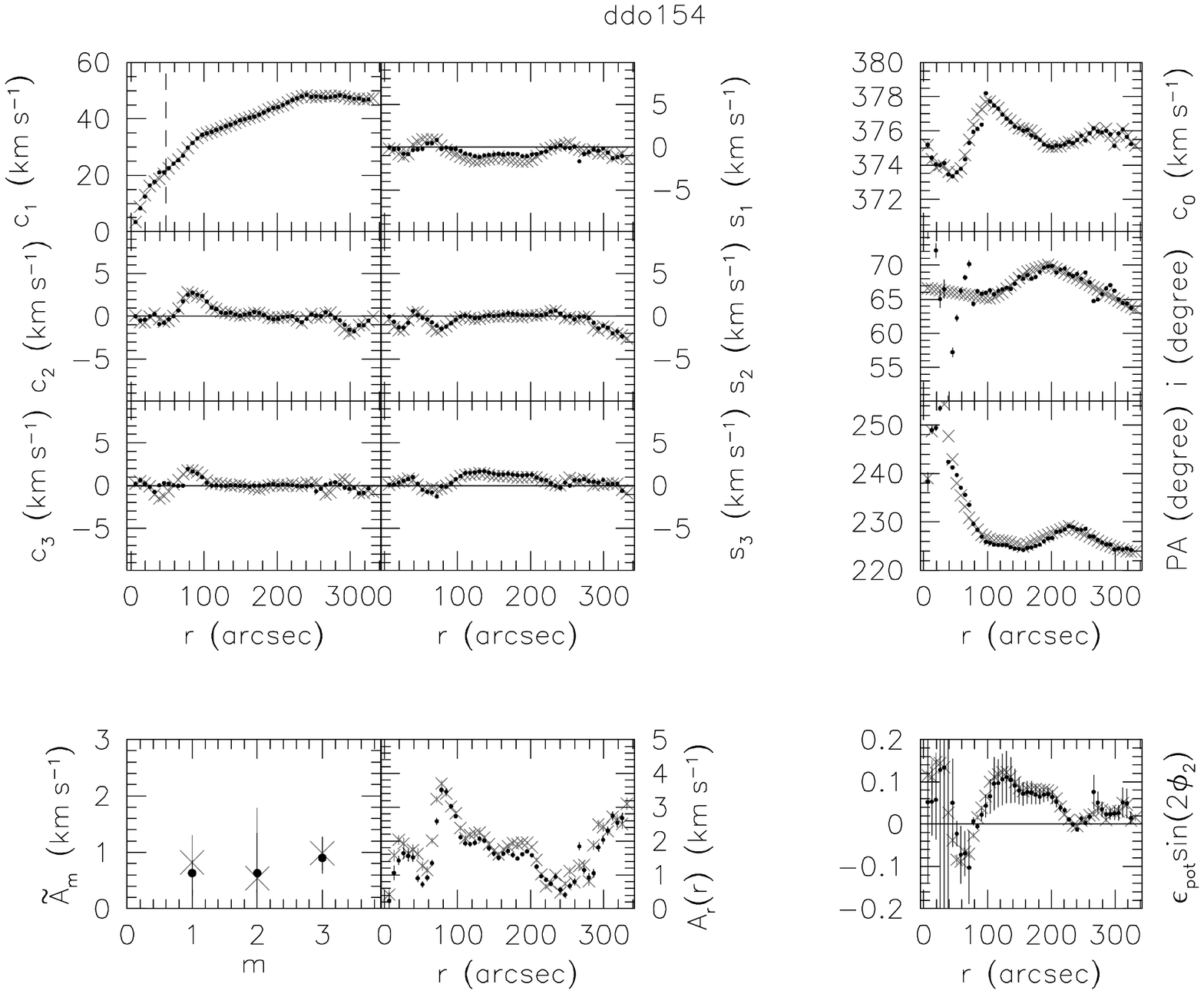}
\caption[Constrained {\it vs.} unconstrained harmonic decomposition of DDO 154]{Like Fig.~\ref{fig:fixed-free-ngc3198} but for DDO\
  154. The differences between the two decompositions are mostly
  negligible. The close-to solid-body rotation of DDO\ 154 makes a  determination of the inclination in the inner parts (in an unconstrained fit) more difficult. The impact of the less well-constrained $i$ is, however, small.} \label{fig:fixed-free-ddo154} 
\end{center}
\end{figure*}

For DDO\ 154 (Fig.~\ref{fig:fixed-free-ddo154}), the differences between the
harmonic components of the two decompositions are generally of the order of $\sim 1\,\kms$.
The inclinations in the unconstrained fit differ inwards of $r\approx 80\arcsec$ from the ones in the constrained fit. This is because of the close-to solid-body
rotation of DDO~154, which makes the simultaneous determination of the correct
inclination 
difficult. The impact on the results, even in the inner 80\arcsec, is
nevertheless small. For instance, the derived rotation velocities are
indistinguishable between the two decompositions. This shows that the magnitude of the non-circular motions is not very sensitive to small variations in PA or inclination.

\subsubsection{Hermite {\it vs}. Intensity weighted mean VF}
Hermite velocity
fields differ from intensity weighted mean (IWM) velocity fields when dealing
with asymmetric (i.e., non-Gaussian) profiles. It was noted, e.g., by \citet{deblok-07}, that hermite velocity fields better reproduce the velocity of the peak intensity of the profile, as their analytical function includes an $h_3$ (skewness) term \citep[see, e.g.,][]{vandermarel-1993}. For a detailed overview of the different types of velocity fields we refer to the discussion in \cite{deblok-07}.

\begin{figure*}[t]
\begin{center}
\includegraphics[angle=0,width=0.65\textwidth,bb=19 235 592 697,clip=]{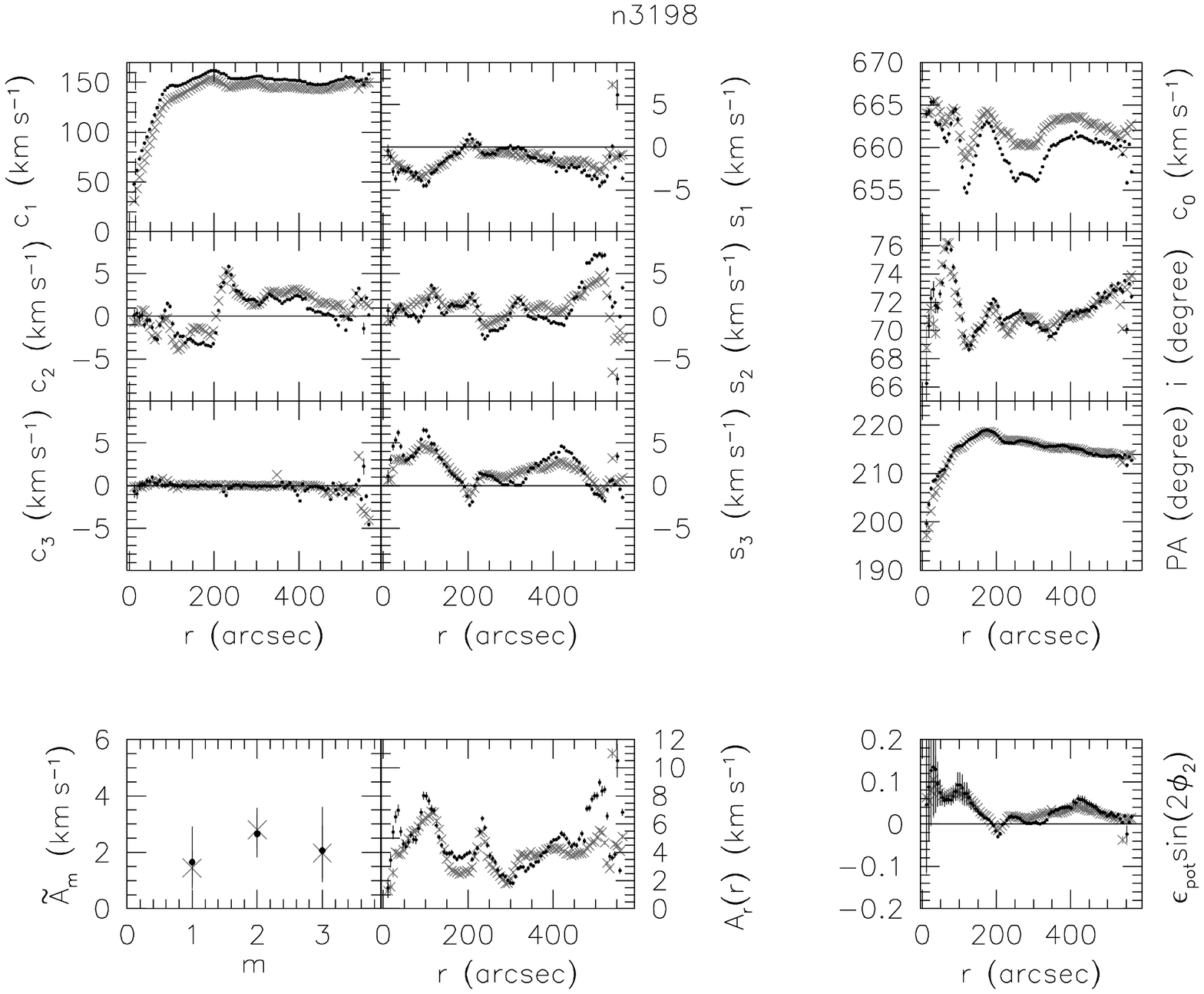}
\caption[Comparing the harmonic decomposition of the HER VF with that of the IWM VF for NGC 3198]{Harmonic analysis of NGC\ 3198 using the hermite velocity field
  (black dots) and the intensity weighted mean velocity field (gray
  crosses). The layout of the figure is identical to that of
  Figs.~\ref{fig:n3198-3} and \ref{fig:n3198-4}, except that we do not show any weighted means in the panels on the right-hand side. The 
  differences between the derived quantities are small. The radial
  variation of the harmonic components seems, however, more pronounced in the hermite velocity field.} \label{fig:iwm-her-ngc3198}
\end{center}
\end{figure*}

\begin{figure*}[t]
\begin{center}
\includegraphics[angle=0,width=0.65\textwidth,bb=19 235 592 697,clip=]{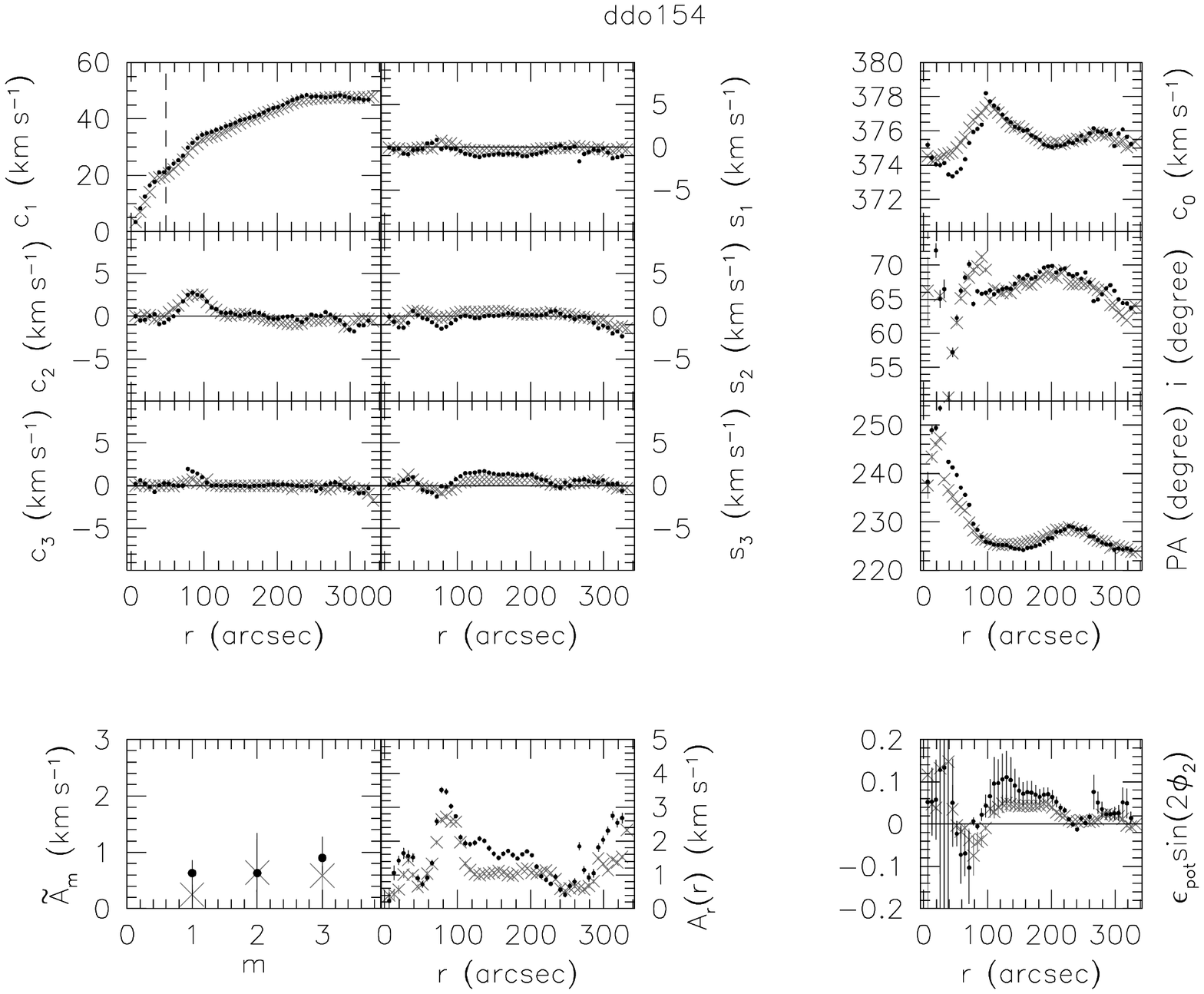}
\caption[Comparing the harmonic decomposition of the HER VF with that of the
IWM VF for DDO 154]{Like Fig.~\ref{fig:iwm-her-ngc3198} but for DDO~154.} 
\label{fig:iwm-her-ddo154}
\end{center}
\end{figure*}

As we are using the hermite velocity fields, an investigation of how using IWM velocity fields affects our analysis, if at all, is warranted. We perform unconstrained harmonic decompositions with \reswri\ using 
identical initial conditions on the hermite and on the IWM velocity
fields of NGC\ 3198 and DDO\ 154. The results are summarized in
Figs.~\ref{fig:iwm-her-ngc3198} and \ref{fig:iwm-her-ddo154}, respectively. 
As small differences between the two decompositions are visible, it is clear that how a velocity field is created does play a role in the analysis. 
A close inspection of the radial variation of the various quantities shown 
in both figures indicates that the radial variations are more
pronounced in the case of the hermite velocity field.
The choice of the velocity field construction method has, however, little impact on the derived quantities (see, e.g., bottom panels of Figs.~\ref{fig:iwm-her-ngc3198} and \ref{fig:iwm-her-ddo154}), and 
therefore does not change our conclusions in any significant way.

\subsubsection{Decomposition under the assumption of an incorrect center}\label{sec:center-offsets}
\begin{figure*}[t]
\begin{center}
\includegraphics[angle=0,width=0.65\textwidth,bb=19 235 592 697,clip=]{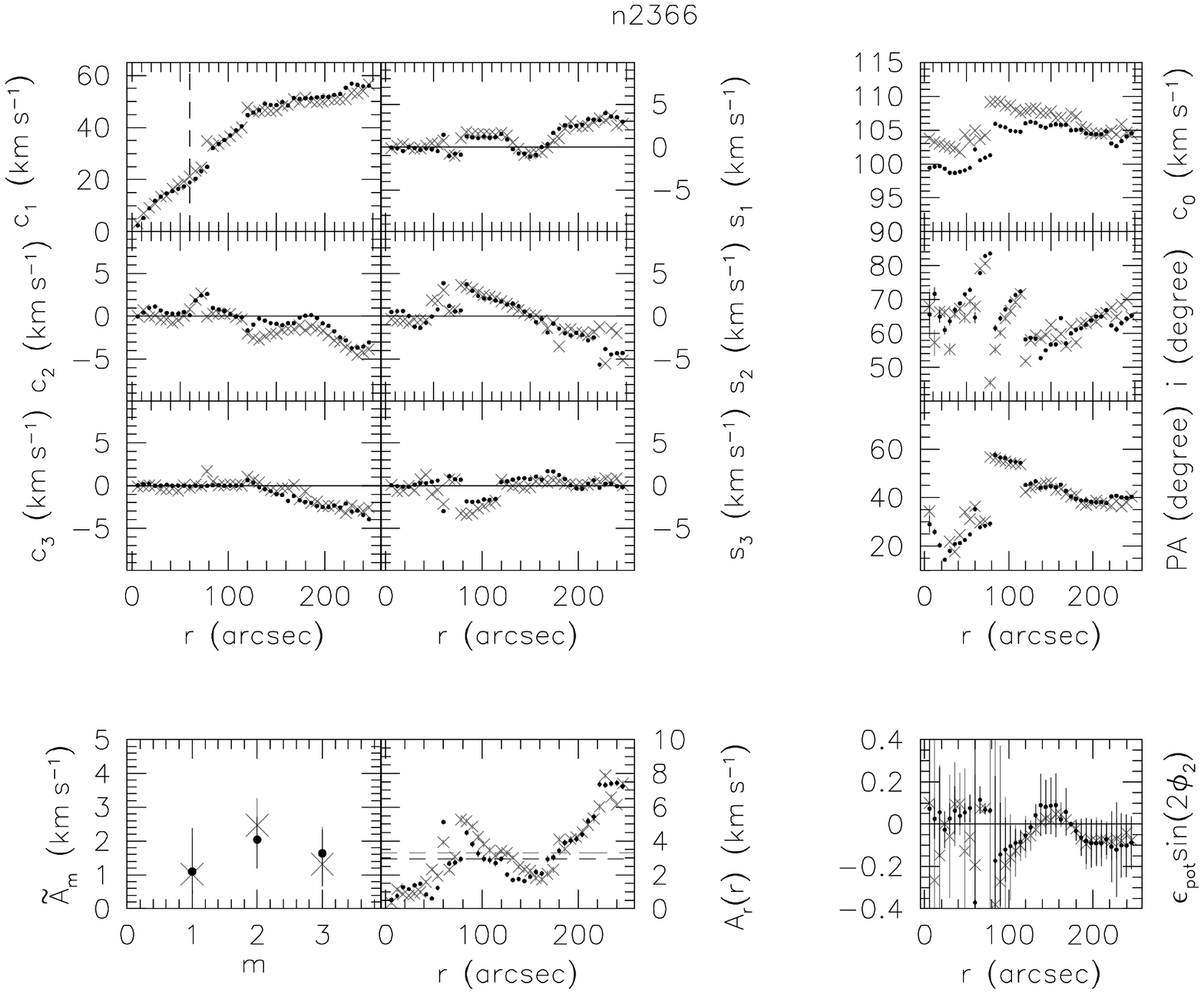}
\caption[How a 10\arcsec\ shift along the major axis affects the harmonic
decomposition of NGC 2366]{Harmonic analysis of NGC\ 2366 using our best center position
  (black) and a center position, offset by 10\arcsec\ along the major axis
  (gray). The layout of the figure is identical to that of
  Figs.~\ref{fig:n3198-3} and \ref{fig:n3198-4}, except that we do not show any weighted means in the panels on the right-hand side. The difference between the
  two decompositions is 
  marginal. The offset center causes the median amplitude of the
  non-circular motions (dashed line in the panel showing the distribution of
  \Ar) to increase from $\A \sim 3 \kms$ to $\A \sim 3.5 \kms$, showing that
  a small offset along the major axis has no significant effect on the derived quantities
  in the case of NGC 2366.} \label{fig:center-offset-n2366}  
\end{center}
\end{figure*}

\begin{figure*}[t]
\begin{center}
\includegraphics[angle=0,width=0.65\textwidth,bb=19 235 592 697,clip=]{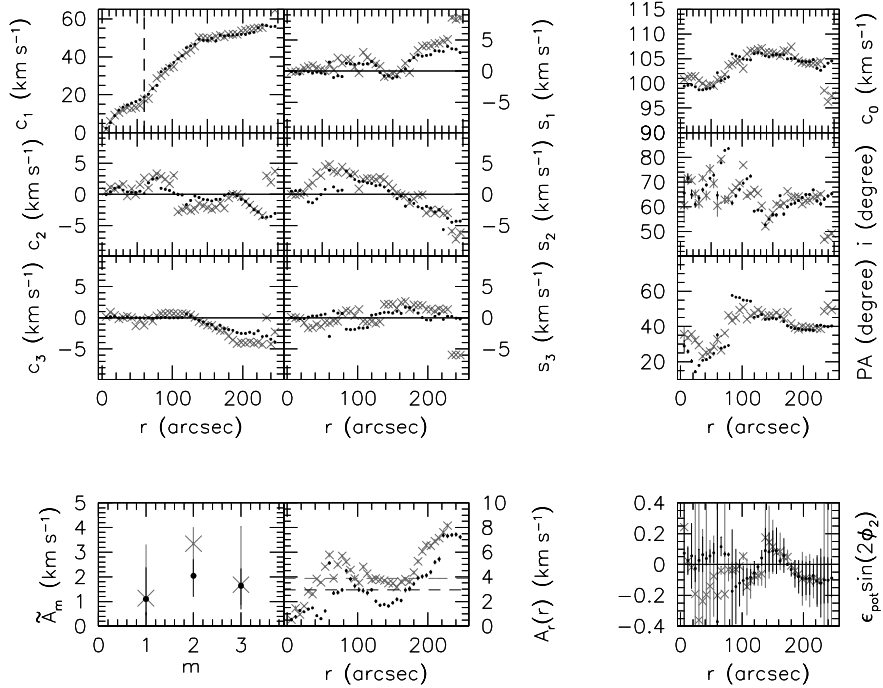}
\caption[How a 10\arcsec\ shift along the minor axis affects the harmonic
decomposition of NGC 2366]{Like Fig.~\ref{fig:center-offset-n2366}, but the harmonic analysis
  shown in gray is using a center position offset by 10\arcsec\ along the
  minor axis. Although being larger than in the case of the
  offset along the major 
  axis, the difference between the two decompositions is
  still small. The offset center causes the median amplitude of the
  non-circular motions (dashed line in the panel showing the distribution of
  \Ar) to increase from $\A \sim 3 \kms$ to $\A \sim 4 \kms$, showing that
  even a small offset along the minor axis does not influence the derived
  quantities in a significant way in the case of a dwarf galaxy like NGC
  2366.} \label{fig:center-offset-n2366-minor} 
\end{center}
\end{figure*}

\begin{figure*}[t]
\begin{center}
\vspace{5mm}
\includegraphics[angle=0,width=0.65\textwidth,bb=19 235 592 697,clip=]{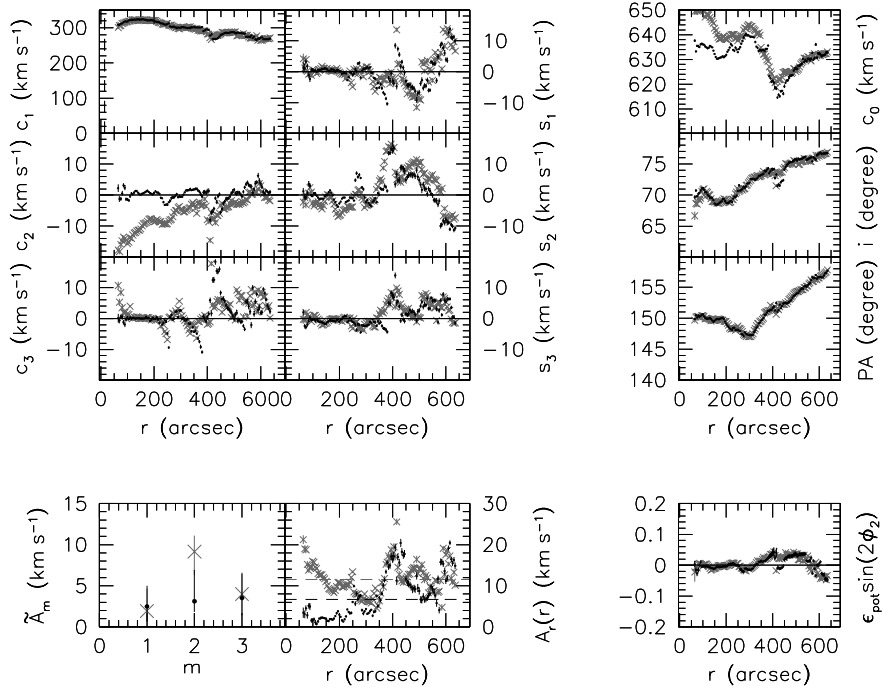}
\caption[How a 10\arcsec\ shift along the major axis affects the harmonic
decomposition of NGC 2841]{Harmonic analysis of NGC\ 2841 using our best center position
  (black) and a center position, offset by 10\arcsec\ along the major axis
  (gray). The layout of the figure is identical to that of
  Figs.~\ref{fig:n3198-3} and \ref{fig:n3198-4}, except that we do not show any weighted means in the panels on the right-hand side. For 
  NGC 2841, the amount of non-circular motions increases clearly 
  when choosing an offset center. The largest differences are visible in the
  $c_2$ term. The median amplitude of the non-circular
  motions (dashed-line in the panel showing the distribution of \Ar) increases
  from $\A \sim 7 \kms$ to $\A \sim 12 \kms$, showing that in the case of NGC 2841, an offset along the major axis will clearly show itself as an increase in the amplitudes of the harmonic
  components.} \label{fig:center-offset-n2841} 
\end{center}
\end{figure*}

\begin{figure*}[t]
\begin{center}
\vspace{5mm}
\includegraphics[angle=0,width=0.65\textwidth,bb=19 235 592 697,clip=]{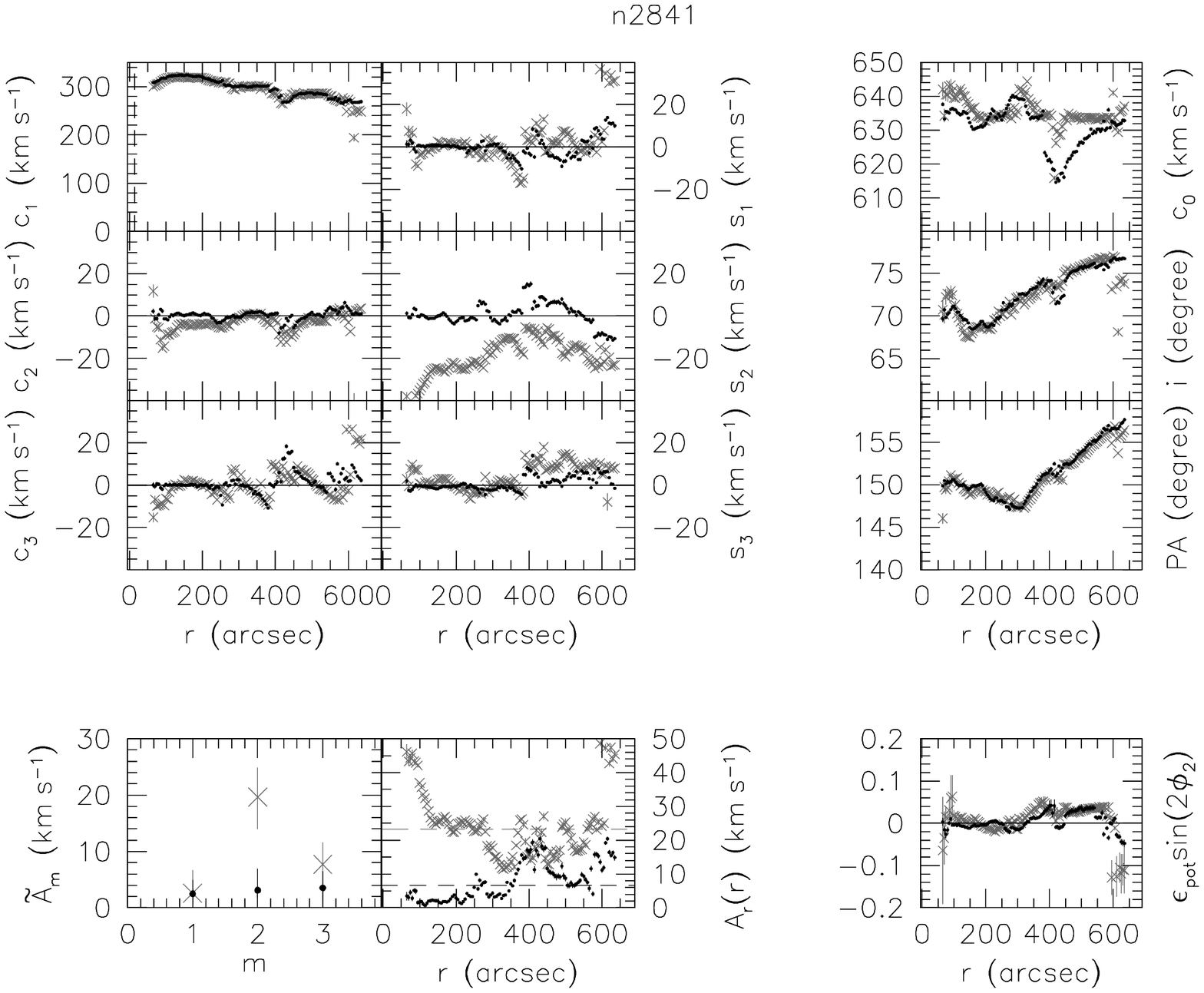}
\caption[How a 10\arcsec\ shift along the minor axis affects the harmonic
decomposition of NGC 2841]{Like Fig.~\ref{fig:center-offset-n2841}, but the harmonic analysis
  shown in gray is using a center position offset by 10\arcsec\ along the
  minor axis. For NGC\ 2841, the amount of non-circular
  motions increases clearly 
  when choosing an offset center. The largest differences are visible in the
  $s_2$ term. \A\ 
(dashed-line in the panel showing the distribution of \Ar) increases
  from $\A \sim 7 \kms$ to $\A \sim 25 \kms$, showing that in the case of NGC 2841, an offset along the minor axis will produce
  significantly different results than if the dynamical center position is chosen.}
  \label{fig:center-offset-n2841-minor}
\end{center}
\end{figure*}

In this section, we test
the influence of an incorrect center position on our results. As test
candidates, we have chosen one galaxy where the position of the center is very
well-defined (NGC\ 2841), and one where the center is less obvious (NGC\
2366). 
For both galaxies, we deliberately shift the center
positions used in \reswri\ by 2\arcsec, 4\arcsec, 6\arcsec, and
10\arcsec\ along their major and minor axes. The
results 
of the most extreme cases (10\arcsec\ offsets, i.e., approximately the beam size) are compared with the results 
from our best center positions in
Figs.~\ref{fig:center-offset-n2366}-\ref{fig:center-offset-n2841-minor}.

For NGC\ 2366, the effect of an incorrectly chosen center is 
small, irrespective of whether the center is shifted along the major (Fig.~\ref{fig:center-offset-n2366}) or along the minor (Fig.~\ref{fig:center-offset-n2366-minor}) axis. 
For offsets along the major (minor) axis, the median
amplitude of the non-circular motions increases from $\A \sim
3 \kms$ for our best center position to $\A \sim 3.5 \kms$ ($\A \sim 4 \kms$).
Our results for the dwarf galaxy NGC 2366 are therefore unaffected by 
small offsets in the center position, indicating that for a rotation curve
showing close-to solid-body rotation, a modest offset from the true
center position is not crucial for the analysis presented here.

For NGC 2841, the situation is different. A 10\arcsec\ offset along the
major axis (Fig.~\ref{fig:center-offset-n2841}) already increases the median
amplitude of the non-circular motions 
from $\A \sim 7 \kms$ to $\A \sim 12 \kms$. In the case of an offset along the
minor axis (Fig.~\ref{fig:center-offset-n2841}), the increment is even larger
($\A \sim 25 \kms$).

Note that the 
difference in the harmonic components is mostly to be found in the $m=2$
term. This is to be expected, as a galaxy will appear kinematically lopsided if a center offset from the dynamical center is chosen for the harmonic decomposition.
Also note that, as expected, it is the cosine coefficient $c_2$ which
shows the largest amplitude in the case of an offset along the major axis, and the
sine coefficient $s_2$ in the case of an offset along the minor axis. 

For NGC 2841 --- a galaxy with a steep and subsequently flat rotation curve --- our results are sensitive to
the chosen center position and an offset center would clearly show itself
by increased non-circular motions. But this example is of course
rather contrived,  
as for galaxies like NGC 2841, the center is usually so well-defined that uncertainties
of $10\arcsec$ as modeled here are unlikely to occur within the THINGS
sample.

We have shown that the results of a harmonic decomposition of galaxies like NGC 2841 (i.e., having a steep rotation curve) are sensitive to offsets in the galaxies' center position.
For NGC 2366, i.e., a dwarf galaxy showing close-to solid-body rotation, our results are mostly insensitive to small offsets in the center position. The less well-defined center for NGC 2366 does therefore not affect our results of that galaxy.

\section{Summary and Conclusions}
We have analyzed \HI\ velocity fields of 19 THINGS disk and dwarf galaxies at the currently
best available spatial and
spectral resolution. The observations, data reduction and analysis of all these
galaxies were done in an identical and homogeneous manner \citep{walter-07, deblok-07}.
We have determined the center positions of these 19
galaxies by comparing the kinematic center estimates with those from the
radio continuum and/or NIR data.

We have derived reliable center
positions and show that most galaxies in our sample do not exhibit significant offsets between 
their kinematic and their optical centers. 
For 18 out of 19 galaxies from our sample, we have performed a harmonic decomposition of their velocity fields with the aim of
quantifying the amount of non-circular motions and deriving a lower limit for
the elongation of the potential. 

Our results show that in the (large) majority of our sample galaxies,
the effects of non-circular
motions are small, irrespective of whether these non-circular motions
are measured in the inner few kpc or over the entire radial range. 
Within our sample, the average amplitude of the non-circular motions is
$\tilde{A}_{r} = 4.8 \pm 4.0\,\kms$ for the
inner 1 kpc and $\tilde{A}_{r} = 6.7 \pm 5.9\,\kms$ when averaged over
the entire radial range.
High non-circular motions in the inner parts of the galaxies are mostly
found in barred, and/or luminous galaxies.
The galaxies in our sample least dominated by baryonic matter (i.e., the low
luminosity galaxies) show the smallest non-circular motions in the sample,
especially in the inner parts, thus not showing the streaming motions one would associate with elliptical distortions or tri-axial halos.

The inner few kpc of the large majority of the galaxies in our sample contain non-circular motions of the order of ten percent of the local rotation velocity.
Averaged over our sample, the median amplitude of the non-circular motions in the inner 1 kpc is $8 \pm 3$ percent of the local rotation velocity.
For three out of the four dwarf galaxies in our sample, the amplitudes of the non-circular motions amount to (at most) ten percent of the local rotation velocity.

Whether the non-circular motions are expressed in absolute terms or if they are normalized by the local rotation velocity does not affect the conclusions:
the amount of non-circular motions detected in the majority of our galaxies is
significantly smaller than what is expected from CDM simulations and they are
far too small to ``hide a cusp in a core'' as proposed, e.g., by \cite{hayashi-2006}. 

The average elongation of the gravitational potential and its scatter, both statistically corrected for the unknown viewing angle $\varphi_2$, is $\langle \epsilon_{pot}\rangle=0.017 \pm 0.020$.
This is significantly lower than the predictions from cosmological simulations.
The individual elongation measurements show that the large majority of the galaxies in our sample have elongation measurements which are systematically below the CDM predictions. The elongation of all galaxies in our sample is consistent with a round potential, although some galaxies have large enough uncertainties to make them also (marginally) consistent with the lower end of the predicted range for CDM halos. 
We therefore see no indication for a significant tri-axiality within the \HI\ disks of the THINGS galaxies.

\acknowledgments
CT gratefully acknowledges the help and support from Ralf-J\"urgen Dettmar.
The work of CT is supported by the German Ministry
for Education and Science (BMBF) through grant 05 AV5PDA/3.
The work of WJGdB is based upon research supported by the South
African Research Chairs Initiative of the Department of Science and
Technology and National Research Foundation. 
EB gratefully acknowledges financial support through 
an EU Marie Curie International Reintegration Grant 
(Contract No. MIRG-CT-6-2005-013556). 
This research has made use of the NASA/IPAC Extragalactic Database (NED) which is operated by the Jet Propulsion Laboratory, California Institute of Technology, under contract with the National Aeronautics and Space Administration.
We acknowledge the usage of the HyperLeda database (http://leda.univ-lyon1.fr).
\bibliographystyle{apj}
\bibliography{bib}

\begin{thebibliography}{47}
\expandafter\ifx\csname natexlab\endcsname\relax\def\natexlab#1{#1}\fi

\bibitem[{{Battaglia} {et~al.}(2006){Battaglia}, {Fraternali}, {Oosterloo}, \&
  {Sancisi}}]{battaglia-2006}
{Battaglia}, G., {Fraternali}, F., {Oosterloo}, T., \& {Sancisi}, R. 2006,
  \aap, 447, 49

\bibitem[{{Braun} {et~al.}(1994){Braun}, {Walterbos}, {Kennicutt}, \&
  {Tacconi}}]{braun-1994}
{Braun}, R., {Walterbos}, R.~A.~M., {Kennicutt}, Jr., R.~C., \& {Tacconi},
  L.~J. 1994, \apj, 420, 558

\bibitem[{{Capuzzo-Dolcetta} {et~al.}(2007){Capuzzo-Dolcetta}, {Leccese},
  {Merritt}, \& {Vicari}}]{capuzzo-2007}
{Capuzzo-Dolcetta}, R., {Leccese}, L., {Merritt}, D., \& {Vicari}, A. 2007,
  \apj, 666, 165

\bibitem[{{de Blok}(2004)}]{deblok-2004}
{de Blok}, W.~J.~G. 2004, in IAU Symposium, Vol. 220, Dark Matter in Galaxies,
  ed. S.~{Ryder}, D.~{Pisano}, M.~{Walker}, \& K.~{Freeman}, 69

\bibitem[{{de Blok}(2005)}]{deblok-2005}
{de Blok}, W.~J.~G. 2005, \apj, 634, 227

\bibitem[{{de Blok} \& {Bosma}(2002)}]{deblok-2002}
{de Blok}, W.~J.~G., \& {Bosma}, A. 2002, \aap, 385, 816

\bibitem[{{de Blok} {et~al.}(2003){de Blok}, {Bosma}, \&
  {McGaugh}}]{deblok-2003}
{de Blok}, W.~J.~G., {Bosma}, A., \& {McGaugh}, S. 2003, \mnras, 340, 657

\bibitem[{{de Blok} {et~al.}(2001{\natexlab{a}}){de Blok}, {McGaugh}, {Bosma},
  \& {Rubin}}]{deblok-2001a}
{de Blok}, W.~J.~G., {McGaugh}, S.~S., {Bosma}, A., \& {Rubin}, V.~C.
  2001{\natexlab{a}}, \apjl, 552, L23

\bibitem[{{de Blok} {et~al.}(2001{\natexlab{b}}){de Blok}, {McGaugh}, \&
  {Rubin}}]{deblok-2001b}
{de Blok}, W.~J.~G., {McGaugh}, S.~S., \& {Rubin}, V.~C. 2001{\natexlab{b}},
  \aj, 122, 2396

\bibitem[{{de Blok} {et~al.}(2008){de Blok}, {Walter}, {Brinks},
  {Trachternach}, {Oh}, \& {Kennicutt}}]{deblok-07}
{de Blok}, W.~J.~G., {Walter}, F., {Brinks}, E., {Trachternach}, C., {Oh},
  S.-H., \& {Kennicutt}, Jr., R.~C. 2008, submitted to AJ

\bibitem[{{Dubinski}(1994)}]{dubinski-1994}
{Dubinski}, J. 1994, \apj, 431, 617

\bibitem[{{Franx} \& {de Zeeuw}(1992)}]{franx-92}
{Franx}, M., \& {de Zeeuw}, T. 1992, \apjl, 392, L47

\bibitem[{{Frenk} {et~al.}(1988){Frenk}, {White}, {Davis}, \&
  {Efstathiou}}]{frenk-1988}
{Frenk}, C.~S., {White}, S.~D.~M., {Davis}, M., \& {Efstathiou}, G. 1988, \apj,
  327, 507

\bibitem[{{Gentile} {et~al.}(2005){Gentile}, {Burkert}, {Salucci}, {Klein}, \&
  {Walter}}]{gentile-2005}
{Gentile}, G., {Burkert}, A., {Salucci}, P., {Klein}, U., \& {Walter}, F. 2005,
  \apjl, 634, L145

\bibitem[{{Gentile} {et~al.}(2007){Gentile}, {Salucci}, {Klein}, \&
  {Granato}}]{gentile-2007}
{Gentile}, G., {Salucci}, P., {Klein}, U., \& {Granato}, G.~L. 2007, \mnras,
  375, 199

\bibitem[{{Gentile} {et~al.}(2004){Gentile}, {Salucci}, {Klein}, {Vergani}, \&
  {Kalberla}}]{gentile-2004}
{Gentile}, G., {Salucci}, P., {Klein}, U., {Vergani}, D., \& {Kalberla}, P.
  2004, \mnras, 351, 903

\bibitem[{{Hayashi} \& {Navarro}(2006)}]{hayashi-2006}
{Hayashi}, E., \& {Navarro}, J.~F. 2006, \mnras, 373, 1117

\bibitem[{{Hayashi} {et~al.}(2004){Hayashi}, {Navarro}, {Jenkins}, {Frenk},
  {Power}, {White}, {Springel}, {Stadel}, {Quinn}, \& {Wadsley}}]{hayashi-2004}
{Hayashi}, E., {Navarro}, J.~F., {Jenkins}, A., {Frenk}, C.~S., {Power}, C.,
  {White}, S.~D.~M., {Springel}, V., {Stadel}, J., {Quinn}, T., \& {Wadsley},
  J. 2004, ArXiv Astrophysics e-prints

\bibitem[{{Hayashi} {et~al.}(2007){Hayashi}, {Navarro}, \&
  {Springel}}]{hayashi-2007}
{Hayashi}, E., {Navarro}, J.~F., \& {Springel}, V. 2007, \mnras, 248

\bibitem[{{Kasun} \& {Evrard}(2005)}]{kasun-2005}
{Kasun}, S.~F., \& {Evrard}, A.~E. 2005, \apj, 629, 781

\bibitem[{{Kennicutt} {et~al.}(2003){Kennicutt}, {Armus}, {Bendo}, {Calzetti},
  {Dale}, {Draine}, {Engelbracht}, {Gordon}, {Grauer}, {Helou}, {Hollenbach},
  {Jarrett}, {Kewley}, {Leitherer}, {Li}, {Malhotra}, {Regan}, {Rieke},
  {Rieke}, {Roussel}, {Smith}, {Thornley}, \& {Walter}}]{SINGS}
{Kennicutt}, Jr., R.~C., {Armus}, L., {Bendo}, G., {Calzetti}, D., {Dale},
  D.~A., {Draine}, B.~T., {Engelbracht}, C.~W., {Gordon}, K.~D., {Grauer},
  A.~D., {Helou}, G., {Hollenbach}, D.~J., {Jarrett}, T.~H., {Kewley}, L.~J.,
  {Leitherer}, C., {Li}, A., {Malhotra}, S., {Regan}, M.~W., {Rieke}, G.~H.,
  {Rieke}, M.~J., {Roussel}, H., {Smith}, J.-D.~T., {Thornley}, M.~D., \&
  {Walter}, F. 2003, \pasp, 115, 928

\bibitem[{{Kuzio de Naray} {et~al.}(2008){Kuzio de Naray}, {McGaugh}, \& {de
  Blok}}]{naray-2008}
{Kuzio de Naray}, R., {McGaugh}, S.~S., \& {de Blok}, W.~J.~G. 2008, \apj, 676,
  920

\bibitem[{{Kuzio de Naray} {et~al.}(2006){Kuzio de Naray}, {McGaugh}, {de
  Blok}, \& {Bosma}}]{naray-2006}
{Kuzio de Naray}, R., {McGaugh}, S.~S., {de Blok}, W.~J.~G., \& {Bosma}, A.
  2006, \apjs, 165, 461

\bibitem[{{Maoz} {et~al.}(1996){Maoz}, {Filippenko}, {Ho}, {Macchetto}, {Rix},
  \& {Schneider}}]{maoz-1996}
{Maoz}, D., {Filippenko}, A.~V., {Ho}, L.~C., {Macchetto}, F.~D., {Rix}, H.-W.,
  \& {Schneider}, D.~P. 1996, \apjs, 107, 215

\bibitem[{{Marchesini} {et~al.}(2002){Marchesini}, {D'Onghia}, {Chincarini},
  {Firmani}, {Conconi}, {Molinari}, \& {Zacchei}}]{marchesini-2002}
{Marchesini}, D., {D'Onghia}, E., {Chincarini}, G., {Firmani}, C., {Conconi},
  P., {Molinari}, E., \& {Zacchei}, A. 2002, \apj, 575, 801

\bibitem[{{Matthews} \& {Gallagher}(2002)}]{matthews-2002}
{Matthews}, L.~D., \& {Gallagher}, III, J.~S. 2002, \apjs, 141, 429

\bibitem[{{McGaugh} \& {de Blok}(1998)}]{mcgaugh-1998b}
{McGaugh}, S.~S., \& {de Blok}, W.~J.~G. 1998, \apj, 499, 41

\bibitem[{{Moore} {et~al.}(2004){Moore}, {Kazantzidis}, {Diemand}, \&
  {Stadel}}]{moore-2004}
{Moore}, B., {Kazantzidis}, S., {Diemand}, J., \& {Stadel}, J. 2004, \mnras,
  354, 522

\bibitem[{{Navarro} {et~al.}(1996){Navarro}, {Frenk}, \&
  {White}}]{navarro-1996}
{Navarro}, J.~F., {Frenk}, C.~S., \& {White}, S.~D.~M. 1996, \apj, 462, 563

\bibitem[{{Navarro} {et~al.}(1997){Navarro}, {Frenk}, \&
  {White}}]{navarro-1997}
---. 1997, \apj, 490, 493

\bibitem[{{Oh} {et~al.}(2008){Oh}, {de Blok}, {Walter}, {Brinks}, \&
  {Kennicutt}}]{se-heon}
{Oh}, S.-H., {de Blok}, W.~J.~G., {Walter}, F., {Brinks}, E., \& {Kennicutt},
  Jr., R.~C. 2008, submitted to AJ

\bibitem[{{Rhee} {et~al.}(2004){Rhee}, {Valenzuela}, {Klypin}, {Holtzman}, \&
  {Moorthy}}]{rhee-2004}
{Rhee}, G., {Valenzuela}, O., {Klypin}, A., {Holtzman}, J., \& {Moorthy}, B.
  2004, \apj, 617, 1059

\bibitem[{{Schoenmakers}(1999)}]{schoenmakers-thesis}
{Schoenmakers}, R.~H.~M. 1999, PhD thesis, Univ. Groningen

\bibitem[{{Schoenmakers} {et~al.}(1997){Schoenmakers}, {Franx}, \& {de
  Zeeuw}}]{schoenmakers-1997}
{Schoenmakers}, R.~H.~M., {Franx}, M., \& {de Zeeuw}, P.~T. 1997, \mnras, 292,
  349

\bibitem[{{Simon} {et~al.}(2003){Simon}, {Bolatto}, {Leroy}, \&
  {Blitz}}]{simon-2003}
{Simon}, J.~D., {Bolatto}, A.~D., {Leroy}, A., \& {Blitz}, L. 2003, \apj, 596,
  957

\bibitem[{{Simon} {et~al.}(2005){Simon}, {Bolatto}, {Leroy}, {Blitz}, \&
  {Gates}}]{simon-2005}
{Simon}, J.~D., {Bolatto}, A.~D., {Leroy}, A., {Blitz}, L., \& {Gates}, E.~L.
  2005, \apj, 621, 757

\bibitem[{{Spano} {et~al.}(2008){Spano}, {Marcelin}, {Amram}, {Carignan},
  {Epinat}, \& {Hernandez}}]{spano-2007}
{Spano}, M., {Marcelin}, M., {Amram}, P., {Carignan}, C., {Epinat}, B., \&
  {Hernandez}, O. 2008, \mnras, 383, 297

\bibitem[{{Spergel} {et~al.}(2007){Spergel}, {Bean}, {Dor{\'e}}, {Nolta},
  {Bennett}, {Dunkley}, {Hinshaw}, {Jarosik}, {Komatsu}, {Page}, {Peiris},
  {Verde}, {Halpern}, {Hill}, {Kogut}, {Limon}, {Meyer}, {Odegard}, {Tucker},
  {Weiland}, {Wollack}, \& {Wright}}]{spergel-2007}
{Spergel}, D.~N., {Bean}, R., {Dor{\'e}}, O., {Nolta}, M.~R., {Bennett}, C.~L.,
  {Dunkley}, J., {Hinshaw}, G., {Jarosik}, N., {Komatsu}, E., {Page}, L.,
  {Peiris}, H.~V., {Verde}, L., {Halpern}, M., {Hill}, R.~S., {Kogut}, A.,
  {Limon}, M., {Meyer}, S.~S., {Odegard}, N., {Tucker}, G.~S., {Weiland},
  J.~L., {Wollack}, E., \& {Wright}, E.~L. 2007, \apjs, 170, 377

\bibitem[{{Swaters} {et~al.}(2003{\natexlab{a}}){Swaters}, {Madore}, {van den
  Bosch}, \& {Balcells}}]{swaters-2003a}
{Swaters}, R.~A., {Madore}, B.~F., {van den Bosch}, F.~C., \& {Balcells}, M.
  2003{\natexlab{a}}, \apj, 583, 732

\bibitem[{{Swaters} {et~al.}(2003{\natexlab{b}}){Swaters}, {Verheijen},
  {Bershady}, \& {Andersen}}]{swaters-2003b}
{Swaters}, R.~A., {Verheijen}, M.~A.~W., {Bershady}, M.~A., \& {Andersen},
  D.~R. 2003{\natexlab{b}}, \apjl, 587, L19

\bibitem[{{Valenzuela} {et~al.}(2007){Valenzuela}, {Rhee}, {Klypin},
  {Governato}, {Stinson}, {Quinn}, \& {Wadsley}}]{valenzuela-2007}
{Valenzuela}, O., {Rhee}, G., {Klypin}, A., {Governato}, F., {Stinson}, G.,
  {Quinn}, T., \& {Wadsley}, J. 2007, \apj, 657, 773

\bibitem[{{van der Hulst} {et~al.}(1992){van der Hulst}, {Terlouw}, {Begeman},
  {Zwitser}, \& {Roelfsema}}]{vanderhulst-1992}
{van der Hulst}, J.~M., {Terlouw}, J.~P., {Begeman}, K.~G., {Zwitser}, W., \&
  {Roelfsema}, P.~R. 1992, in Astronomical Society of the Pacific Conference
  Series, Vol.~25, Astronomical Data Analysis Software and Systems I, ed. D.~M.
  {Worrall}, C.~{Biemesderfer}, \& J.~{Barnes}, 131

\bibitem[{{van der Marel} \& {Franx}(1993)}]{vandermarel-1993}
{van der Marel}, R.~P., \& {Franx}, M. 1993, \apj, 407, 525

\bibitem[{{Walter} {et~al.}(2008){Walter}, {Brinks}, {de Blok}, {Bigiel},
  {Kennicutt}, \& {Thornley}}]{walter-07}
{Walter}, F., {Brinks}, E., {de Blok}, W.~J.~G., {Bigiel}, F., {Kennicutt},
  Jr., R.~C., \& {Thornley}, M. 2008, submitted to AJ

\bibitem[{{Walter} {et~al.}(1998){Walter}, {Kerp}, {Duric}, {Brinks}, \&
  {Klein}}]{walter-1998}
{Walter}, F., {Kerp}, J., {Duric}, N., {Brinks}, E., \& {Klein}, U. 1998,
  \apjl, 502, L143+

\bibitem[{{Wong} {et~al.}(2004){Wong}, {Blitz}, \& {Bosma}}]{wong-2004}
{Wong}, T., {Blitz}, L., \& {Bosma}, A. 2004, \apj, 605, 183

\bibitem[{{Zackrisson} {et~al.}(2006){Zackrisson}, {Bergvall}, {Marquart}, \&
  {{\"O}stlin}}]{zackrisson-2006}
{Zackrisson}, E., {Bergvall}, N., {Marquart}, T., \& {{\"O}stlin}, G. 2006,
  \aap, 452, 857

\end{thebibliography}

\FloatBarrier

\begin{appendix}
\section{Description of individual galaxies}\label{sec:indiv-gal}

In this section, we present for each galaxy in our sample a detailed description of \emph{(a)} the derivation of the center position, and \emph{(b)} the results of the harmonic decomposition.
The galaxies are presented in order of increasing right ascension. 
The results are presented in graphical form in
Figs.~\ref{fig:ngc-925}-\ref{fig:ngc-7793}. Channel maps, optical images and moment maps for the galaxies in our sample are given in \cite{walter-07}.
The different center
estimates are summarized in Table~\ref{table:center-pos}, where our adopted
best center positions are shown in bold face. The results from the harmonic
decomposition are summarized in Table~\ref{table:harm-decomp}.

\subsection[NGC 925]{NGC 925 (Fig.~\ref{fig:ngc-925})}\label{sec:n925}
\subparagraph{a) Center estimates \\[1.5ex]}
NGC 925 is a barred late-type spiral galaxy whose \HI\ disk looks rather
clumpy. The 3.6\,$\mu$m IRAC image
shows a weak central bar and two spiral arms in the outer region.
NGC 925 lacks a clear center in the
IRAC image as well as in the radio continuum. 
We used {\sc ellfit} at varying isophote levels to derive a central position
from the IRAC image. This center is offset from the pointing center by
$\dx=-0\farcs 8$ and $\dy=-2\farcs 6$ (i.e., towards the south-east) and was
used as an initial estimate for
a \rotcur\ run with all parameters left free.
Averaging the values for \dx, \dy\ over the region not too heavily
affected by spiral arms and asymmetries
($r\le 121\arcsec$, see top row of Fig.~\ref{fig:ngc-925}) results in a center
position offset from the pointing center by $\dx=4\farcs 4 \pm 8\farcs 3$ and $\dy=-1\farcs 5 \pm
4\farcs 2$. As this position is consistent with the
center from the IRAC image (see row 2 of Fig.~\ref{fig:ngc-925}), we adopt the 
former one as our best center for NGC 925.

\subparagraph{b) Harmonic expansion \\[1.5ex]}
The fitted values for the inclination show a lot of scatter in the inner
$100\arcsec$. This is partially caused by the near 
solid-body rotation in that region, which makes determining a kinematic
inclination more difficult. Note that in the case of pure solid-body rotation,
\VROT\ and $\sin(i)$ are degenerate and cannot be fitted
simultaneously. However, we seldom have to deal with {\it pure}
solid-body rotation, but more usually with near solid-body rotation, which
gives the algorithm some handle on the fitted values.
Although the amplitudes of the individual components are not constant over the 
radial range, the median amplitudes of all harmonic orders are relatively small 
($\Am \le 5\ \mathrm{km\,s^{-1}}$ when averaged over the
entire galaxy). The median amplitudes 
are slightly higher if one averages over the inner 1 kpc only, which might be
due to the stellar bar. \A\ is $\sim
6.3\,\kms$ if averaged over the entire range and $\sim 9.5 \, \kms$ if
averaged over the inner 1 kpc only (see also Table~\ref{table:harm-decomp}).
The elongation of the potential is unconstrained inwards of
$r\approx 100\arcsec$, mainly because of the large uncertainties for the
derived inclination (which enters into the uncertainty in $\epsilon_{\mathrm{pot}}$ as a fourth
  power). 
Its weighted mean is consistent with a round potential ($
\langle \epsilon_{\mathrm{pot}}\,\sin(2\varphi_2) \rangle =0.000 \pm 0.046$,
see again Table~\ref{table:harm-decomp}).
The median of the absolute residual velocity field is 3.0\,\kms, 
showing that 
a harmonic decomposition up to third order was able to capture most
non-circular motions present in NGC~925.

\subsection[NGC 2366]{NGC 2366 (Fig.~\ref{fig:ngc-2366})}\label{sec:n2366}
\subparagraph{a) Center estimates \\[1.5ex]}
NGC 2366 is classified as a dwarf irregular and belongs, like NGC 2403, to the
M81 group. As for all the other dwarf galaxies in the sample presented
here, the 3.6\,$\mu$m IRAC image and the radio continuum map do not show a
nuclear source. We derived the center of the emission in the
IRAC image by using {\sc ellfit} at varying isophote levels. The 
IRAC center is offset from the pointing center by $\dx=7\farcs 8 \pm 0\farcs 6$ and 
$\dy=-18\farcs 0 \pm 1\farcs 8$.
This center estimate was used for 
a \rotcur\ run with all parameters left free. We averaged the derived
values for \dx, \dy\ for $r\le 200\arcsec$ (see row 1 in 
Fig.~\ref{fig:ngc-2366}), but excluded the discrepant data
points between $70\arcsec \le r \le 100\arcsec$. The resulting dynamical
center is offset from the pointing center by $\dx=5\farcs 2 \pm 3\farcs 5$ and
$\dy=-19\farcs 2 \pm 7\farcs 7$, and within the uncertainties consistent with
the center position derived from the IRAC
image (see row 2 of Fig.~\ref{fig:ngc-2366}). Deriving a kinematic center by
averaging over all data points out to $r=200\arcsec$ gives a consistent center
estimate, though with a larger scatter ($\dx=2\farcs 9\pm 6\farcs 5$, 
$\dy=-18\farcs 5 \pm 7\farcs 9$). 
Therefore, we adopt the former kinematic estimate 
as our final center position.

\subparagraph{b) Harmonic expansion \\[1.5ex]}
NGC 2366 shows no clear signs of spiral
structure. Its rotation curve is clearly dominated by near solid-body
rotation, 
making it difficult to determine the rotation velocity and $i$
simultaneously. 
At $r\sim 100\arcsec$, the
systemic velocity $c_0$ rises from $\sim 100\ \mathrm{km\,s^{-1}}$
to $\sim 105\ \mathrm{km\,s^{-1}}$. 
At the same radius, the amplitude of $s_2$ ``jumps'' from $\sim 0\, \kms$ to $\sim 4\, \kms$ and decreases then to $-4 \kms$ in the outskirts of NGC 2366.
The distribution of \Ar\
shows that the largest amount of 
non-circular motions is to be found in the outer parts of NGC\ 2366, and that
the inner 1 kpc exhibit only minor non-circular motions.
The weighted mean elongation is $ \langle \epsilon_{\mathrm{pot}}\,\sin(2\varphi_2) \rangle =0.003 \pm 0.067$ and
the median of the absolute residual velocity field is 2.4
$\mathrm{km\,s^{-1}}$, once again showing that the galaxy has a round
potential and that there are no large non-circular motions present in the
$m>3$ terms.

\subsection[NGC 2403]{NGC 2403 (Fig.~\ref{fig:ngc-2403})}\label{sec:n2403}
\subparagraph{a) Center estimates \\[1.5ex]}
NGC 2403 belongs to the M81 group, and is a
late-type Sc spiral. Its 3.6\,$\mu$m IRAC image shows multiple
spiral arms and a bright central component. However, neither the IRAC image,
nor 
the radio continuum map show a nuclear source. Therefore, we used
3.6\,$\mu$m isophote fits to determine center coordinates and used
the resulting center position (\dx=$-$4\farcs 7, \dy=$-$13\farcs 6) as an initial estimate for an unconstrained \rotcur\ run.
As can be seen in the first row of Fig.~\ref{fig:ngc-2403}, \dx\
and \dy\ stay fairly constant over a large part of the galaxy. We
determine the 
dynamical center by averaging the values for $r\le 220\arcsec$, thus
excluding the radii where \dx\ changes considerably, probably due to the spiral arm
located at that radius.
The offset from the pointing center is $\dx=-5\farcs 2 \pm 5\farcs 2$ to the
east and $\dy=-12\farcs 8 \pm 4\farcs 1$ to the south. Averaging
\dx\ and \dy\ over a larger part of the galaxy gives comparable results
but with a larger scatter. Given 
the agreement between the dynamical center estimate and the center as derived using the IRAC image,
we adopt the $r\le 220\arcsec$ kinematic center estimate as our best center position.

\subparagraph{b) Harmonic expansion \\[1.5ex]}
The radial variation of the inclination is only $\sim 5\degree$, except at a few inner radii. 
The PA shows a steep inner rise, a flat
outer part and a dip at $r\approx 200\arcsec$,
which coincides with a high amplitude in the $c_2$ and $s_2$ components. 
Inspection of the integrated \HI\ map shows a 
spiral arm crossing at this radius. The $s_1$ and $s_3$ components also show
the characteristic wiggles caused by spiral arms. Additionally, $s_3$
has a 
small offset from zero, which through Eq.~\ref{eq:epot} can indicate a
slightly elongated potential.
The average elongation of NGC 2403, however, is 
$ \langle \epsilon_{\mathrm{pot}}\,\sin(2\varphi_2) \rangle =-0.022 \pm 0.025$,
and therefore consistent with a round potential.
The distribution of \Am\ shows that the individual harmonic orders
contribute only $\sim
2$ percent to \vmax, indicating that non-circular motions play a 
minor role in NGC 2403. The latter can also be seen in the distribution of \Ar.

\subsection[NGC 2841]{NGC 2841 (Fig.~\ref{fig:ngc-2841})}\label{sec:n2841}
\subparagraph{a) Center estimates \\[1.5ex]}
NGC 2841 is an early-type spiral galaxy. Its \HI\ distribution has a central
hole and the IRAC 3.6\,$\mu$m image shows a distinctive bulge and a flocculent
spiral structure. 
The radio continuum map of NGC 2841 shows a strong nuclear source coinciding within
0\farcs 2 with the position of the central source in the 
3.6\,$\mu$m IRAC image.
We used the position of the radio continuum source (\dx=$-$6\farcs 3,
\dy=0\farcs 5) as input for
\rotcur\ and made a fit with all parameters left free. We averaged the
values for \dx\ and \dy\ between $75\arcsec<r<220\arcsec$ (see first row
in Fig.~\ref{fig:ngc-2841}). The lower
limit excludes the two innermost rings, as these are sparsely filled. The
upper limit restricts our averaging to a region which is unaffected by spiral
arms or by the warp.
The dynamical center derived in such a way ($\dx=-6\farcs 5 \pm 0\farcs 6$,
$\dy=0\farcs 4 \pm 0 \farcs 9$) deviates  by less than 0\farcs 2 from
the other center estimates. Therefore, we choose the radio
continuum center as the best center position. 

\subparagraph{b) Harmonic expansion \\[1.5ex]}
The inner 200\arcsec-250\arcsec\ of NGC 2841 are in unperturbed circular rotation and 
show only small harmonic
terms. For the inner 1 kpc, we have no data as the \HI\ surface density in
this region falls below the $3\sigma$ detection limit imposed on the
velocity fields. Most of the parameters show
 a large change at $r \approx 350\arcsec-400\arcsec$: the systemic velocity
drops by about 20 km\,s$^{-1}$, and most of the harmonic components show their
highest 
amplitude at these radii. Looking at the total \HI\ intensity map, one can see
that this radius coincides with the location of a strong spiral arm. The dip in the
fitted inclination at those radii causes a non-zero $c_3$ term, induced by the
spiral arm. The radial variation of the PA and
inclination indicates that the outer disk of NGC 2841 is
warped. All harmonic components are similar in amplitude, having median
amplitudes of $3\le \Am \le 4$ km\,s$^{-1}$, or $\le 2\,$ percent of \vmax. \Ar\ is
small in the inner parts, but $> 10\ 
\mathrm{km\,s^{-1}}$ in the outer parts of NGC\ 2841. Its median value is
$\A \sim 7\, \kms$. The elongation of the potential 
is fairly constant over the radius and its weighted average is consistent with
a round potential (see Table~\ref{table:harm-decomp}).

\subsection[NGC 2903]{NGC 2903 (Fig.~\ref{fig:ngc-2903})}\label{sec:n2903}
\subparagraph{a) Center estimates \\[1.5ex]}
NGC 2903 is a barred galaxy with tightly wound spiral arms. Its dominant bar
can be seen both in the IRAC image and in the radio
continuum map. The radio continuum additionally shows a strong nuclear source,
coinciding within 1\arcsec\ 
with the center of the bulge in the 3.6\,$\mu$m image. 
The central position as
given by the radio continuum (\dx=$-$1\farcs 2, \dy=2\farcs 4 ) was used to
make a \rotcur\ run with 
all parameters left free. To derive the kinematic center, we averaged the
\dx\ and \dy\ values over the stable parts of the fit, i.e., for
$100\arcsec \leq r \leq 300\arcsec$ (see top row in
Fig.~\ref{fig:ngc-2903}). The resulting kinematic center ($\dx=0\farcs 4
\pm 2\farcs 4$, $\dy=0\farcs 6 \pm 2 \farcs 2$) agrees within
the uncertainties and within one beam with both the IRAC and the radio
continuum center (cf. second row in Fig.~\ref{fig:ngc-2903}).  
We therefore adopt the center position as derived from the radio continuum as our 
final central position.

\subparagraph{b) Harmonic expansion \\[1.5ex]}
NGC 2903 shows high amplitudes in all harmonic components in the inner
$125\arcsec$. These non-circular motions are presumably caused
by the bar in NGC 2903. The innermost 4-5 data points show the largest
deviations from the general trends, most clearly visible in the distributions
of the PA, inclination and $c_0$. 

The pronounced shape in the distribution
of $s_2$ could be caused by an $m=1$ or $m=3$ term in the gravitational
potential. Given that NGC~2903 is not kinematically
lopsided (there is, e.g., no significant difference in the rotation curves of 
the approaching and receding sides), the contribution from an $m=1$ term is
likely to be small.
Inspection of the total \HI\ maps presented in \citet{walter-07} shows
that a spiral arm is located at the same radii as the pronounced
variation in $s_2$. 
The wiggles in the radial distributions of $s_1$ and
$s_3$ at $r\sim 100\arcsec$ and $r\sim 350\arcsec$ also coincide with the
locations of spiral arms. The median amplitudes of the individual
harmonic components are small when averaged over the entire galaxy ($\Am
\le 4\ \mathrm{km\,s^{-1}}$, or 2-3 percent of \vmax), but increase towards the
center of the galaxy. Within the inner kpc, the $m=2$ component of \Am\ has a
median value of about $14\,\kms$. These high non-circular motions, which can
also be seen in the distribution of \Ar, are likely to be caused by the
strong bar, and the associated streaming motions. 
Like some of the $c_i$, $s_i$, and thus also \Ar, the elongation of the
potential shows some variations with 
radius. Its mean value is nevertheless consistent with a round potential (see
Table~\ref{table:harm-decomp}).

\subsection[NGC 2976]{NGC 2976 (Fig.~\ref{fig:ngc-2976})}\label{sec:n2976}
\subparagraph{a) Center estimates \\[1.5ex]}
The 3.6\,$\mu$m image of NGC 2976 shows no sign of a bar or spiral
arms. Nevertheless, the image shows two enhanced star forming regions at
either end of the disk which coincide with density enhancements in the
total \HI\ map.
Additionally, the IRAC image contains a nuclear source in the central parts of the galaxy,
presumably a nuclear star cluster, which has, however, no counterpart in the
radio continuum. To test whether the position of this central source
agrees with the dynamical center, we perform an unconstrained \rotcur\
run and use the position of the nuclear star cluster (\dx= 0\farcs 2, \dy=
0\farcs 1) as an
initial center estimate. By averaging \dx, \dy\ over $r<50\arcsec$ (see
Fig.~\ref{fig:ngc-2976}, top row), we derive a dynamical center which is offset
from 
the pointing center by $\dx=2\farcs 2 \pm 3\farcs 0$ and $\dy=0\farcs 8 \pm
1\farcs 4$
towards the north-west. Averaging over the entire galaxy gives consistent
results ($\dx =2\farcs 5 \pm 3\farcs 0$, $\dy=1\farcs 6 \pm 3\farcs 9$). As the
dynamical center and the 
position of the nuclear star cluster agree within the uncertainties
and to within one beam (cf. second row in Fig.~\ref{fig:ngc-2976}), we
adopt the position of the nuclear star cluster as the best
center position of NGC 2976.

\subparagraph{b) Harmonic expansion \\[1.5ex]}
The harmonic components of the velocity field of NGC 2976 show a regular
behavior. Both $s_1$ and $s_3$ scatter within $\sim 1\, \kms$ around zero, indicating that
the elongation of the gravitational potential is very small. Its weighted mean 
is consistent with zero ($ \langle
\epsilon_{\mathrm{pot}}\,\sin(2\varphi_2) \rangle =-0.010 \pm 0.018$).

The $c_2$ component is slightly offset from zero and shows
sine-like variations with radius, especially for $r>80\arcsec$, where
the two \HI\ density enhancements are located. According to
\citet{schoenmakers-thesis}, a non-zero $m=2$ harmonic component in the velocity field can be caused by an $m=1$ or $m=3$ term in the potential. NGC 2976 does not seem to be kinematically lopsided \citep[cf. the negligible differences in the rotation curves of the approaching and receding sides in][]{deblok-07}. Therefore,
there is most 
likely only a small contribution of an $m=1$ term in the potential to the $m=2$
harmonic term in the velocity field, and the significant contribution should
come from an $m=3$ term in the potential.

Beyond $r\approx 80\arcsec$, the PA stays  
constant. The median amplitudes of all harmonic component are  
small ($\Am \le 3$ percent \vmax),
irrespective of whether one averages over the entire galaxy or only over the
inner 1 kpc. The distribution of \Ar\ is also small for most
radii. Only a few data points at 
extreme radii show large amplitudes, but these are associated with large
uncertainties.
The mean elongation of the potential is, despite being unconstrained in the
inner 30\arcsec, again consistent with a round
potential (see Table~\ref{table:harm-decomp}).

\subsection[NGC 3031]{NGC 3031 (Fig.~\ref{fig:ngc-3031})}\label{sec:n3031}
\subparagraph{a) Center estimates \\[1.5ex]}
NGC 3031 (better known as M 81) is a grand design spiral. Its two well-defined
spiral arms 
are easily visible in the IRAC 3.6\,$\mu$m image as well as in the \HI\ intensity
map. 
NGC 3031 contains a strong central radio source. The 3.6\,$\mu$m image also
shows a well-defined 
central component with a central minimum which coincides with the position of
the central radio-continuum source (\dx= $-$202\farcs 7, \dy= $-$371\farcs 9).
We used the position of the radio
continuum source as input for a \rotcur\ run with all parameters
left free. As can be seen in the top row of Fig.~\ref{fig:ngc-3031},
\dx\ and \dy\ show a clear break at $r\sim400\arcsec$, probably
because of the prominent spiral arm which is located at approximately that
radius. We 
therefore estimated the dynamical center by averaging \dx\ and \dy\ over
$200\arcsec \leq r \leq 385\arcsec$, thus omitting the innermost data
points whose tilted-rings are only sparsely filled and have correspondingly
large 
uncertainties. The resulting dynamical center ($\dx 
=-205\farcs 1 \pm 3\farcs 1$, $\dy=-374\farcs 6 \pm 3\farcs 9$) coincides
within the uncertainties and within 
less than one beam with the other center estimates (second row of 
Fig.~\ref{fig:ngc-3031}) and we therefore adopt the position of the radio
continuum source as the best center position of M 81. 

\subparagraph{b) Harmonic expansion \\[1.5ex]}
The PA, $c_0$, and the inclination of M\ 81 rise steadily beyond $r \approx
400\arcsec$, indicating that the outer disk is warped or
disturbed, which is not unexpected given that M\ 81 interacts with M\ 82 and
NGC\ 3077.
The radii with the highest gradient in
$s_1$ and $s_3$ ($r\approx 350\arcsec-400\arcsec$) correspond with the location
of a prominent spiral arm. Between $400\arcsec \le r \le 700\arcsec$, $c_2$
and $s_2$ change from
values around $+5\ \mathrm{km\,s^{-1}}$ to $<-10\ \mathrm{km\,s^{-1}}$. Note that 
if we were to perform a harmonic decomposition with an unconstrained  
center position, the center would vary at those radii in order to
minimize the $c_2$ and $s_2$ term, thus underestimating the amount of
non-circular motions present in M~81. However, as we keep the center position
fixed, we are able to quantify and detect these non-circular motions.
The median amplitudes \Am\ of the $m=2$ and $m=3$ component are roughly
equally high ($\sim 5\ \kms$), whereas that of the $m=1$ component is slightly smaller ($\sim 2\ \kms$). As M81 has a
central \HI\ minimum, we have no data for the inner 1 kpc. The radial
distribution of the non-circular motions varies between $3\le \Ar \le 15$
km\,s$^{-1}$ 
and has a median of $\A \sim 9\, \kms$ ($<$5 percent of \vmax), making the
amount of non-circular 
motions quite small, despite the prominent spiral arms. 
The elongation of the potential shows distinct radial variation which is due to
the spiral arms. Its weighted average is again consistent
with zero (see Table~\ref{table:harm-decomp}). 

\subsection[NGC 3198]{NGC 3198 (Figs.~\ref{fig:n3198-1}-\ref{fig:n3198-4})} \label{sec:n3198}
\subparagraph{a) Center estimates \\[1.5ex]}
NGC\ 3198 is classified as a SBc spiral. Its 3.6\,$\mu$m IRAC image shows two
well-defined spiral arms, 
emanating from a prominent bulge. The central component has a nuclear point
source embedded, which has a counterpart in the radio continuum. The
center estimates from the IRAC image and the continuum map agree to within 
1\arcsec. We make a \rotcur\
run with the position of the radio continuum 
center (\dx= $-1$\farcs 4, \dy= $-0$\farcs 1) as an initial estimate.
Fig.~\ref{fig:n3198-1} shows that the outer parts ($r \ge
150\arcsec$) of 
NGC 3198 are strongly affected by the spiral arms. For the derivation of the
dynamical center, we therefore restrict the averaging of \dx\ and \dy\ to
radii with $r\le 100\arcsec$. The derived dynamical center ($\dx
= -1\farcs 4 \pm 1\farcs 7$, $\dy=-0\farcs 2 \pm 2\farcs 1$) agrees
well with the other estimates (see Fig.~\ref{fig:n3198-2}). We
therefore adopt the 
position of the continuum source as the best center position for NGC 3198.

\subparagraph{b) Harmonic expansion \\[1.5ex]}
The PA of NGC 3198 shows a steep increase within the inner $200\arcsec$, and then
declines slowly. The inclination varies in the inner parts
in a range of about five degrees, but shows a steady increase beyond
$r\sim 450\arcsec$, 
indicating that the outer disk is warped. The $c_3$ term is 
small, meaning that the fitted inclination is close to the intrinsic inclination of the disk. Although there is no global offset from zero for $c_2$ and
$s_2$, they show small deviations from zero for some radii which are
coinciding with the locations of spiral arms in the 3.6\,$\mu$m image or the
total \HI\ map presented in \cite{walter-07}.
The distributions of $s_1$ and $s_3$ are best described as wiggles caused by spiral arms on top of
a small offset. The global elongation of the potential is
with $ \langle \epsilon_{\mathrm{pot}}\,\sin(2\varphi_2)
\rangle =0.017 \pm 0.020$ consistent with zero.

The median amplitudes of the individual harmonic
components are
similar in amplitude, and of order $\Am \sim 2-3$ km\,s$^{-1}$, 
or $\le$ 2 percent of \vmax. 
The distribution of \Ar\ shows that the amplitude of the
non-circular motions is $\le 8\ \mathrm{km\,s^{-1}}$ for most 
radii. Its median is $\A \sim 4.5\,\kms$ when averaged over the
entire radial range, and $\A \sim 1.5\, \kms$ for the inner 1 kpc. The median
value of the absolute residual velocity field is $\sim 2.6\,\kms$, showing
that a harmonic decomposition up to third order was capable of capturing most
non-circular motions.

\subsection[IC 2574]{IC 2574 (Fig.~\ref{fig:ic-2574})}\label{sec:ic2574}
\subparagraph{a) Center estimates \\[1.5ex]}
IC 2574 has neither a central radio continuum source, nor
a clear nuclear source in the 3.6\,$\mu$m IRAC image. The dynamics of this
galaxy show clear evidence of random non-circular motions, showing themselves
as kinks in the iso-velocity contours. For the center determination, we make use of a velocity
field where most non-circular features were removed \citep[as presented in][]{se-heon}. 
We make a \rotcur\ 
run with all parameters free, using the center position derived by
\cite{se-heon} as an initial center estimate.  
The results of the \rotcur\ fit are shown in row 1 of 
Fig.~\ref{fig:ic-2574}. It can be seen that the values for \dx\ and
\dy\ do not change much over a large part of the galaxy. Averaging \dx,
\dy\ over $r\le 400\arcsec$ results in a dynamical center which is offset by
$\dx=8\farcs 7 \pm 14\farcs 9$ and $\dy=18\farcs 6 \pm 10\farcs 4$ towards
the north-west of the pointing center. This corresponds to
$\alpha_{2000}=10^h28^m27.5^s$, $\delta_{2000}=+68\arcdeg 24\arcmin 58\farcs7$ and is in
excellent agreement with the estimate derived by \cite{se-heon} ($\alpha_{2000}=10^h28^m27.7^s$,
$\delta_{2000}=+68\arcdeg 24\arcmin 59\farcs4$). 

\subparagraph{b) Harmonic expansion \\[1.5ex]}
Although we made use of the bulk velocity field from \cite{se-heon} for the
center estimate, we use the hermite velocity field for the harmonic
decomposition, as we aim to quantify non-circular motions. 
The results of the harmonic decomposition of the hermite velocity field of
IC\ 2574 are shown in the third row of Fig.~\ref{fig:ic-2574}.

At a radius of $r\sim 250\arcsec$, the PA and inclination drop significantly, and the harmonic
components jump to more extreme values. The distribution of \Ar\ has its
maximum at this radius, and $c_1$ also 
shows a clear break.
The feature which is most likely responsible for the major non-circular
motions is the 
supergiant shell in the north-east of IC\ 2574 \citep[see][]{walter-1998}. 
Looking only at the inner 1 kpc, the amplitudes of the individual harmonic components are all 
$\le1$ km\,s$^{-1}$.
If one measures the
amplitudes over the entire radial range of the galaxy, the $m=2$ component is
the dominant one, 
mainly because of the large amplitudes in $c_2$ and $s_2$ at $r\sim
250\arcsec$.

The median amplitude of the quadratically added non-circular motions is $\A
\sim 3.8\,\kms$ when averaged over the entire radial range, and $\A \sim
1.4\,\kms$ for the inner 1 kpc. This shows that although IC~2574 contain a significant 
amount of {\it chaotic} non-circular motions \citep[see][]{se-heon}, the
effect of {\it systematic} (potential induced) non-circular motions is small.
For some parts of the galaxy, the elongation of the potential is 
unconstrained. Its weighted mean is consistent with a round potential, although it
has a large uncertainty associated
($ \langle \epsilon_{\mathrm{pot}}\,\sin(2\varphi_2) \rangle =0.015 \pm
0.045$). 

\subsection[NGC 3521]{NGC 3521 (Fig.~\ref{fig:ngc-3521})}\label{sec:n3521}
\subparagraph{a) Center estimates \\[1.5ex]}
NGC 3521 appears in the 3.6\,$\mu$m IRAC image as well as in the \HI\ map as a
multi-armed, flocculent spiral galaxy. The IRAC image shows an
inner disk, in
which a strong nuclear point source (without a counterpart in the radio
continuum) is embedded. We use the position of the nuclear point
source (\dx= $-$8\farcs 8, \dy= 5\farcs 7) as input for a \rotcur\
fit with all parameters left free. 
The top row of Fig.~\ref{fig:ngc-3521} shows that the values
of \dx\ and \dy\
stay constant for radii smaller than 220\arcsec. Beyond that
radius, the rotation curves for the approaching and receding sides start to
differ \citep[see][]{deblok-07} and this most likely also affects the derived
central positions. 
In order to derive a dynamical center, we average \dx\ and \dy\ for $r \le$
220\arcsec. The resulting center ($\dx =-8\farcs 9 \pm 0\farcs 8$,
$\dy=6\farcs 6 \pm 1\farcs 4$) agrees with the central position 
found in the IRAC image (see second row of Fig.~\ref{fig:ngc-3521})
and we adopt the IRAC position as the best center estimate.

\subparagraph{b) Harmonic expansion \\[1.5ex]}
As can be seen, e.g., in the
radial distribution of PA, $i$, and \Ar, the inner few data points clearly deviate from the rest. 
Note that the $c_3$ component is
higher for these points, meaning that their inclination could not
be fitted correctly in an unconstrained fit. Some of these inner data points
show large non-circular motions, while others (e.g., the
inner two points) do not. This becomes even clearer if one looks at the
distribution of \Am. Although the median of the $m=3$
component in the inner 1 kpc is small, its upper quartile
is rather large, as the inner 1 kpc contain only three data points, one of
which has a large amplitude. 
The $s_1$ and $s_3$ components are both slightly offset from zero 
and superimposed on a wiggle at $r\approx 200\arcsec$. The wiggle coincides
with 
a density enhancement in the total \HI\ intensity map (in this case a spiral
arm). However, as $s_1$ is offset to negative values, and $s_3$ to positive
ones, their offsets almost cancel out for the derivation of the elongation of
the potential (cf. Eq.~\ref{eq:epot}), whose weighted mean is within the
uncertainties consistent with zero ($ \langle \epsilon_{\mathrm{pot}}\,\sin(2\varphi_2) \rangle =0.017 \pm 0.019$).

The effect of the spiral arms is clearly visible in the $c_2$ and $s_2$ components. NGC\ 3521 has the highest median value in the absolute residual velocity field of our
sample (4.5 $\mathrm{km\,s^{-1}}$).

\subsection[NGC 3621]{NGC 3621 (Fig.~\ref{fig:ngc-3621})}\label{sec:n3621}
\subparagraph{a) Center estimates \\[1.5ex]}
The 3.6\,$\mu$m IRAC image of NGC 3621 is dominated by a flocculent
spiral structure, which contains a central point source. The prominent
star forming region in the south-west is connected to the center of
the galaxy by a spiral arm which is clearly visible in the total \HI\
map and the IRAC image. While
this star forming region has a counterpart in the radio continuum, the
IRAC nuclear point source has not.
We use the coordinates of the nuclear point source in the IRAC image (\dx=
 $-$4\farcs 2, \dy= $-$1\farcs 1) as an initial estimate for a \rotcur\ fit
 with all parameters left free. The results (top row of
 Fig.~\ref{fig:ngc-3621}) show that the derived central positions
are fairly constant for radii smaller than 450\arcsec\ and we therefore
derive the dynamical center by averaging over all points with
$r<450\arcsec$, with the additional exclusion of the innermost data point. The
resulting dynamical center ($\dx =-6\farcs 0 \pm 2\farcs 
0$, $\dy=1\farcs 4 \pm 6\farcs 5$) coincides with the
coordinates of the point source both within the uncertainties and the size of
the beam (Fig.~\ref{fig:ngc-3621}, second row). Therefore, we adopt the
position of the central point source as our best center estimate for NGC 3621.

\subparagraph{b) Harmonic expansion \\[1.5ex]}
The data points beyond $r = 600\arcsec$ are not used for our analysis
due to their sparsely filled tilted-rings.
The PA of NGC 3621 is, except for the innermost two data points, 
constant to within a few degrees. Apart from a few data points at
extreme radii, the amplitude of $s_1$ is almost negligible. 
$c_3$ is close to zero, 
except for $r\approx 300\arcsec$. This position also coincides with the radius
of the highest amplitude in $c_2$, and $s_2$.
Inspection of the total \HI\ intensity map shows that there is a ring-like \HI\
density enhancement at which a spiral arm emerges in the south-east. The $s_3$
component shows clear wiggles, indicating spiral arms, and a steep rise
beyond $r\approx 500\arcsec$, where the tilted-rings are less filled. The
elongation of the potential again shows the kinematic signature of 
the spiral arms. Its weighted average is consistent with zero
(see Table~\ref{table:harm-decomp}).
The median amplitudes for each individual harmonic component are 
fairly small ($\Am <\,$3 km\,s$^{-1}$, or $\le 2$ percent of \vmax). The
distribution of \Ar\ is
$\le 5\ \mathrm{km\,s^{-1}}$ for the majority of radii and has a median
value of $\A \sim 3.4\ \mathrm{km\,s^{-1}}$. 

\subsection[NGC 3627]{NGC 3627 (Fig.~\ref{fig:ngc-3627})}\label{sec:n3627}
\subparagraph{a) Center estimates \\[1.5ex]}
The 3.6\,$\mu$m IRAC image shows that NGC 3627 is a barred galaxy with two
asymmetric spiral arms, which are also 
clearly visible in the total \HI\ map. The central parts of NGC 3627 contain a nuclear point 
source which is 
visible in the IRAC image as well as in the radio continuum. Using the 
position of the point source ($\dx=0\farcs 1$, $\dy=0\farcs 1$) as input
for an unconstrained 
\rotcur\ fit results in a slightly different center estimate (see row
1 and 2 of 
Fig.~\ref{fig:ngc-3627}). We exclude some data points at extreme radii
because of their large uncertainties, and
average \dx, \dy\ for $45\arcsec \le r \le 160\arcsec$, resulting in a
dynamical center which is offset  
from the pointing center by $\dx=-4\farcs 5 \pm 3\farcs 1$ and
$\dy=-6\farcs 8 \pm 4\farcs 8$. These values deviate
from the radio continuum source by approximately $1.5 \sigma$. Given the
asymmetric appearance
of NGC\ 3627, it is to be expected that the results from \rotcur\ are
somehow affected by it. However, the difference between
the kinematic center and the center from the radio continuum is only one
beam. We therefore use the coordinates of the central radio continuum
source as the best center position for NGC 3627.

\subparagraph{b) Harmonic expansion \\[1.5ex]}
NGC 3627 has a central hole in the \HI\ distribution. Therefore, we see only the
flat part of the rotation curve and have no data within the inner kpc. The
systemic velocity changes from $\sim$730 km\,s$^{-1}$ at $r\le 60\arcsec$ to
$\sim$705 km\,s$^{-1}$ for $r\geq 90\arcsec$. All harmonic components
($c_2,\ c_3$, $s_1,\ s_2,\ s_3$) show 
high amplitudes. The galaxy is kinematically and 
morphologically lopsided \citep[cf.,][]{deblok-07}, which is also indicated
by the non-zero values for $c_2$ and $s_2$. 
Both the $s_1$ and the $s_3$
component show a ``dip'' at $r\approx 70\arcsec$, followed by a steep rise
beyond that radius. In the case of $s_3$, the rise is followed by a
second decline. The radius of that first ``dip'' coincides with the location of
the large arm which extends to the south of NGC 3627. The median
amplitudes of the individual 
harmonic components are the highest in our sample, reaching from $\Am \sim 10$
km\,s$^{-1}$ for $m=3$ to $\Am \sim 17$ km\,s$^{-1}$ for $m=2$. The
distribution of \Ar\ shows that the amplitude of non-circular motions is large
at all radii. The median amplitude is $\A \sim 28.5 \ \mathrm{km\,s^{-1}}$, or
up to 14.7 percent of \vmax. Given the 
morphology of NGC 3627, this is not unexpected. Despite
the high non-circular motions present here, the weighted mean elongation
of the potential is still small, and within its uncertainty consistent with
a round potential (cf. Table~\ref{table:harm-decomp}), although there are 
large errors 
associated with the data points in the outer parts and in the region at
$r\approx 70\arcsec$. 

\subsection[NGC 4736]{NGC 4736 (Fig.~\ref{fig:ngc-4736})}\label{sec:n4736}
\subparagraph{a) Center estimates \\[1.5ex]}
The 3.6\,$\mu$m IRAC image of NGC 4736 is dominated by a ring of star formation
which is also prominently visible in the
radio continuum and the total \HI\ map. The center is well-defined
both in the 3.6\,
$\mu$m image and in the radio continuum. The former shows a small central
\emph{minimum} in the flux which coincides within 1\farcs 3 with
the central radio-continuum \emph{maximum}. We use the position of the
radio continuum source ($\dx=0\farcs 2$, $\dy=-0\farcs 4$) for a
\rotcur\ run with all parameters left free. The fitted center positions
are stable within the inner 100\arcsec, but have an increasingly large scatter
towards the outer parts, most likely due to the marginally filled tilted-rings 
beyond $r\sim 100\arcsec$, and also because of the low inclination of NGC 4736. 
We therefore average the estimates for \dx\ and \dy\ over the inner
100\arcsec, resulting in a dynamical center ($\dx=0\farcs 5 \pm 1\farcs 8$,
$\dy=0\farcs 6 \pm 2\farcs 0$) whose deviation from the other center estimates
slightly exceeds the uncertainties (1.2$\sigma$ deviation for \dy). Though, to
put things into perspective, this is far smaller than one beam (see second row
of Fig.~\ref{fig:ngc-4736}). Thus, we will use
the position of the radio continuum source as the best center estimate for NGC 4736.

\subparagraph{b) Harmonic expansion \\[1.5ex]}
The star forming ring at $r\sim 80\arcsec$ causes a lot of confusion in
the inner parts of NGC 4736. Here the unconstrained fit by \reswri\
produces an apparent rotation velocity which rises to twice the
value in the outer parts. The other harmonic components also show high
amplitudes. 
The median amplitudes of each harmonic component (averaged over the inner 1
kpc) are negligible for the $m=1$ and $m=3$ component ($\Am \sim 1$
km\,s$^{-1}$), but much higher for the $m=2$ component ($\Am \sim 7$
km\,s$^{-1}$). The
amplitude of the quadratically added non-circular motions, \Ar, is
$\sim 10$ km\,s$^{-1}$ for most radii.

NGC 4736 is both
kinematically and morphologically lopsided, which can be seen not only in the
differences between the rotation curves of the approaching and receding sides
\citep[see][]{deblok-07}, but also in the high amplitudes in the $c_2$ and
$s_2$ terms of our harmonic expansion. The highest amplitudes in the 
harmonic terms are found for radii between $r\approx200\arcsec$ and
$r\approx 300\arcsec$, i.e., those radii where the large northern spiral arm is
located.
The weighted average of the 
elongation of the potential is $ \langle
\epsilon_{\mathrm{pot}}\,\sin(2\varphi_2) \rangle =-0.055 \pm 0.149$. 

\subsection[DDO 154]{DDO 154 (Fig.~\ref{fig:ddo-154})}\label{sec:ddo154}
\subparagraph{a) Center estimates \\[1.5ex]}
DDO 154 is a quiescent, gas-rich dwarf galaxy with a warp in the outer
parts. It does not have a compact, well-defined center in the 3.6\,$\mu$m IRAC
image, or in the radio continuum. 
We attempted to derive a center position by fitting ellipses at a few representative intensity levels to the
IRAC image, and the total \HI\ map. The resulting central positions showed a variation of up to 
15\arcsec, demonstrating that deriving an unambiguous {\it photometric} center
for DDO 154 is not straightforward. 
We used the center which was derived using the total \HI\ map as a initial estimate for a \rotcur\ fit with all parameters left free.
The central positions as derived by \rotcur\ are fairly stable over
the radial range of the galaxy and the kinematic center was derived by
averaging the \dx, \dy\ values over $100\arcsec\le r \le 260\arcsec$, thus
omitting the slightly more 
unstable inner and outer regions. The
resulting center position is given in Table \ref{table:center-pos}. 
Including the inner data points gives a similar center position, but with two
times larger scatter.
We therefore adopt the dynamical center as averaged over
$100\arcsec\le r \le 260\arcsec$ as the best center position of DDO 154.

\subparagraph{b) Harmonic expansion \\[1.5ex]}
The $c_3$ term is close to zero for the entire radial range, normally
indicating that the inclination could be well-determined.
However, due to the nearly solid-body rotation in the innermost parts of
DDO~154, the inclination is not very well-constrained in this region, as can
be seen in the large 
scatter of the individual data points for $r\le 70\arcsec$. 
This is studied in more detail in Section~\ref{sec:sanity-checks}. 
The $c_2$ component shows a maximum at around $r\approx 80\arcsec$ which
coincides with a minimum in the $s_2$ component and a change in the
systemic velocity $c_0$, providing evidence for a small kinematic
lopsidedness as supported by the variation of the center
position 
at these radii and the differences
between the approaching and receding sides of the rotation curve \citep[see][]{deblok-07}. However, the effects described above are small.
Despite the small-scale structure present in the velocity field, the amplitudes
of all harmonic components --- both for the entire galaxy and for the innermost 1
kpc --- are $\le 1\ \mathrm{km\,s^{-1}}$, or $\sim 2$ percent of \vmax. The
distribution of \Ar\ shows that the amplitudes of non-circular motions in
DDO\ 154 are small for all radii. The mean elongation
of the potential is small, $\langle
\epsilon_{\mathrm{pot}}\,\sin(2\varphi_2) \rangle =0.024 \pm 0.033$, and
consistent with zero. The median of the absolute residual velocity field
is 1.2 $\mathrm{km\,s^{-1}}$ (the smallest in our sample).

\subsection[NGC 4826]{NGC 4826 (Fig.~\ref{fig:ngc-4826})}\label{sec:n4826}
\subparagraph{a) Center estimates \\[1.5ex]}
NGC 4826 shows no sign of spiral structure in the 3.6\,$\mu$m IRAC
image, but has two counter-rotating gas disks \citep{braun-1994}. The center of
NGC 4826 is well-defined by the central, compact source visible in
the radio continuum, and the IRAC 3.6\,$\mu$m image, whose coordinates agree to
within 1\arcsec. We use the position of the radio continuum source
($\dx=2\farcs 4$, $\dy=8\farcs 5$) as an
initial estimate for an unconstrained \rotcur\ fit. 
We derive the dynamical center by
averaging the \dx, \dy\ values inwards of 60\arcsec, i.e., over the radial
range of the inner disk. The kinematic center derived in such a way
($\dx=2\farcs 2 \pm 1\farcs 6$, $\dy=7\farcs 5 \pm 0\farcs 8$)
agrees well with the other center estimates and we therefore assume the
position of the radio continuum source to be the best center for NGC 4826.

\subparagraph{b) Harmonic expansion \\[1.5ex]}
Because of the sparsely filled tilted-rings, and the additional
complication of the two
counter-rotating disks, we have not attempted to derive a harmonic decomposition for this galaxy.

\subsection[NGC 5055]{NGC 5055 (Fig.~\ref{fig:ngc-5055})}\label{sec:n5055}
\subparagraph{a) Center estimates \\[1.5ex]}
NGC 5055 is classified as an Sbc galaxy and has a flocculent
structure in the 3.6\,$\mu$m IRAC image. The IRAC image shows a
compact nucleus with a 
well-defined center, whose position coincides to within $1\arcsec$ with a
faint central radio continuum source. We use the position of the latter 
($\dx=8\farcs 8$, $\dy=0\farcs 4$) as the
input center for a \rotcur\ run with all parameter left free. 

We derive the dynamical center by averaging the
\dx, \dy\ values for radii smaller than 450\arcsec\ to exclude the parts
with sparsely filled tilted-rings beyond $r\sim 450\arcsec$. This center ($\dx=8\farcs
2 \pm 4\farcs 8$, $\dy=0\farcs 4 \pm 2\farcs 0$) 
coincides within the uncertainties and to within a beam size with the other
center estimates. We therefore adopt the position of the radio continuum
source as the best estimate for the central position of NGC 5055.

The \HI\ distribution and kinematics of NGC 5055 were recently
analyzed by \citet{battaglia-2006}. They define a dynamical center
coinciding with their optical center at 
$\alpha_{2000}=13^h15^m49.25^s$, $\delta_{2000}=+42\arcdeg 01\arcmin 49\farcs3$, which 
is $4\arcsec$ to the north of
our choice. As our choice was also guided by the position of the
central source in the 3.6 $\mu$m image, this would imply that the
respective photometric centers must be shifted by 4\arcsec\ with respect to
each other. The optical center position listed in \citet{battaglia-2006}
is ultimately based on a listing in \citet{maoz-1996} of the position of
the central UV source in NGC 5055 determined using an HST FOC image.
On examination of the relevant image in the Hubble Space Telescope
archive, we find the central source to be located at 
$\alpha_{2000}=13^h15^m49.3^s$, $\delta_{2000}=+42\arcdeg 01\arcmin 46\farcs2$, which is in much closer
agreement (0\farcs8) with our position.

\subparagraph{b) Harmonic expansion \\[1.5ex]}
For the harmonic analysis of NGC 5055, a distinction between the inner and the
outer parts has to be made. NGC 5055 has a well-defined \HI\ disk which
extends to about $450\arcsec$. Beyond that radius, the \HI\ column density is
lower and generally falls below the 3\,$\sigma$ column density limit which 
we imposed during the construction of the velocity fields
\citep[see][]{deblok-07}. Therefore, the uncertainties
increase rapidly and we 
considered only data points with radii
$r<450\arcsec$ for the radial averaging of the parameters.
The median amplitudes of the individual harmonic components (averaged over
$r<450\arcsec$) are all fairly low ($\Am < 3\ 
\mathrm{km\,s^{-1}}$ or $\sim 2$ percent of \vmax). The median amplitudes averaged
over the inner 1 kpc contain only three data points and are also low for
the $m=1$ and $m=3$ components, but quite high ($\sim 8\ \mathrm{km\,s^{-1}}$)
for the $m=2$ component. 
The latter has a large scatter associated with it, mainly
because of the innermost data point which shows high non-circular motions, but
is derived from an only partially filled tilted-ring.
The global elongation of the potential is well-constrained in the inner part
of NGC 5055. The weighted mean 
elongation within 450\arcsec\ is again consistent with a round
potential (see Table~\ref{table:harm-decomp}). 

\subsection[NGC 6946]{NGC 6946 (Fig.~\ref{fig:ngc-6946})}\label{sec:n6946}
\subparagraph{a) Center estimates \\[1.5ex]}
NGC 6946 is a late-type spiral galaxy with well-defined multiple
spiral arms. The IRAC 3.6\,$\mu$m image shows a point
source in the center of the galaxy which coincides to better than
1\arcsec\ with the central radio continuum source. We 
use the position of the latter ($\dx=-1\farcs 8$,
$\dy=14\farcs 5$) as input for a \rotcur\ fit
with all parameters left free and determine the dynamical center by averaging
the \dx, \dy\ values for $22\arcsec\le r \le 290\arcsec$, thus excluding
the two innermost points which are affected by sparsely filled 
tilted-rings. The 
resulting dynamical center ($\dx=-3\farcs 3 \pm 4\farcs 4$,
$\dy=11\farcs 8 \pm 5\farcs 8$) coincides within the uncertainties and to
within one beam with the other center estimates (see
Fig.~\ref{fig:ngc-6946}, row 2). The large uncertainties in the
kinematic estimate are caused
by the relatively low inclination of NGC 6946, which makes an unconstrained
tilted-ring fit of the velocity field more difficult. We adopt the
position of the central radio continuum source as the best center for
NGC 6946.

\subparagraph{b) Harmonic expansion \\[1.5ex]}
NGC 6946 is the galaxy in our sample with the lowest inclination. This
introduces larger uncertainties in the rotation curve and the harmonic
decomposition. 
For our analysis of the median amplitudes,
and for the derivation of the average elongation,
we regard only the inner $420\arcsec$, i.e., the region where the
unconstrained apparent circular velocity $c_1$ does not rise to a value 
two times that in the flat part. This radius 
also coincides with the location from which, when moving outwards, the
approaching and receding sides of the 
velocity field start to differ substantially \citep[see][]{deblok-07}. The
median amplitudes averaged over $r<420\arcsec$ are small for each harmonic
component ($\Am \le 5$ km\,s$^{-1}$).
Due to the deficiency of \HI\ in the 
center of NGC 6946, we can not study the kinematics within the inner 1 kpc. 
Again averaging only over $r<420\arcsec$, the median of the quadratically
added amplitudes of all non-circular 
components is $\A \approx 7\ \mathrm{km\,s^{-1}}$.
The steep rise in $s_2$ and $s_3$ at $r\approx 120\arcsec$ is probably caused by the
spiral arm which crosses at that radius. 
Due to the low inclination of NGC\ 6946, and the associated large error in the
derivation of the 
fitted parameters, the elongation of the potential is not well-constrained
for $r\le 150\arcsec$ and $r\geq 420\arcsec$. Its weighted mean (again
averaged over $r\le 420\arcsec$), however, is consistent with a round
potential. 

\subsection[NGC 7331]{NGC 7331 (Fig.~\ref{fig:ngc-7331})}\label{sec:n7331}
\subparagraph{a) Center estimates \\[1.5ex]}
The 3.6\,$\mu$m IRAC image of NGC 7331 shows prominent spiral arms, and
a well-defined nucleus which has no counterpart in the radio continuum. 
We use the center from the 3.6\,$\mu$m image ($\dx=0\farcs 1$, $\dy=0\farcs
3$) as input for an unconstrained fit with \rotcur. 
Ignoring the two innermost points as their tilted-rings are only
sparsely filled, we
average the values for \dx\ and \dy\ over the stable region with $r\le 
185\arcsec$. This results in a
dynamical center of $\dx=0\farcs 0 \pm 0\farcs 6$, 
$\dy=-1\farcs 8 \pm 2\farcs 2$, which agrees within the uncertainties and to
within one beam with the IRAC center. Therefore, we 
adopt the IRAC center as our best center position for NGC 7331.

\subparagraph{b) Harmonic expansion \\[1.5ex]}
Although PA and inclination do not change much over the radial range
($\sim$4-6 degrees), 
they both increase outwards, indicating a warping of the disk. The 
small $c_3$ values denote that the inclination could be well-determined 
for most radii. The $s_1$ and $s_3$ values both show a wiggle
at $r\approx 150\arcsec$, coinciding 
with the location of a spiral arm visible in the total intensity \HI\ map and in
the IRAC 3.6\,$\mu$m image. The variation of $c_0$, $c_2$, and $s_2$ shows that
NGC 7331 is kinematically lopsided, as shown by the
differences in the rotation curves of the receding and 
approaching sides of the velocity field \citep[cf.,][]{deblok-07}.
The median amplitudes of the individual harmonic components are all
small ($\Am < 5\ \mathrm{km\,s^{-1}}$).
Because of the central \HI\ deficiency of NGC 7331, there is 
no kinematic data for the inner 1 kpc. The distribution of \Ar\ varies between
$\Ar \sim 2\ \mathrm{km\,s^{-1}}$ and $\Ar \sim 10\ \mathrm{km\,s^{-1}}$ and
has a median of $\A \sim 6\ \mathrm{km\,s^{-1}}$.
The elongation of the potential is
well-determined and close to zero for most radii. The wiggle at $r\approx
150\arcsec$ is again an indication of the aforementioned spiral arm. The
weighted mean elongation is $ \langle \epsilon_{\mathrm{pot}}\,\sin(2\varphi_2)
\rangle =-0.003 \pm 0.017$, and therefore consistent with a round potential.

\subsection[NGC 7793]{NGC 7793 (Fig.~\ref{fig:ngc-7793})}\label{sec:n7793}
\subparagraph{a) Center estimates \\[1.5ex]}
NGC 7793 is a flocculent spiral whose 3.6\,$\mu$m IRAC image shows a
well-defined central source without a counterpart in the radio continuum. We
adopt the position of this source 
($\dx=-9\farcs 1$, $\dy=-0\farcs 9$) as the initial estimate for an
unconstrained \rotcur\ fit, and average the \dx, \dy\ values 
over the stable part of the galaxy ($r\le 180\arcsec$). The resulting
kinematic center 
($\dx=-9\farcs 9 \pm 2\farcs 6$, $\dy=1\farcs 8 \pm 2\farcs 1$) agrees
reasonably well with the IRAC center (1.5$\sigma$ deviation). However, the
deviation between these two center estimates is $\sim 6$ times smaller than the size of
one beam and we therefore use the IRAC center as our best center position.

\subparagraph{b) Harmonic expansion \\[1.5ex]}
The PA and the inclination vary continuously with radius, indicating
that the disk of NGC 7793 might be warped. The systemic velocity, $c_0$,
stays fairly constant over radius. 
The offset from zero in the $s_2$ component might
indicate a slight kinematic lopsidedness, which is also visible in the
difference between the receding and approaching sides of the rotation curve in
\cite{deblok-07}. 
The $s_3$ term is
significantly larger than the $s_1$ term, which is typical for galaxies with
low inclination \citep{schoenmakers-thesis}. In fact, the $s_3$ component
seems to be offset from zero, indicating through Eq.~\ref{eq:epot} an
elongation of the potential. The weighted mean elongation is
$ \langle \epsilon_{\mathrm{pot}}\,\sin(2\varphi_2) \rangle =-0.067 \pm 0.085$,
which is the largest elongation we 
have measured in our sample. It is due to its large uncertainty consistent
with both the CDM predictions by \cite{hayashi-2007} and with a round potential. 
The large uncertainty for the elongation arises from the relatively large
error bars of the inclination values, which enter into the uncertainty in
$\epsilon_{\mathrm{pot}}$ as a fourth power (cf. Eq.~\ref{eq:epot}).
The median amplitude of each 
harmonic component is fairly low ($\Am <4\ \mathrm{km\,s^{-1}}$), both for the
entire 
galaxy and for the inner 1 kpc. The
distribution of \Ar\ shows that the amplitude of non-circular motions is
small, especially in the inner parts of 
NGC\ 7793. Its median value (averaged over the entire galaxy) is $\A \sim 5$
km\,s$^{-1}$. 
The median of the absolute residual velocity field is
2.2 $\mathrm{km\,s^{-1}}$, again showing that a harmonic
expansion up to third order did cover most of the non-circular motions.

\section{The atlas}\label{sec:first-figure}
\begin{figure*}[t!]
\begin{center}
\includegraphics[angle=0,width=0.50\textwidth,bb=18 144 592 520,clip=]{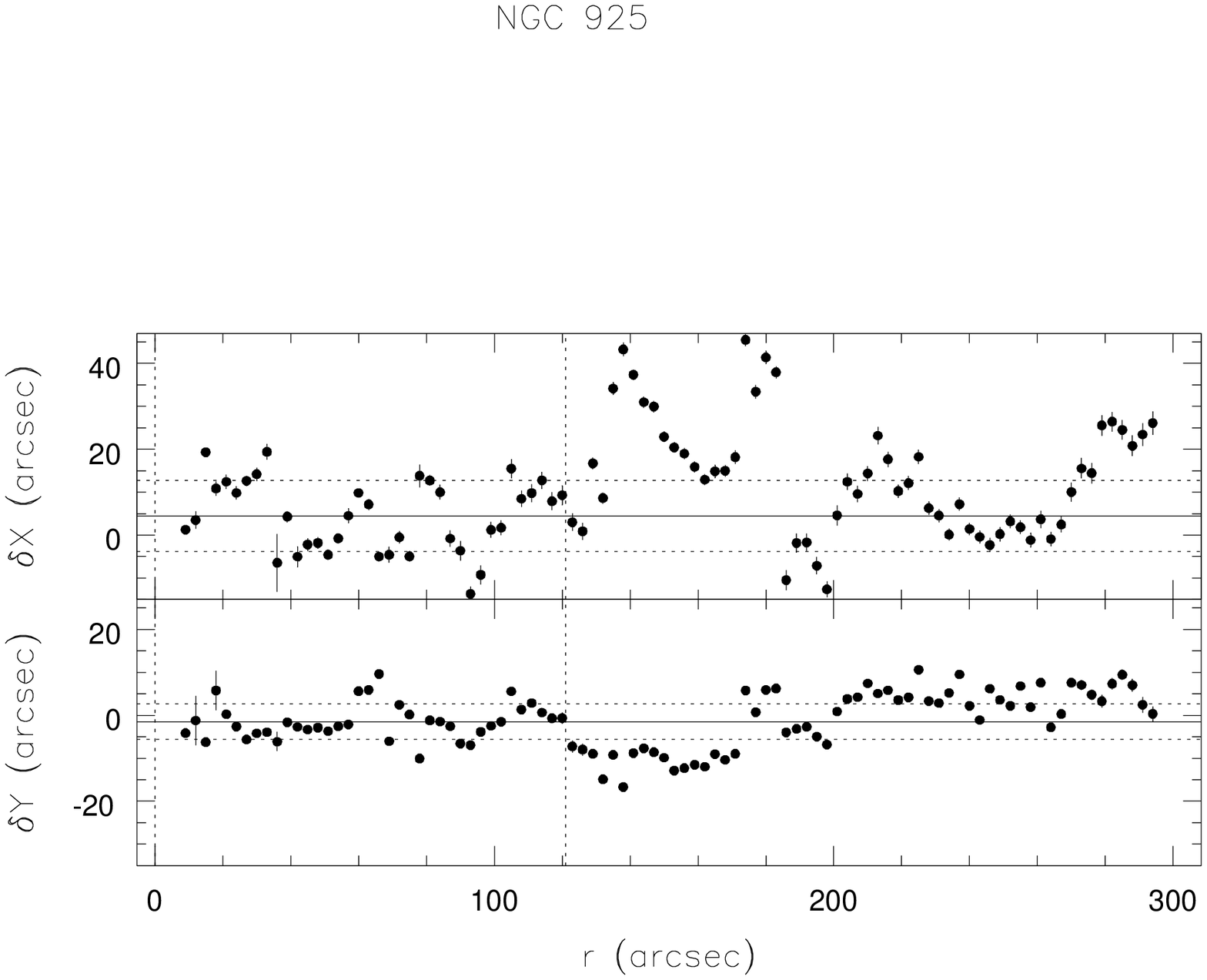}
\includegraphics[angle=0,width=0.55\textwidth,bb=18 79 520 520,clip=]{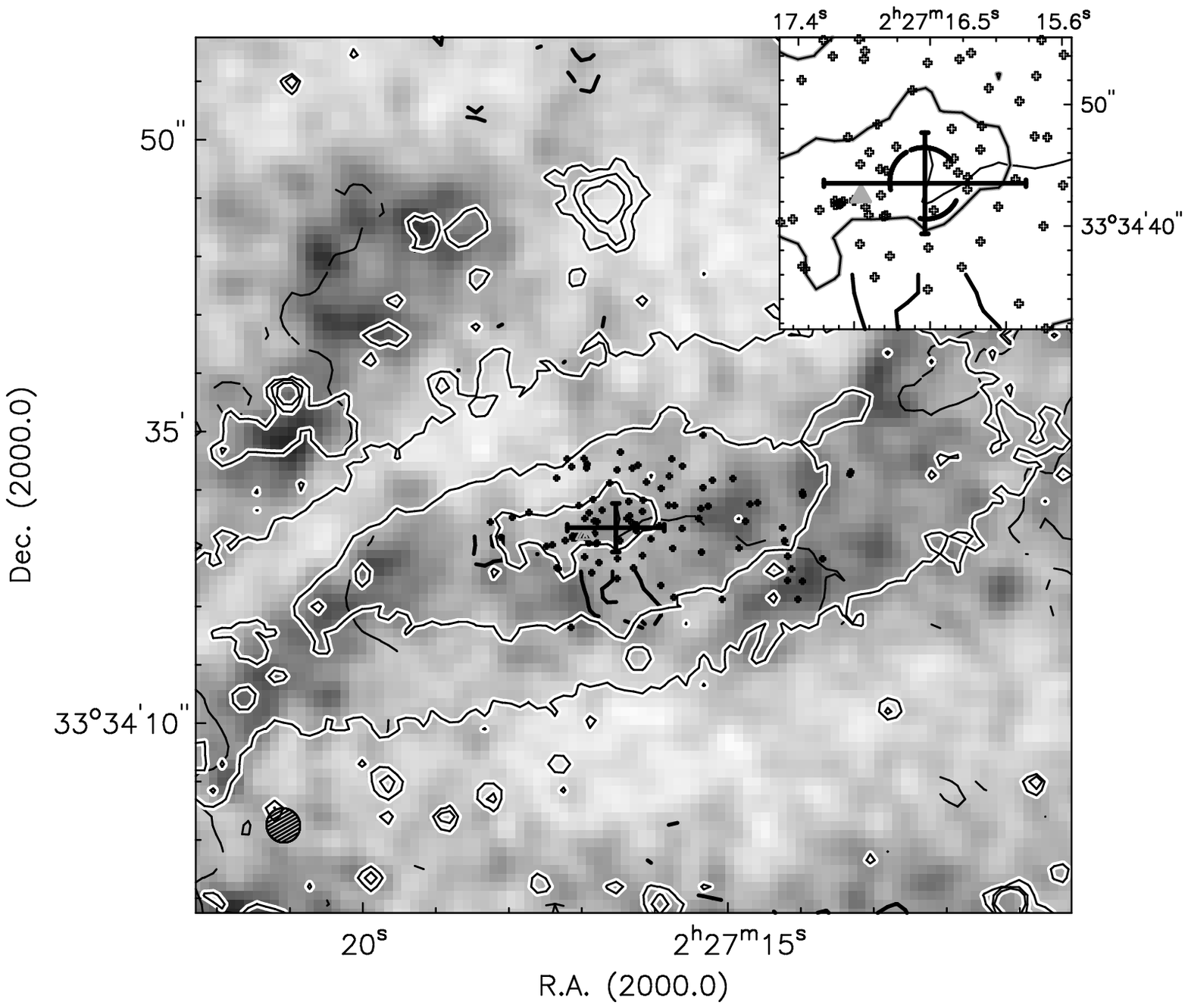}\\
\includegraphics[angle=0,width=0.65\textwidth,bb=19 235 592 697,clip=]{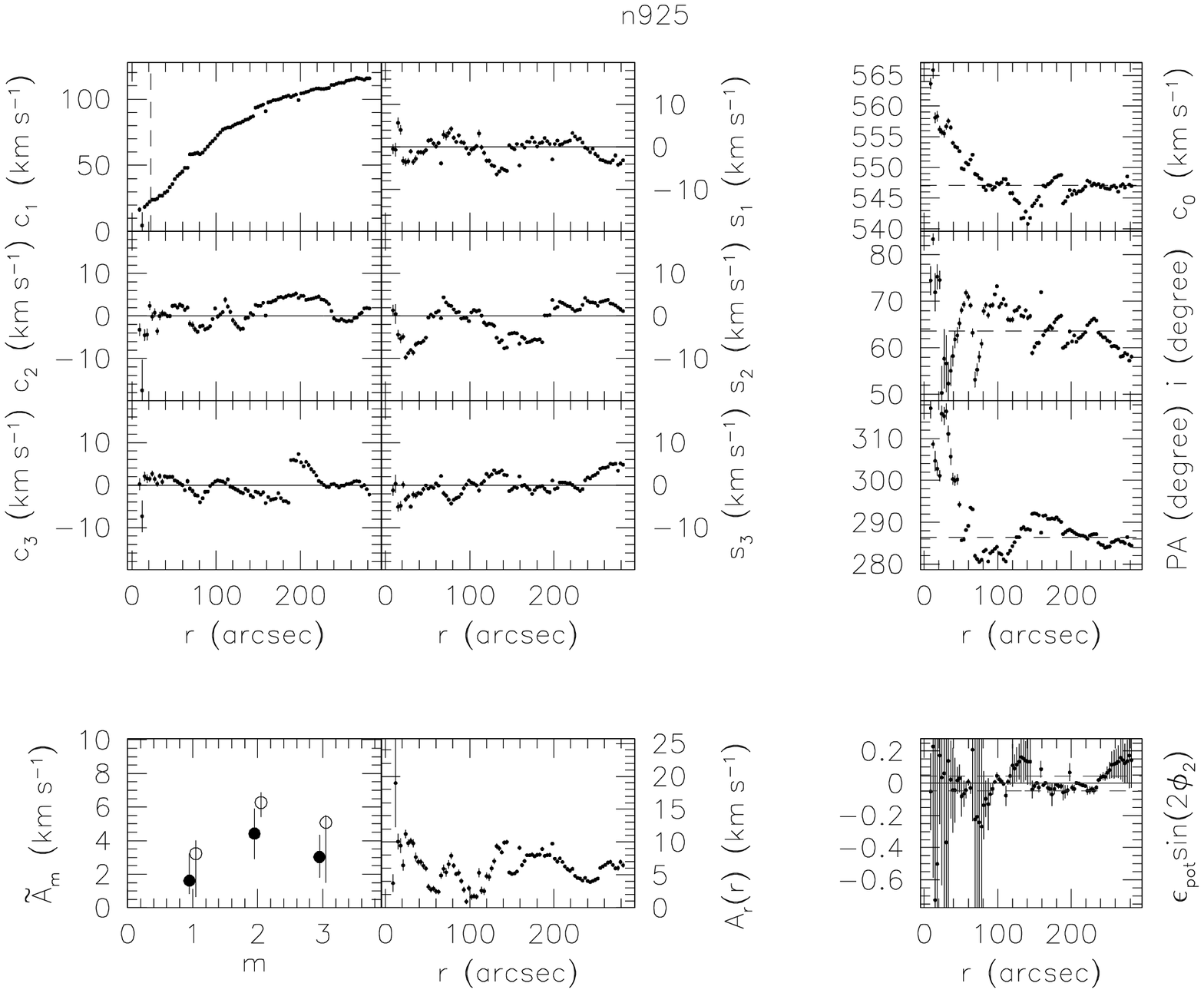}
\caption[Summary panel for NGC 925]{Summary panel for NGC 925. Lines and symbols are described in the text, Appendix~\ref{sec:first-figure}. The IRAC
  center shown was derived using {\sc ellfit}. We
  omitted the IRAC contours at the 2\% and 5\% levels
  for clarity reasons. See Appendix \ref{sec:n925} for a discussion of this galaxy.} \label{fig:ngc-925}
\end{center}
\end{figure*}

\begin{figure*}[t!]
\begin{center}
\includegraphics[angle=0,width=0.50\textwidth,bb=18 144 592 520,clip=]{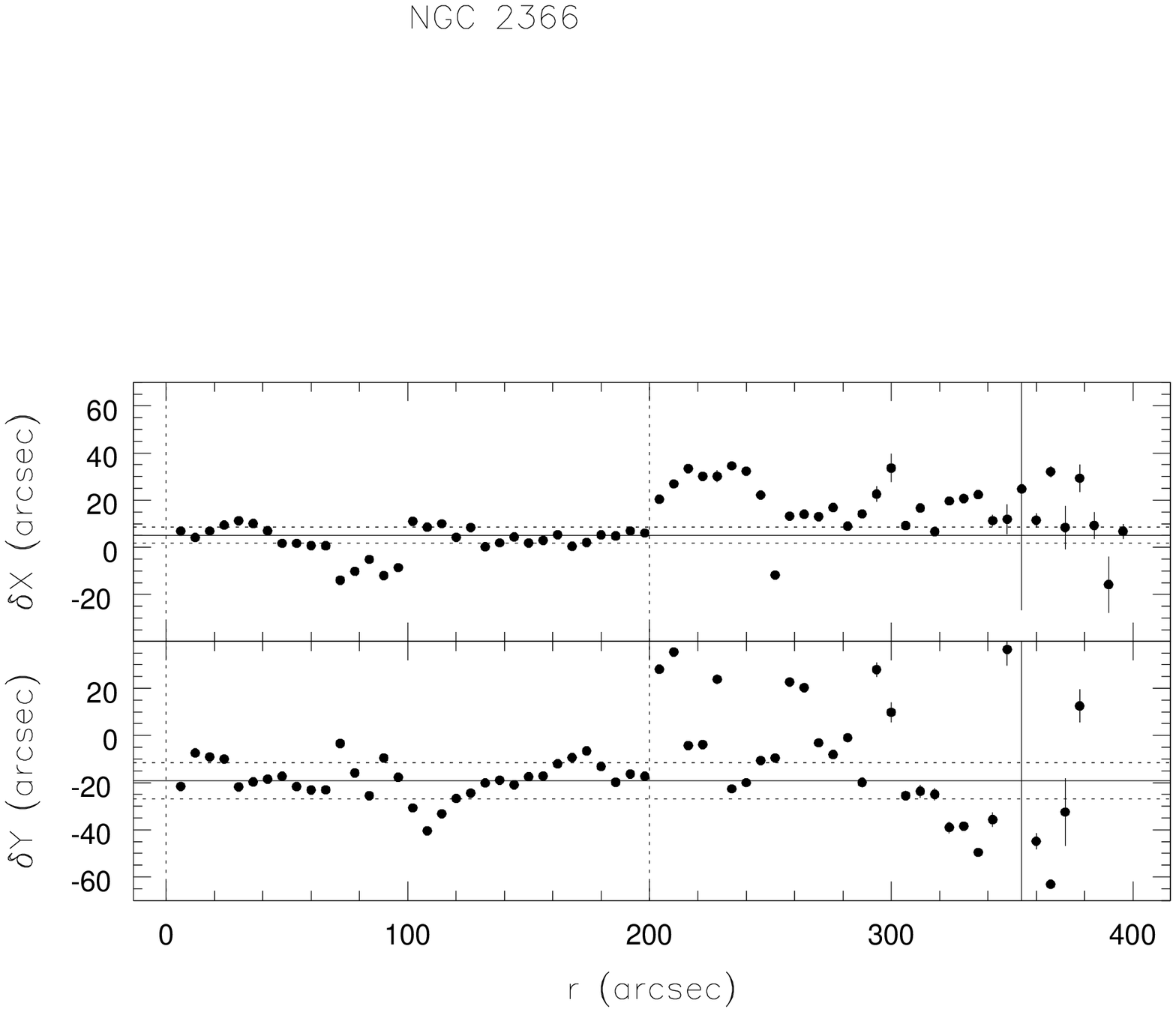}
\includegraphics[angle=0,width=0.55\textwidth,bb=18 79 520 520,clip=]{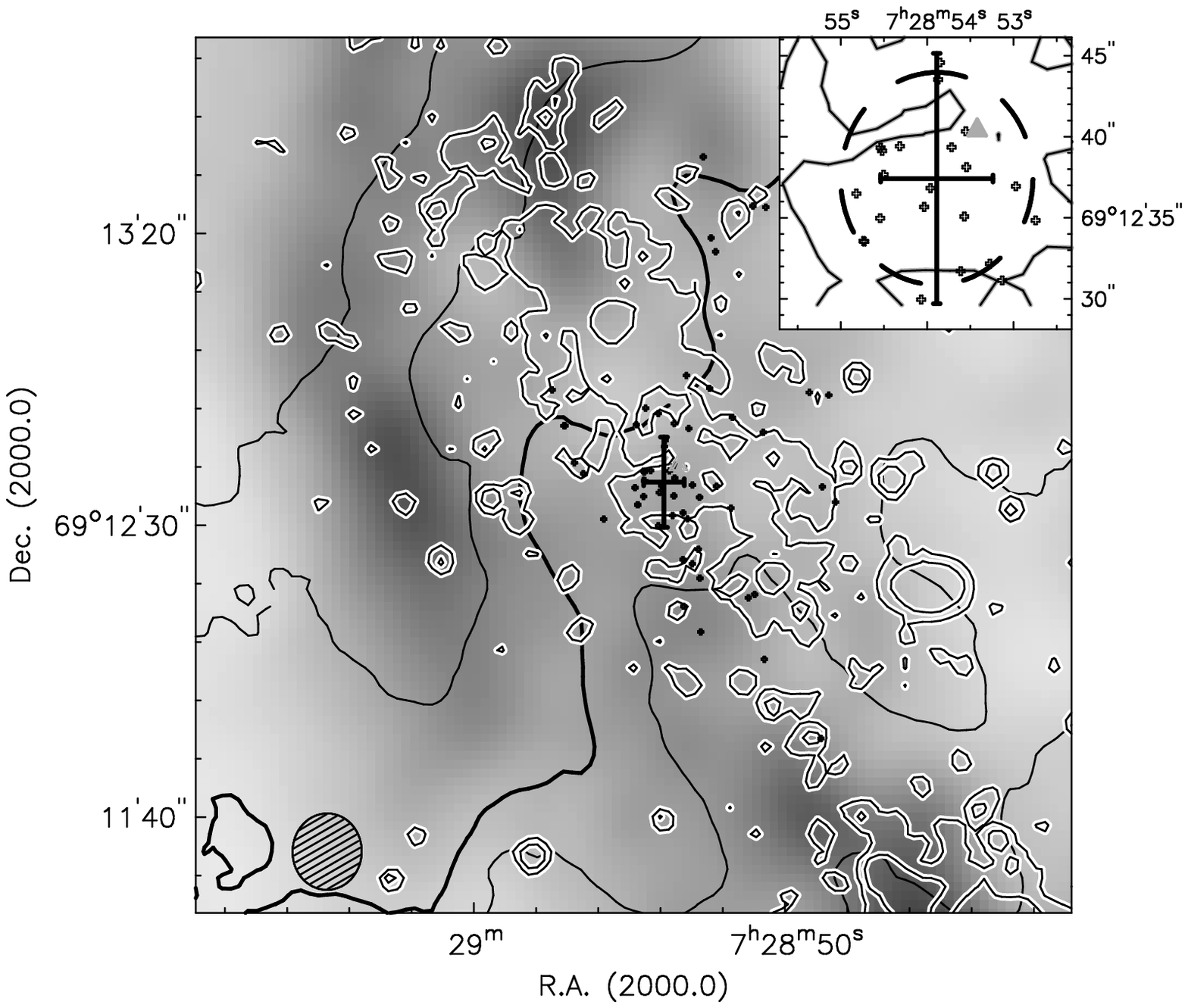}\\
\includegraphics[angle=0,width=0.65\textwidth,bb=19 235 592 697,clip=]{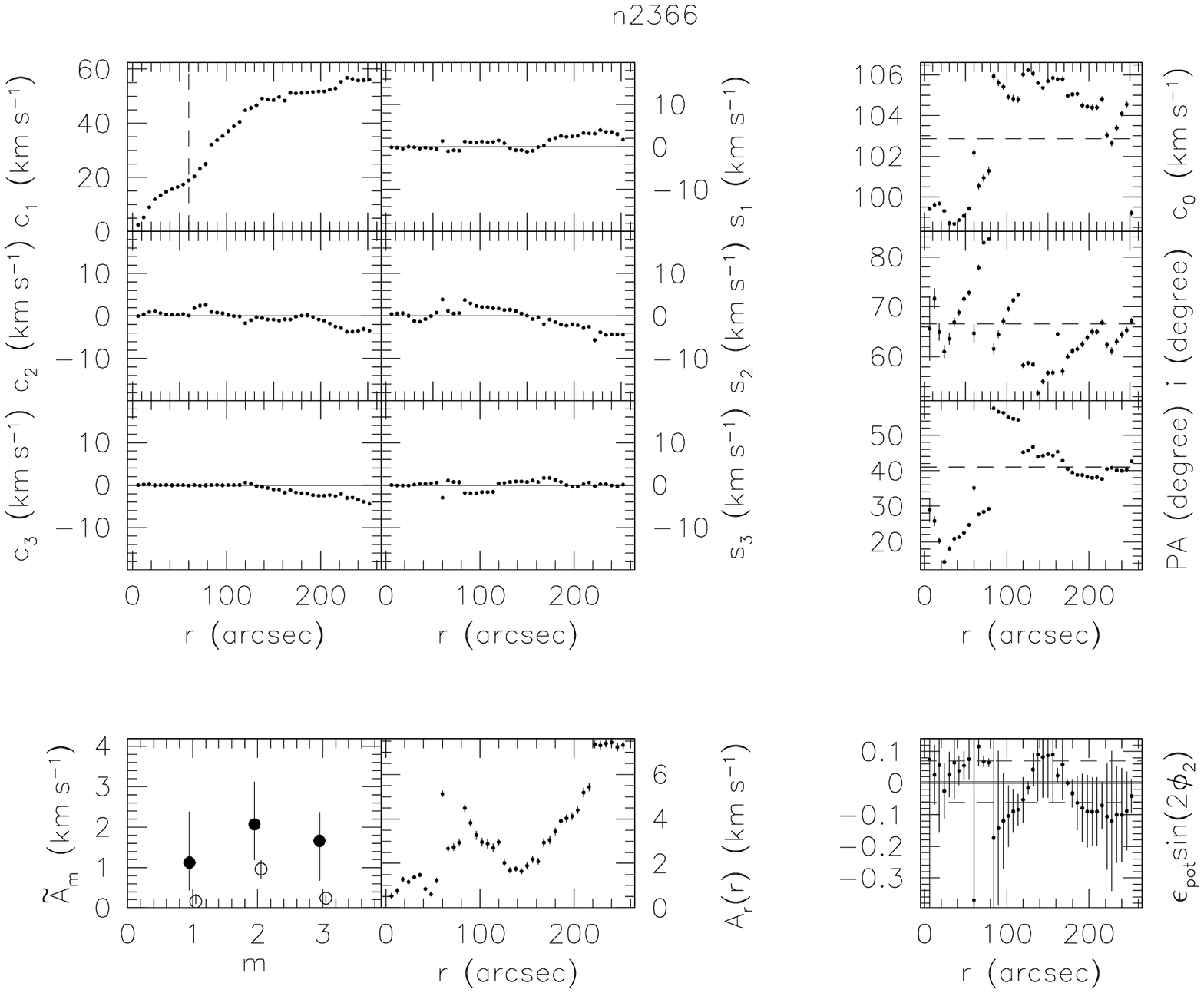}
\caption[Summary panel for NGC 2366]{Summary panel for NGC 2366. Lines and symbols are described in the text, Appendix~\ref{sec:first-figure}. The IRAC
  center shown was derived using {\sc ellfit}, and the contours of the IRAC image are given at the 20\% and 50\% level of the maximum intensity. See Appendix \ref{sec:n2366} for a discussion of this
  galaxy.} \label{fig:ngc-2366} 
\end{center}
\end{figure*}

\begin{figure*}[t!]
\begin{center}
\includegraphics[angle=0,width=0.50\textwidth,bb=18 144 592 520,clip=]{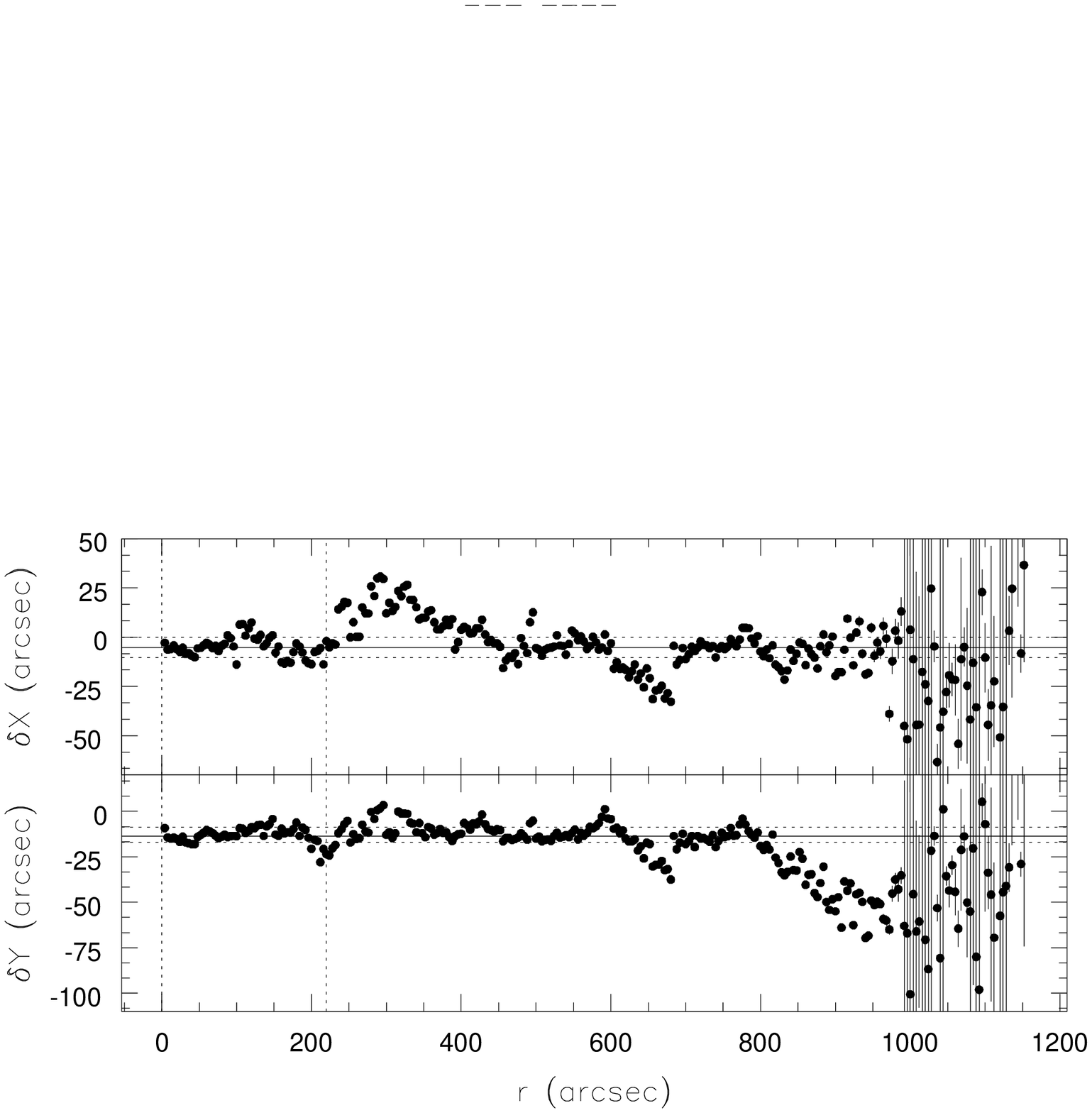}
\includegraphics[angle=0,width=0.55\textwidth,bb=18 79 520 520,clip=]{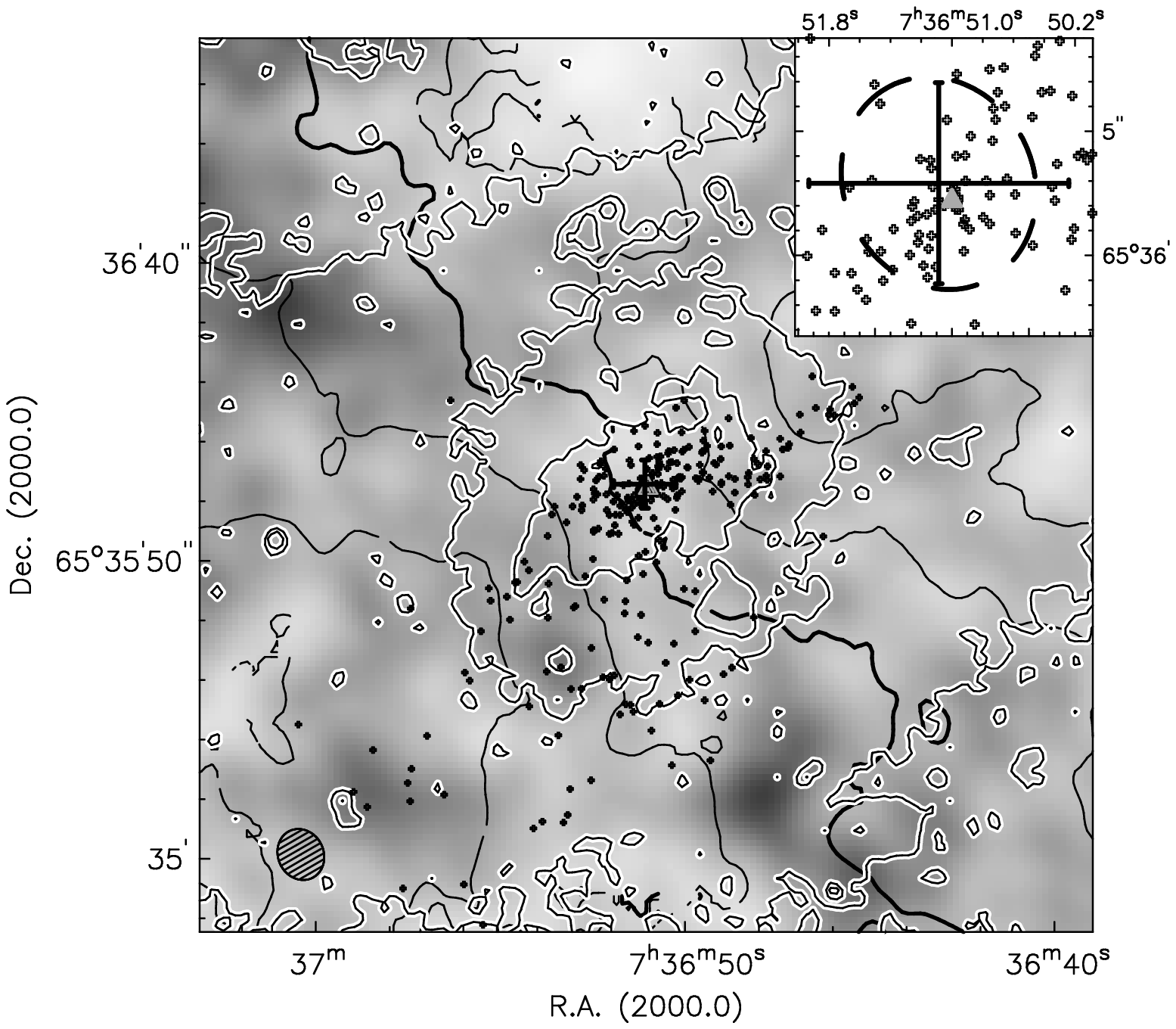}\\
\includegraphics[angle=0,width=0.65\textwidth,bb=19 235 592 697,clip=]{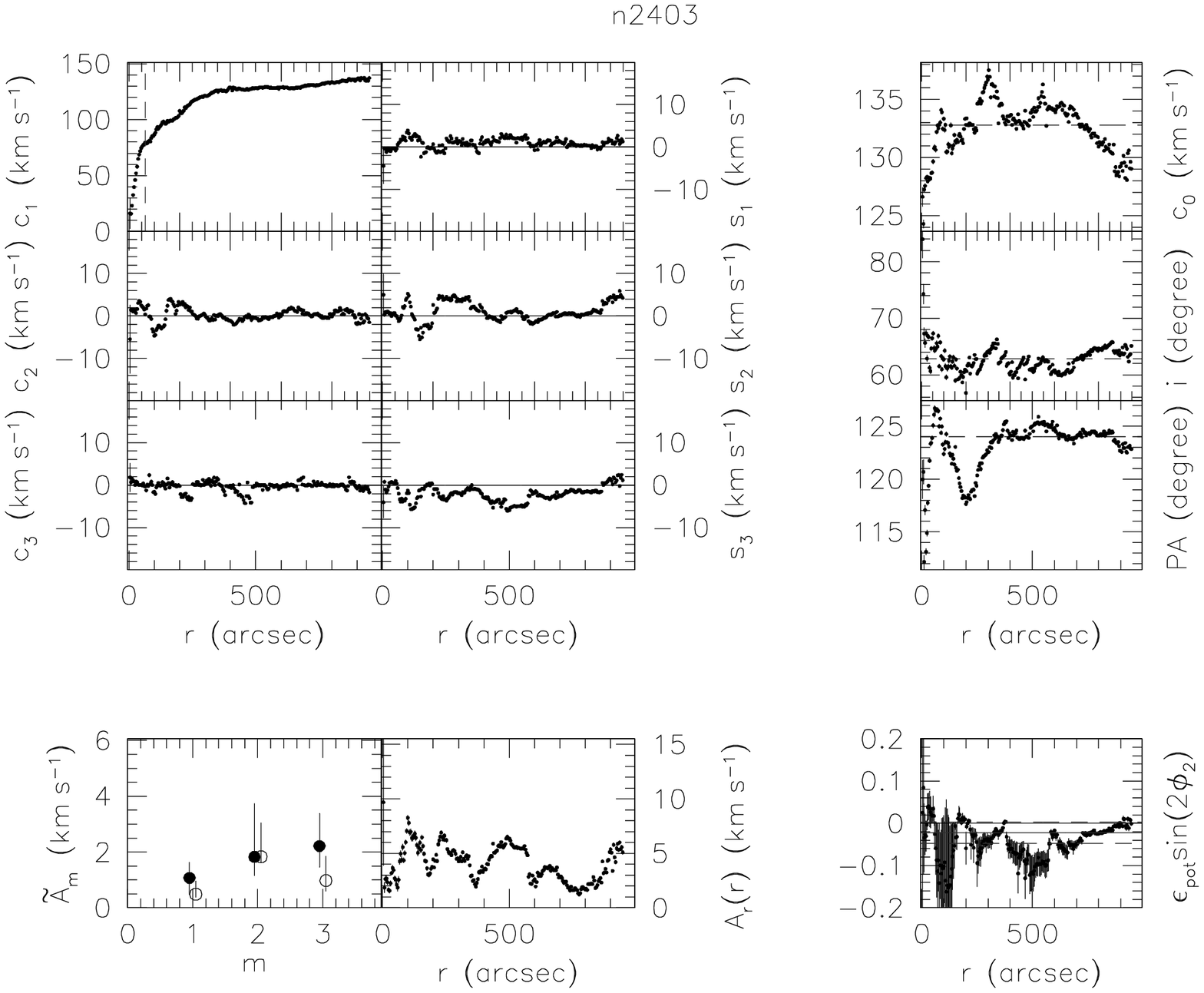}
\caption[Summary panel for NGC 2403]{Summary panel for NGC 2403. Lines and symbols are described in the text, Appendix~\ref{sec:first-figure}. The IRAC
  contours are given at the 20, 50, and 80\% level of the maximum intensity. See Appendix \ref{sec:n2403} for a discussion of this galaxy.} \label{fig:ngc-2403}
\end{center}
\end{figure*}

\begin{figure*}[t!]
\begin{center}
\includegraphics[angle=0,width=0.50\textwidth,bb=18 144 592 520,clip=]{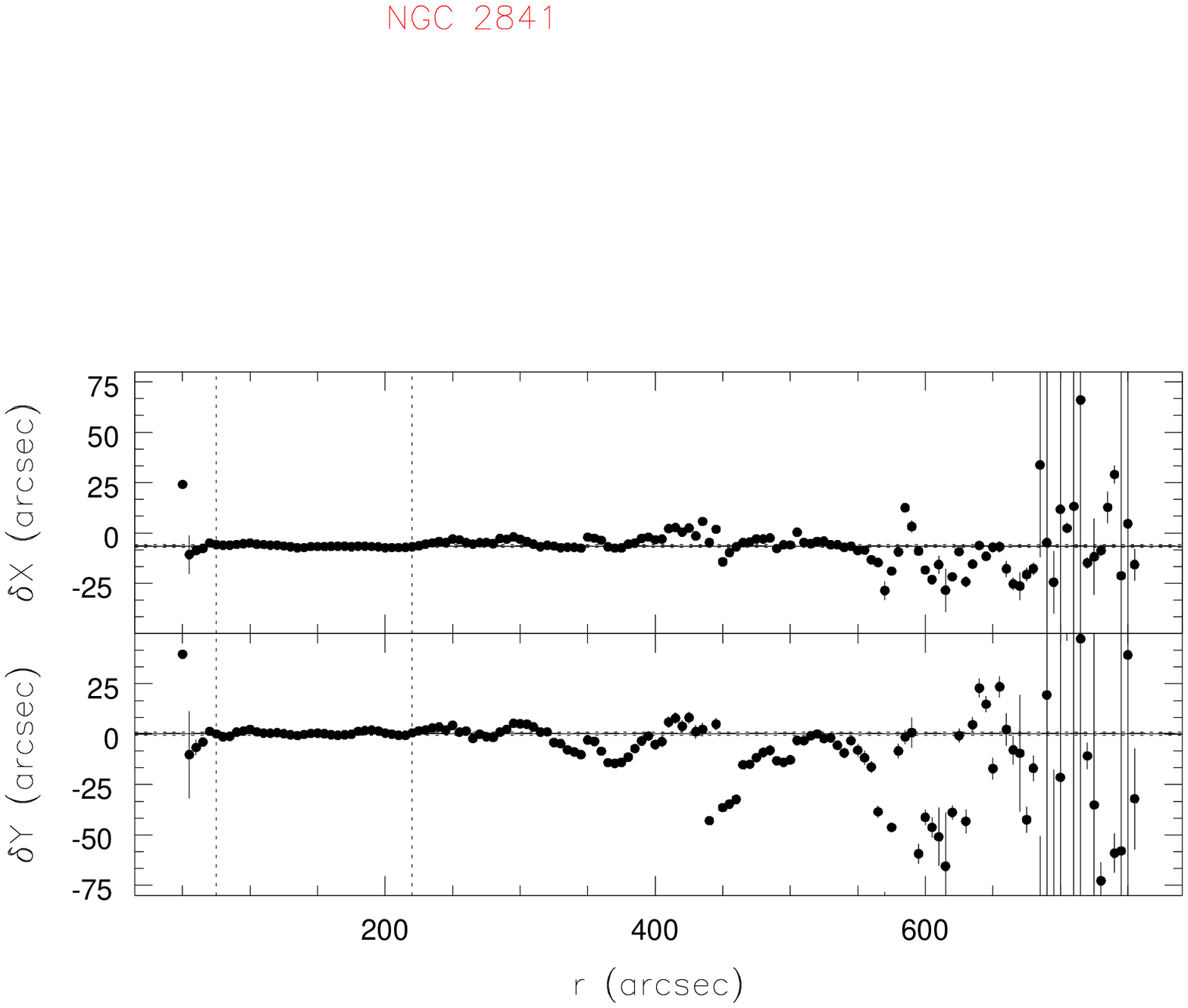}
\includegraphics[angle=0,width=0.55\textwidth]{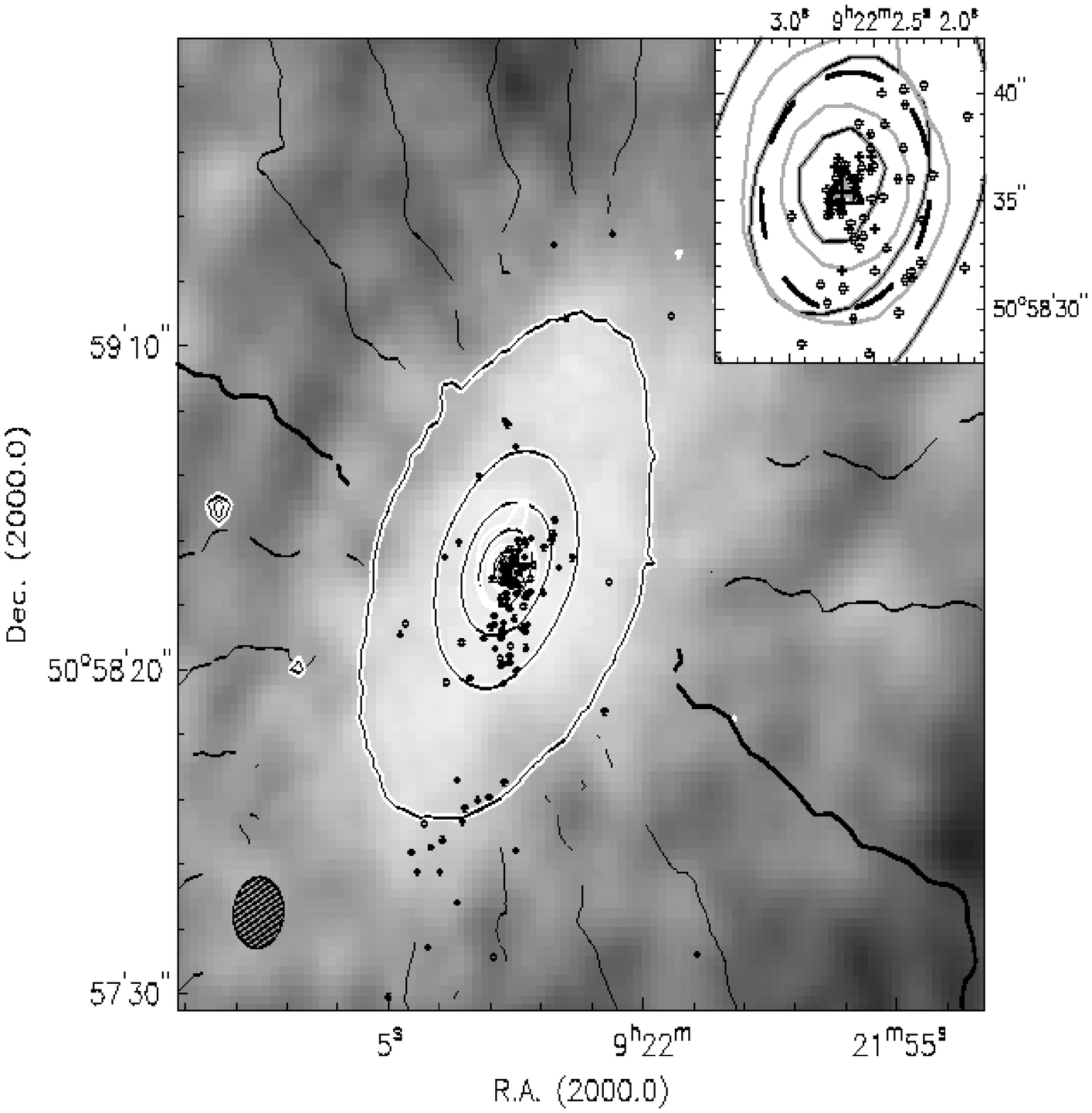}\\
\includegraphics[angle=0,width=0.65\textwidth,bb=19 235 592 697,clip=]{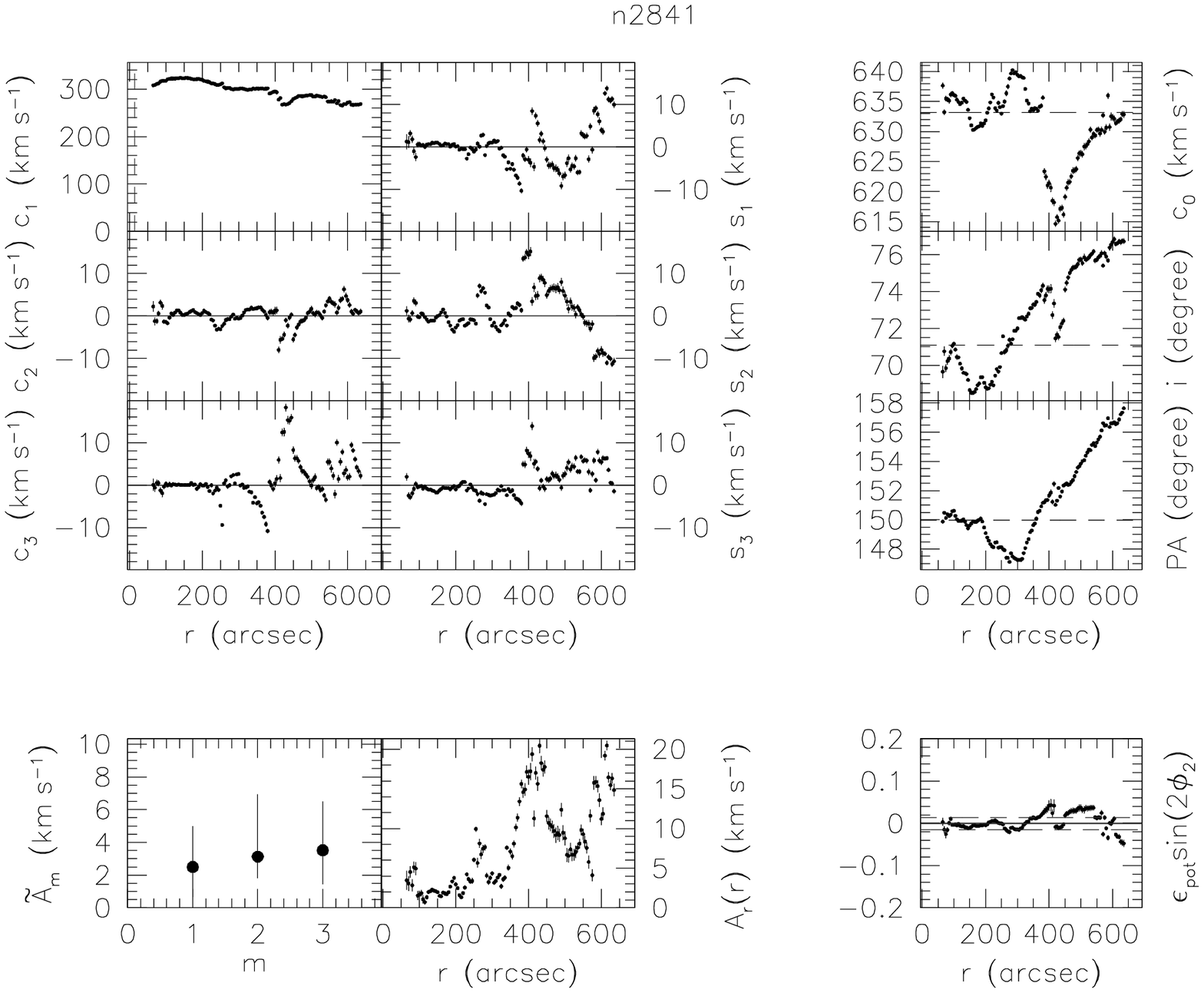}
\caption[Summary panel for NGC 2841]{Summary panel for NGC 2841. Lines and symbols are described in the text, Appendix~\ref{sec:first-figure}. See Appendix \ref{sec:n2841} for a discussion of this galaxy.} \label{fig:ngc-2841}
\end{center}
\end{figure*}

\begin{figure*}[t!]
\begin{center}
\includegraphics[angle=0,width=0.50\textwidth,bb=18 144 592 520,clip=]{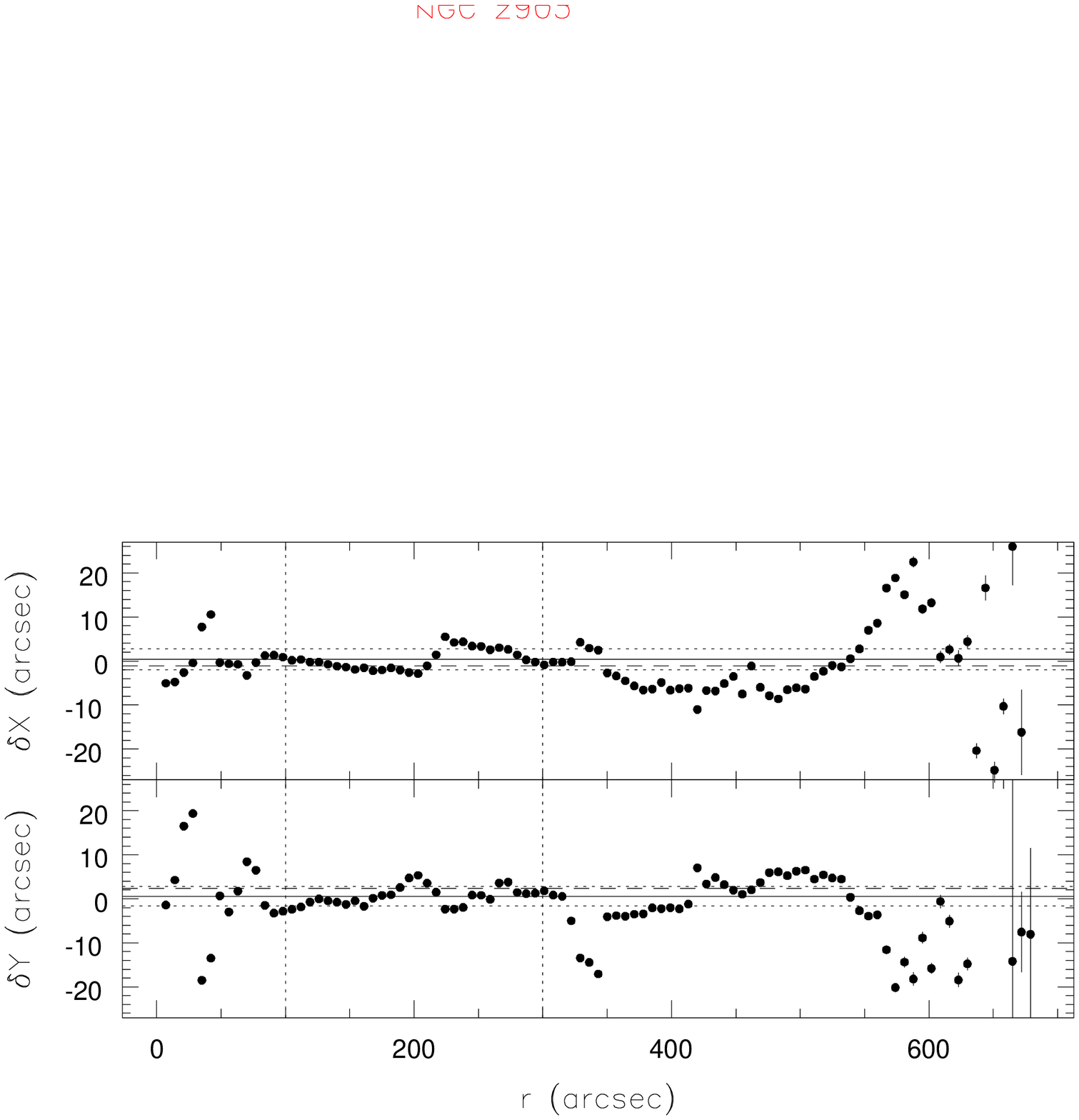}
\includegraphics[angle=0,width=0.55\textwidth,bb=18 79 520 520,clip=]{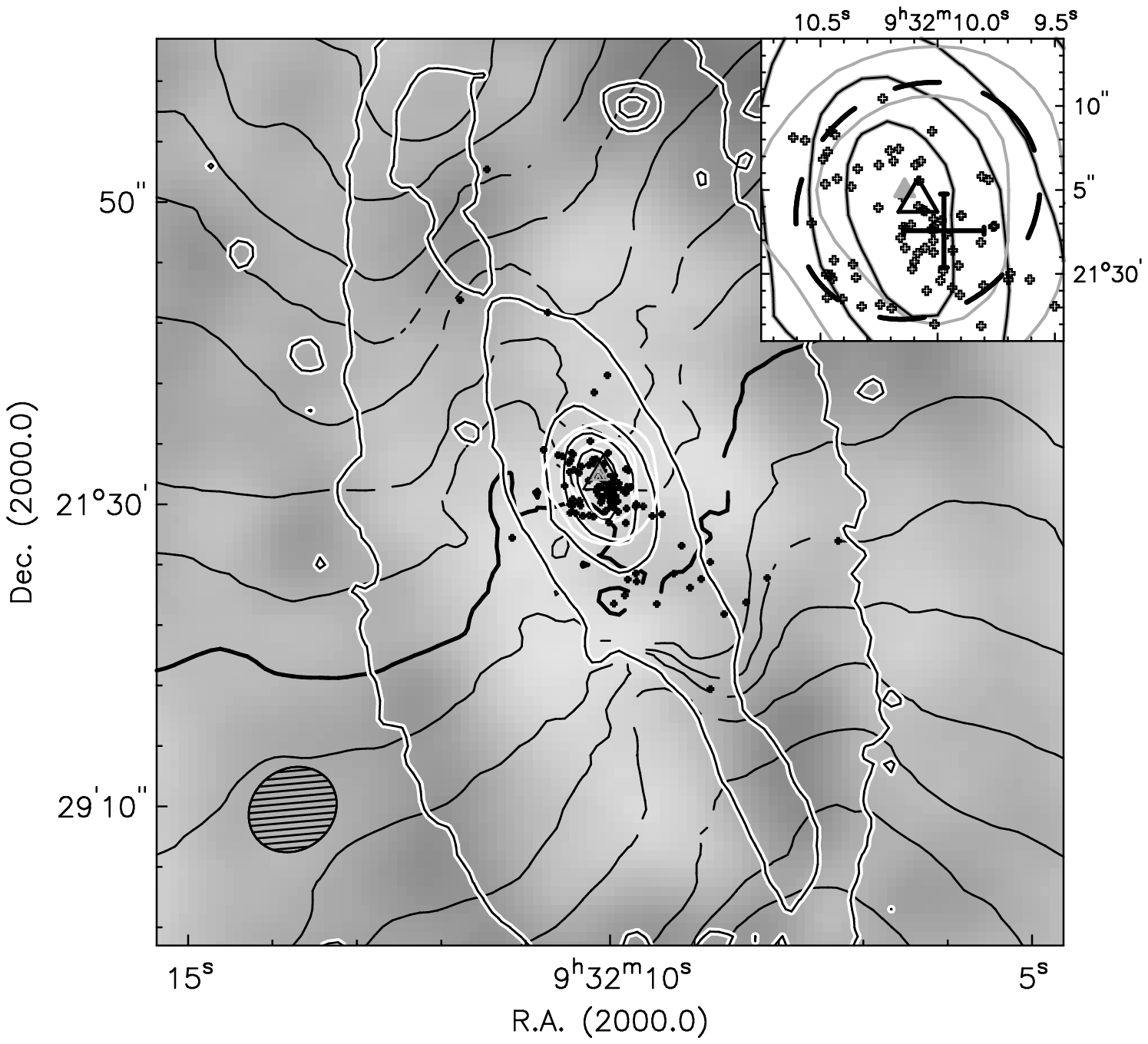}\\
\includegraphics[angle=0,width=0.65\textwidth,bb=19 235 592 697,clip=]{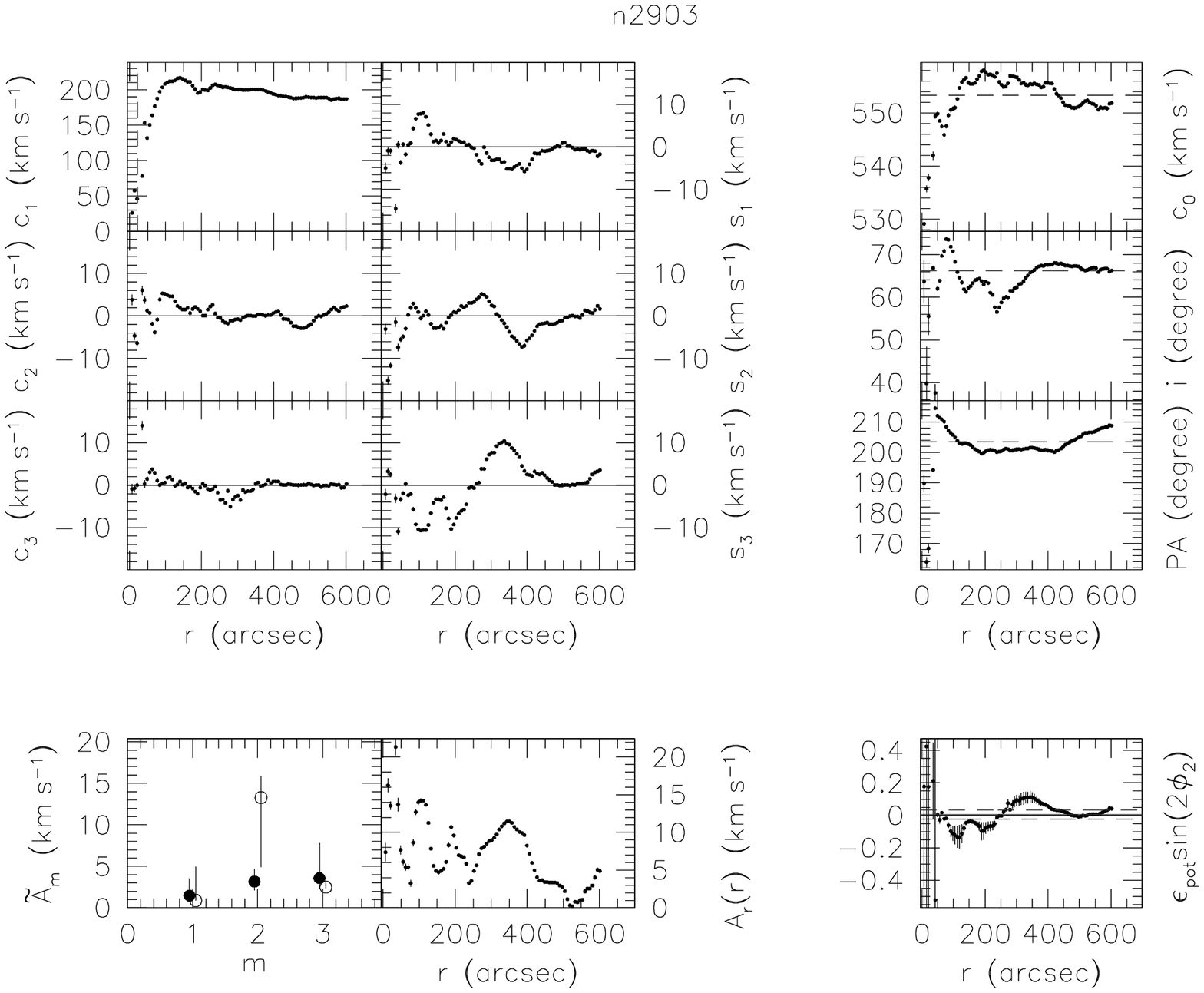}
\caption[Summary panel for NGC 2903]{Summary panel for NGC 2903. Lines and symbols are described in the text, Appendix~\ref{sec:first-figure}. See Appendix \ref{sec:n2903} for a discussion of this galaxy.} \label{fig:ngc-2903}
\end{center}
\end{figure*}

\begin{figure*}[t!]
\begin{center}
\includegraphics[angle=0,width=0.50\textwidth,bb=18 144 592 520,clip=]{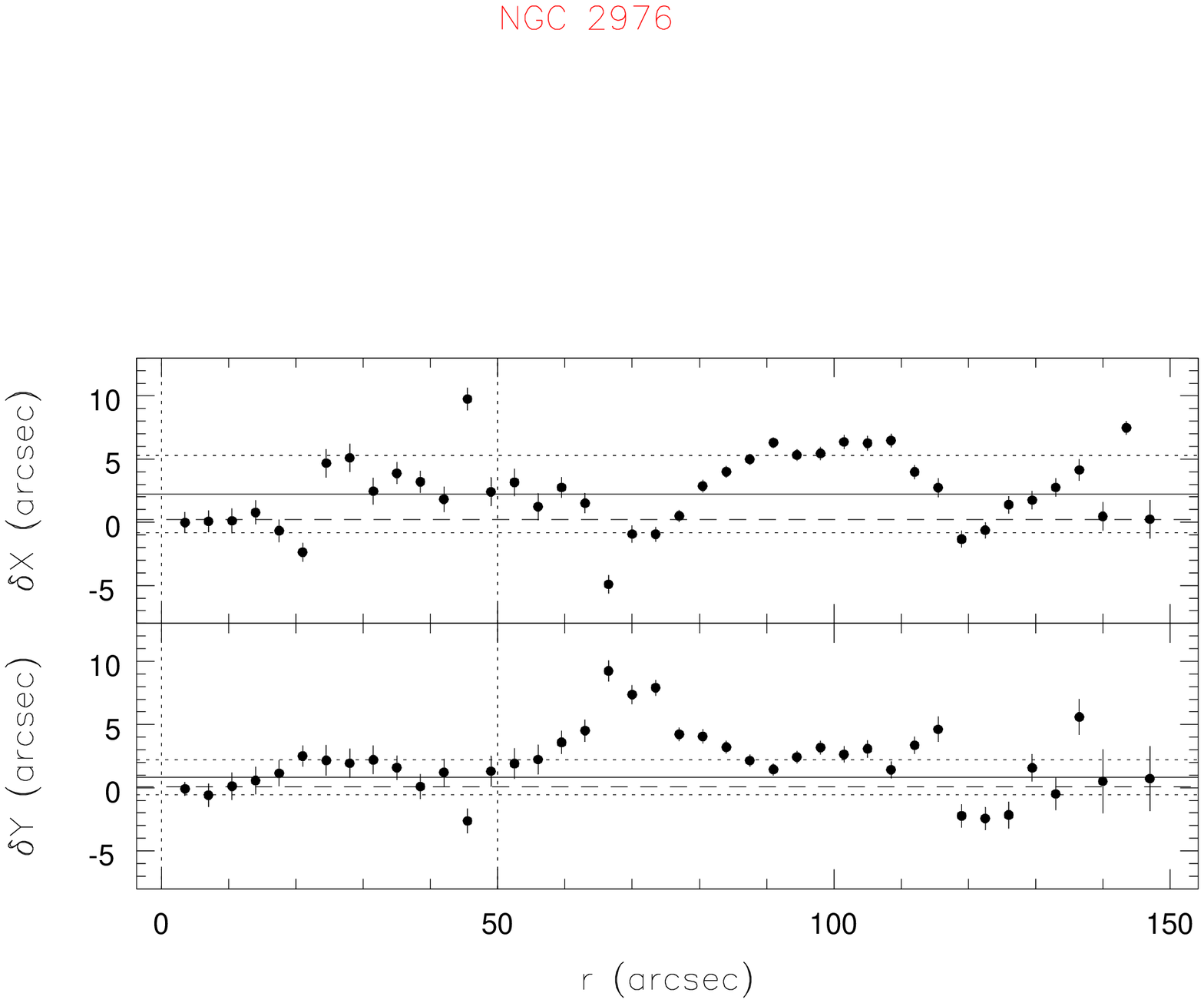}
\includegraphics[angle=0,width=0.55\textwidth,bb=18 79 520 520,clip=]{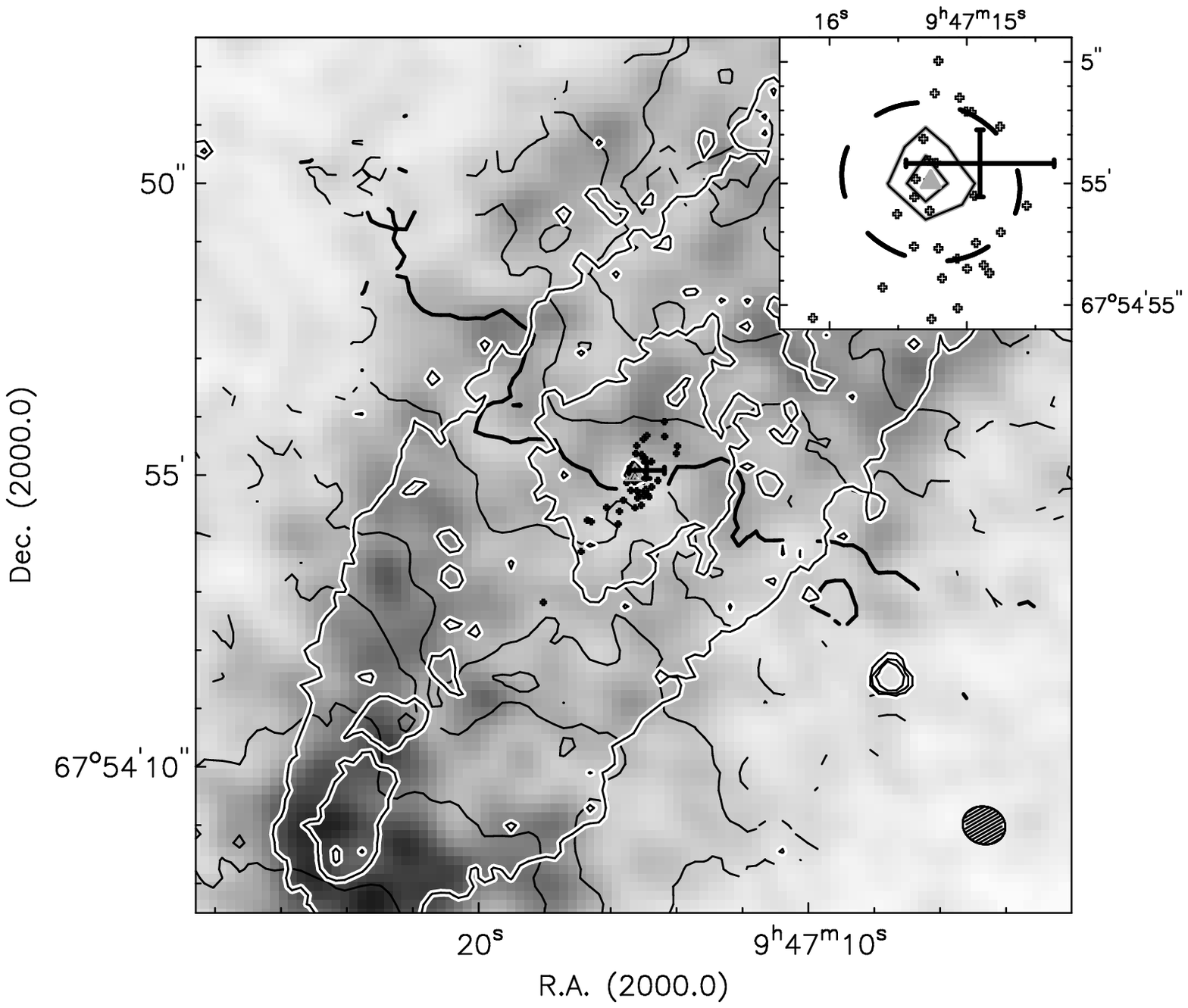}\\
\includegraphics[angle=0,width=0.65\textwidth,bb=19 235 592 697,clip=]{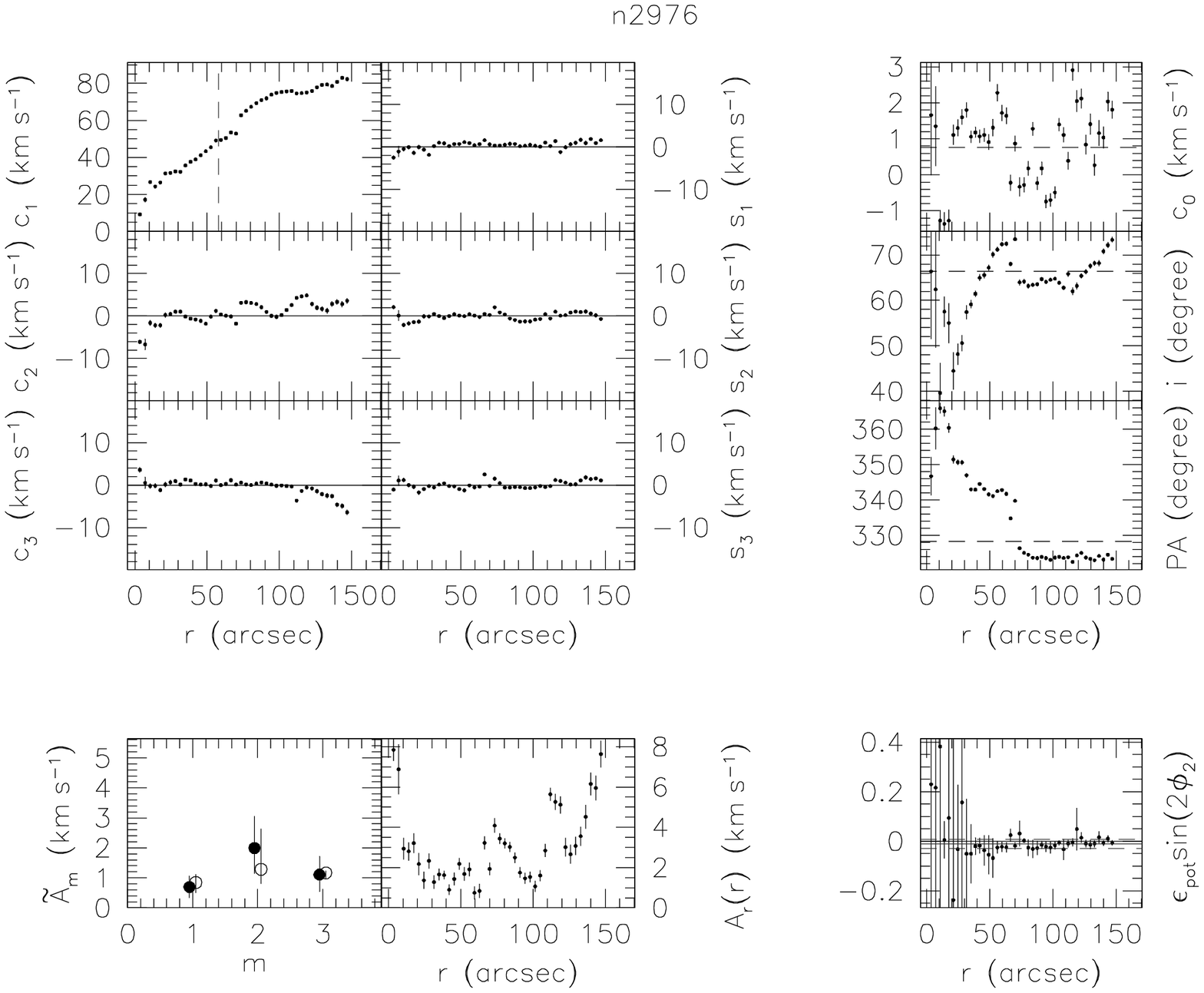}
\caption[Summary panel for NGC 2976]{Summary panel for NGC 2976. Lines and symbols are described in the text, Appendix~\ref{sec:first-figure}. The IRAC
  contours are given at the 10, 20, 50, and 80\% level of the maximum intensity. See Appendix \ref{sec:n2976} for a discussion of this galaxy.} \label{fig:ngc-2976}
\end{center}
\end{figure*}

\begin{figure*}[t!]
\begin{center}
\includegraphics[angle=0,width=0.50\textwidth,bb=18 144 592 520,clip=]{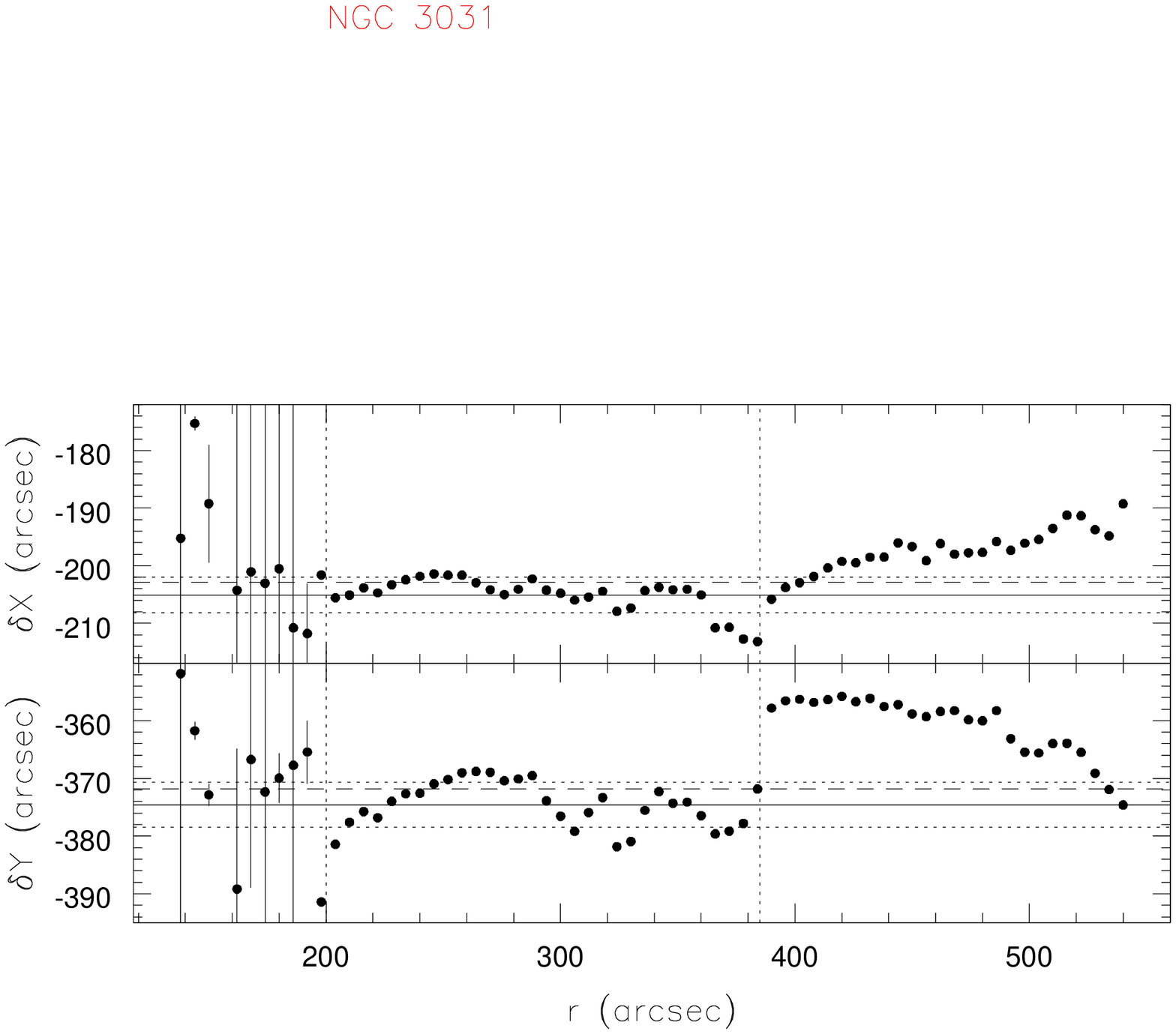}
\includegraphics[angle=0,width=0.55\textwidth,bb=18 79 520 520,clip=]{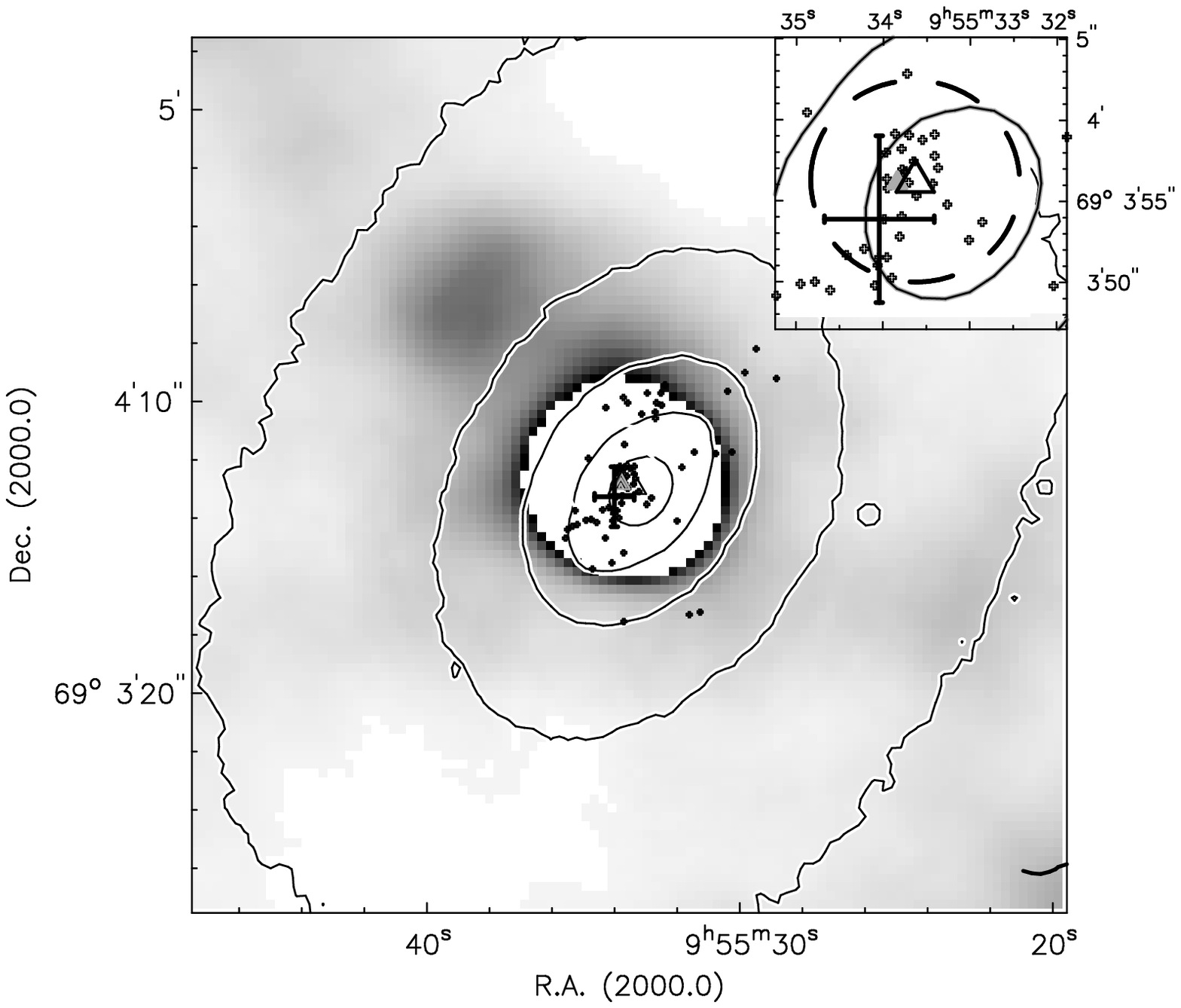}\\
\includegraphics[angle=0,width=0.65\textwidth,bb=19 235 592 697,clip=]{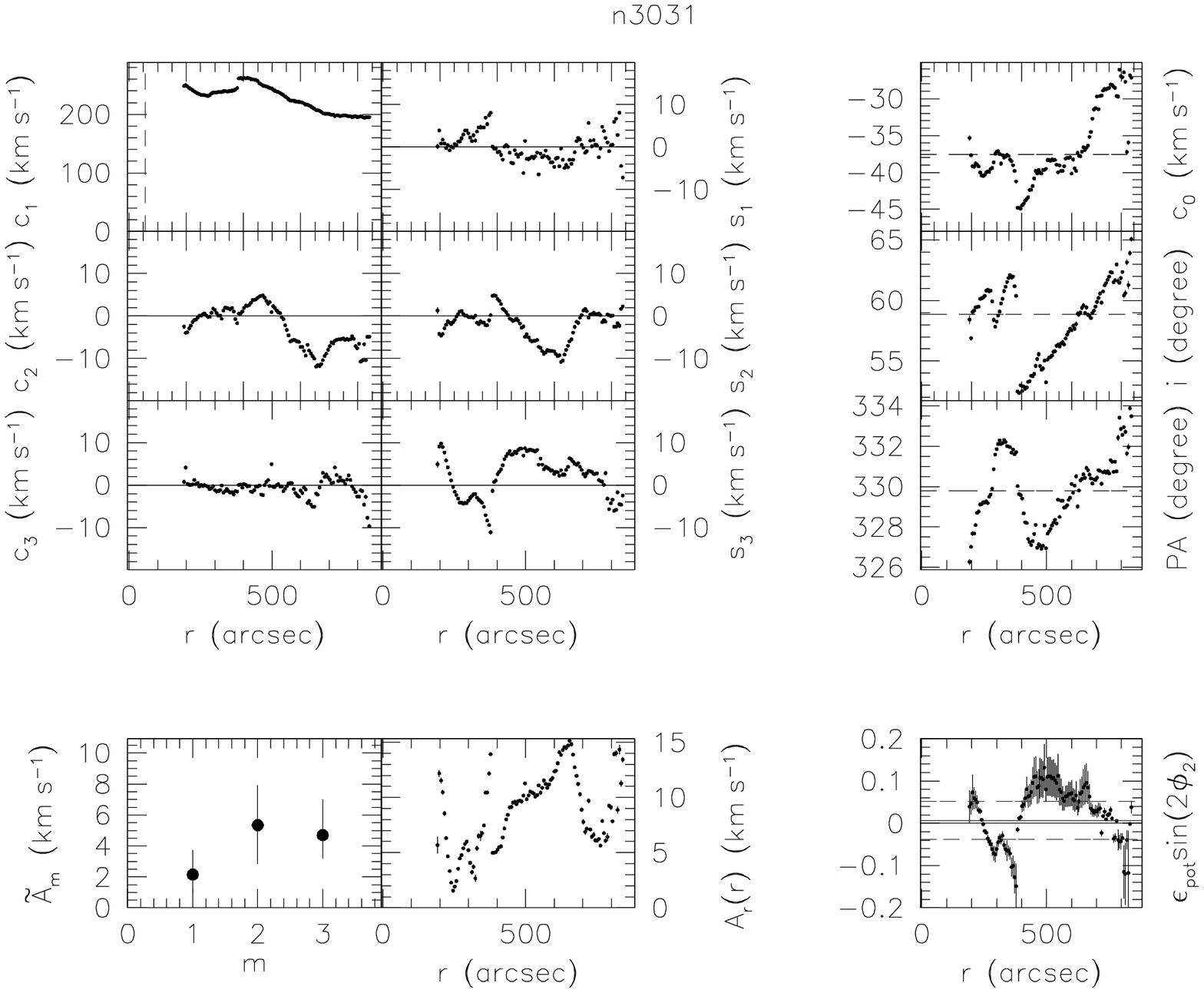}
\caption[Summary panel for NGC 3031]{Summary panel for NGC 3031. Lines and symbols are described in the text, Appendix~\ref{sec:first-figure}. The bright radio continuum source which is still
  visible in the total intensity \HI\ map was blanked out for clarity reasons. Due to the central \HI\ deficiency, no velocity contours are visible. See Appendix \ref{sec:n3031} for a discussion of this galaxy.} \label{fig:ngc-3031}
\end{center}
\end{figure*}

\begin{figure*}[t!]
\begin{center}
\includegraphics[angle=0,width=0.50\textwidth,bb=18 144 592 520,clip=]{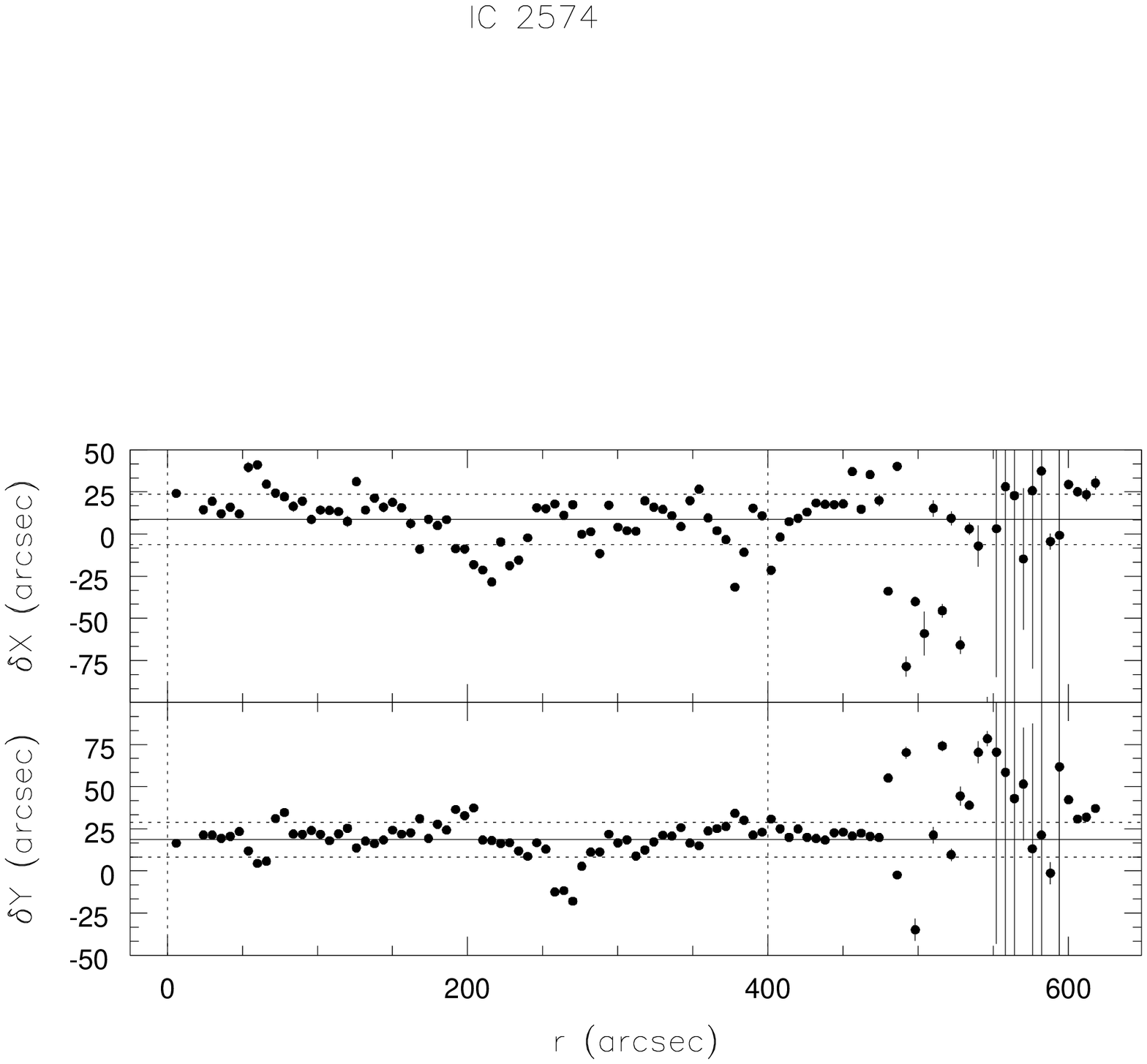}
\includegraphics[angle=0,width=0.55\textwidth,bb=18 79 520 520,clip=]{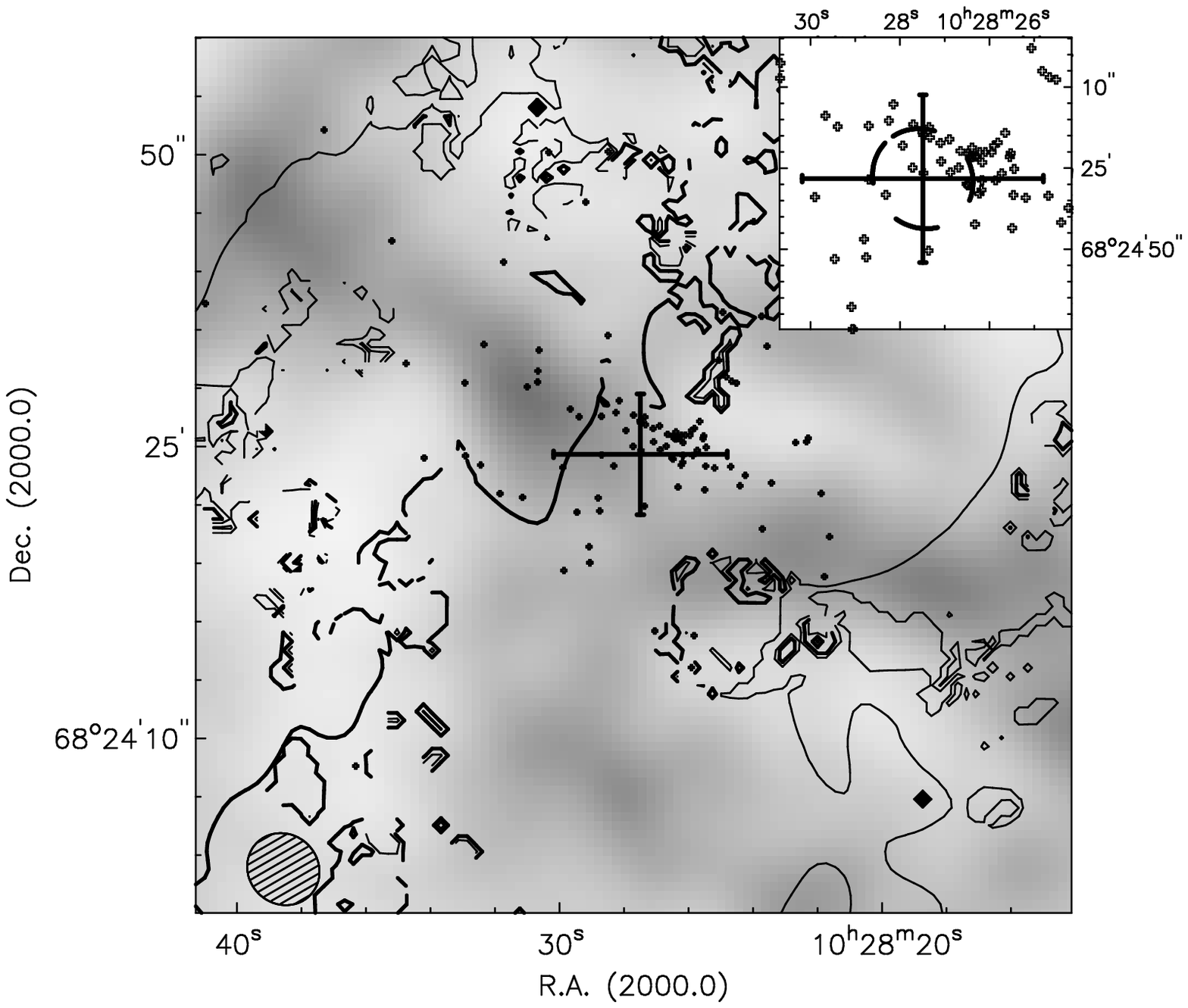}\\
\includegraphics[angle=0,width=0.65\textwidth,bb=19 235 592 697,clip=]{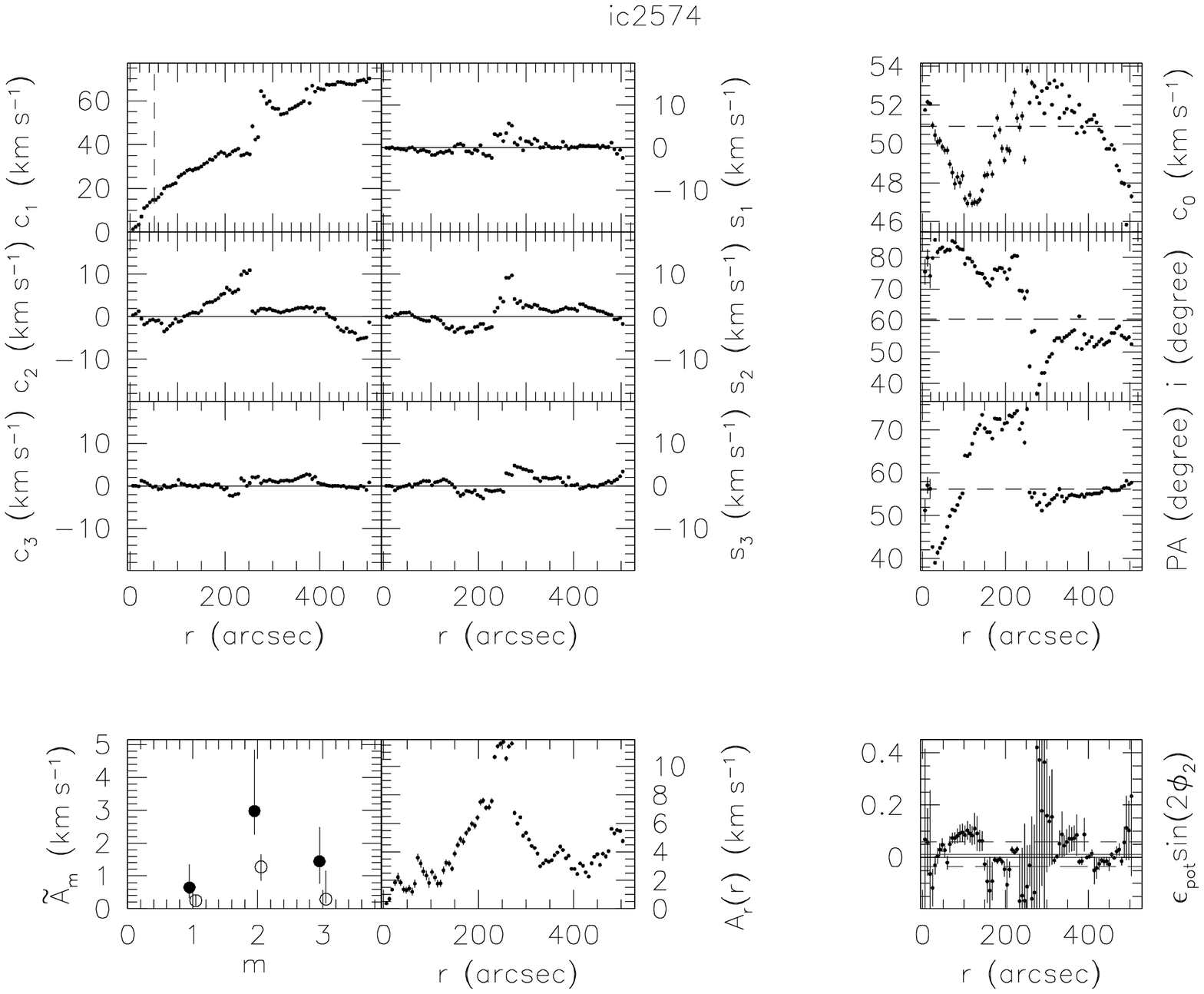}
\caption[Summary panel for IC 2574]{Summary panel for IC 2574. Lines and symbols are described in the text, Appendix~\ref{sec:first-figure}. No estimate could be derived from the
  IRAC and radio continuum images, and their respective contours are therefore
  not shown. See Appendix \ref{sec:ic2574} for a discussion of this
  galaxy.} \label{fig:ic-2574}
\end{center}
\end{figure*}

\begin{figure*}[t!]
\begin{center}
\includegraphics[angle=0,width=0.50\textwidth,bb=18 144 592 520,clip=]{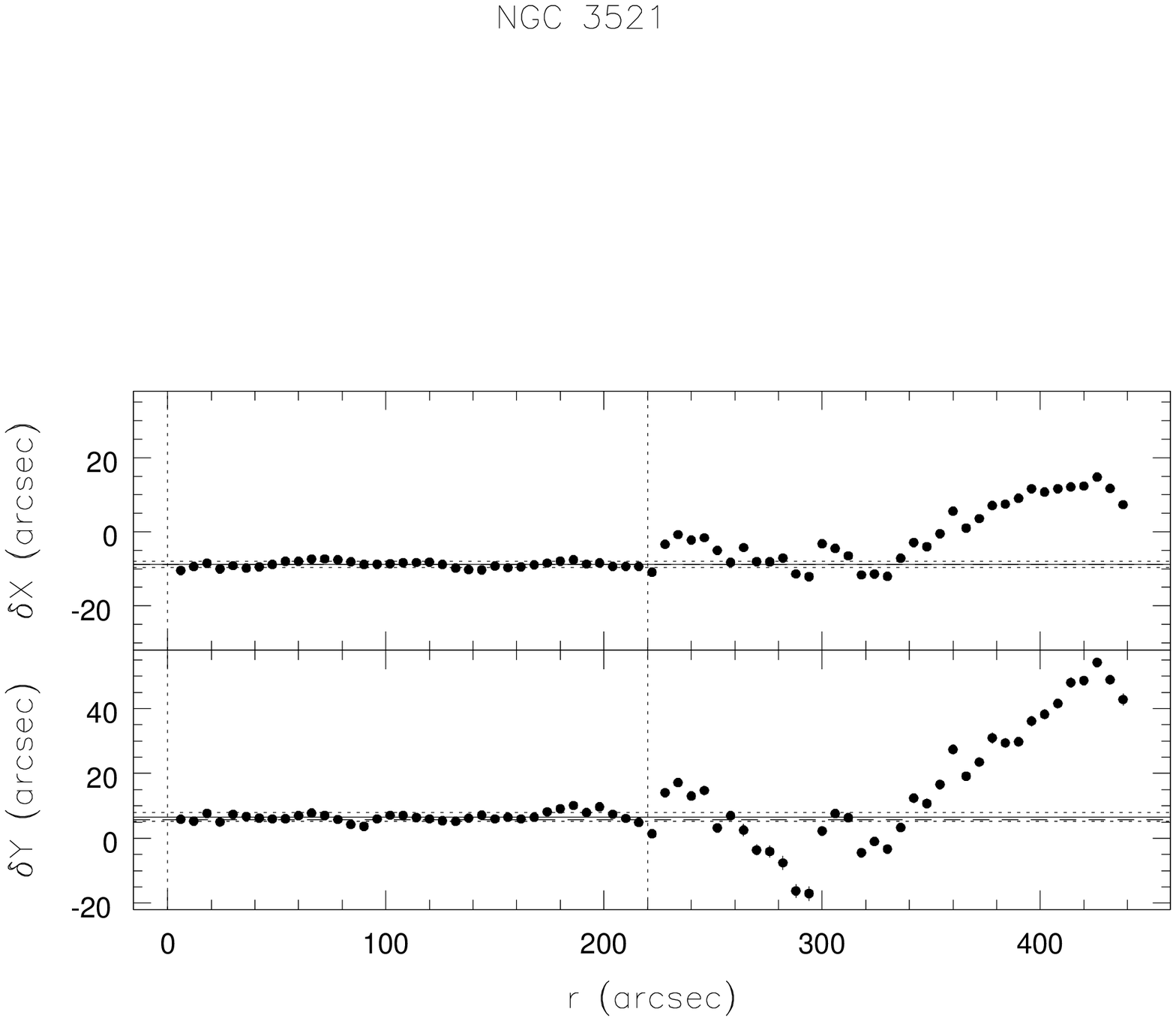}
\includegraphics[angle=0,width=0.55\textwidth,bb=18 79 520 520,clip=]{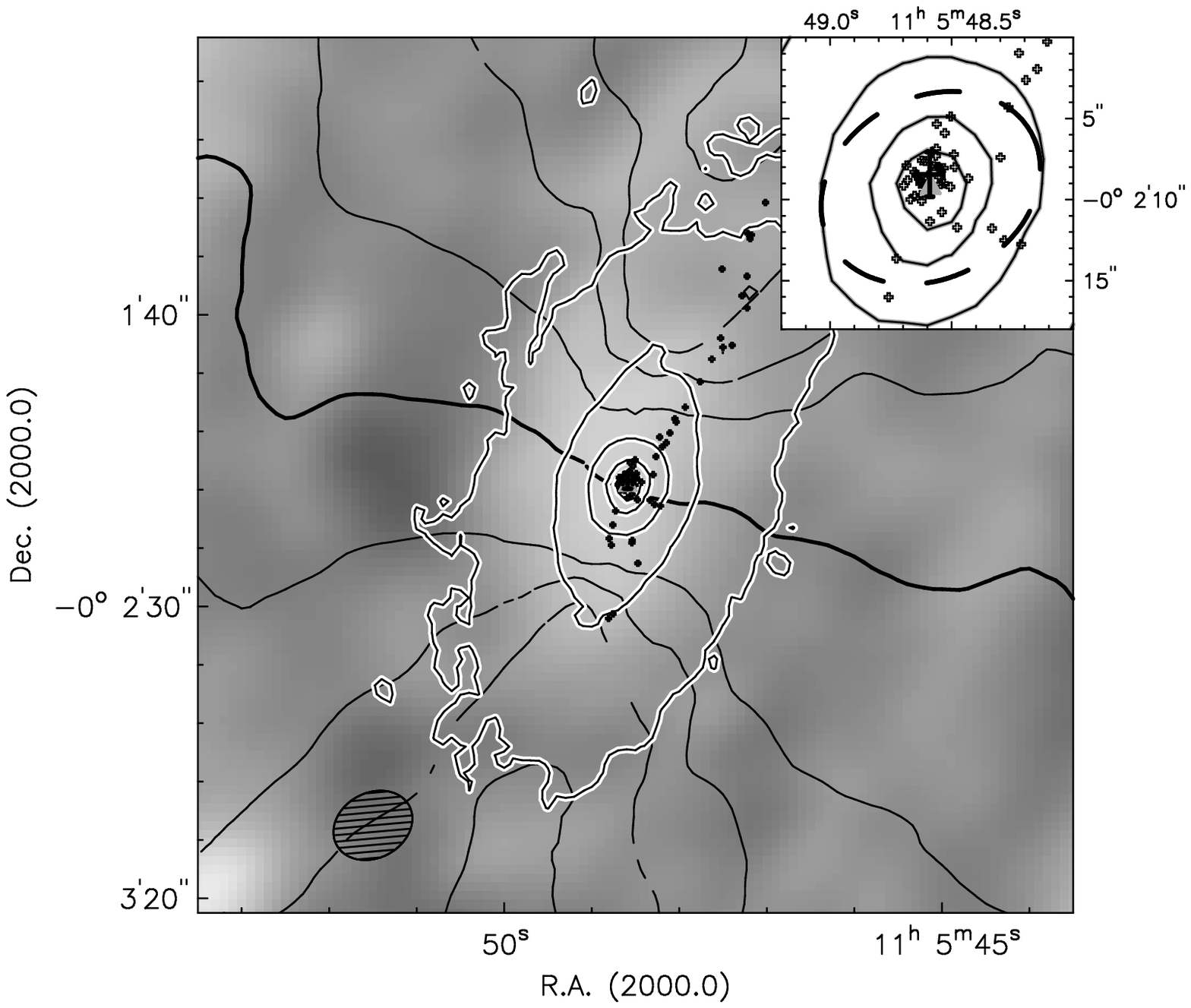}\\
\includegraphics[angle=0,width=0.65\textwidth,bb=19 235 592 697,clip=]{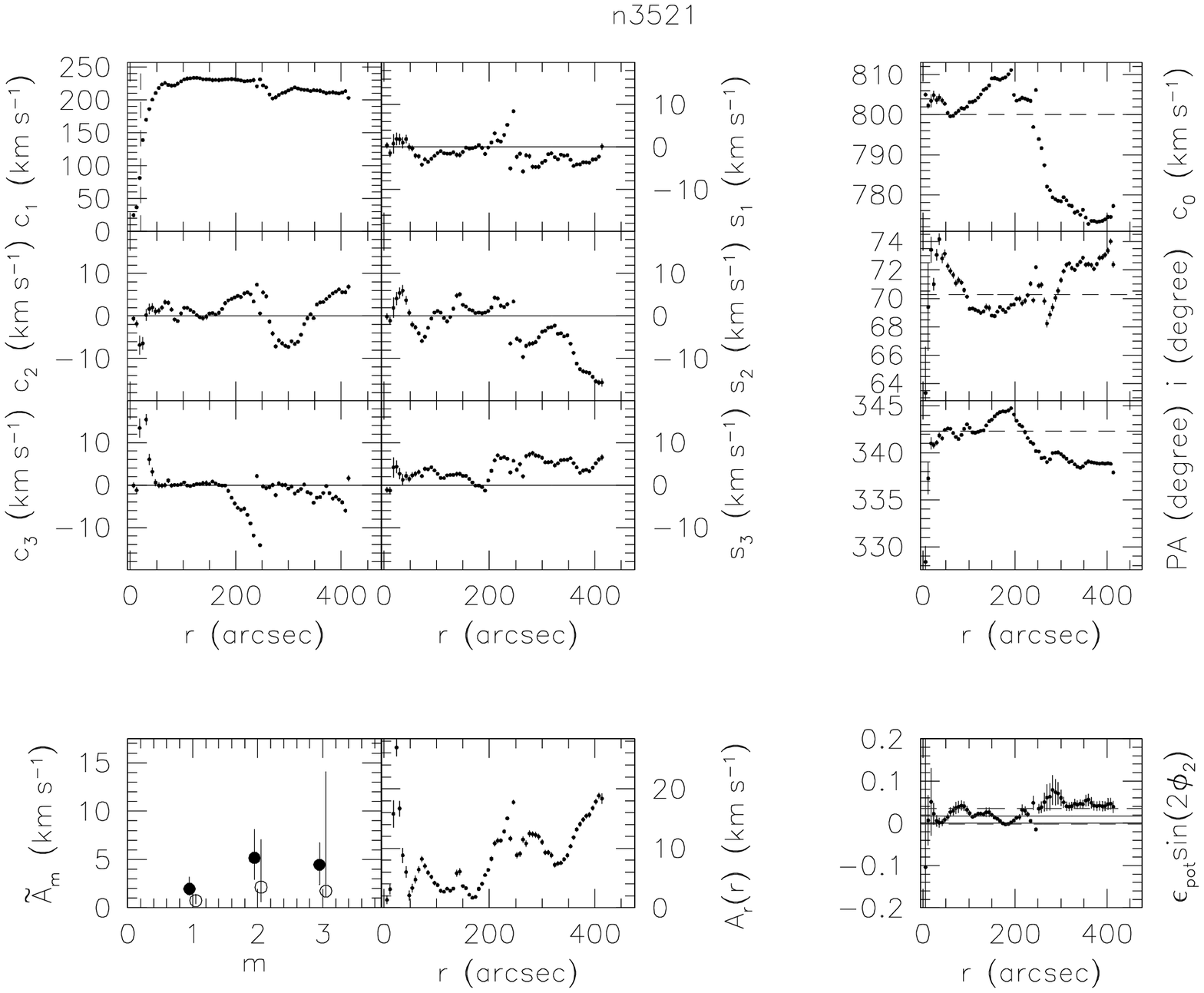}
\caption[Summary panel for NGC 3521]{Summary panel for NGC 3521. Lines and symbols are described in the text, Appendix~\ref{sec:first-figure}. See Appendix \ref{sec:n3521} for a discussion of this galaxy.} \label{fig:ngc-3521}
\end{center}
\end{figure*}

\begin{figure*}[t!]
\begin{center}
\includegraphics[angle=0,width=0.50\textwidth,bb=18 144 592 520,clip=]{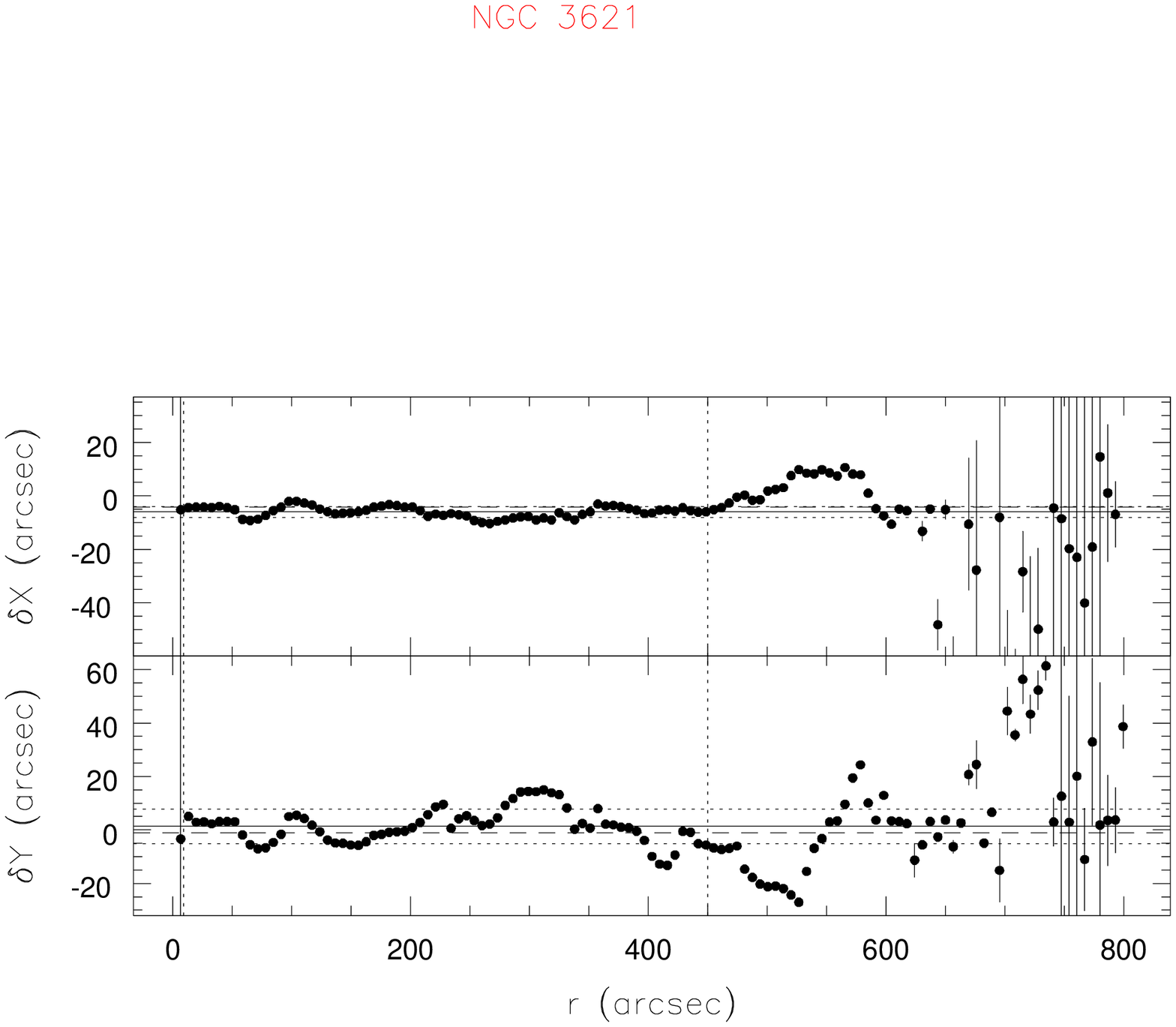}
\includegraphics[angle=0,width=0.55\textwidth,bb=18 79 520 520,clip=]{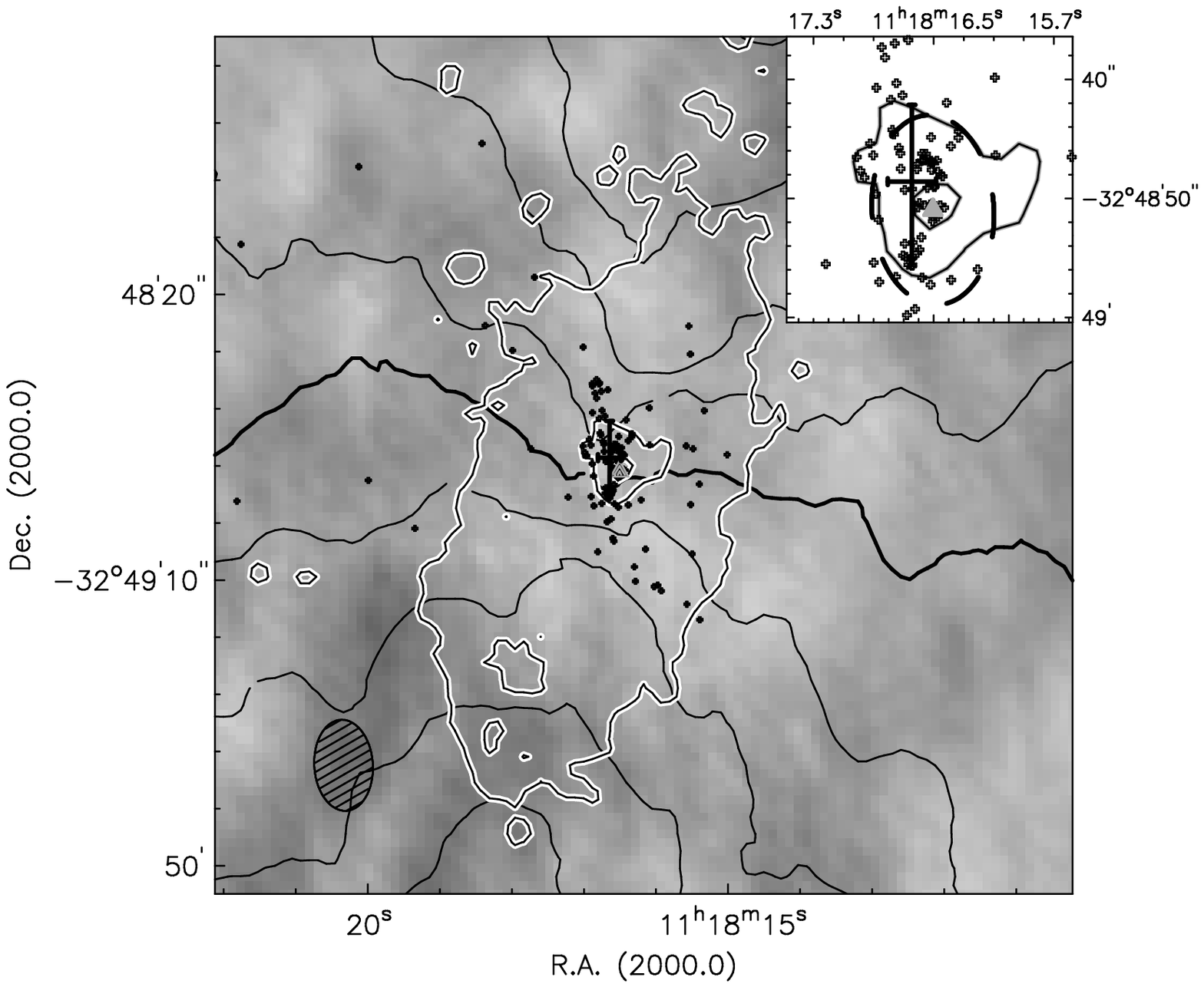}\\
\includegraphics[angle=0,width=0.65\textwidth,bb=19 235 592 697,clip=]{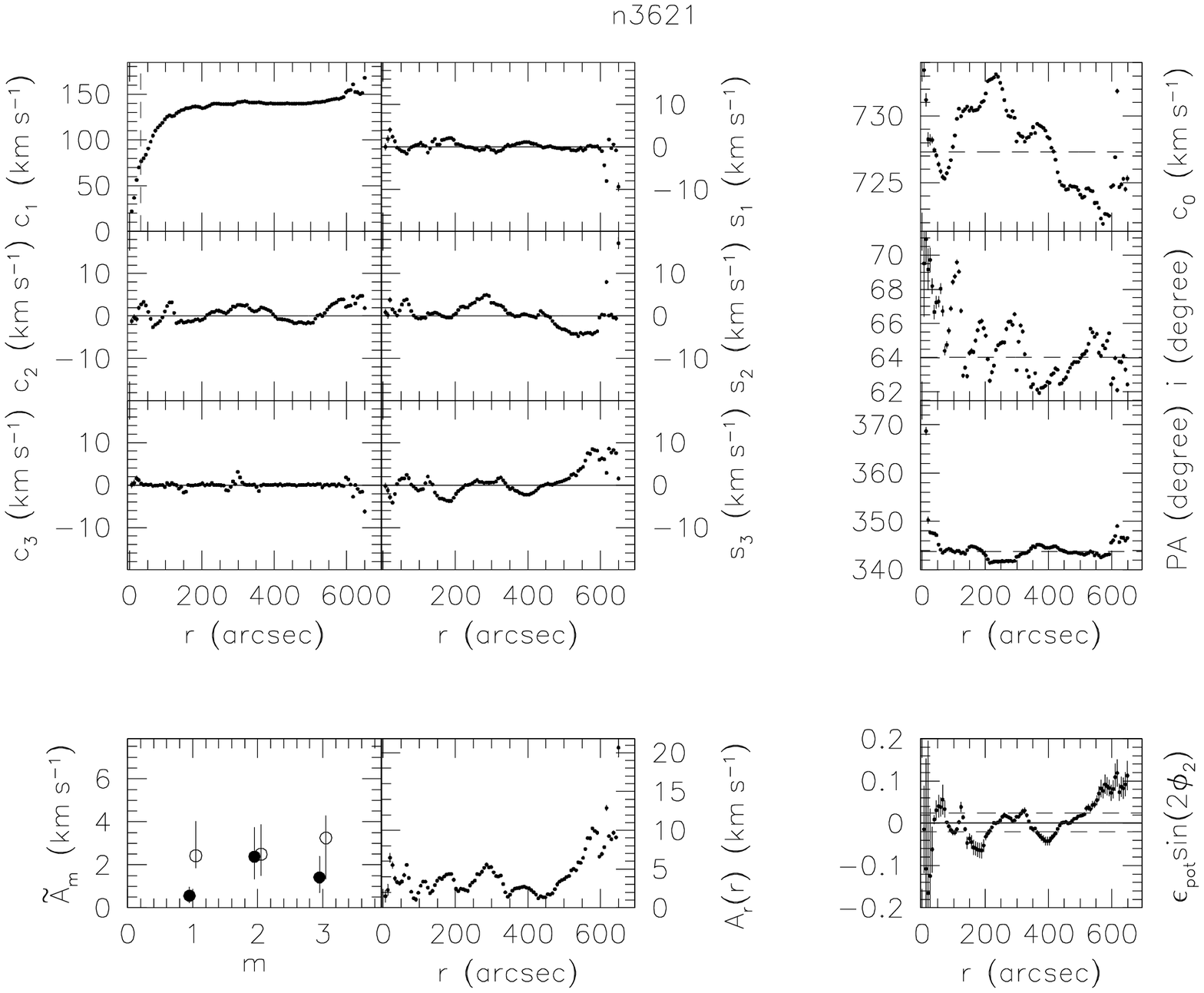}
\caption[Summary panel for NGC 3621]{Summary panel for NGC 3621. Lines and symbols are described in the text, Appendix~\ref{sec:first-figure}. The IRAC contours are given at the 20, 50, and 80\% level of the maximum intensity. Only data with $r\le 600\arcsec$ was used for radial averaging of \Am\ or $\epsilon_{\mathrm{pot}}$.
See Appendix \ref{sec:n3621} for a discussion of this galaxy.} \label{fig:ngc-3621}
\end{center}
\end{figure*}

\begin{figure*}[t!]
\begin{center}
\includegraphics[angle=0,width=0.50\textwidth,bb=18 144 592 520,clip=]{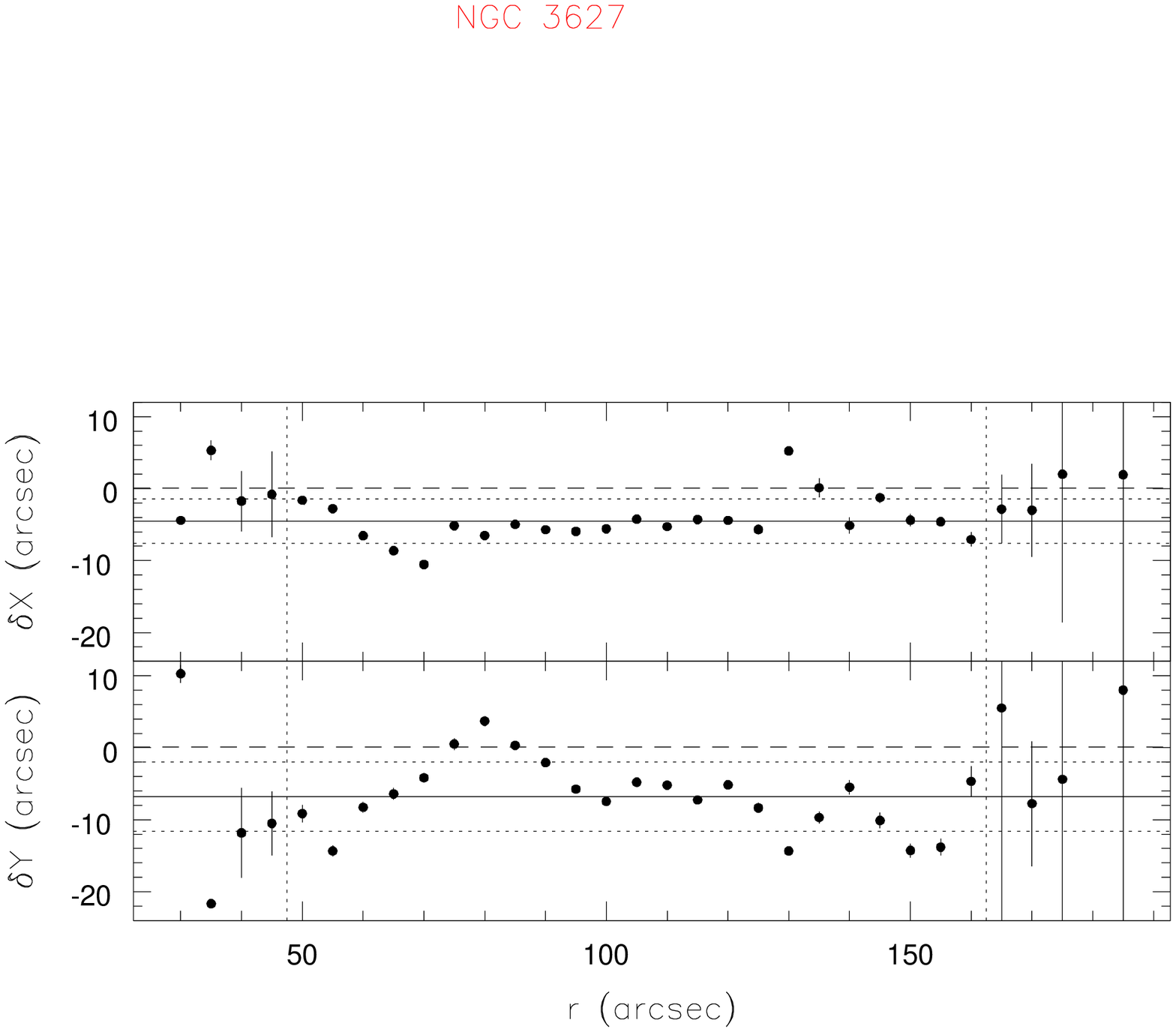}
\includegraphics[angle=0,width=0.55\textwidth,bb=18 79 520 520,clip=]{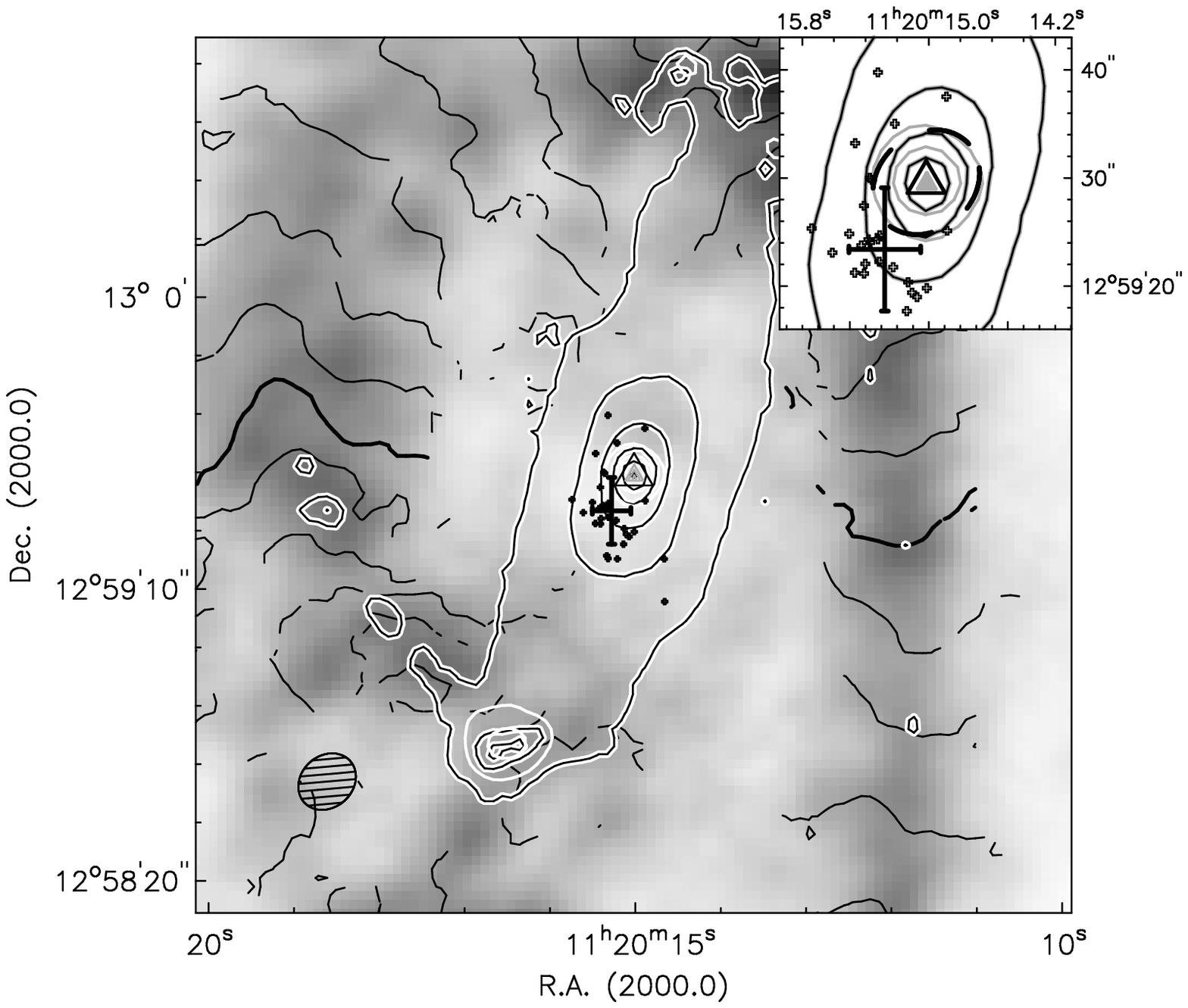}\\
\includegraphics[angle=0,width=0.65\textwidth,bb=19 235 592 697,clip=]{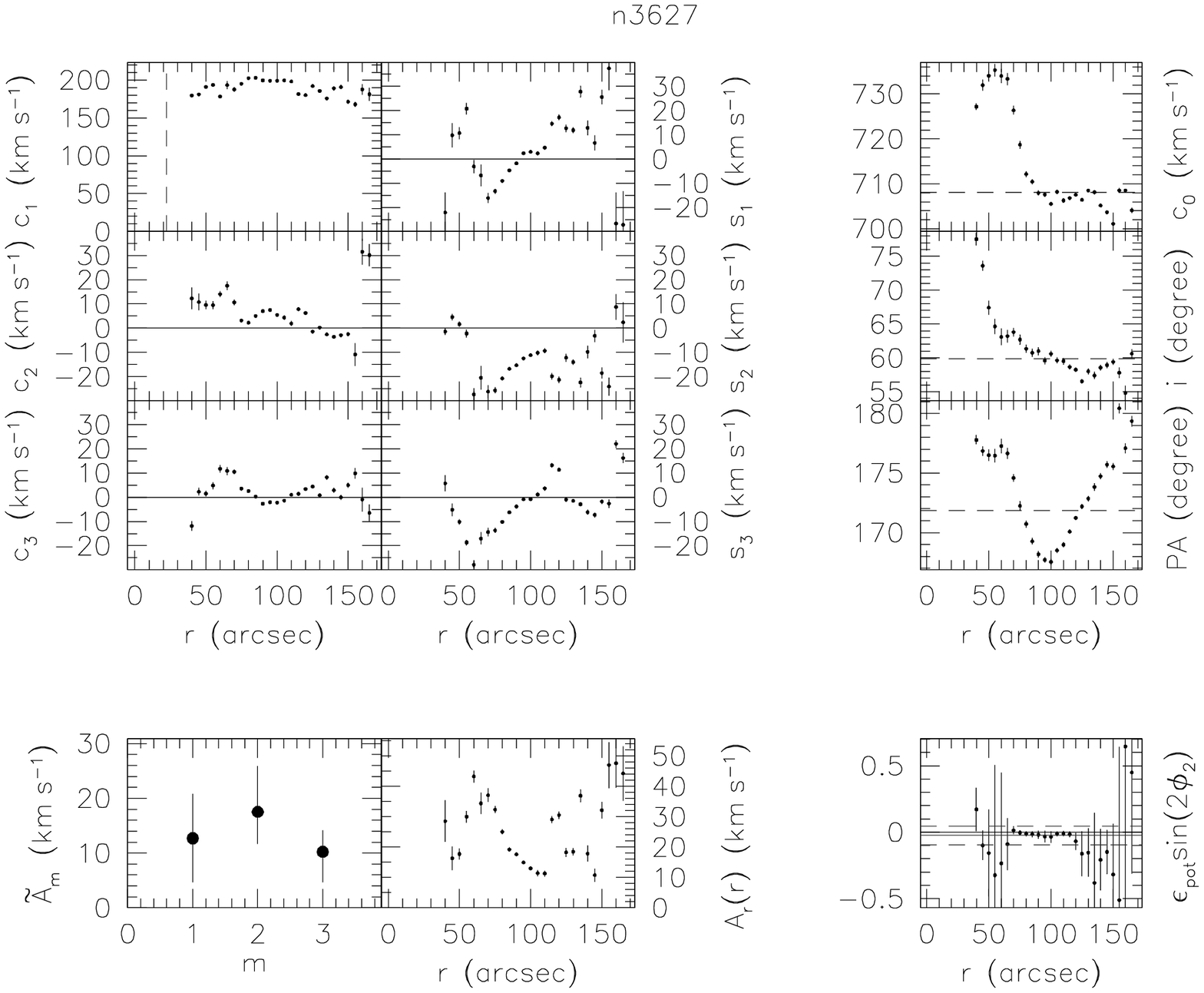}
\caption[Summary panel for NGC 3627]{Summary panel for NGC 3627. Lines and symbols are described in the text, Appendix~\ref{sec:first-figure}. The axis scale on the panels showing
  the non-circular components runs from $-30$ to $40$ \kms. See Appendix \ref{sec:n3627} for a discussion of this galaxy.} \label{fig:ngc-3627}
\end{center}
\end{figure*}

\begin{figure*}[t!]
\begin{center}
\includegraphics[angle=0,width=0.50\textwidth,bb=18 144 592 520,clip=]{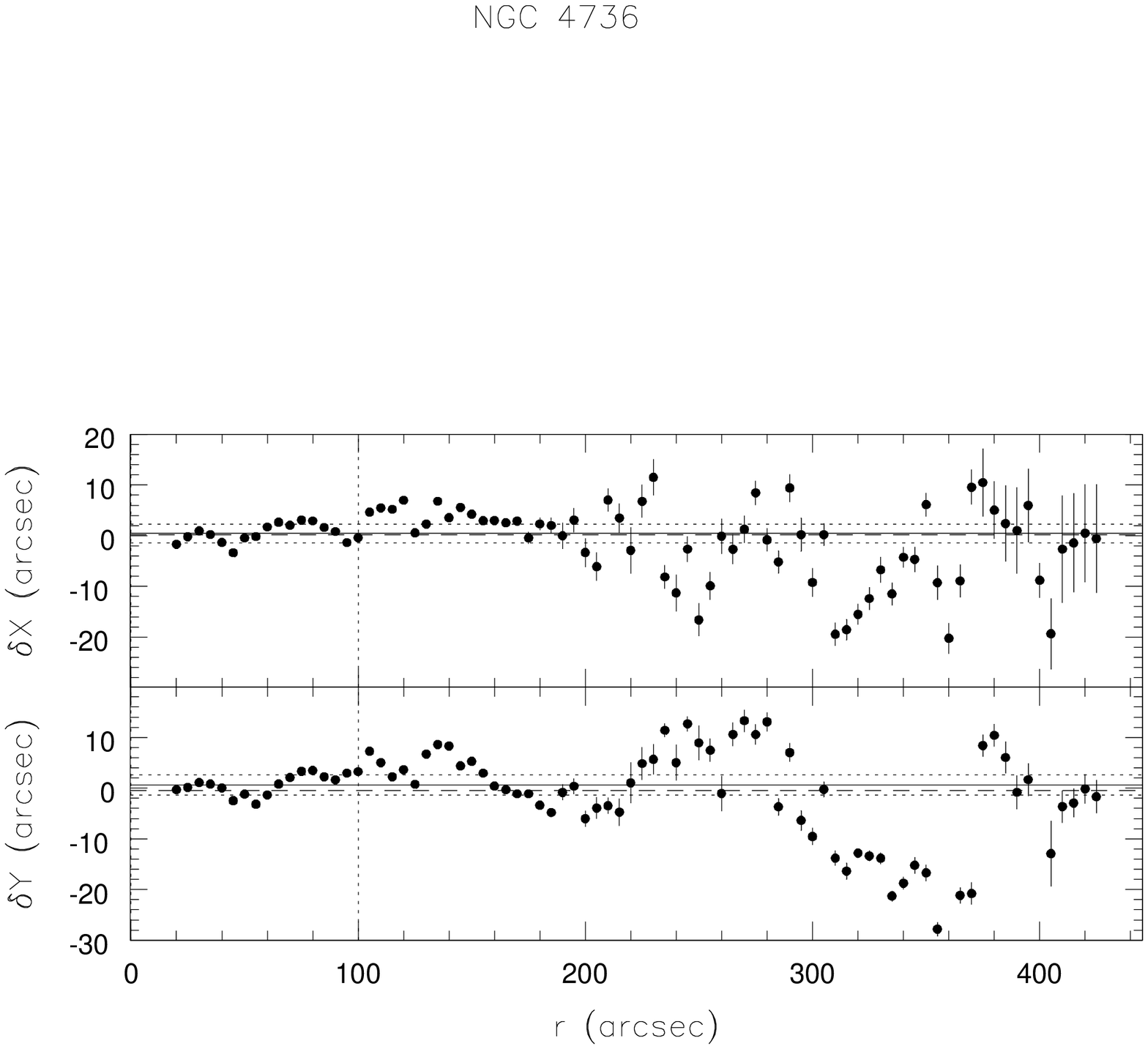}
\includegraphics[angle=0,width=0.55\textwidth,bb=18 79 520 520,clip=]{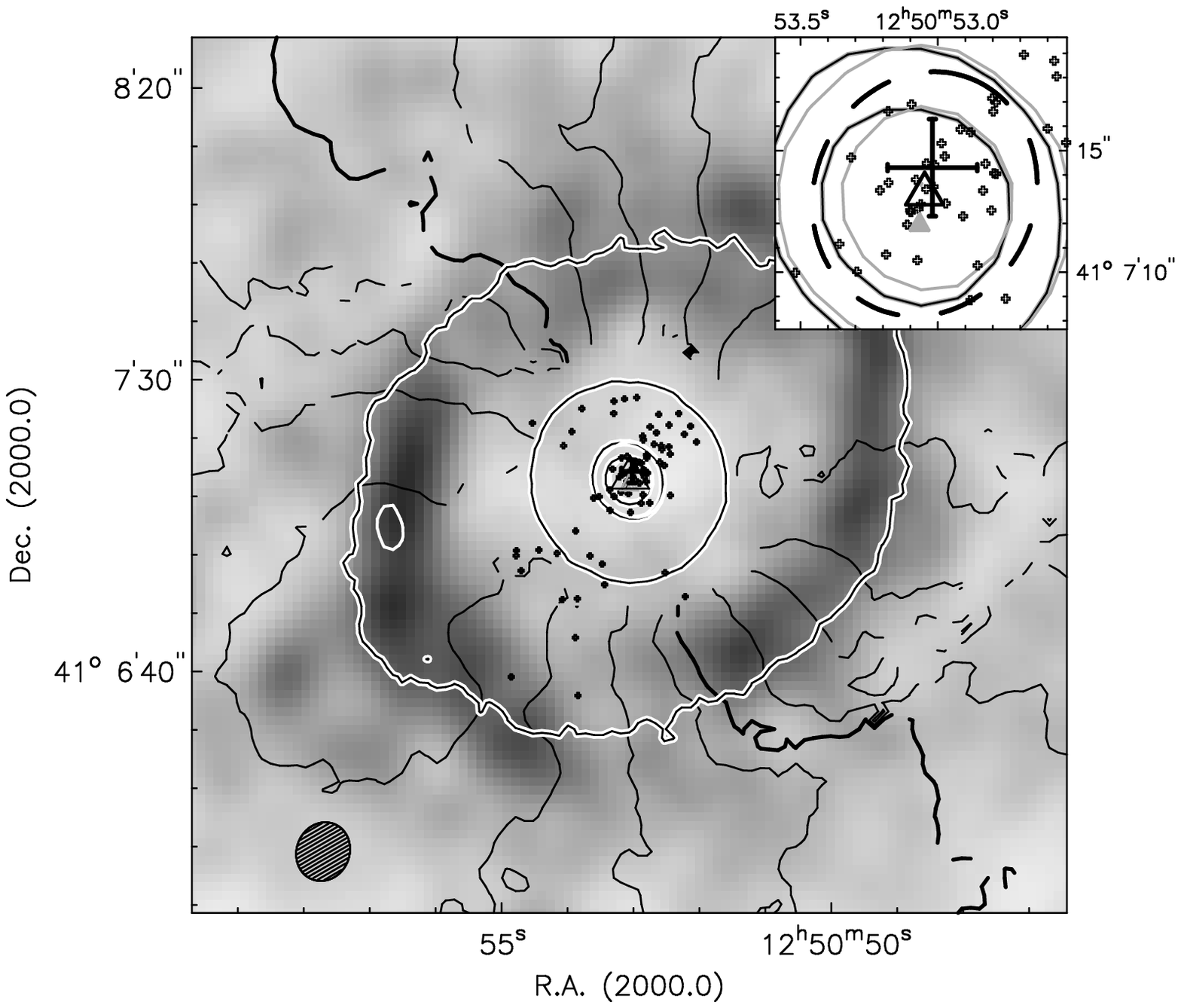}\\
\includegraphics[angle=0,width=0.65\textwidth,bb=19 235 592 697,clip=]{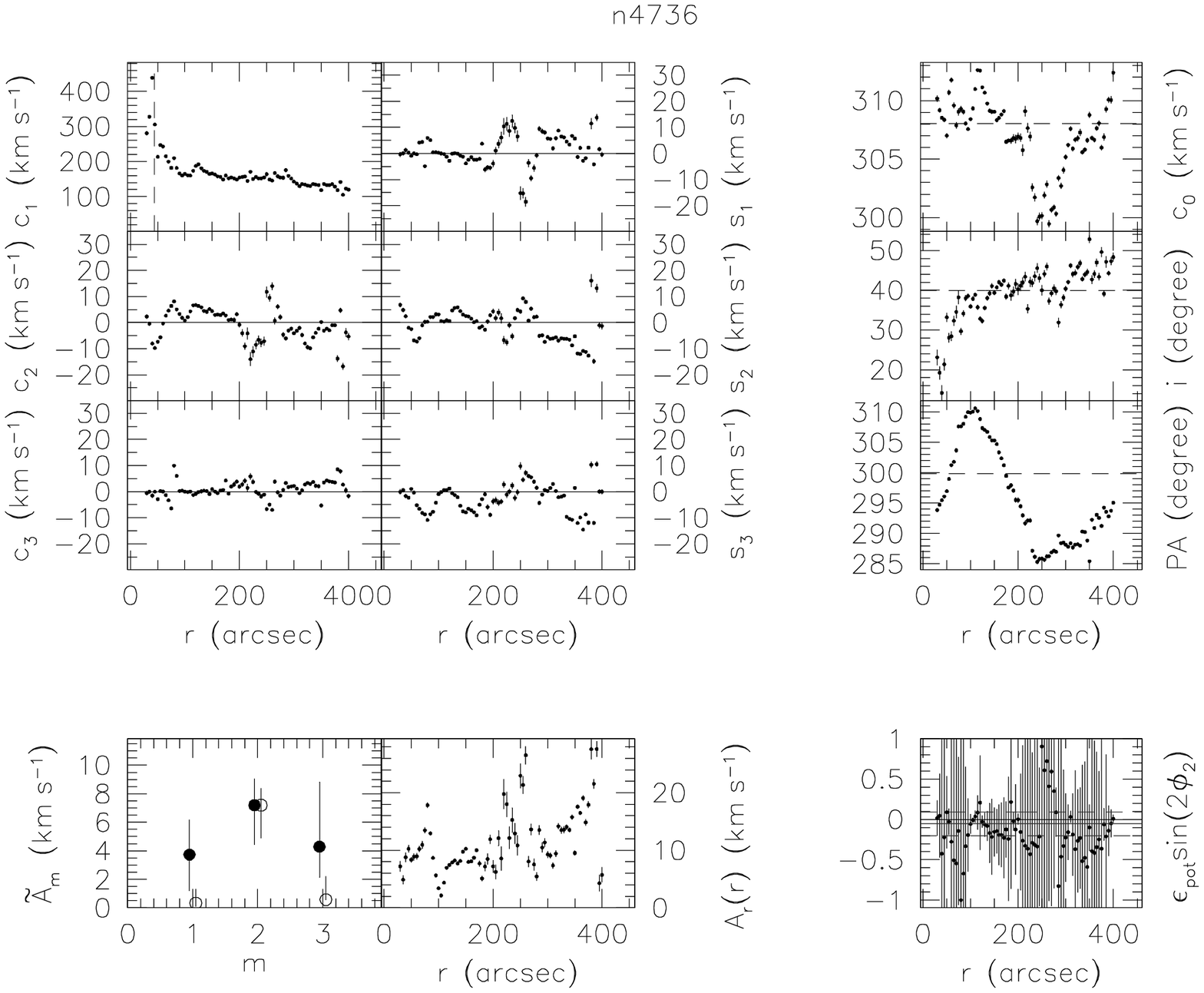}
\caption[Summary panel for NGC 4736]{Summary panel for NGC 4736. Lines and symbols are described in the text, Appendix~\ref{sec:first-figure}. The IRAC
  contours are given at the 2, 10, 50, and 80\% level of the maximum intensity. The axis scale on the panels showing
  the non-circular components runs from $-30$ to $35$ \kms. See Appendix \ref{sec:n4736} for a discussion of this galaxy.} \label{fig:ngc-4736}
\end{center}
\end{figure*}

\begin{figure*}[t!]
\begin{center}
\includegraphics[angle=0,width=0.50\textwidth,bb=18 144 592 520,clip=]{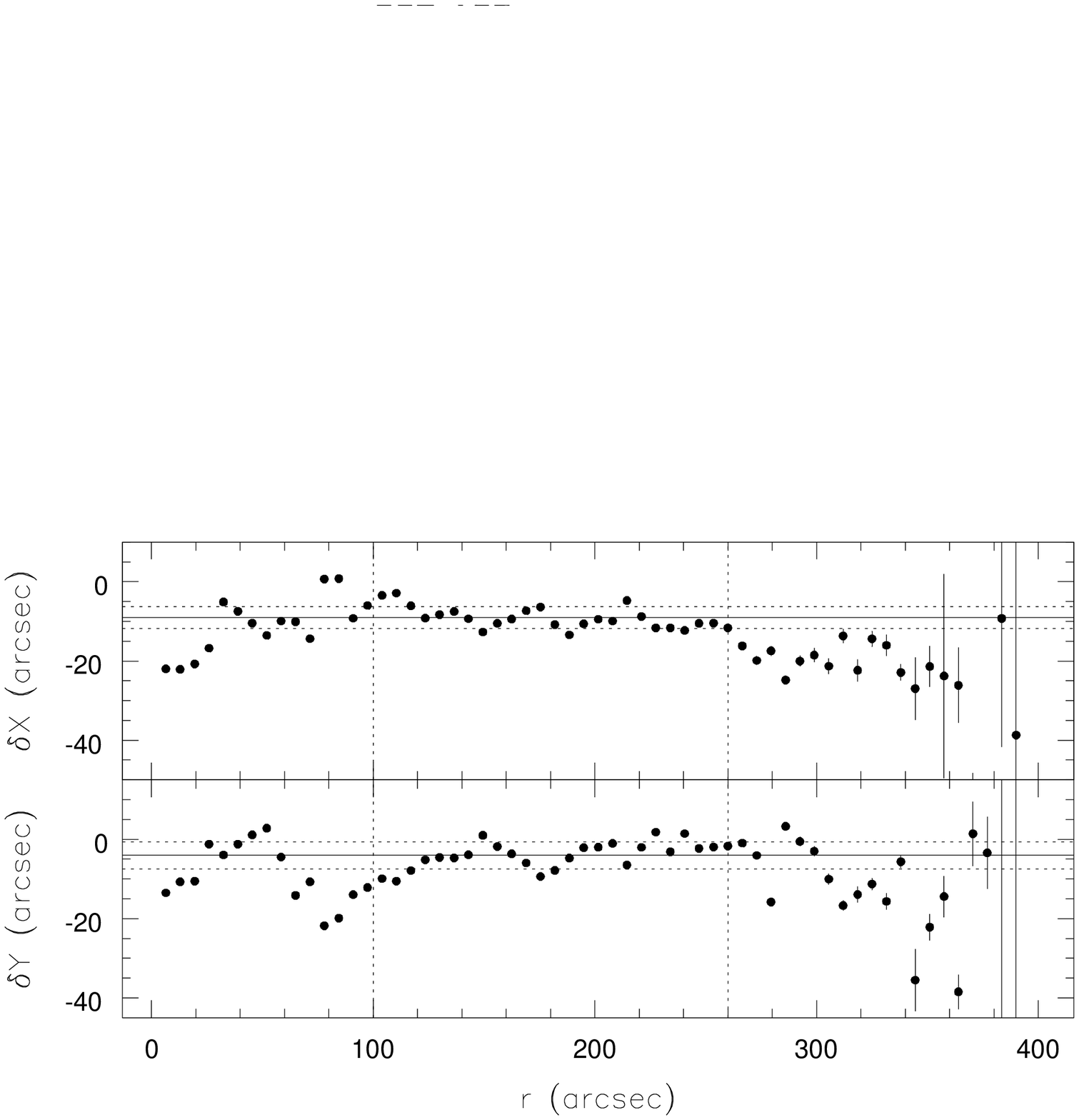}
\includegraphics[angle=0,width=0.55\textwidth,bb=18 79 520 520,clip=]{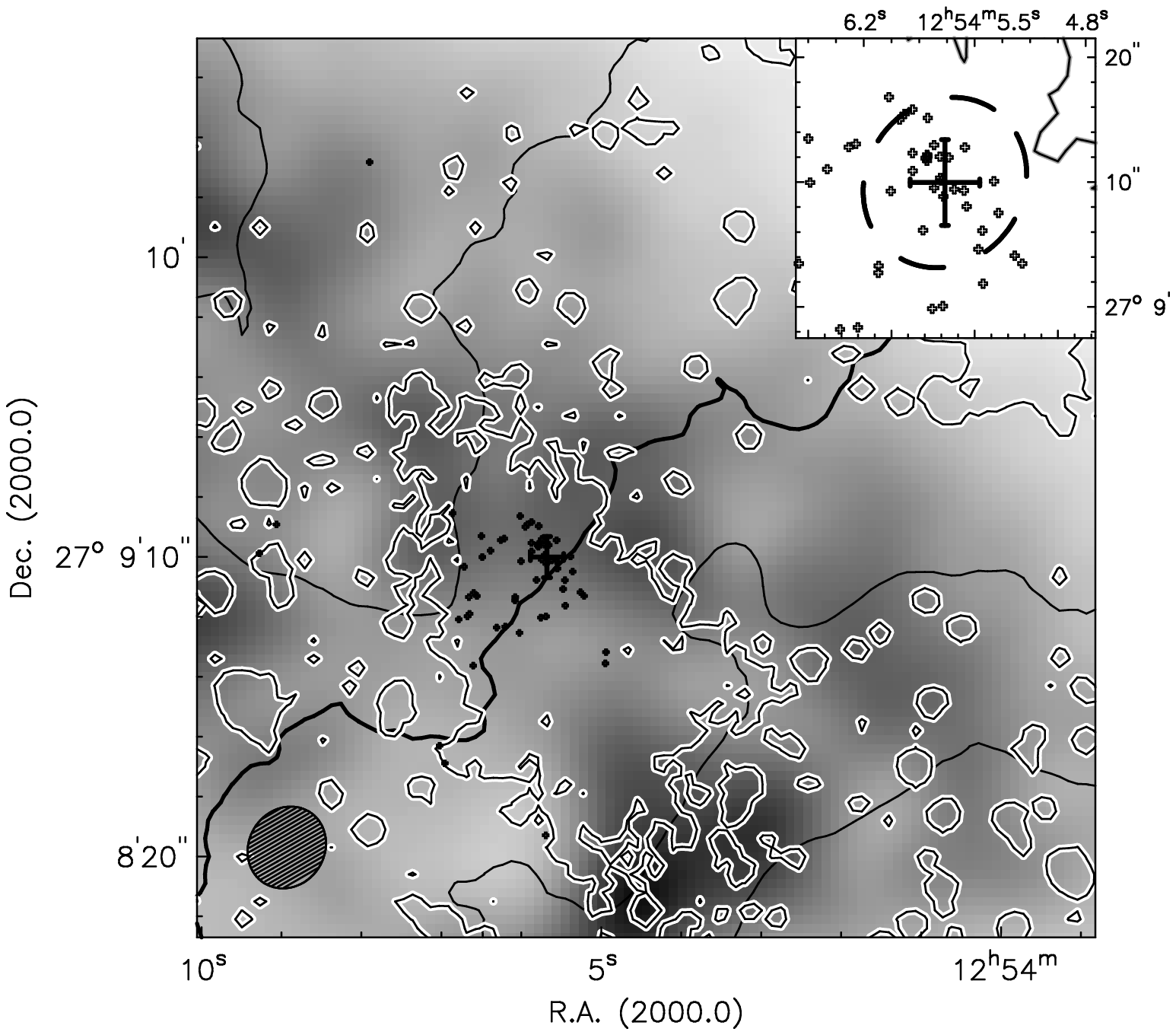}\\
\includegraphics[angle=0,width=0.65\textwidth,bb=19 235 592 697,clip=]{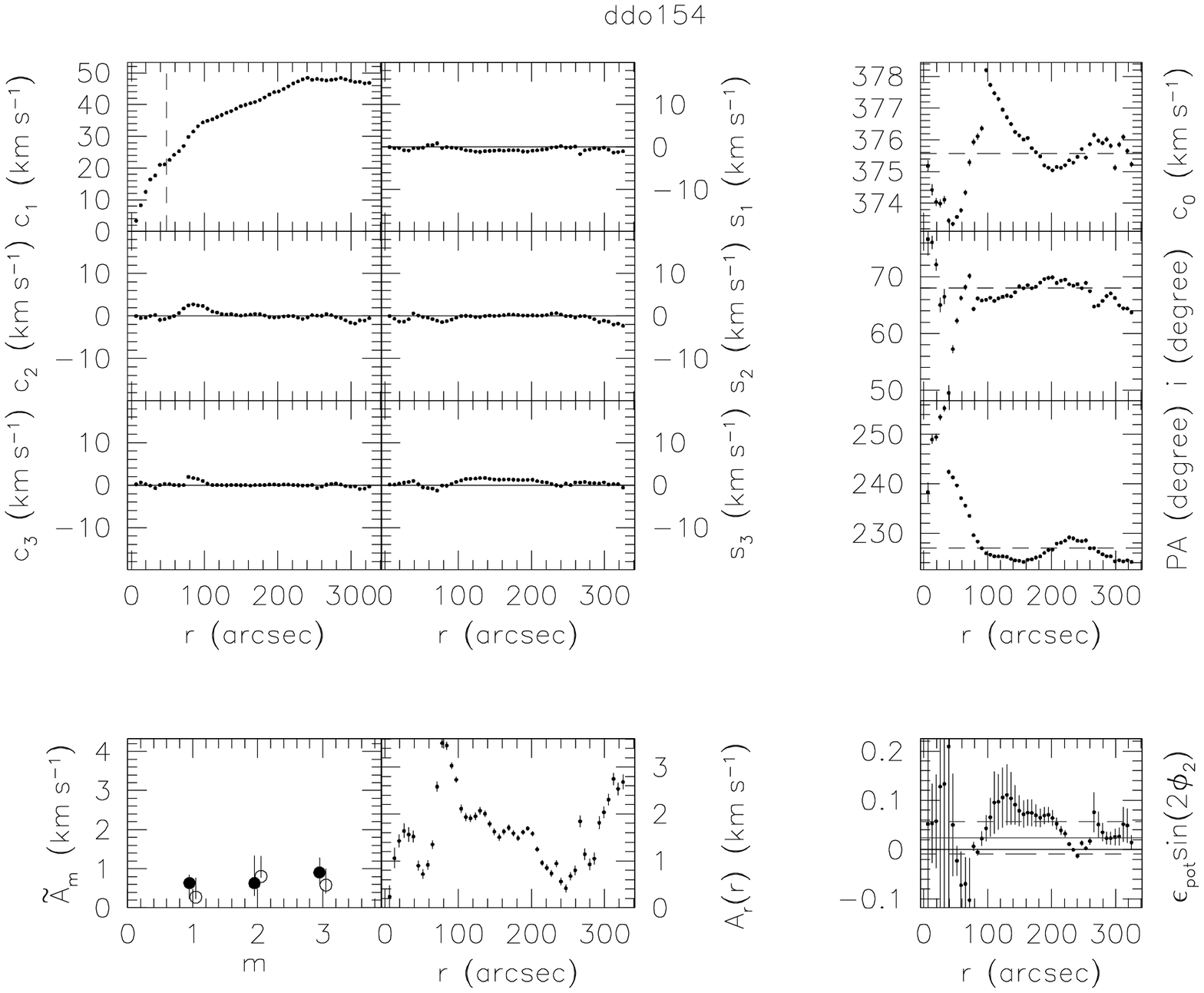}
\caption[Summary panel for DDO 154]{Summary panel for DDO 154. Lines and symbols are described in the text, Appendix~\ref{sec:first-figure}. No estimate could be derived from the
  IRAC and radio continuum images, and their respective contours are therefore
  not shown. See Appendix \ref{sec:ddo154} for a discussion of this galaxy.} \label{fig:ddo-154}
\end{center}
\end{figure*}

\begin{figure*}[t!]
\begin{center}
\includegraphics[angle=0,width=0.50\textwidth,bb=18 144 592 520,clip=]{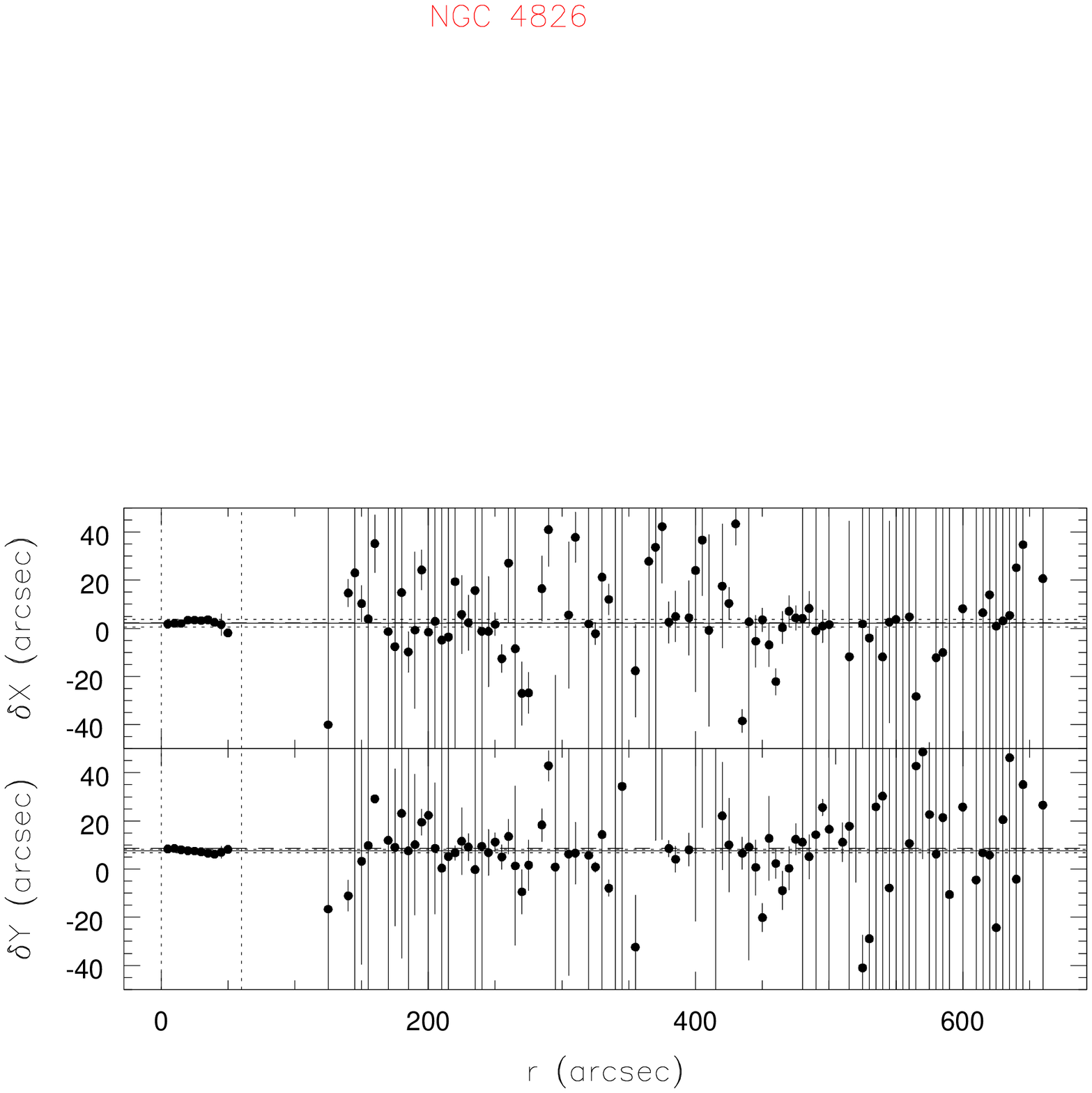}
\includegraphics[angle=0,width=0.55\textwidth,bb=18 79 520 520,clip=]{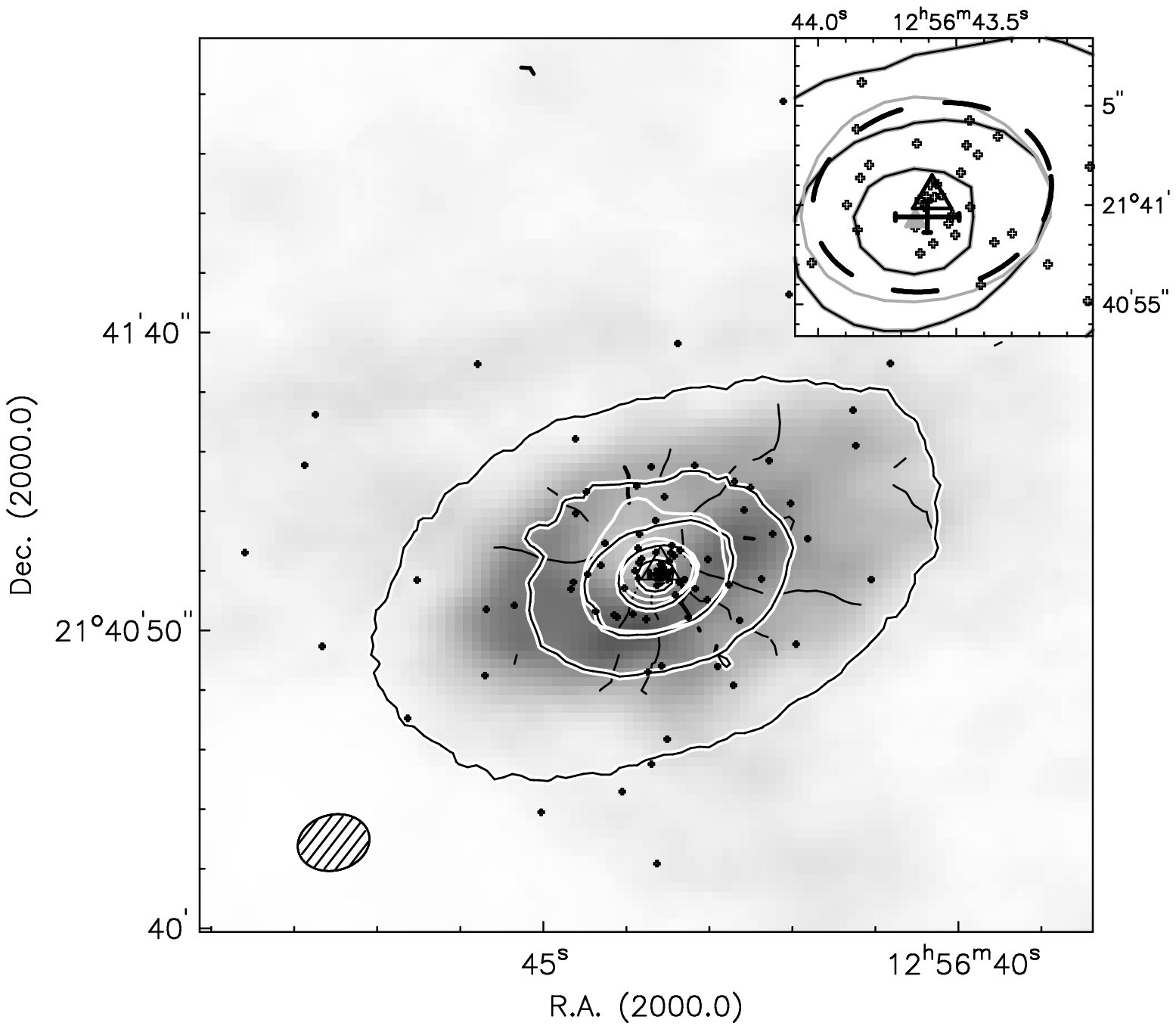}\\
\caption[Summary panel for NGC 4826]{Summary panel for NGC 4826. Lines and symbols are described in the text, Appendix~\ref{sec:first-figure}. The lower panel is missing given that
  we were not able to derive a meaningful harmonic decomposition of the
  velocity field. See Appendix \ref{sec:n4826} for a discussion of this galaxy.} \label{fig:ngc-4826}
\end{center}
\end{figure*}

\begin{figure*}[t!]
\begin{center}
\includegraphics[angle=0,width=0.50\textwidth,bb=18 144 592 520,clip=]{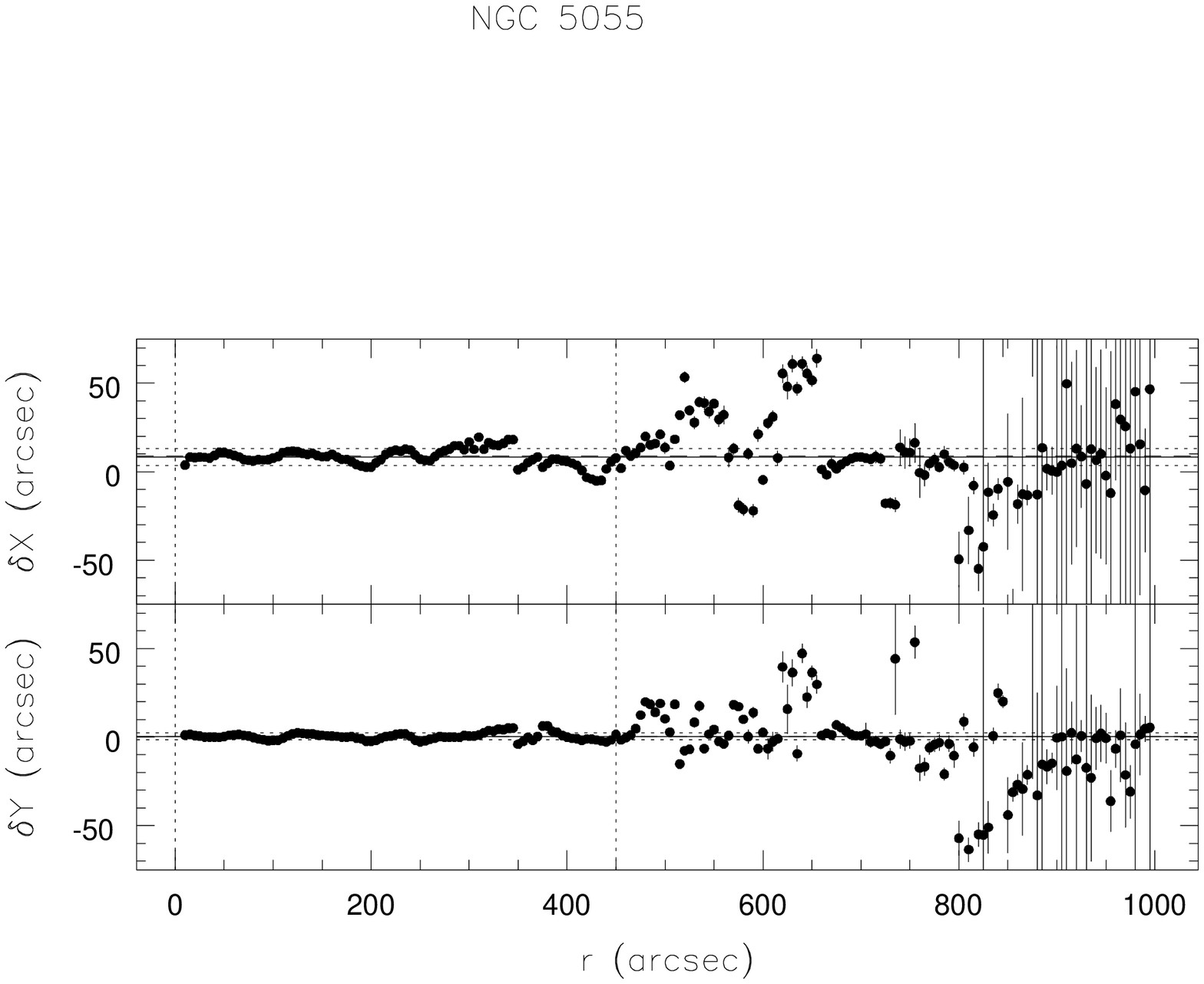}
\includegraphics[angle=0,width=0.55\textwidth,bb=18 79 520 520,clip=]{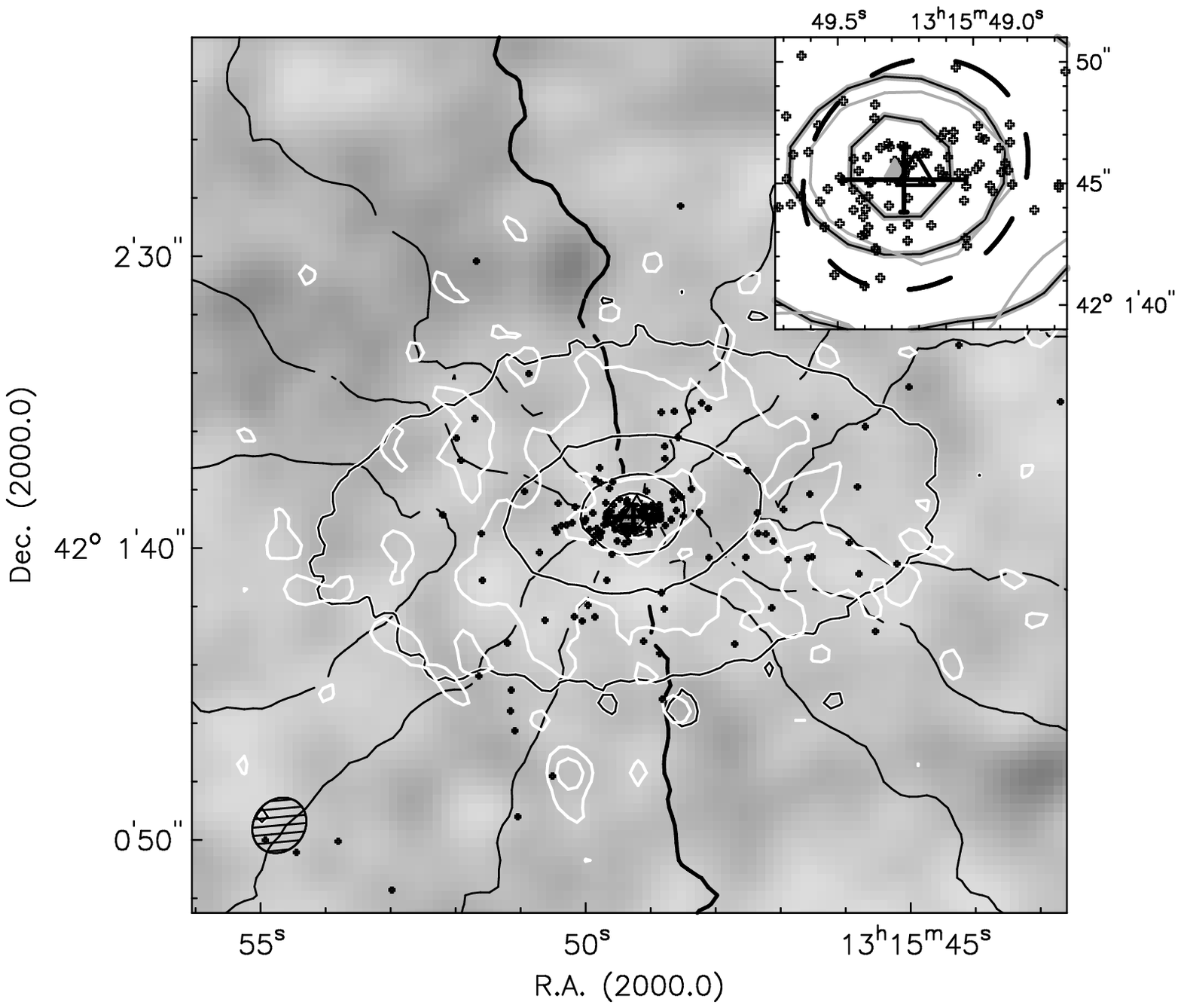}\\
\includegraphics[angle=0,width=0.65\textwidth,bb=19 235 592 697,clip=]{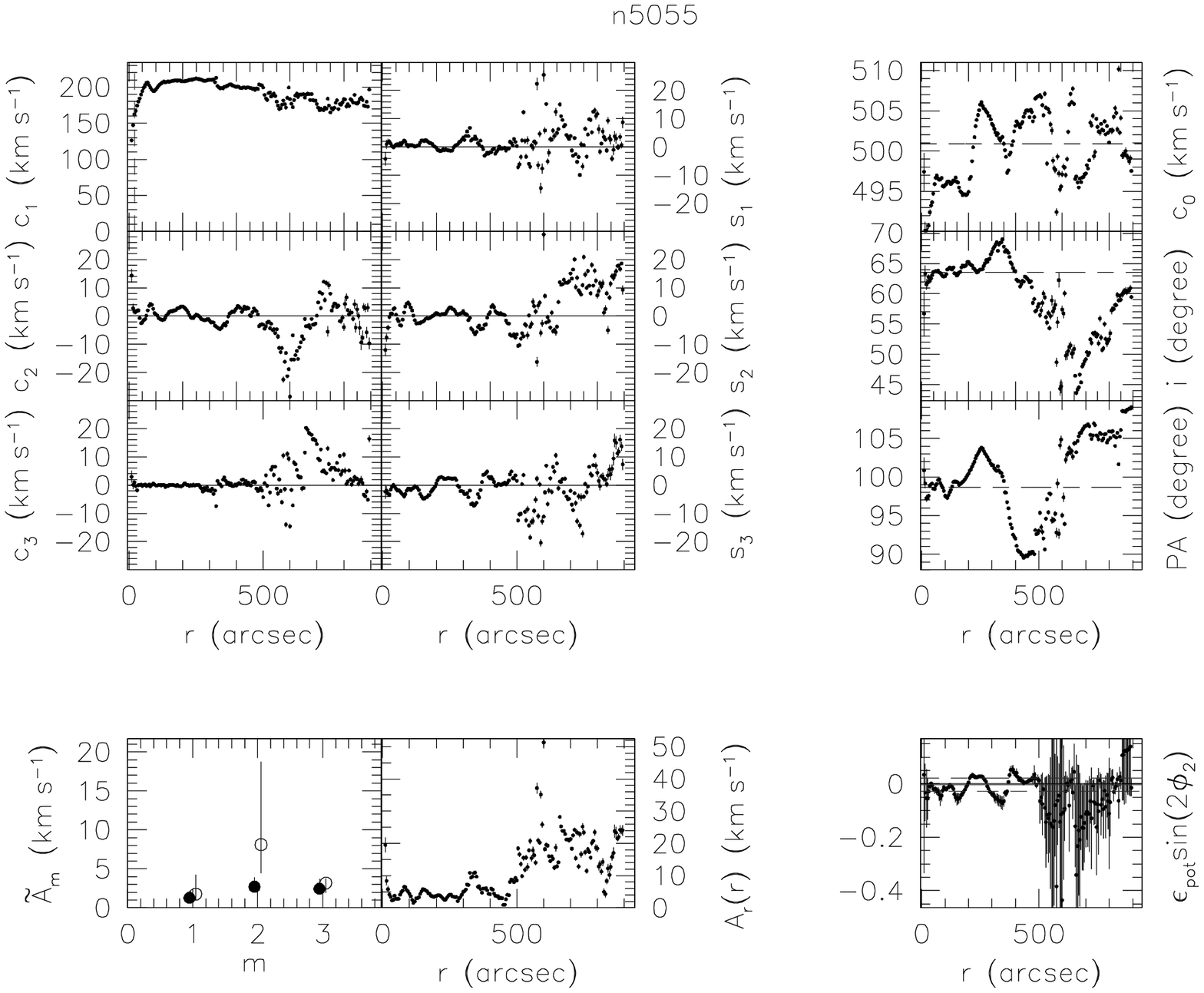}
\caption[Summary panel for NGC 5055]{Summary panel for NGC 5055. Lines and symbols are described in the text, Appendix~\ref{sec:first-figure}. The axis scale on the panels showing
  the non-circular components runs from $-30$ to $30$ \kms. Only data with $r\le 450\arcsec$
  was used for radial averaging of \Am\ or $\epsilon_{\mathrm{pot}}$. See Appendix \ref{sec:n5055} for a discussion of this galaxy.} \label{fig:ngc-5055}
\end{center}
\end{figure*}

\begin{figure*}[t!]
\begin{center}
\includegraphics[angle=0,width=0.50\textwidth,bb=18 144 592 520,clip=]{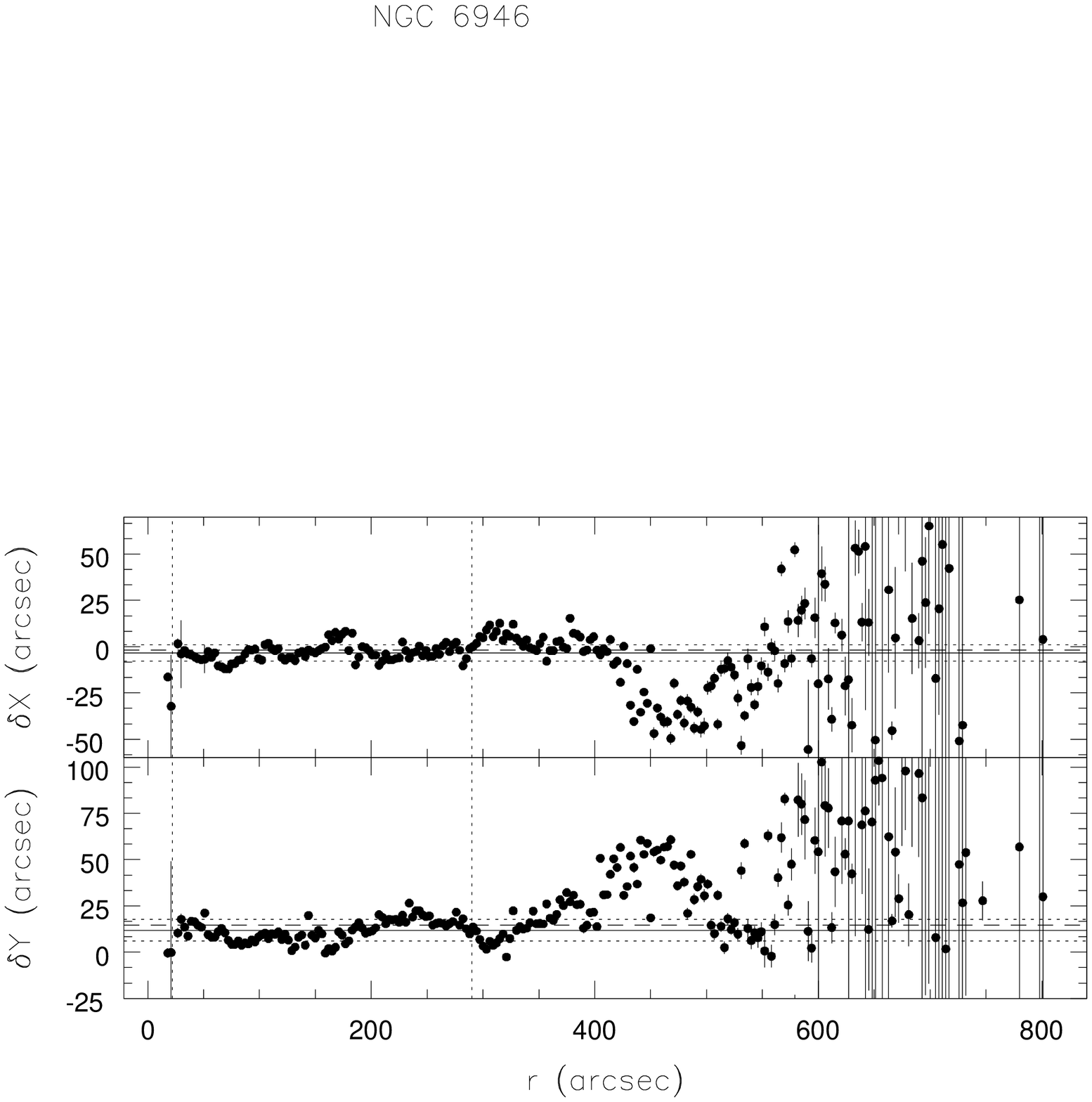}
\includegraphics[angle=0,width=0.55\textwidth,bb=18 79 520 520,clip=]{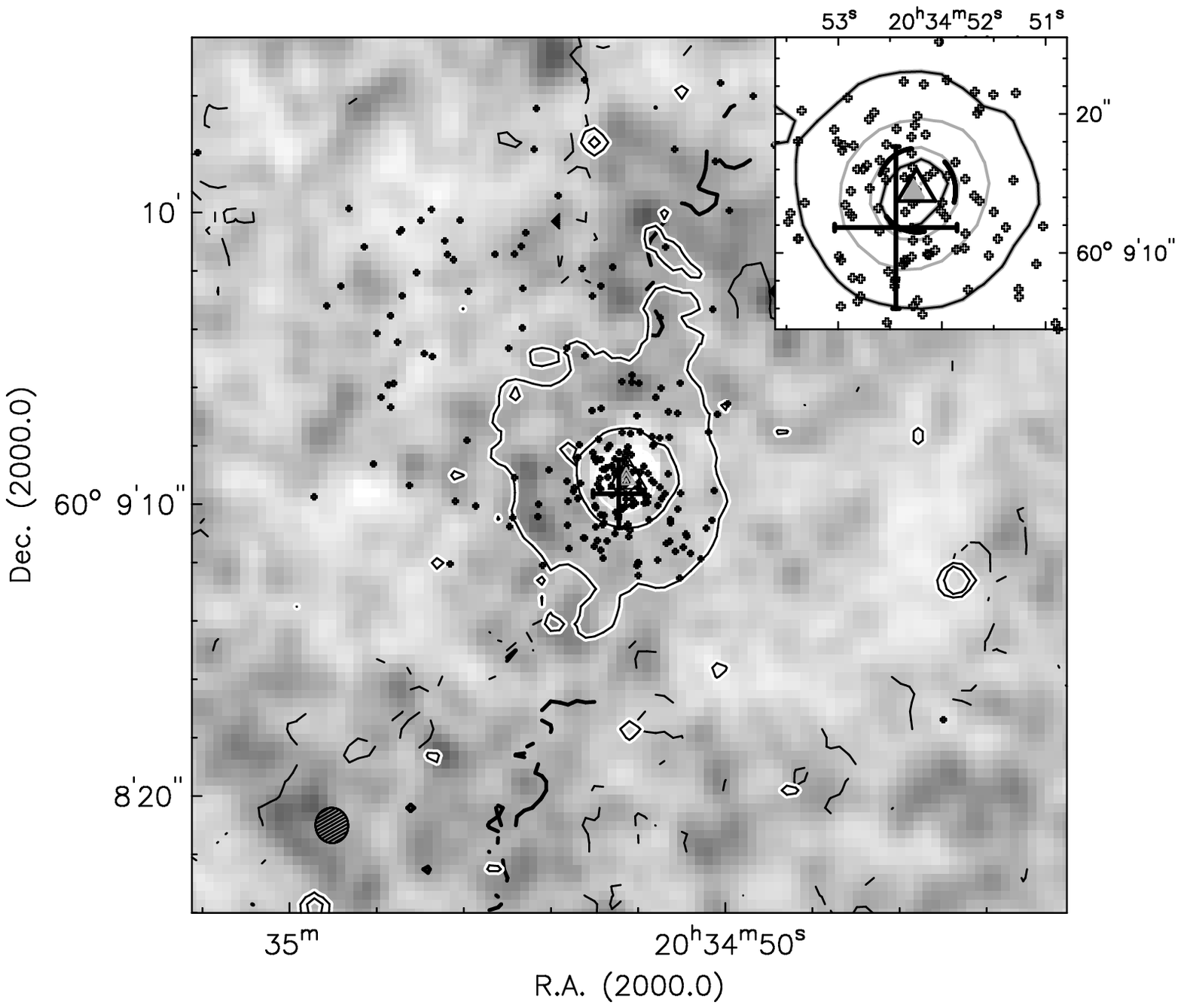}\\
\includegraphics[angle=0,width=0.65\textwidth,bb=19 235 592 697,clip=]{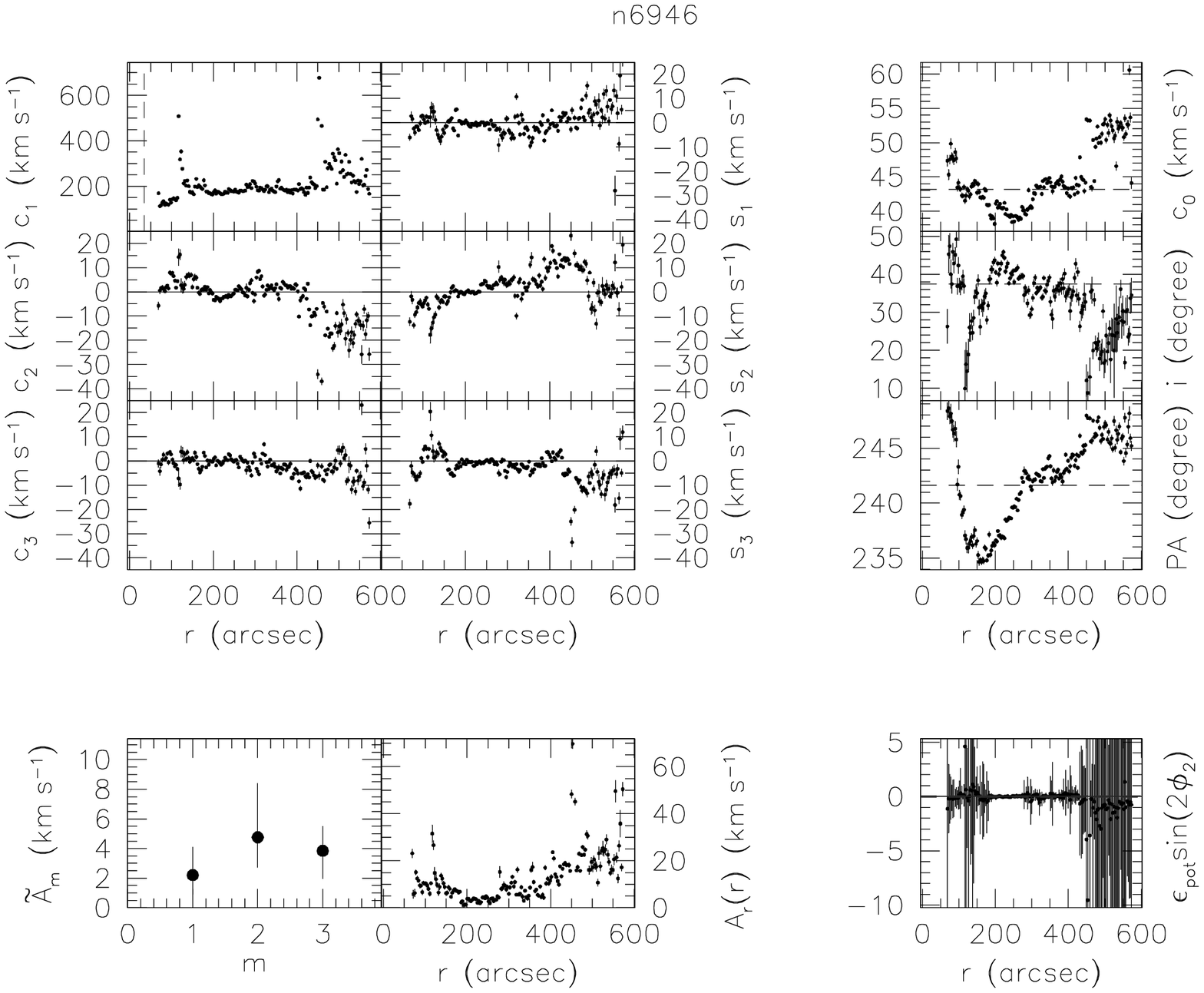}
\caption[Summary panel for NGC 6946]{Summary panel for NGC 6946. Lines and symbols are described in the text, Appendix~\ref{sec:first-figure}. The IRAC contours are given at the 2, 5,  and 50\% level of the maximum intensity. The axis scale on the panels showing the non-circular components runs from $-45$ to $25$ \kms. Only data with $r\le 420\arcsec$
  was used for radial averaging of \Am\ or $\epsilon_{\mathrm{pot}}$. See Appendix \ref{sec:n6946} for a discussion of this galaxy.} \label{fig:ngc-6946}
\end{center}
\end{figure*}

\begin{figure*}[t!]
\begin{center}
\includegraphics[angle=0,width=0.50\textwidth,bb=18 144 592 520,clip=]{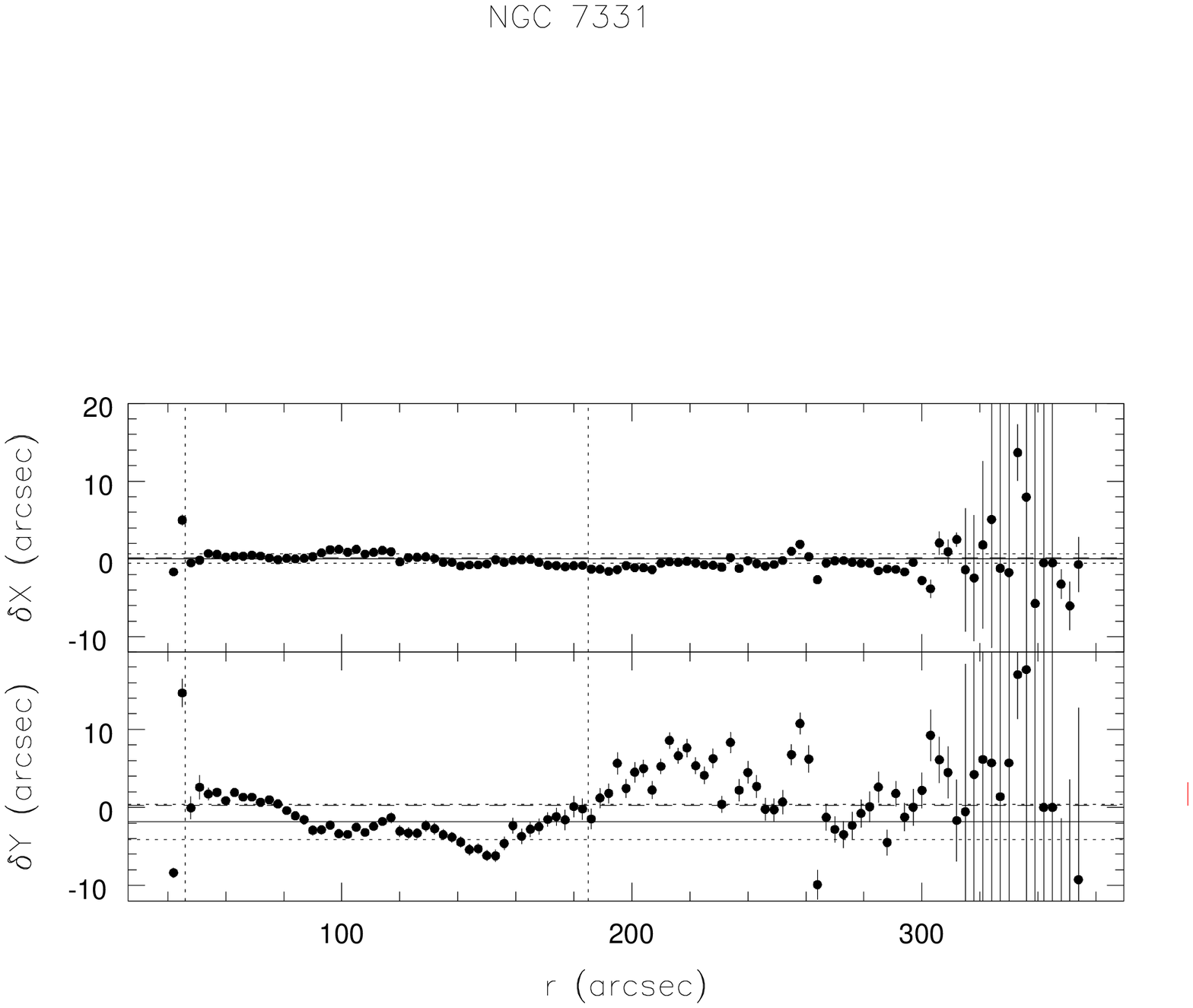}
\includegraphics[angle=0,width=0.55\textwidth,bb=18 79 520 520,clip=]{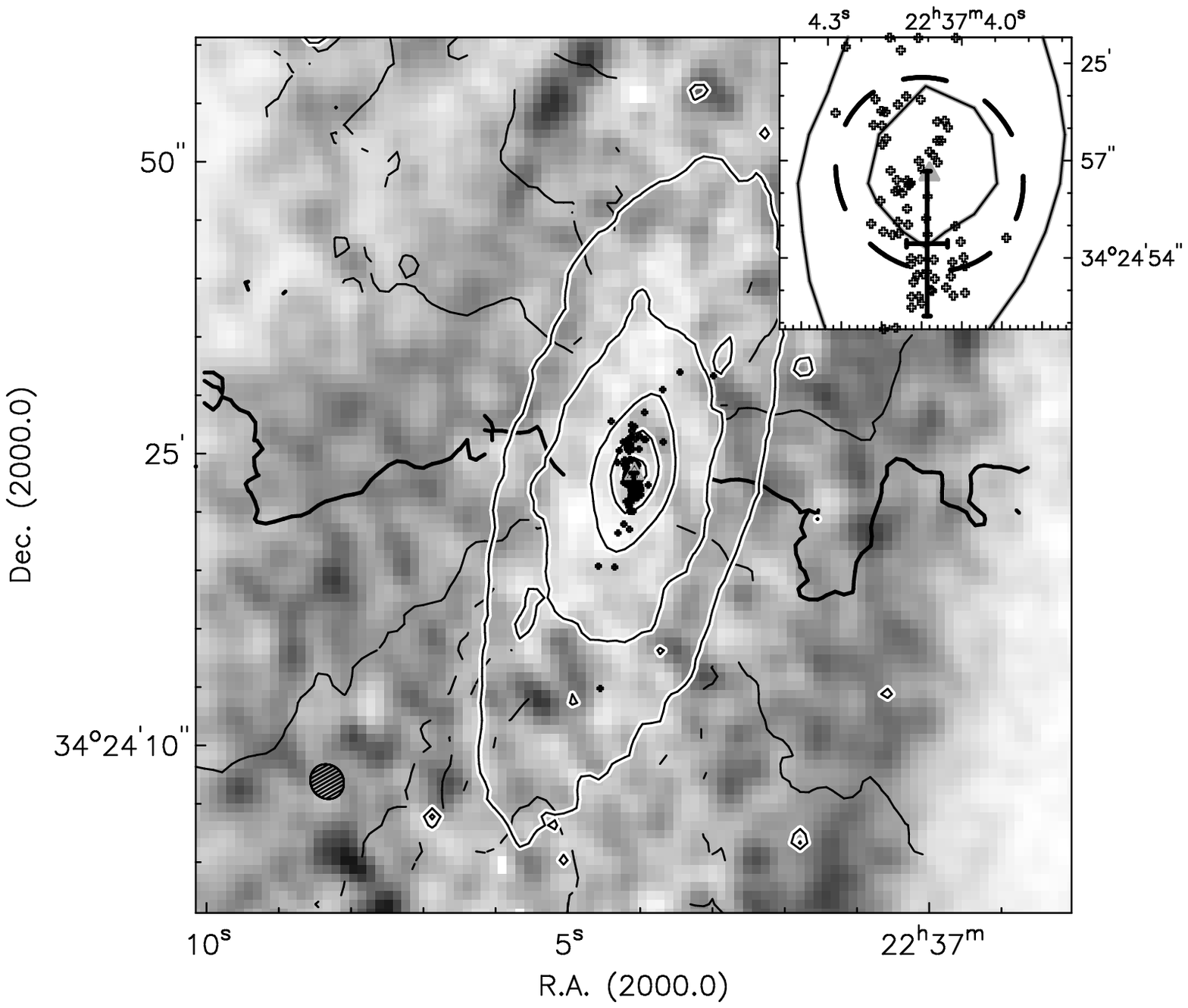}\\
\includegraphics[angle=0,width=0.65\textwidth,bb=19 235 592 697,clip=]{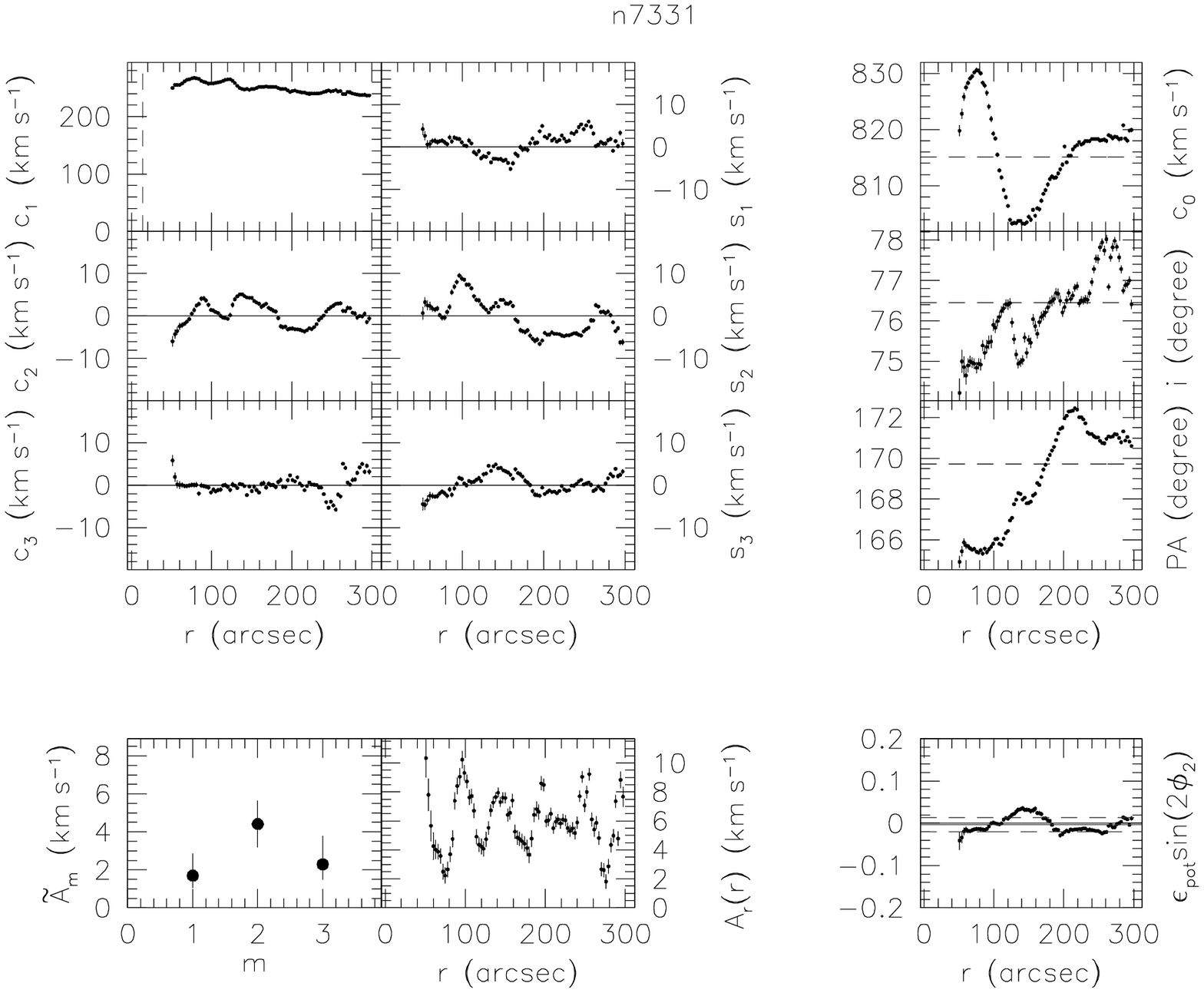}
\caption[Summary panel for NGC 7331]{Summary panel for NGC 7331. Lines and symbols are described in the text, Appendix~\ref{sec:first-figure}. See Appendix \ref{sec:n7331} for a discussion of this galaxy.} \label{fig:ngc-7331}
\end{center}
\end{figure*}

\begin{figure*}[t!]
\begin{center}
\includegraphics[angle=0,width=0.50\textwidth,bb=18 144 592 520,clip=]{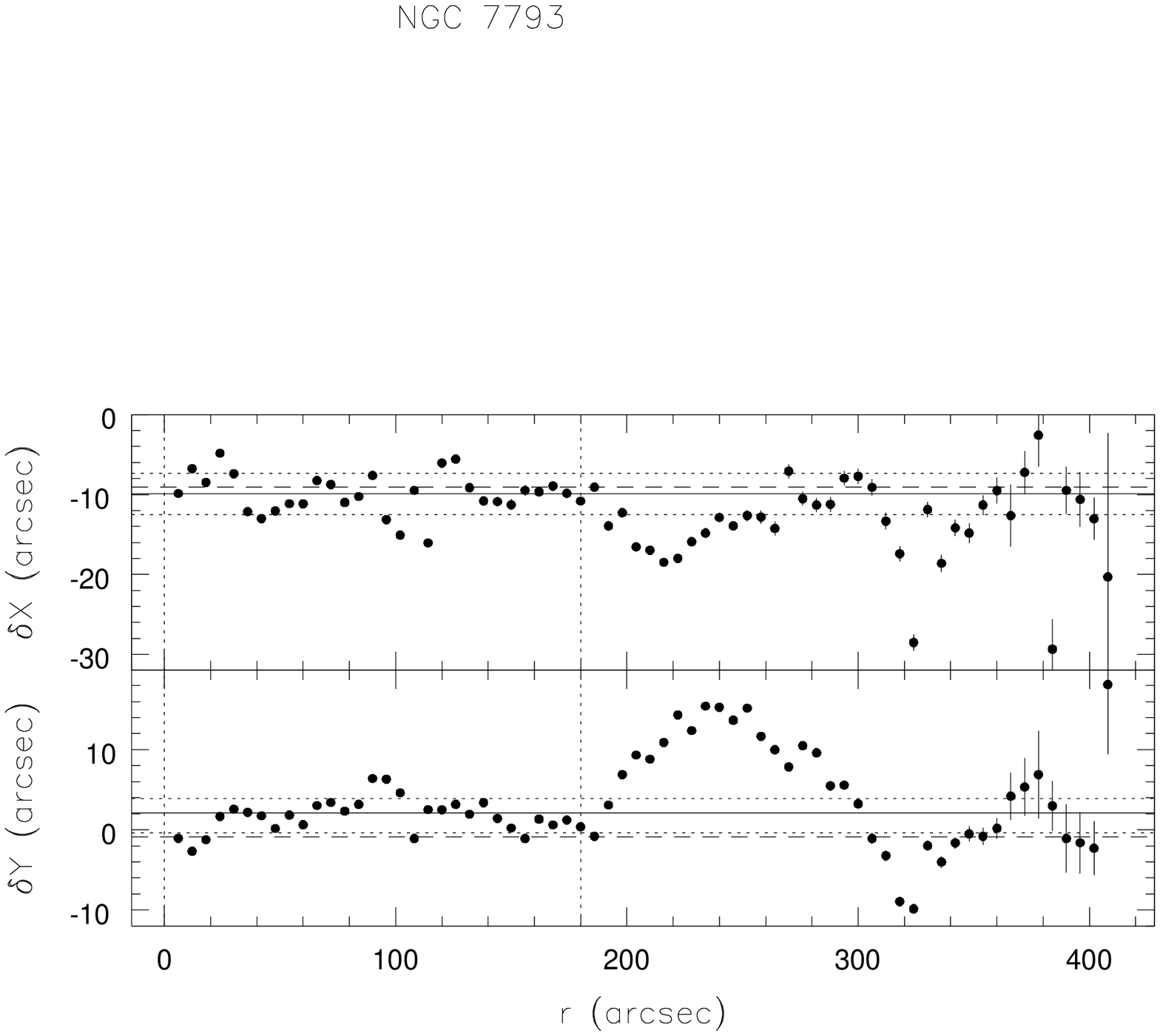}
\includegraphics[angle=0,width=0.55\textwidth,bb=18 79 520 520,clip=]{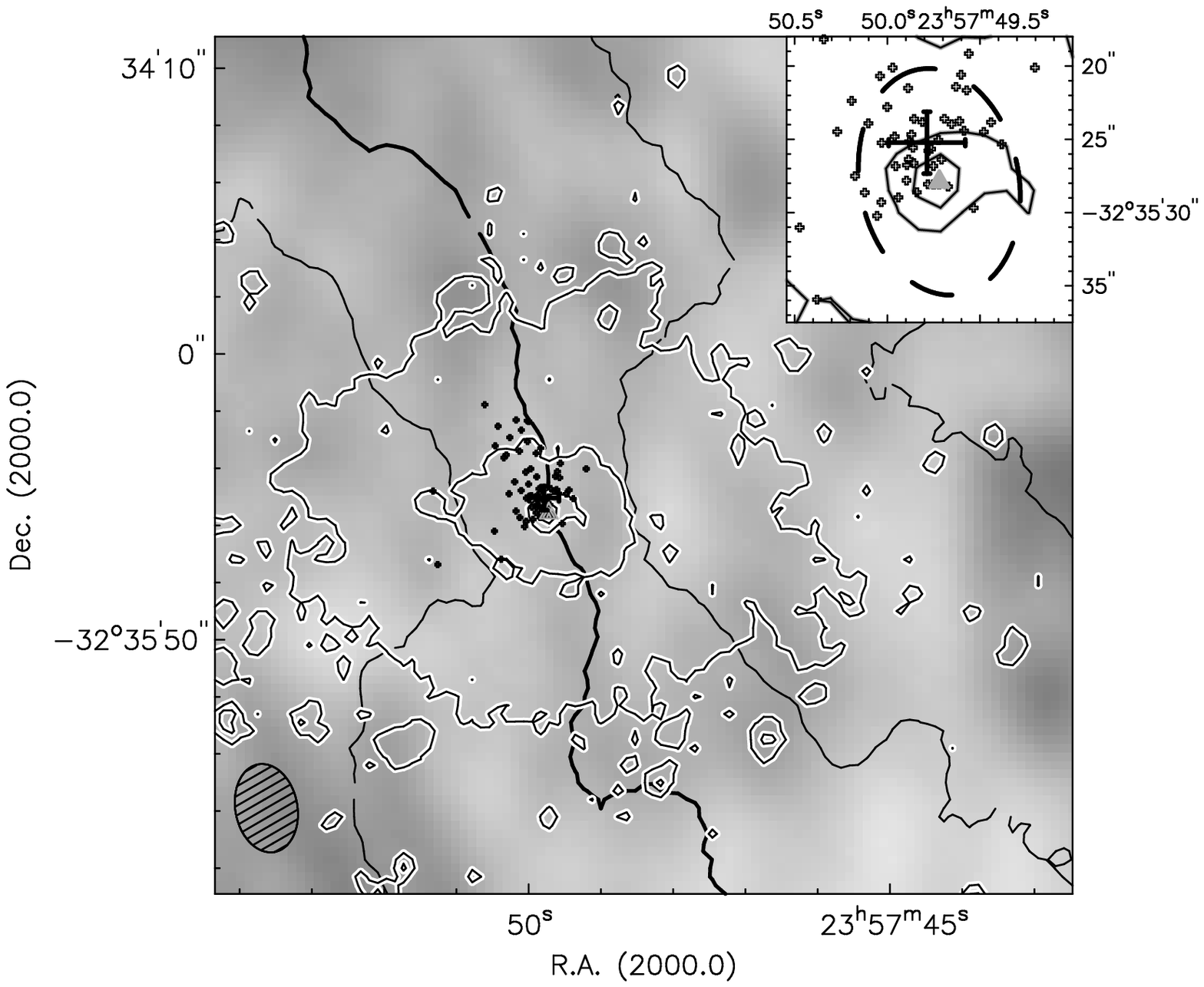}\\
\includegraphics[angle=0,width=0.65\textwidth,bb=19 235 592 697,clip=]{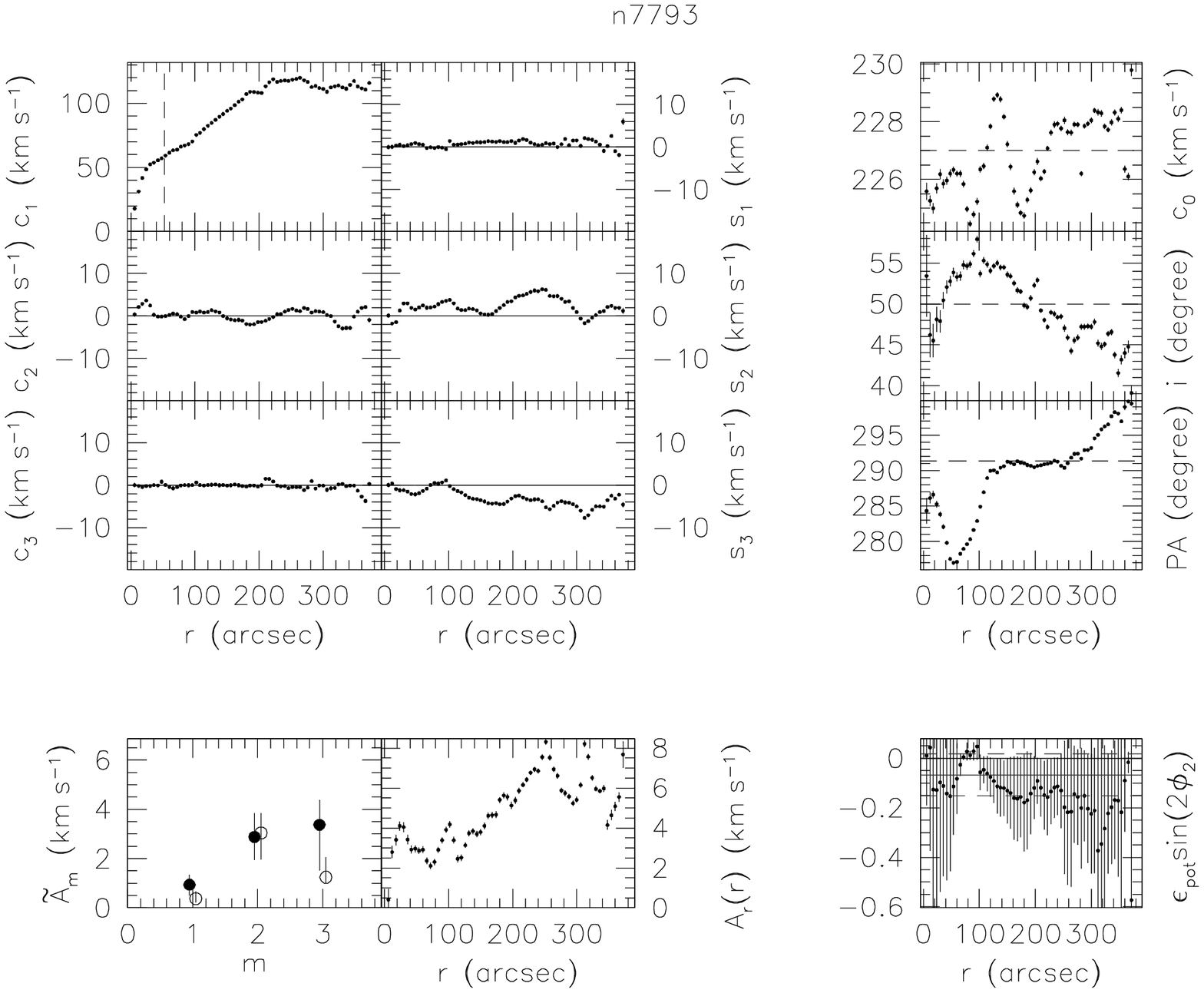}
\caption[Summary panel for NGC 7793]{Summary panel for NGC 7793. Lines and symbols are described in the text, Appendix~\ref{sec:first-figure}. The 2\% IRAC contour was
  omitted for clarity reasons. See Appendix \ref{sec:n7793} for a discussion of this galaxy.} \label{fig:ngc-7793}
\end{center}
\end{figure*}

The center estimates and the harmonic decompositions are shown in
Figs.~\ref{fig:ngc-925}-\ref{fig:ngc-7793}. Following is a
description of the content and layout of the figures:
Each figure consists of four rows. From top to bottom, they contain:

{\bf Row 1:} The radial variation of the center position. Shown is the offset 
from the pointing center in arcsecond. The top panel shows the offsets in the X 
(or right ascension) direction, with positive \dx\ to the west; the bottom panel 
those in the Y (declination) direction, with positive \dy\ to the north.
The filled circles represent the individual center positions
derived using \rotcur, and the error bars indicate the
formal uncertainties derived by
\rotcur. The two dotted, vertical lines indicate the
radial range over which the positions were averaged in order to derive a
kinematic center. The kinematic center is indicated by the solid
horizontal line, and the $1\sigma$ standard deviation by the dotted,
horizontal lines. For those cases where the kinematic center was not used as
our best center position, we additionally show the best estimate with a
dashed, horizontal line.

{\bf Row 2:} The inner $150\arcsec\,\times\,150\arcsec$ of the galaxy. The
total intensity \HI\ map is shown in grayscales, and the beam size is indicated in the
bottom-left corner. The black contours are drawn from the
velocity field which was used for the estimate of the dynamical center (bulk
velocity field from \citealt{se-heon} for IC\ 2574, hermite velocity field from
\citealt{deblok-07} for the other galaxies). The thick line represents the
systemic velocity as derived in \cite{deblok-07}. The thin black
contours overlaid on the thick white
contours belong to the 3.6\,$\mu$m IRAC
image and are usually given at a 2, 5, 10, 20, and 50 percent level of the
maximum flux in the image.
The white contours are taken from the THINGS radio continuum
maps and are usually given at a 10, 20, and 50 percent level of the maximum
flux. The black, filled circles indicate the individual center positions from
\rotcur\ and the black cross represents the chosen dynamical center and
its uncertainty. The derived center from the 3.6\,$\mu$m image is shown as a
gray, filled triangle, and the one from the radio continuum as a
black, open triangle.\\
\emph{Inset:} To better highlight the different
center estimates, we additionally show in the upper-right corner an inset with
only the central few arcseconds. For clarity, we 
omit the total \HI\ map and the velocity field contours in the inset. The
contours from the 3.6\,$\mu$m image are shown in black and are given at the same
intensity levels as in the main plot. The same holds for the radio continuum
contours, which are shown here in gray. The individual center estimates from
\rotcur\ are shown as small crosses.
In the inset, the beam size is indicated by the thick dashed ellipse, which is centered on
our best center position.

{\bf Row 3:} The error bars of all
plots in this row are the formal uncertainties as derived by \reswri.\\
\emph{Left panel:} This panel consists of six sub-panels, all of
which are plotted against radius in arcsecond.  The upper-left
sub-panel shows $c_1$, the 
amplitude of the circular velocity. The other five sub-panels show the amplitudes of the
non-circular components, namely the second and third order component of the
cosine term ($c_2, c_3$) and the first, second, and third order
component of the sine term ($s_1, s_2, s_3$). The axis scale of these  
five sub-panels is generally $-20$ to $+20$ \kms. The solid, vertical line in
the panel containing $c_1$ indicates the radius corresponding to 1 kpc. All
quantities shown in Row 3 are corrected for inclination effects.\\
\emph{Right panel:} From top to bottom: systemic velocity $c_0$; inclination
angle $i$; and position angle PA, all plotted against radius in
arcsecond. The dashed, horizontal line in all three sub-panels indicates the weighted mean, using the inverse square of the uncertainties as weight.

{\bf Row 4:} \emph{Left panel:} This panel consists of two sub-panels. The
left sub-panel shows \Am, the median 
amplitudes for each harmonic order $m$. The values are calculated by
radially averaging the quadratically added amplitudes as described by
Eqs.~\ref{eq:A(m)1} and \ref{eq:A(m)}. We show the median as determined for
the entire extent of the galaxy (filled circles) as well as using only the
inner 1 kpc (open circles). The error bars indicate the lower and upper
quartile respectively. The right 
sub-panel shows \Ar, the radial distribution of the amplitude of all non-circular
motions derived following Eq.~\ref{eq:A(r)}. The error bars
of \Ar\ are derived with formal error propagation assuming a Gaussian error
distribution. Both \Am\ and 
\Ar\ are corrected for inclination effects.\\
\emph{Right panel:} This panel contains the elongation of the potential
$ \langle \epsilon_{\mathrm{pot}}\,\sin(2\varphi_2) \rangle $ (cf.,
Eq.~\ref{eq:epot}) {\it vs.} radius. The elongation is also corrected for
inclination. The solid,
black line indicates the zero level, the dotted gray one represents the
weighted mean elongation, and the dashed, gray ones its standard
deviation. The error bars shown here are derived using a formal error
propagation assuming a Gaussian error distribution.

\clearpage

\clearpage

\clearpage


\clearpage

\clearpage

\end{appendix}
\end{document}